\documentclass[%
  reprint,
  aps,
  pra,
  floatfix,
  superscriptaddress,
]{revtex4-2}

\usepackage[T1]{fontenc}
\usepackage{lmodern}
\usepackage{microtype}

\usepackage{amsmath}
\usepackage{mathtools}
\usepackage{amssymb}
\usepackage{amsthm}
\usepackage{physics}

\usepackage{graphicx}
\usepackage[table]{xcolor}
\usepackage{tikz}
\usetikzlibrary{decorations.pathmorphing, math}
\tikzset{snake it/.style={decorate, decoration=snake}}
\usepackage{qcircuit}

\definecolor{ibmOrange}{HTML}{FFB000}
\definecolor{ibmRed}{HTML}{FE6100}
\definecolor{ibmMagenta}{HTML}{DC267F}
\definecolor{ibmViolet}{HTML}{785EF0}
\definecolor{ibmBlue}{HTML}{648FFF}
\definecolor{armygreen}{HTML}{009F6B}
\definecolor{terracotta}{HTML}{A55B4B}

\usepackage{float}
\usepackage{placeins}
\usepackage{booktabs}
\usepackage{multirow}
\usepackage{adjustbox}
\newcolumntype{P}[1]{>{\raggedright\arraybackslash}p{#1}}

\newcommand{\subpanel}[3]{%
  \begin{minipage}[t]{#1}%
    \vspace{0pt}\textbf{#2}\par\centering #3%
  \end{minipage}}

\usepackage[inline]{enumitem}
\usepackage{appendix}
\usepackage[normalem]{ulem}     %
\usepackage{comment}
\usepackage{csquotes}

\usepackage{hyperref}
\hypersetup{
  colorlinks=true,
  linkcolor=blue,
  citecolor=blue,
  urlcolor=blue,
}
\usepackage{cleveref}
\crefname{section}{Sec.}{Secs.}
\crefname{figure}{Fig.}{Figs.}
\crefname{appendix}{Appendix}{Appendices}
\crefname{equation}{Eq.}{Eqs.}
\crefname{table}{Table}{Tables}
\Crefname{section}{Section}{Sections}
\Crefname{figure}{Figure}{Figures}
\Crefname{appendix}{Appendix}{Appendices}
\Crefname{equation}{Equation}{Equations}
\Crefname{table}{Table}{Tables}

\makeatletter
\def\namelabel#1#2{\@bsphack
  \begingroup
  \UseHookWithArguments{label}{1}{#1}%
  \protected@write\@auxout{}%
    {\string\newlabel{#1}{{\@currentlabel}{\thepage}%
     {#2}{\@currentHref}{\@kernel@reserved@label@data}}}%
  \endgroup
  \@esphack}
\makeatother

\newif\ifshowcomments
\showcommentstrue

\newcommand{\makeauthorcomment}[3]{%
  \newcommand{#1}[1]{\ifshowcomments{\color{#2}[#3: ##1]}\fi}}

\makeauthorcomment{\MC}{red}{MC}
\makeauthorcomment{\EC}{blue}{EC}
\makeauthorcomment{\AN}{green}{AN}
\makeauthorcomment{\FMG}{orange}{FMG}
\makeauthorcomment{\MK}{purple}{MK}
\makeauthorcomment{\FA}{brown}{FA}
\makeauthorcomment{\CM}{teal}{CM}
\makeauthorcomment{\am}{magenta}{am}
\makeauthorcomment{\RS}{cyan}{RS}
\makeauthorcomment{\SF}{olive}{SF}
\makeauthorcomment{\MHG}{pink}{MHG}

\begin{document}

\title{Thermalization Across Scales on a Digital Quantum Processor}

\newcommand{\PCuk}{Phasecraft Ltd, London, UK}
\newcommand{\PCus}{Phasecraft Inc, Washington DC, USA}
\newcommand{\VT}{Virginia Tech}
\newcommand{\Bristol}{University of Bristol}

\author{Faisal Alam}\affiliation{\PCuk}
\author{Marcos Crichigno}\thanks{Corresponding author.}\affiliation{\PCus}
\author{Elizabeth Crosson}\affiliation{\PCus}
\author{Steven T. Flammia}\affiliation{\PCus}\affiliation{\VT}
\author{Filippo~Maria~Gambetta}\affiliation{\PCuk}
\author{Max Hunter Gordon}\affiliation{\PCuk}
\author{Michael Kreshchuk}\affiliation{\PCus}
\author{Ashley Montanaro}\affiliation{\PCuk}\affiliation{\Bristol}
\author{Alberto~Nocera}\affiliation{\PCuk}
\author{Raul A. Santos}\affiliation{\PCuk}

\date{\today}

\begin{abstract}
  
Thermalization of isolated quantum many-body systems normally unfolds via dynamical spreading of entanglement across progressively increasing spatial scales. The speed of this dynamical process, however, can be affected by the presence of local disorder. 
Here we study the crossover between slow and fast thermalization in a disordered two-dimensional Heisenberg Floquet model using digital quantum simulations with up to 100 superconducting qubits. 
Our experiment measures subsystem purities and $Z$-basis marginal collision entropies for patches of up to $3\times 3$ qubits. We resolve the onset of thermalization across scales by tracking each quantity's approach to its asymptotic value. In regimes where the system is moderately entangled, we validate our calibrated quantum simulations against state-of-the-art classical simulations. Near the crossover to fast thermalization, the classical methods no longer reliably converge due to volume-law growth of entanglement.
By exploring this computationally hard regime, our work establishes a frontier of applications for digital quantum computers in probing complex many-body thermalization.

\end{abstract}

\maketitle

\addtocontents{toc}{\protect\SMtocHide}

When quantum systems thermalize under information-preserving coherent dynamics, local observables describing an initial microstate relax by spreading the information they once contained into many non-local degrees of freedom~\cite{Nandkishore-Huse-Review,Alessio03052016}. Periodically driven (Floquet) systems undergo these coherent dynamics without energy conservation,
which means they can generate intricately entangled global quantum pure states alongside asymptotically featureless (infinite temperature) local observables~\cite{DAlessio2014,Lazarides2014}.   However, even for these driven systems the presence of disorder can reduce the rate of equilibration, and may prevent the system from exploring the full Hilbert space at late times, slowing down the dynamical thermalization process~\cite{Ponte2015,abanin2019colloquium}. 

\begin{figure*}
    \centering
    \subpanel{0.28\textwidth}{a}{\includegraphics[width=\linewidth]{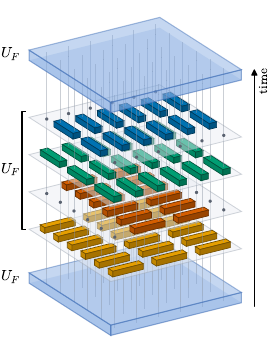}}%
    \hfill
    \begin{minipage}[t]{0.7\textwidth}\vspace{0pt}%
        \subpanel{\linewidth}{b}{\includegraphics[width=\linewidth]{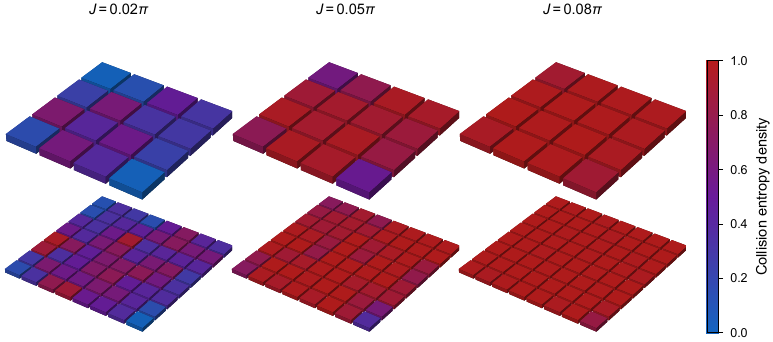}}\par\vspace{2mm}%
        \subpanel{\linewidth}{c}{\includegraphics[width=\linewidth]{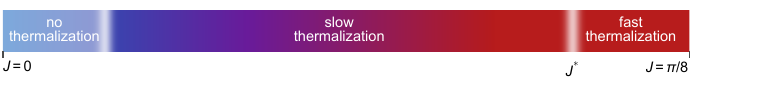}}%
    \end{minipage}
    \caption{\textbf{Thermalization in a Floquet model.}
  \textbf{a}, The periodic quantum circuits we consider consist of a 2D brickwork of gates.
  Each blue slab covering all qubits represents one application of $U_F$, a single Floquet layer.
  The sequence of underlying two-qubit gates is shown for one such layer, forming a densely packed brickwork.
  Each gate consists of the entangling unitary $e^{iJ(X_iX_j+Y_iY_j+Z_iZ_j)}$ followed by two random single-qubit rotations $e^{i(h_iZ_i+h_jZ_j)}$, with $h_i$ drawn uniformly from $[-\pi/2,\pi/2]$ on the first Floquet layer and then repeated periodically in time.
  \textbf{b}, A visualization created from the results of an $8\times 8 = 64$ qubit simulation on an IBM Quantum Nighthawk-family device, illustrating the crossover from slow to fast thermalization in the Heisenberg Floquet system as a function of the coupling $0 \le J < \pi/4$.
  At each of the three values of $J$ we exhibit the collision entropy of the $64$ single-qubit marginals, as well as the collision entropy of the $16$ non-overlapping $2\times 2$ subsystems.
  All patches of up to size $3\times 3$ can be similarly obtained but are not shown here.
  \textbf{c}, Schematic of the dynamical regimes at a fixed, finite system size. At weak coupling a single resonant spin exchange takes longer than the Heisenberg time and the system does not thermalize at any time~\cite{sierant2023stability}; this window is expected to shrink with system size. At intermediate coupling marginals thermalize slowly, and above the crossover coupling $J^\star$ they reach their random-state values after a number of cycles set by the linear size of the lattice. Our experiments resolve the slow-to-fast crossover at early times. Boundaries are schematic.}
  \label{fig:overview}
\end{figure*}

The Heisenberg Floquet model with local disorder is a standard model for these questions (\cref{fig:overview}c). At any finite size, sufficiently weak coupling leaves the system non-thermal at all times, because a single resonant spin exchange takes longer than the Heisenberg time~\cite{sierant2023stability}; whether this regime survives as the system grows is the subject of the many-body localization debate~\cite{suntajs2020quantum,sierant2023stability,PhysRevB.109.094207,Sierant_2025}. At stronger coupling the system thermalizes, first slowly and then, above a crossover coupling $J^\star$, on time scales set by ballistic spreading, analogous to the subdiffusive and diffusive regimes of the Hamiltonian model~\cite{agarwal2015anomalous,Luitz_2017}. We study this slow-to-fast crossover. Quantum hardware currently reaches only early times, but these constrain late times: a marginal that has thermalized after $n_F$ cycles remains thermal, so the early-time $J^\star$ upper-bounds the coupling at which thermalization sets in at late times. We are motivated by the open question of whether this crossover persists as the system size increases, or whether the onset of fast thermalization drifts to weaker coupling, for example through rare regions of disorder~\cite{gopalakrishnan2016griffiths} or seeding avalanches~\cite{de2017stability,thiery2018many,Morningstar_2022}.

These key questions about thermalization remain open because they lie at a formidable computational frontier. Classical tensor-network simulations are the leading numerical approach, but the rapid growth of entanglement with the coupling causes their cost to rise sharply precisely as the system approaches the dynamical crossover.  Quantum simulation on the other end offers a natural platform to generate these rapidly entangling dynamics~\cite{schreiber2015observation,Mi2021,kim2023evidence,andersen2025thermalization,shtanko2025uncovering,alam2025programmable}, but one must extract the physics from highly entangled global states using measurement samples.   %

 To track the spreading of information into higher order correlations, we measure subsystem purities (quantum R\'enyi-2 entropy) as well as the $Z$-basis marginal collision entropy. By balancing the exponential sampling overhead of higher-order correlations against the statistical costs of hardware noise calibration, we push this analysis to mesoscopic subsystems of up to $3 \times 3$ qubits. 

A basic intuition is that local scrambling dynamically spreads quantum information into growing mesoscopic regions or patches and only affects the entire system at late times. Inspired by the inverse participation ratio (IPR) of the physics literature~\cite{RevModPhys.80.1355}, encoding the \emph{collision} probability (drawing the same bitstring from two statistically independent samples) of the whole system in a given basis, we here adapt this definition to mesoscopic patches of the lattice, calling it the marginal IPR. The key observation is that the purity of a subset $A$ of qubits is lower bounded by the collision probability in the $Z$ basis. This follows from the Pauli expansions of the purity $P_2[\rho_A]$ and of the $Z$ basis collision probability $C_2[\rho_A]$ of the marginal density matrix $\rho_A$ of $A$,
\begin{align}
P_2[\rho_A] &\equiv \textrm{Tr} \rho^2_A  = 2^{-|A|} \sum_{p \in \{I,X,Y,Z\}^{|A|}} \langle p\rangle^2 \\ 
C_2[\rho_A] &\equiv \sum_{x \in \{0,1\}^{|A|}} \langle x | \rho_A |x\rangle^2  = 2^{-|A|} \sum_{p \in \{I,Z\}^{|A|}} \langle p\rangle^2 \label{eq:parseval}
\end{align}
where $|A|$ is the number of qubits in $A$.  Therefore $C_2[\rho_A] \leq P_2[\rho_A]$, and the collision probability would be the remaining purity after dephasing of $\rho_A$ in the $Z$ basis.   While
the purity tracks the scrambling over all Pauli strings in the
marginal, the collision probability only tracks the
Pauli Z strings.  However, an advantage to considering the marginal collision probability is that it can be estimated to \emph{relative error} $\epsilon$ using $\mathcal{O}(2^{|A|/2}/\epsilon^2)$ samples~\cite{SampleComplexityRenyi}, while $\mathcal{O}(2^{|A|/2} /\epsilon^2)$ samples are optimal for determining the purity to within \emph{additive error}  $\epsilon$ (from randomized measurements on a single copy of the system at a time)~\cite{Elben2023toolbox,chen2022exponential}.  Since both quantities approach $2^{-|A|}$ in the ergodic regime, the relative precision of the marginal IPR is an exponential improvement over the additive precision of the purity at the same sample cost. From the purity and the marginal IPR we define the quantum R\'enyi-2 entropy $S_{2}[A] = -\log P_2[A]$ and the marginal collision entropy $S_{2,Z}[A] = -\log C_2[A]$, which we evaluate for patches of up to $3\times 3$ qubits.  

\paragraph*{The Heisenberg Floquet system.}  

We consider the Floquet version of the disordered Heisenberg model,  $H= -J\sum_{\langle i,j \rangle} \left[X_{i}X_j+Y_{i}Y_j+  Z_{i}Z_j\right] -\sum_{i} h_i Z_i$, the paradigmatic model for the sub-ergodic to ergodic crossover in $U(1)$ conserving Hamiltonian dynamics~\cite{Nandkishore-Huse-Review,ZhangKhemaniHuse2016,Ponte2015,PhysRevB.75.155111,pal2010many}.  The quantum circuit consists of a brickwork lattice of 2-qubit unitary gates (see \cref{fig:overview}(a)) parametrised by a coupling strength $J\in [0,\pi/4]$: 

\begin{equation}
\label{Gateuijmain}
    U_{i,j}= e^{i J (X_i X_j+Y_iY_j +Z_iZ_j)}e^{i(h_i Z_i +h_j Z_j)}\, , \, U_F = \prod_{\langle i,j\rangle} U_{i,j}
\end{equation}
where $J=0$ is an identity while $J=\frac{\pi}{4}$ is a SWAP up to single qubit rotations.  To construct the Floquet operator $U_F$ these gates are applied across the 2D lattice with the disorder fields $h_i, h_j \in [-\pi/2,\pi/2]$ sampled independently at each gate location (see Supplementary Material~\cref{sec:The Heisenberg Floquet operator} for additional details).  The full time evolution is given by repeating this layer $U_F$ for $n_F$ cycles, $U =(U_F)^{n_F}$, and so time is naturally measured in units of cycles $n_F$ for this ensemble of unitary operators that are random in space and periodic in time~\cite{PhysRevX.8.041019}.   

\begin{figure*}[htbp]
  \centering
  \subpanel{0.32\textwidth}{a}{%
    \includegraphics[width=\linewidth]{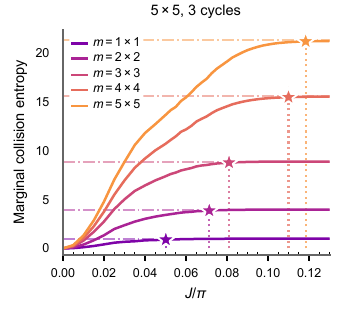}}%
  \hfill
  \subpanel{0.32\textwidth}{b}{%
    \includegraphics[width=\linewidth]{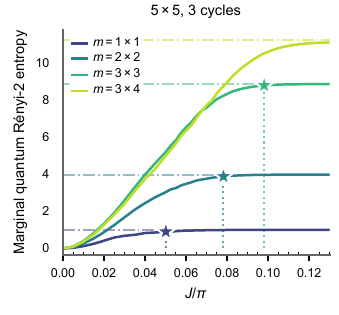}}%
  \hfill
  \subpanel{0.32\textwidth}{c}{%
    \includegraphics[width=\linewidth]{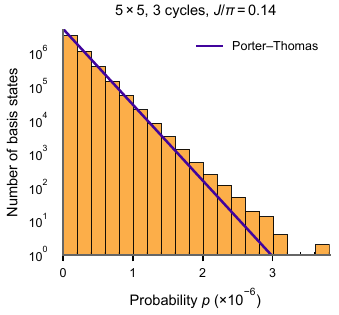}}%
  \caption{\textbf{Exact simulation of $5\times 5$ system.}
    \textbf{a}, Marginal collision entropies $S_{2,Z}[A]$ and \textbf{b}, R\'enyi-2 entropies $S_2[A]$ for \emph{central} patches in the $5\times5$ system at $n_F=3$ cycles, computed via exact state vector simulation, averaged over $256$ disorder realizations.
    The stars ($\bigstar$) indicate the couplings $J^\star$ at which entropy reaches a plateau within $0.1$ of the corresponding Haar values.
    In \textbf{b}, subsystems are shown up to $3\times 4$ ($12$ qubits, closest to half the system size, since $S_2[A]$ of a pure state equals that of its complement).
    Unlike the interior $3\times 3$ patch, a $3\times 4$ patch cannot fit in the bulk of $5\times 5$ and has one edge on the open boundary, cutting $11$ rather than $12$ bonds; at small $J$, where bipartite entanglement is generated locally across the cut, $S_2[3\times 4]$ therefore lies slightly below $S_2[3\times 3]$ before overtaking it in the volume-law regime.
    \textbf{c}, We also find that the distribution of computational-basis probabilities $P(x) = \lvert\langle x|\psi(t)\rangle\rvert^2$ after $n_F=3$ and $J/\pi=0.14$ already closely follows the Porter–Thomas form~\cite{porter1956fluctuations,boixo2018characterizing}.
    }
  \label{fig:marginal_ipr_5x5_main_text}
\end{figure*}

\paragraph*{Behavior of the model.} Using exact state-vector simulation of up to $5\times 5$ qubits, we observe a smooth crossover in both the purity and the marginal collision entropies as $J$ increases (\cref{fig:marginal_ipr_5x5_main_text}). At small $J$ the entropies remain well below their Haar-random values even on time scales of the order of the Heisenberg time, as supported by spectral analysis of the Floquet operator. Above a crossover coupling $J^\star$, the entropies of all marginals reach their Haar values after a number of cycles set by the linear size of the system, consistent with ballistic spreading of correlations; at larger $J$ the system is fully ergodic by random-matrix spectral diagnostics~\cite{PhysRevB.75.155111,atas2013distribution} and by the Porter--Thomas form of its output distribution~\cite{porter1956fluctuations,boixo2018characterizing}. The value of $J^\star$ decreases slowly with evolution time, consistent with the upper bound above. %

The question of whether this behavior persists as the system size increases faces a computational bottleneck. Beyond state-vector simulations, our matrix product state~\cite{schollwock2011} (MPS) simulations with bond dimension up to $\chi=4096$, and projected entangled pair state simulations with belief propagation~\cite{tindall2023gauging} (BP-PEPS) up to $\chi=32$, are reliable at weak coupling but lose stability under bond-dimension refinement at $J/\pi\simeq0.035$--$0.06$, before the crossover, as the entanglement of the evolved states grows towards a volume law (Methods and Supplementary \cref{sec:Classical algorithms}).

\subsection*{Simulation on a superconducting quantum processor}

We perform quantum simulations of the system on an IBM Quantum's Nighthawk-family QPU (\texttt{ibm\_miami}), which consists of a rectangular grid of $n=10\times 12$ superconducting qubits. 
We run simulations at different system sizes, starting with small systems to validate the experimental results against classical simulations, and going up to a $10\times 10$ system. 
We run each at a number of Floquet cycles $n_F$ that is half the linear size of the system, or fix the number of cycles to be $n_F = 6$ when probing convergence to the thermodynamic limit.
This ensures that the system has sufficient time to become fully entangled when the value of $J$ is sufficiently high.  Throughout this section we always initialize the qubits in the $Z$ basis N\'eel state, a product state with a check-board pattern of spin up and spin down. This is done to ensure that the experiment captures that largest symmetry sector of the Heisenberg Floquet system. Then, a number of Floquet unitaries \cref{Gateuijmain} are applied (see \cref{fig:overview}(a)). Finally, all qubits are measured in the computational basis.  Raw experimental results are mitigated by Pauli weight-dependent empirical calibration factors performed on the least entangled circuits at $J = 0.005\pi$, which we refer to as Low Entanglement Calibration (LEC), followed by an additional $J$ dependent debasing step. Extensive details of the actual procedure are outlined in (Methods).

In \cref{fig:overview}(b) we visualize the emergence of thermalization across scales using experimental data at system size $8\times8$. Each spatial patch is color-coded according to its measured collision entropy, providing a heatmap of local thermalization for $1\times1$ and $2\times2$ marginals. At very small $J$, many patches of both sizes remain far from thermal. At intermediate $J$, most $1\times1$ patches appear thermalized, but some $2\times2$ regions are not. Only at sufficiently large $J$ patches of all sizes do approach the thermal value, indicating ergodic behaviour across scales.

\begin{figure*}[!htbp]
    \centering
    \includegraphics[width=\textwidth]{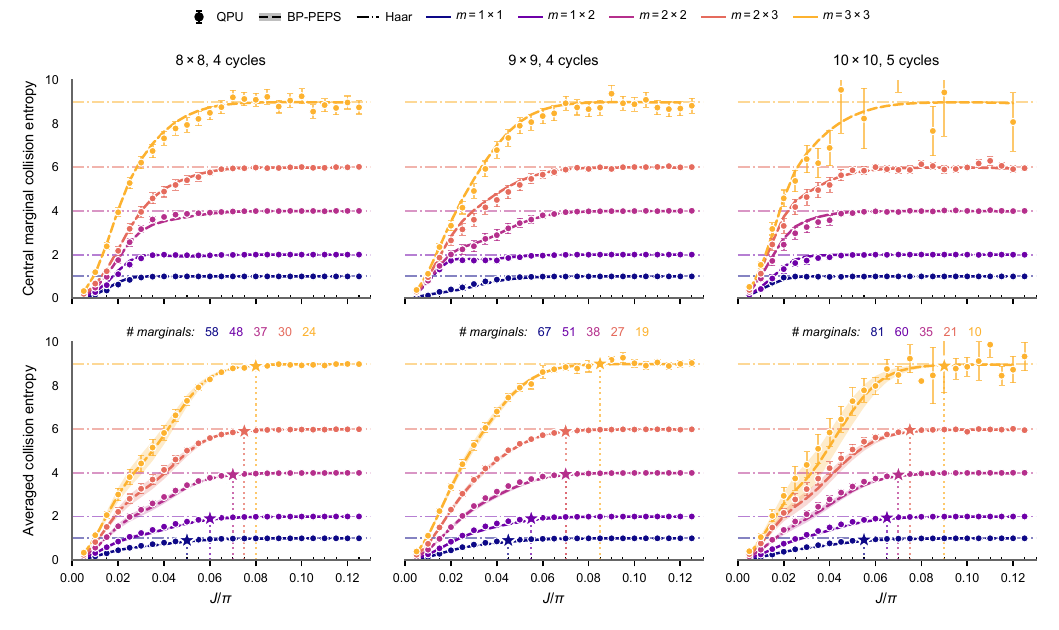}
    \caption{\textbf{Onset of thermalization.} Collision entropy $S_{2,Z}[A]$ for subsystems $A$ of shape $m$, for a range of
system sizes beyond exact diagonalization. 
Top row: the single subsystem at the centre of the lattice. 
Bottom row: averaged over subsystems of each shape. 
Points are from IBM Nighthawk-family QPUs, calibrated with LEC and corrected for residual bias. 
Dashed lines are BP-PEPS at $\chi = 32$, with a fourth-order cumulant expansion for marginals larger than $2\times2$ (Methods); the shaded bands show the variance of the PEPS subsystem average. Their stability under bond-dimension refinement is shown in \cref{fig:classical_cost_main}a and Supplementary \cref{fig:reliability}. 
Error bars combine shot noise with the uncertainty of the bias correction and, for the averages, the variance over subsystems.
The stars ($\star$) mark $J^\star[A]$, the first $J$ at which the averaged $S_{2,Z}[A]$ comes within $0.1$ of its Haar value. 
Numbers above each lower panel state how many subsystems of each shape avoid broken couplers and enter the average.}
    \label{fig:entropy_large_systems}
\end{figure*}

In \cref{fig:entropy_large_systems} we show central marginal collision entropies as well as spatially averaged marginal collision entropies, for systems of size $8\times8$, $9\times9$, and $10\times10$ (evolved for $n_F=4$, $n_F=4$, and $n_F=5$ cycles, respectively). These results demonstrate that the crossover behaviour in exact state-vector simulations of small systems (see  \cref{fig:marginal_ipr_5x5_main_text}) persists at these larger system sizes. Our results establish that there is a region in $J$ for which the system does not thermalize (at the times we consider) and that a smooth crossover transition to ergodicity for each marginal size exists.  The symbol $\bigstar$ indicates the value of $J^\ast$ at which the measured, spatially-averaged entropy lies within $\varepsilon=0.1$ of the Haar value; since the behaviour appears smooth numerically, there is no sharply defined onset of ergodicity, and these markers are shown only for reference to indicate trends. 
The mitigated data agree with the classical simulations where the latter can be trusted: on $8\times8$ the $3\times3$ marginal IPR changes by less than $0.5\%$ under the largest refinement of the bond dimension, for both MPS and BP-PEPS, up to $J/\pi=0.04$, and by $0.7$--$3.5\%$ for $0.05\le J/\pi\le0.09$ (\cref{fig:classical_cost_main}a). At intermediate $J$ and for the larger marginals the tensor-network entropies lie systematically below the mitigated ones; whether this deviation comes from the error mitigation or from the truncation of the entanglement in the simulations remains open.

The results in \cref{fig:entropy_large_systems} are obtained using \emph{spatial averaging} over translations of marginals within a single disorder instance. Because the system self-averages as it approaches the thermodynamic limit in system size~\cite{schiulaz2020self}, spatial averaging captures ensemble properties while allowing for higher statistical resolution of the crossover. 
Focusing on a single disorder instance at each system size also lets us push the experiment and the MPS and BP-PEPS simulations to their limits on the same circuits. However, as observed when probing the convergence of these entropies with system size, residual effects from the particular disorder instance can still be present in spatially averaged quantities in moderately sized systems.

The results of \cref{fig:entropy_large_systems}  by themselves do not determine the value of $J^\ast$ for the full system at these sizes (which would come from computing the global IPR), but they do provide lower bounds: if a marginal of size $m<n$ has not thermalized, then the full system cannot be thermalized either. Since noise tends to increase entropy, these lower bounds remain valid even in the presence of residual unmitigated noise.

\begin{figure}[t]
    \centering
    \includegraphics[width = 0.99\columnwidth]{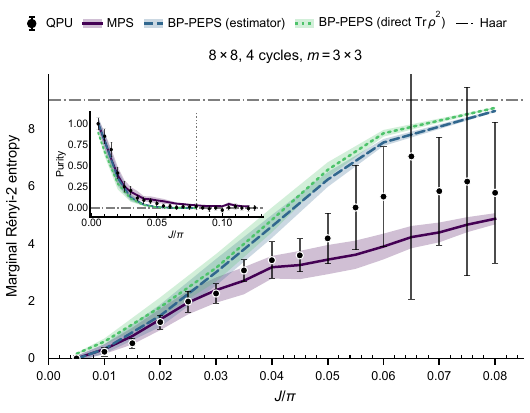}
    \caption{\textbf{Measurement of the R\'enyi-2 entropy.} Marginal R\'enyi-2 entropy $S_2[A]=-\log_2P_2[A]$ of $3\times3$ subsystems $A$ of an $8\times8$ system after $n_F=4$ cycles, obtained from the purity $P_2[A]$ averaged over four subsystems and four disorder realizations. QPU (points): error-mitigated data from IBM Quantum Nighthawk-family devices, where the purity is measured in a random basis, implemented by two additional cycles with $J=\pi/8$ restricted to each subsystem, and estimated from the $Z$-basis outcomes resolved by magnetization sector. MPS: the same protocol simulated with a matrix product state at $\chi=2048$. BP-PEPS (estimator): purity obtained by applying the two scrambling cycles exactly to the BP-PEPS reduced density matrix $\rho_A$; BP-PEPS (direct $\textrm{Tr}\,\rho^2$): the purity of $\rho_A$ itself. BP-PEPS uses $\chi=8$ for $J/\pi\le0.03$ and $\chi=16$ for $J/\pi\ge0.04$. Dash-dotted line: value for a Haar-random state at fixed total magnetization. Error bars and shaded bands give one standard deviation of the averaged purity, propagated to $S_2$. Inset: purity over the full range of $J$; to the right of the vertical line the signal-to-noise ratio of the QPU data is too low to convert the purity into an entropy.}
    \label{fig:purity_experiment}
\end{figure}

The collision entropy probes only the $Z$-basis structure of each marginal. To access the full Rényi-2 entropy we also measured the purity of $3\times3$ subsystems of the $8\times8$ system, using a randomized-measurement protocol in which the random unitaries are generated by the Floquet circuit itself, restricted to each patch and resolved by magnetization sector~\cite{Elben2018,Brydges2019,Vitale2022,Elben2023toolbox} (Methods). \Cref{fig:purity_experiment} shows that the entropy rises from zero to its random-state value over a similar range of $J$ to that in which the $3\times3$ collision entropies saturate (\cref{fig:entropy_large_systems}). At small $J$ the mitigated results agree with both MPS and BP-PEPS simulations. At larger $J$ the two classical methods separate, and the mitigated results lie between them, above MPS but below BP-PEPS. Since both methods are least reliable in this regime of rapid entanglement growth, the comparison does not determine whether the device under- or overestimates the entropy there. As $J$ increases and the purity approaches zero, the signal-to-noise ratio becomes too low to resolve the entropy accurately, so the inset shows the purity over the full range of $J/\pi$. These results show that the Rényi-2 entropy, the purity and the marginal collision entropy locate the crossover consistently, and the Rényi-2 entropy confirms that the marginals approaching the Haar collision entropy are close to maximally entangled with the rest of the system, not merely dephased in the $Z$ basis.

\begin{figure*}[!htbp]
    \centering
    \includegraphics[width=\textwidth]{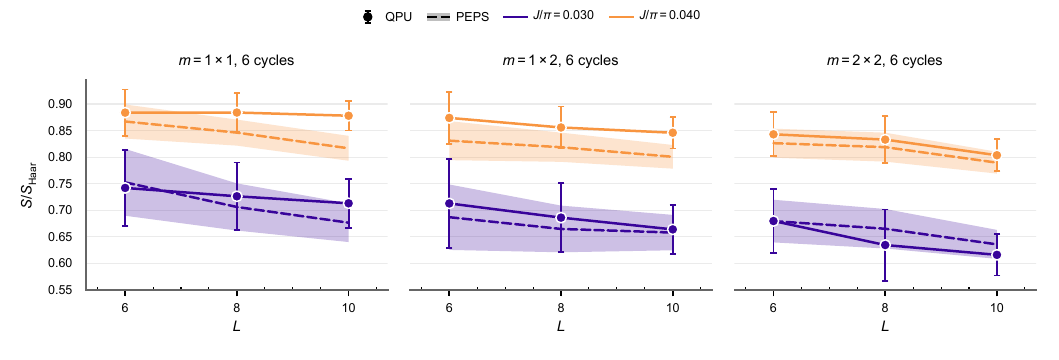}
\caption{\textbf{Approach to the thermodynamic limit.} 
Spatial averaged collision entropy $S_{2,Z}[A]$ after 6 cycles, normalized by its Haar value $S_{\rm Haar}$ for each system size and subsystem shape $m$, against the linear size $L$ of the lattice, at $J/\pi = 0.030$ and $0.040$. 
Points are from IBM Nighthawk-family QPUs, calibrated with LEC and corrected for residual bias.
Dashed lines are tensor-network results: MPS at $\chi = 4096$ for $L = 6$ and BP-PEPS at $\chi = 32$ for $L \geq 8$. At these couplings the averaged BP-PEPS IPRs change by at most $0.3\%$ between $\chi=32$ and $48$ on $6\times6$ and $8\times8$ (on $10\times10$, by about $1\%$ at $J/\pi=0.05$).
Averages run over subsystems of each shape that do not touch the lattice boundary, with experimental data additionally excluding subsystems that include broken couplers.
Error bars combine shot noise, the uncertainty of the bias correction and the variance over subsystems.
}
    \label{fig:TDL_main}
\end{figure*}
When the number of Floquet cycles $n_F$ is held fixed as $L\to\infty$, every finite subsystem has a finite backward light cone of gates, so its reduced state in the thermodynamic limit is reproduced exactly by any finite lattice that contains that light cone. At $n_F=6$ this causal light cone is larger than all of the lattice sizes we simulate, so agreement between system sizes is not guaranteed by causality alone. In the slow-thermalization regime, however, disorder suppresses the spreading of correlations~\cite{burrell2007bounds}, and the region on which a marginal effectively depends can be much smaller than its causal light cone.

To test whether this convergence occurs at $n_F=6$, we compare the normalized, spatially averaged collision entropy $S/S_{\rm Haar}$ of the $1\times1$, $1\times2$ and $2\times2$ marginals in the $6\times6$, $8\times8$ and $10\times10$ systems (\cref{fig:TDL_main}), excluding subsystems that touch the boundary. We focus on $J/\pi=0.03$ and $0.04$, which lie in the crossover regime and are therefore most sensitive to finite-size effects. If the thermodynamic limit had been reached, $S/S_{\rm Haar}$ at fixed $J$ would be independent of $L$. Instead the classical and quantum results both show a slight downward trend. We attribute this trend to the specific disorder instance probed rather than to a property of the system: for each system size we use a single disorder realization, and at these sizes the average over its bulk marginals does not fully self-average to the ensemble value.

\subsection*{Classical and quantum costs at the crossover regime}

\begin{figure*}[t]
    \centering
    \subpanel{0.40\textwidth}{a}{\includegraphics[width=\linewidth]{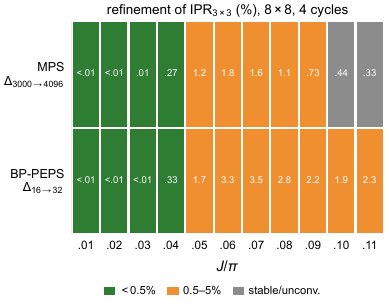}}%
    \hfill
    \subpanel{0.29\textwidth}{b}{\includegraphics[width=\linewidth]{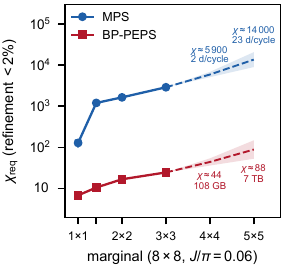}}%
    \hfill
    \subpanel{0.29\textwidth}{c}{\includegraphics[width=\linewidth]{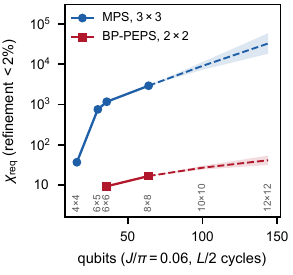}}%
    \caption{\textbf{Classical cost of Tensor Network simulations.} \textbf{a}, Reliability of the IPR of the central $3\times3$ marginal of $8\times8$ after four cycles for MPS and BP-PEPS. Each cell gives the change of the IPR (in \%) under the largest refinement performed: bond dimension $\chi=3000\to4096$ (MPS) and $16\to32$ (BP-PEPS). Green: below $0.5\%$; orange: $0.5$--$5\%$; grey: below $0.5\%$ beyond the coupling at which the refinement peaks, where the marginal approaches its random-state value (stable but not converged); see Supplementary \cref{fig:reliability} for all lattices. \textbf{b}, Bond dimension $\chi_{\rm req}$ for which the refinement of the marginal IPR falls below $2\%$, against the size of the marginal of $8\times8$ at $J/\pi=0.06$ (for MPS the $1\times1$ value is an upper bound). The $4\times4$ and $5\times5$ values are extrapolated (dashed lines; the bands span an exponential growth with the number of sites and with the linear size of the marginal) and annotated with the corresponding cost: wall-clock time of one Floquet cycle on four cores for MPS, and largest intermediate tensor of the exact contraction of a $2\times3$ block for BP-PEPS. \textbf{c}, $\chi_{\rm req}$ at $J/\pi=0.06$ against the number of qubits, after $L/2$ cycles, for the $3\times3$ (MPS) and $2\times2$ (BP-PEPS) marginals, extrapolated to $10\times10$ and $12\times12$. Details in Methods and Supplementary \cref{fig:methods_complexity}.}
    \label{fig:classical_cost_main}
\end{figure*}

\emph{Tunable classical hardness.} The coupling $J$ controls the cost of tensor-network simulations (\cref{fig:classical_cost_main}a). At weak coupling all of the tensor-network simulations are reliable; their stability under increasing bond dimension (a necessary condition for convergence) decreases in the range $J/\pi\simeq0.035$--$0.06$, depending on method and system size, prior to the thermalization crossover.   The required bond dimension grows with the size of the lattice (\cref{fig:classical_cost_main}c) as well as the scale of the probe: at fixed $J$ the required resources grow exponentially with the size of the marginal (\cref{fig:classical_cost_main}b); at $J/\pi=0.06$ and for a refinement below $2\%$, a $4\times4$ marginal of $8\times8$ would require $\chi\approx5{,}000$--$7{,}000$ with MPS, i.e.\ $1.5$--$3.5$ days per Floquet cycle, and intermediate tensors of $30$--$370$\,GB with BP-PEPS to execute tensor contractions at larger bond dimensions.  The scale-resolved entropies therefore provide a natural ladder of benchmarks at a fixed large system size.  

\emph{Scaling of the quantum experiment.} On the quantum processor the circuit at every $J$ has the same structure and depth, so the wall-clock time of each data point is effectively independent of $J$, and all marginal collision entropies of a circuit are estimated from the same measurement record. In the absence of noise, the number of samples to estimate a collision probability to relative error $\epsilon$ scale as $2^{|A|/2}/\epsilon^2$, i.e.\ a factor of $4$ for each additional pair of qubits in the patch.   Therefore in a noiseless system  the collision entropy of $4\times4$ marginals could be estimated using the $10^4$--$10^5$ samples in \cref{fig:entropy_large_systems}. In the noisy experiment each $Z$ string is damped by the noisy gates that its Heisenberg-evolved operator spreads over, so that larger marginals and larger $J$ incur more error (Methods). This does not change the contrast with the classical scaling described above: the quantum resources grow with the patch size and the noise, while the classical resources grow with the patch size and the entanglement, which $J$ controls.

\emph{Classical simulability beyond tensor networks.} Building on an earlier version of this work, Rudolph et al.~\cite{rudolph2026pushing} developed randomized Pauli propagation~\cite{suchsland2026quantum} and applied it to the central 3×3 collision entropy of the 8×8 system at four couplings, combining a replica estimator for the squared expectation values with a weight-dependent allocation of samples; they report agreement with the tensor-network results in the localized regime, and estimates between MPS and BP-PEPS near the crossover, where they note that their results are not fully converged. We also developed a cluster discrete truncated Wigner approximation~\cite{schachenmayer2015many,wurtz2018cluster} for our circuits (Methods and Supplementary Sec.~\ref{app:dtwa}), whose $3\times3$ collision entropies on $8\times8$ after six cycles lie within $2\%$ of BP-PEPS for $0.02\le J/\pi\le0.08$, at a cost that grows exponentially with the cluster size, which also controls its accuracy and bounds the size of the marginals it can describe reliably. The scale-resolved entropies introduced here provide a well-defined, experimentally accessible family of quantities for testing the performance across the growing catalog of approximate classical simulation methods.  

\subsection*{Open questions and outlook.} Our experiments do not settle whether the crossover persists in the thermodynamic limit or drifts to weaker coupling with system size: over the range accessible here, residual disorder-instance effects dominate any size dependence (\cref{fig:TDL_main}). Resolving that question requires larger lattices, longer evolutions and larger patches, precisely the directions in which the classical cost above grows fastest and in which the quantum cost grows only through the noise. As error rates fall, experiments of this kind, validated where classical calculations are controlled and extended where they are not, offer a route to many-body thermalization at and beyond the limits of classical simulation.

\clearpage
\section{Methods}
\label{sec:methods}

We briefly summarize the experimental implementation, error mitigation procedures, and classical simulation methods, as well as the algorithm used to measure subsystem purities. Full details are provided in the Supplementary Material.

\subsection{Experimental details}

Our experiments were performed on \texttt{ibm\_miami} and \texttt{ibm\_berlin}, the first Nighthawk-family QPUs released by IBM Quantum, between January 2026 and July 2026. Each device has 120 superconducting qubits arranged on a $12\times 10$ square lattice (see \cref{fig:error_rates}). The disordered Heisenberg Floquet circuit was compiled into native single-qubit rotations and CZ gates, with each Heisenberg interaction decomposed into three $\mathrm{CZ}$ gates~\cite{vatan2004optimal}. 

Depending on system size and number of Floquet cycles, circuits reached up CZ depth of 72 and as many as 3240 two-qubit gates, with 40k-120k shots allocated per parameter point. Pauli twirling of two-qubit gates~\cite{Wallman_2016_noise} and measurements~\cite{vandenberg2022model} was used to suppress coherent errors. Dynamical decoupling~\cite{Viola_1998_dynamical} was also used to suppress noise. Additional details on hardware specifications, compilation, calibration data and execution parameters are provided in Sec.~\ref{sec:Experimental details} of the Supplementary Material.

\subsection{Error mitigation}
All collision entropies from QPU data reported in \cref{fig:entropy_large_systems,fig:TDL_main} are mitigated in three steps: marginals containing broken couplers are discarded, a calibration against tensor-network simulations at the smallest value of $J$ is applied to each marginal, and the residual bias of that calibration is corrected by an error model fitted to experiments with a trusted classical reference.

\paragraph*{Avoiding broken couplers.} Our experimental data reveals that certain qubits in each run return measurement outcomes showing they are fully decohered at every value of $J$. We identify these \textit{dead qubits} using the criterion that their IPR at the smallest coupling $J_0$, where the ideal state is still close to the initial product state, lies within $0.02$ of the maximally mixed value of $1/2$. 

\begin{figure*}[h!]
    \centering
    \subpanel{0.5\textwidth}{a}{\includegraphics[width=\linewidth]{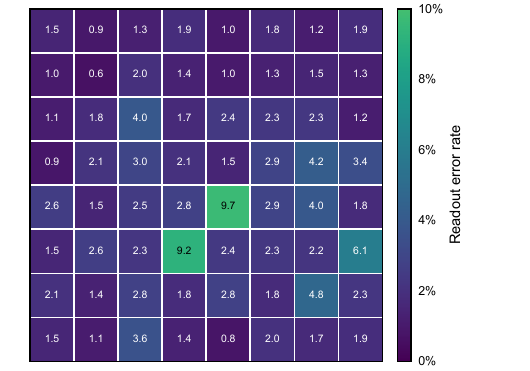}}%
    \subpanel{0.5\textwidth}{b}{\includegraphics[width=\linewidth]{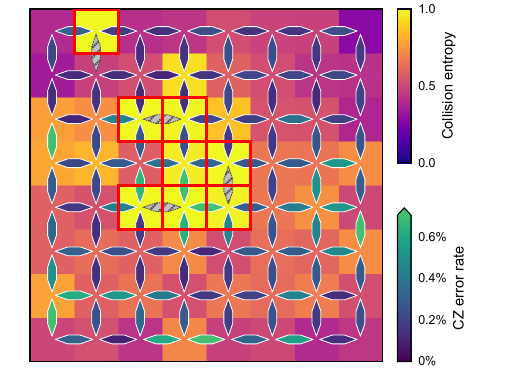}}%
    \caption{\textbf{Error rates on \texttt{ibm\_berlin}.}
    \textbf{a}, Readout error rates, $(p_{01}+p_{10})/2$, where $p_{ab}$ is the probability that a qubit prepared in $\ket{a}$ is measured as $b$, from calibration circuits run at the start of each experiment.
    \textbf{b}, Single-qubit collision entropies measured in our experiment at the smallest coupling, $J = 0.005\pi$ (tiles), and CZ error rates from IBM Quantum's calibration data (bonds). Dead qubits, whose IPR is within shot noise of the maximally mixed value of $1/2$, are outlined in red. Faulty CZ gates, which the calibration data lists with an error rate of $100\%$, are drawn in grey. Locations of dead qubits are highly correlated with faulty CZ gates.
    Data are from an $8\times 8$ experiment on \texttt{ibm\_berlin} on 26 June 2026.}
    \label{fig:error_rates}
\end{figure*}

The locations of dead qubits vary between devices and between experiments run on the same device on different dates. \cref{fig:error_rates}b indicates that their locations are strongly correlated with faulty $\mathrm{CZ}$ gates, which are marked in IBM Quantum's calibration data with error rates of 100\%. Therefore, we hypothesize that a broken coupler leads to excessive decoherence on one or both of the qubits it touches. Marginals that include dead qubits are discarded from the subsystem averages in \cref{fig:entropy_large_systems,fig:TDL_main}. 

\paragraph*{Low entanglement calibration.} The calibration finds a scaling factor that matches the experimental results at the smallest available coupling $J_0$ to tensor-network simulations, since this regime is weakly entangled and classically accessible at all system sizes we study (here $J_{0} = 0.005\pi$). Once determined, the scaling is applied to all other values of $J$. We refer to this procedure as Low Entanglement Calibration (LEC)~\cite{czarnik2021error,montanaro2021error,Mi2021,kim2023evidence,shtanko2025uncovering}.

For each marginal $A$ whose IPR we estimate, LEC is applied to the expectation values $z_s:=\langle Z_s\rangle$, $s \subseteq A$, that make up the IPR through the Parseval \cref{eq:parseval}, with one multiplicative factor per Pauli weight:
\begin{align}
z_s^\textrm{LEC}(J) &= R_{|s|}\, z_s^\textrm{exp}(J)\,, \label{eq:LECdefinitionmain}\\
R_k &= \arg\min_R \sum_{|s|=k}\big(z_s^\textrm{exact} - R\,z_s^\textrm{exp}\big)^2\Big|_{J=J_0} \nonumber\\
    &= \frac{\sum_{|s|=k} z_s^\textrm{exact}\, z_s^\textrm{exp}}{\sum_{|s|=k} \big(z_s^\textrm{exp}\big)^2}\bigg|_{J=J_0}\,, \nonumber
\end{align}
where the sums run over the $Z$ strings of weight $k$ supported on $A$. Each $R_k$ is thus the least-squares scaling factor matching the strings of that weight to the reference at $J=J_0$; the objective is quadratic in $R$, and setting its derivative to zero gives the closed form in the third line. 

The IPR is then reconstructed from \cref{eq:parseval} with each $(z_s^\textrm{LEC})^2$ computed from the scaling factor and the unbiased estimator of $(z_s^\textrm{exp})^2$ from the measured shots.

The form of this ansatz for the calibrated expectation values can be motivated by considering local two-qubit depolarizing noise on each gate, with a rate that does not depend on $J$. Such a noise model implies that each $Z$ string is damped once for every noisy gate in its backward lightcone, a number set mainly by its weight $|s|$. See \cref{sec:depolarizingnoiseinversion} of the Supplementary Material for details. 

\paragraph*{Removing residual J dependent bias.} LEC reproduces the classical value of the entropy at $J_0$, but, on average, overestimates the entropy at larger $J$. This effect worsens with the depth of the circuit. We attribute this to operator spreading~\cite{nahum2018operator,suchsland2026quantum}. Although noise does not depend on $J$ explicitly, the operator seen by each noisy gate does. In the Heisenberg picture a measured $Z$ string evolves backward through the circuit. Each time the evolved Pauli sum encounters a noisy gate, its norm diminishes. The strength of this effect depends on the fraction of the Pauli sum (by squared coefficients) that is incident on the gate. At small $J$, where the LEC scaling factor is computed, the fraction incident on each noisy gate is smaller than at larger $J$. As a result, LEC tends to underestimate the effect of noise on expectation values at larger $J$. See \cref{sec:depolarizingnoiseinversion} for a detailed explanation. 

Writing $\Delta Q:=\text{IPR}_{2,Z}[A]-2^{-n_A}$ for the contribution of the nontrivial expectation values to the IPR, we correct this residual damping by
\begin{align}
\Delta Q_\textrm{corr} &= e^{2\delta}\,\Delta Q_\textrm{LEC}\,, \label{eq:errormodel}\\
\delta &= A\alpha\,G\,\big[e^{-(n_F J_0/x_0)^2} - e^{-(n_F J/x_0)^2}\big]\,, \nonumber
\end{align}
where $G$ is the number of two-qubit gates in the backward light cone of the marginal and $n_F$ the number of Floquet cycles. The damping vanishes at $J_0$, grows quadratically with the accumulated coupling $n_F J$ as the operators spread, and saturates at $\delta \approx A\alpha\,G$ once they have spread over the whole light cone. The corrected entropy carries a systematic error 
\begin{align}
\sigma_A=[(\sigma n_A)^2+(\eta\,\Delta S^\kappa)^2]^{1/2}, \label{eq:sys_error}
\end{align}
added in quadrature to the shot noise, where the first term accounts for the error of LEC itself, which shares one factor between all strings of a given weight, and the second for the uncertainty of the correction, which grows with its size $\Delta S$ in bits. The five parameters ($A, x_0, \sigma, \eta, \kappa$) are fitted to the classically simulable data points from 20 experiments ($4\times4$ to $10\times10$, 2 to 12 cycles). 

To test the effectiveness of bias correction, we hold out each experiment in turn from the training data and find that the resulting correction (\cref{eq:errormodel}) removes about two thirds of the bias on the held out experiment. The error bars computed using \cref{eq:sys_error} contain the classical reference 82\% of the time (median over held out experiments). Details can be found in \cref{sec:wplecerrormodel} of the Supplementary Material.

\paragraph*{Symmetry based error mitigation} Because our circuits conserve the Hamming weight, fitting the output Hamming-weight distribution with a single per-qubit bit-flip probability $p$ and inverting the resulting attenuation mitigates the data without classical simulation (\cref{sec:mitigationbrief}). While this is less accurate than LEC, it has a diagnostic value: $p$ empirically grows with $J$, which indicates the existence of the $J$-dependent residual bias that we correct using \cref{eq:errormodel}. 

\subsection{Estimating collision probabilities}

\paragraph*{Sample-optimal estimator.} We estimate the marginal IPR through the Parseval expansion, \cref{eq:parseval}, rather than by counting collisions directly, so that each Pauli weight receives its own calibration. Before any correction the two estimators coincide: with unbiased estimators for each $\langle Z_s\rangle^2$, the Parseval sum over all strings $s\subseteq A$ equals the coincidence (collision) estimator identically, because $\sum_{s\subseteq A} z_s(x)z_s(y)=2^{|A|}\delta_{xy}$ for any two bit strings $x,y$. The Pauli-weight representation thus loses no statistical efficiency by itself. It costs variance only once mitigation amplifies the high-weight strings, and in return it gives the noise model the structure it actually has, since the damping of a string depends on its weight through its light cone. This combination of a sample-optimal estimator with weight-resolved correction is what made $3\times3$ marginals accessible on current hardware, and we expect it to apply to other sample-based diagnostics of scrambling.

\paragraph*{Shot cost of noisy strings.} In the noisy experiment, each $Z$ string expectation value is damped by the noisy gates that its Heisenberg-evolved operator passes through. At small $J$ the string remains on its own support and is damped only by the gates acting there; as $J$ increases it spreads over its backward light cone and passes through more noisy gates, so that the measured signal decays roughly exponentially in the product of the error rate and the number of gates it has spread over, at most the number in its light cone. If every CZ in the backward light cone ($G$) of a string acted as a depolarizing channel at the median reported error $r=2.7\times10^{-3}$, a weight-9 $Z$-string spanning a $3\times3$ marginal of the $8\times8$ lattice ($G\approx975$ CZs after four cycles) would retain $(1-4r/3)^{G}\approx3\%$ of its ideal expectation value. Since the shot cost of resolving such a string scales as $(1-4r/3)^{-2G}$, a threefold reduction of the CZ error would lower the shots required for $3\times3$ marginals at this depth by roughly two orders of magnitude.

\subsection{Subsystem purity from symmetry-resolved randomized measurements.}
\label{sec:purity_methods}
 
The purities in \cref{fig:purity_experiment} are obtained with randomized measurements~\cite{vanEnk2012,Elben2018,Brydges2019,Elben2023toolbox}, adapted to the $U(1)$ symmetry of the model so that the random unitaries can be generated by the native Floquet circuit itself, in the spirit of the random-quench protocol of Ref.~\cite{Elben2018} and of symmetry-resolved randomized measurements~\cite{Vitale2022}.
 
\emph{Block structure.} Every gate in \cref{Gateuijmain} commutes with the total magnetization $\sum_i Z_i$, and the initial N\'eel state lies in the half-filling sector, so the evolved state $|\psi\rangle$ has a definite total Hamming weight $n/2$. Writing $|\psi\rangle=\sum_k |\psi_k\rangle_A|\phi_{n/2-k}\rangle_{\bar A}$, where $k$ is the Hamming weight of the bit string on patch $A$, the reduced state is block diagonal,
\begin{equation}
\rho_A=\bigoplus_{k=0}^{|A|} w_k\,\rho_k,\quad
w_k=\textrm{Tr}\,\Pi_k\rho_A,\quad
P_2[A]=\sum_{k} w_k^2\,P_k,
\label{eq:purity_blocks}
\end{equation}
where $P_k=\textrm{Tr}\,\rho_k^2$ and $\Pi_k$ projects onto the $d_k=\binom{|A|}{k}$-dimensional weight-$k$ subspace of $A$ and $\rho_k$ is normalized. For a $3\times3$ patch the sector dimensions are $d_k=1,9,36,84,126,126,84,36,9,1$.
 
\emph{Scrambling within sectors.} After $n_F$ cycles of the global evolution, the lattice is partitioned into non-overlapping $3\times3$ patches and each patch receives two further Floquet cycles at $J=\pi/8$ (the $\sqrt{\textrm{SWAP}}$ point of \cref{Gateuijmain}, midway between the non-entangling endpoints $J=0$ and $J=\pi/4$), with gates crossing patch boundaries removed and fresh disorder fields drawn independently for each patch. All patches are scrambled in parallel in the same circuit, after which every qubit is measured in the $Z$ basis. 
The scrambler $V$ conserves the Hamming weight of each patch, so $V=\bigoplus_k V_k$. If $V_k$ is drawn from a unitary 2-design on the weight-$k$ subspace, the collision probability $C_{2,k}=\sum_{x:|x|=k}\langle x|V_k\rho_kV_k^\dagger|x\rangle^2$ of the conditional output distribution satisfies $\mathbb{E}_V\,C_{2,k}=(1+P_k)/(d_k+1)$ ~\cite{Elben2018,Vitale2022}, and so $P_k=(d_k+1)\,\mathbb{E}_V\,C_{2,k}-1 $ and
\begin{equation}
P_2[A]=\sum_k\Big[(d_k+1)\sum_{x:|x|=k}\mathbb{E}_V\,p(x)^2-w_k^2\Big],
\label{eq:purity_estimator}
\end{equation}
where $p(x)$ is the probability of outcome $x$ on $A$. The second form uses only joint (unconditioned) probabilities. From $M$ shots with outcome counts $n_x$ and sector counts $M_k$ we use the unbiased estimators $\widehat{p(x)^2}=n_x(n_x-1)/[M(M-1)]$ and $\widehat{w_k^2}=M_k(M_k-1)/[M(M-1)]$.
 
\emph{A single scrambler realization.} We use one realization of the scrambling disorder per patch and per disorder instance of the evolution, replacing $\mathbb{E}_V$ in \cref{eq:purity_estimator} by a single draw. For Haar-random $V_k$ the single-draw collision probability concentrates around its mean, with fluctuations in the inferred $P_k$ of order $d_k^{-1/2}$ (for a pure $\rho_k$ the standard deviation is $\simeq 2/\sqrt{d_k}$; quantifying this requires fourth moments, i.e.\ a 4-design). Two features of the problem keep the resulting error small. First, the sector weights $w_k$ concentrate on the largest sectors ($k=4,5$, $d_k=126$), and each sector enters \cref{eq:purity_blocks} with weight $w_k^2$, so the poorly concentrated small sectors contribute negligibly. Second, the scrambler disorder is independent across patches and disorder instances, so these fluctuations average down in the reported mean.
 
\emph{Finite-depth scrambler.} Two cycles do not form an exact 2-design within each sector, and the resulting bias depends on the state being measured. It vanishes exactly in the ergodic limit: if $\rho_k=\Pi_k/d_k$, then $C_{2,k}=1/d_k$ for any sector-preserving $V$ and \cref{eq:purity_estimator} returns the exact value $P_k=1/d_k$. The two-cycle bias therefore overestimates purity (underestimates entropy) but this bias is subleading at the target level of statistical accuracy as shown in \cref{fig:purity_experiment}.
 
\emph{Noise.} Because the estimator only accesses $\Delta(\rho_A)=\sum_k\Pi_k\rho_A\Pi_k$, any coherence between sectors created by symmetry-breaking errors is discarded, and since this pinching is a unital channel, $\textrm{Tr}\,\Delta(\rho_A)^2\le\textrm{Tr}\,\rho_A^2$. Errors that do not conserve magnetization still redistribute weight between sectors and are reflected in the measured $w_k$. The purity data in \cref{fig:purity_experiment} are mitigated with the low entanglement calibration (LEC) described below, in its simplest form: rather than calibrating each Pauli weight separately, for each marginal and disorder realization $\textrm{Tr}\,\Delta(\rho_A)^2-P_\textrm{mm}$ is rescaled by a single factor, fixed at $J_0=0.005\pi$ by matching MPS simulations ($\chi=2048$). Here $P_\textrm{mm}=\sum_k d_kD_k^2/D^2$ (with $D_k$ and $D$ as defined below) is the purity of the maximally mixed half-filling state reduced to $A$, the fixed point of depolarizing noise in the presence of the U(1) symmetry. No post-selection on the total Hamming weight of the full register is applied.

\emph{Reference value.} For a random state of the full $8\times8$ system at half filling, the average $3\times3$ purity is $\sum_k d_k D_k(d_k+D_k)/[D(D+1)]\approx1.97\times10^{-3}$, where $D_k=\binom{55}{32-k}$ and $D=\binom{64}{32}$; at this scale this symmetry-constrained value is indistinguishable from the unconstrained $2^{-9}$, and it is the Haar line shown in \cref{fig:purity_experiment}.

\subsection{Classical simulations.}

We complement the experiments with classical simulations using matrix product states (MPS), projected entangled pair states contracted with belief propagation (BP-PEPS) and the cluster discrete truncated Wigner approximation (cluster DTWA). They validate the mitigated hardware data where their results can be considered reliably converged, and show where, and at which cost, such calculations lose control as $J$ approaches the crossover to ergodicity. Full details are given in Secs.~\ref{sec:Classical algorithms} and~\ref{app:dtwa} of the Supplementary Material.

\textit{Matrix product states.} In our two-dimensional MPS simulations, we found the best compromise between performance and accuracy using a hybrid scheme, where we evolve the first Floquet cycle with time-evolving block decimation (TEBD)~\cite{vidal2004efficient} and the subsequent cycles with the \emph{local} time-dependent variational principle (TDVP)~\cite{haegeman2016unifying,sander2025quantum} (inset of Supplementary \cref{fig:methods_complexity}(b)). We simulated systems of up to $8\times8$ qubits with bond dimensions up to $\chi=4096$, and obtained the marginal IPRs by exact contraction of the MPS, except for some $6\times6$ values at $\chi=4096$, estimated from $10^4$ samples of the state.

\textit{Belief-propagation PEPS.} We evolve the state as a PEPS gate by gate with the simple update algorithm~\cite{jiang2008accurate}, and compute contractions with belief propagation~\cite{tindall2023gauging}, with bond dimensions up to $\chi=32$. The IPR of a marginal follows from the Parseval expansion (Supplementary \cref{ParsevalIPR}) in terms of expectation values of $Z$ strings, each obtained by contracting exactly the smallest lattice region that contains the string with the converged messages entering it (inset of Supplementary \cref{fig:methods_complexity}(e)). For marginals larger than $2\times2$ we truncate the corresponding cumulant expansion~\cite{KuboCumulant} at fourth order.

\textit{Cluster DTWA.} In the discrete truncated Wigner approximation every spin is represented by a classical vector sampled from a discrete Wigner distribution. We use a variant of its cluster version~\cite{wurtz2018cluster}, in which the lattice is tiled with clusters of up to $4\times4$ qubits whose quantum state is evolved exactly, while every gate between two clusters is replaced by a single-qubit rotation about a classical spin sampled from the neighbouring cluster. The marginal distributions are averaged over trajectories and over the tilings of the lattice.

\textit{Convergence and cost.} We assess the tensor-network calculations by their stability under bond-dimension refinement, and classify each calculation and coupling as reliable (change below $0.5\%$), borderline ($0.5$--$5\%$) or uncontrolled (above $5\%$) (Supplementary \cref{fig:reliability}). Because the bond dimension is not the only source of error in BP-PEPS, stability does not guarantee accuracy: on $4\times4$ the BP-PEPS IPR becomes independent of $\chi$ while retaining a bias of a few percent with respect to the exact result (Supplementary \cref{fig:conv_bp_4x4}), and on $8\times8$ the MPS IPR at $\chi=4096$ exceeds the BP-PEPS one by $8\%$ to $21\%$ for $0.035\le J/\pi\le0.08$ (Supplementary \cref{fig:MPS_vs_PEPS}). The cost of both tensor-network methods grows with $J$, with the size of the marginal and with the size of the system (Supplementary \cref{fig:methods_complexity}). For MPS on $8\times8$ the bond dimension required for a refinement below $5\%$ rises from $\chi=128$ at weak coupling to $\chi=2048$ for the $3\times3$ marginal, and the time of one Floquet cycle on four CPU cores from about $20$\,s to about $3$ hours ($19$ hours at $\chi=4096$). For BP-PEPS the cost lies in the exact contraction of the $Z$-string regions, which grows as $\chi^8$ for blocks of $2\times3$ sites (Supplementary \cref{fig:bp_contraction_cost}): on a single CPU core the central $3\times3$ marginal takes $22$ minutes at $\chi=32$ and $36$ hours at $\chi=48$ (Supplementary \cref{fig:resource_horizon}). For cluster DTWA the cost grows exponentially with the size of the clusters; with $4\times4$ clusters the $3\times3$ collision entropy stays within $2\%$ of BP-PEPS for $0.02\le J/\pi\le0.08$ (Supplementary \cref{fig:cluster-benchmark}). In all methods the classical calculations are stable under bond-dimension refinement only up to $J/\pi\simeq0.035$--$0.06$, below the crossover to ergodicity; extrapolations to larger marginals and lattices are given in Supplementary Sec.~\ref{sec:tns_complexity}.

\textit{Entanglement.} The growth of the computational resources with $J$ reflects the growth of entanglement in the evolved states. On $8\times8$ after four cycles the R\'enyi-2 entropy $S_2=-\log_2P_2$ of half of the lattice, obtained from the MPS, is independent of $\chi$ up to $J/\pi\simeq0.04$, while for $J/\pi\gtrsim0.05$ it exceeds what can be captured at $\chi=4096$ (Supplementary \cref{fig:mps_entanglement}). With BP-PEPS, patches of up to eight qubits in the bulk of $8\times8$ are within $0.1$ bits of their maximal entropy for $J/\pi\gtrsim0.08$, so that the entropy grows linearly with the number of qubits of the patch, as expected for volume-law entanglement (Supplementary \cref{fig:bp_entanglement}); its scaling for extensive subsystems of the larger lattices is beyond the reach of both methods.

\textit{Classical computing resources.} We ran the tensor-network calculations on a server with an AMD Ryzen Threadripper PRO 7965WX processor ($24$ cores, $503$\,GB of memory) and on E2 virtual machines of Google Cloud Platform, and the cluster DTWA calculations on a laptop with an AMD Ryzen AI 9 HX 370 processor ($12$ cores).

\subsection*{Acknowledgments} The authors gratefully acknowledge David Huse and Toby Cubitt for many fruitful discussions and valuable feedback. We also acknowledge valuable discussions with Adrian Chapman and Joel Klassen and technical support from Jan Lukas Bosse, Brian Flynn, Lana Mineh, Harry McMullan, and Gethin Wright. We thank Cat Mora for valuable feedback on an earlier version of the manuscript. We acknowledge the use of IBM Quantum Credits via the IBM Quantum Startups Program for this work. The views expressed are those of the authors and do not reflect the official policy or position of IBM or the IBM Quantum Platform team.

\subsection*{Author contributions} 
MC conceived and led the project. EC prototyped the study of the marginal collision entropy and the error mitigation methods. MK contributed to the development of the theoretical model and performed exact numerical simulations. RAS coordinated the initial experimental exploration. FMG performed preliminary experiments and tensor-network simulations. AN designed and performed large scale tensor network numerical simulations and the cluster truncated Wigner approach. FA ran the experiments. FA, FMG, MHG and AM performed data analysis and implementation of error mitigation. STF helped guide the initial project direction. All authors discussed the results and contributed to the final manuscript.

\clearpage
\onecolumngrid
\section{Extended data figures and tables}

\begin{table}[H]
    \centering
    \begin{tabular}{llrrrr}
        \toprule
        Lattice & Device & CZ gates & CZ depth & Circuit depth & Shots per $J$ \\
        \midrule
        $8\times 8$, 4 cycles   & \texttt{ibm\_miami}                       & 1{,}344 & 48 & 130 & 40{,}000 \\
        $9\times 9$, 4 cycles   & \texttt{ibm\_miami}                       & 1{,}728 & 48 & 133 & 40{,}000 \\
        $10\times 10$, 5 cycles & \texttt{ibm\_miami}                       & 2{,}700 & 60 & 162 & 40{,}000 \\
        $8\times 8$, 6 cycles   & \texttt{ibm\_berlin}                      & 2{,}016 & 72 & 194 & 80{,}000 \\
        $9\times 9$, 6 cycles   & \texttt{ibm\_miami}, \texttt{ibm\_berlin} & 2{,}592 & 72 & 199 & 120{,}000 \\
        $10\times 10$, 6 cycles & \texttt{ibm\_miami}, \texttt{ibm\_berlin} & 3{,}240 & 72 & 194 & 120{,}000 \\
        \midrule
        $8\times 8$, $4+2$ cycles & \texttt{ibm\_miami}, \texttt{ibm\_berlin} & 1{,}632 & 72 & 200 & $4\times 40{,}000$ \\
        \bottomrule
    \end{tabular}
    \caption{\textbf{Circuit sizes and shot counts for experiments displayed in the main text.}
    The top block lists the experiments that measure collision entropies, and the bottom row lists the R\'enyi-2 entropy experiment. Each experiment runs a single transpiled circuit at a sweep of $J$ values. Each Floquet cycle applies three CZ gates per lattice bond, arranged in 12 CZ layers. The circuit depth counts the layers of native one- and two-qubit gates ($\sqrt{X}$, $X$ and CZ); virtual $R_Z$ rotations and the final measurement are not counted. The R\'enyi-2 entropy experiment adds two scrambling cycles after the four Floquet cycles and uses 40{,}000 shots per disorder instance.}
    \label{tab:experiments}
\end{table}

\clearpage

\setcounter{section}{0}

\numberwithin{equation}{section}

\pagenumbering{arabic}
\setcounter{page}{1}

\begin{center}
\vspace*{4cm}
{\LARGE \textbf{Supplementary Material}}
\end{center}

\makeatletter
\providecommand{\SMtocHide}{\let\SMtoc@contentsline\contentsline\let\contentsline\@gobblefour}
\providecommand{\SMtocShow}{\let\contentsline\SMtoc@contentsline}
\makeatother
\addtocontents{toc}{\protect\SMtocShow}
\tableofcontents
\vspace{1em}
\noindent\rule{\textwidth}{0.4pt}

\section{Background}
\label{sec:Background}

This Supplementary Material provides additional background, definitions, and technical details supporting the results in the main text.
We begin by reviewing information-theoretic quantities used to characterize delocalization and thermalization, with particular emphasis on inverse participation ratios, R\'enyi moments, and their marginal versions.
Subsequent sections collect complementary material on ergodicity and random matrix theory, define the two-dimensional Heisenberg Floquet model studied in this work, present extended numerical and experimental results, and describe both the quantum simulation protocols and the classical algorithms used for comparison.

\subsection{Basic notions of classical and quantum information theory}

We begin by reviewing the classical notion of R\'enyi entropy \cite{renyi1961measures} and its generalization to the context of quantum mechanics.  Given a  classical probability distribution \{$p_i\}_{i=1,\ldots, D}$,  the R\'enyi-$k$ moment is defined as
\begin{equation}
    H_k:= \sum_{i=1}^D {p_i^k}\,.
\end{equation}
It is easy to see that 
\begin{equation}
\frac{1}{D^{k-1}}\leq  H_{k}\leq 1\,.
\end{equation}
The upper bound saturated only when the distribution is peaked in a single outcome, $p_i=1$ for a single $i$. The lower bound is saturated only for the homogeneous distribution, $p_i= \frac{1}{D}$ for all $i$. Thus, the $H_k$ quantify the degree of concentration of the probability distribution. The  classical R\'enyi-$k$ entropy is defined as 
\begin{equation}
    S_{k}:= \frac{1}{1-k}\, \log \sum_{i=1}^{D} p_i^k\,.
\end{equation}
The moments $H_k$ and the corresponding R\'enyi entropies $S_k$ provide a hierarchy of measures quantifying how probability mass is distributed over the sample space. They are a generalization of Shannon entropy, $S_{\rm Shannon}=-\sum_{i=1}^{D}p_i \log p_i$, which is obtained in the limit $S_{\rm Shannon}=\lim_{k\to 1}S_k$.
Unlike the Shannon entropy, which captures average uncertainty, R\'enyi entropies emphasize different parts of the distribution depending on $k$, with larger $k$ increasingly sensitive to high-probability outcomes.

\ 

The classical definitions above extend naturally to the quantum setting by replacing probability distributions with density matrices. Given a density matrix $\rho$, the {\it quantum} R\'enyi-$k$ moment is defined as 
\begin{equation}
    \hat H_{k}(\rho):= \text{Tr} [\rho^k]\,.
\end{equation}
Note these are basis independent quantities. In a basis that diagonalizes the density matrix, $\rho=\sum_{i=1}^D \lambda_i\, \ket{\psi_i}\bra{\psi_i}$, these match the classical R\'enyi moments for the probability distribution $\{\lambda_i\}_{i=1}^D$. Thus, it satisfies the same upper and lower bounds as the classical quantity.  The spectrum of a density matrix $\{\lambda_i\}$ captures the degree of mixedness of a quantum state. The moments $\hat H_{k}(\rho)$ further characterize this mixedness by probing how it is distributed—whether it is concentrated in a few eigenstates or spread across many. The  quantum R\'enyi-$k$ entropies are defined as
\begin{equation}
    S_k(\rho):= \frac{1}{1-k} \,\log \text{Tr}_{\mathcal H} [\rho^k]\,.
\end{equation}
These are a generalization of the von Neumann entropy, $S_{\rm vN}=-\text{Tr}[ \rho  \log \rho]$, which is obtained in the limit $S_{\rm vN}=\lim_{k\to 1}S_k$. 

\ 

Note the R\'enyi entropies all vanish for a pure state $\rho=\ket{\psi}\bra{\psi}$. The R\'enyi entropies are particularly interesting for reduced density matrices, which are typically mixed.  Consider a  bipartite system with Hilbert space $\mathcal H=\mathcal H_A\otimes \mathcal H_B$ and density matrix $\rho$. The reduced density matrix on subsystem $A$ is defined as 
\begin{equation}
    \rho_A := \text{Tr}_{\mathcal H_B}\,[\rho]\,.
\end{equation}
If the original density matrix  is pure, $\rho=\ket{\psi}\bra{\psi}$, the quantum Renyi-$k$ entropy of the reduced density matrix,
\begin{equation}
    S_k(\rho_A):= \frac{1}{1-k} \,\log \text{Tr}_{\mathcal H_A} [\rho_A^k]\,,
\end{equation}
is a measure of entanglement between the subsystems $A$ and $B$.

\subsection{Marginal IPR}
\label{sec:marginal_ipr}

\begin{figure}[t]
\centering
\begin{tikzpicture}[scale=0.9]

\def\s{0.7}

\draw[fill=black!6, draw=black!18, rounded corners=2pt]
  ({0.5*\s},{0.5*\s}) rectangle ({6.5*\s},{6.5*\s});

\draw[fill=black!12, draw=black!25, rounded corners=2pt]
  ({1.5*\s},{1.5*\s}) rectangle ({5.5*\s},{5.5*\s});

\draw[fill=black!30, draw=black!40, rounded corners=2pt]
  ({2.5*\s},{2.5*\s}) rectangle ({4.5*\s},{4.5*\s});

\foreach \x in {0,...,7} {
  \foreach \y in {0,...,6} {
    \draw[line width=0.35pt]
      ({\x*\s},{\y*\s}) -- ({\x*\s},{(\y+1)*\s});
  }
}
\foreach \y in {0,...,7} {
  \foreach \x in {0,...,6} {
    \draw[line width=0.35pt]
      ({\x*\s},{\y*\s}) -- ({(\x+1)*\s},{\y*\s});
  }
}

\foreach \x in {0,...,7} {
  \foreach \y in {0,...,7} {
    \filldraw[fill=white, draw=black, line width=0.4pt]
      ({\x*\s},{\y*\s}) circle (0.07);
  }
}

\end{tikzpicture}
\caption{\textbf{Marginals of a qubit lattice.} Depiction of a set of quantum mechanical degrees of freedom, e.g., qubits or spins and various patches $A$ on which one can define the marginal collision entropies.}
\label{fig:grid_patches_ipr}
\end{figure}

Consider a pure quantum state $\ket{\psi}=\sum_{i=1}^Dc_i \ket{i}$, where $\mathcal B=\{\ket{i},\, i=1,\ldots, D\}$  is some basis for the Hilbert space of dimension $D=\dim \mathcal H$ and the $c_i$ are complex coefficients giving probabilities $p_i=\abs{c_i}^2$ of observing the system in state $\ket{i}$.  A standard measure of thermalization is the ``inverse participation ratio'' (IPR)  and its generalizations \cite{RevModPhys.80.1355,PhysRevLett.84.3690}, defined as
\begin{equation}
    \text{IPR}_{k,\mathcal B}:= \sum_{i}p_i^k\,,
\end{equation}
which is simply the R\'enyi-$k$ moment for the probability distribution $p_i$ associated to the quantum state $\ket{\psi}$ in the basis $\mathcal B$.

\ 

Note for $k=2$ the IPR can be interpreted as the probability of observing the same classical configuration twice, and is thus sometimes referred to as the ``collision probability.'' The collision probability of quantum states captures the  anticoncentration properties of the quantum state and plays a crucial role in the formal arguments of classical hardness of the sampling problems~\cite{Bremner_2010,boixo2018characterizing,bouland2019complexity,dalzell2022random}.  One can also show that the expected value in the Haar-random ensemble is $\Bbb E_{\text{Haar}}[\text{IPR}_{\mathcal B}]=\frac{2}{D}$ for any basis $\mathcal B$. Due to measure concentration~\cite{popescu2006entanglement}, to high precision this is the value for a typical Haar-random state. It is believed that sampling from Haar-random states is classically intractable. 

The IPR is in general notoriously difficult to compute by classical methods, as it requires the computation of an exponential number of probabilities or, via the relation above, the computation of an exponential number of nonlocal expectation values. However, this is also generically inaccessible via a quantum computer since it can be an exponentially small quantity and thus requires an exponential number of samples to be approximated to any good accuracy.  Motivated by the physics of the systems at hand, we introduce a new quantity that is a local measure of delocalization. We consider a decomposition of the Hilbert space $\mathcal H=\mathcal H_A\otimes \mathcal H_B$ with corresponding basis $\mathcal B_A=\{\ket{x},\,x=1,\ldots,d_A\}$ and $\mathcal B_B=\{\ket{y},\,y=1,\ldots,d_B\}$,   and compute the marginal probabilities for the subsystem $A$, i.e.,  
\begin{equation}
   p_A(x):= \sum_{y\in \mathcal B_B} p(x,y)\,.
\end{equation}
Then we define the marginal IPR for the region $A$ as the IPR for the marginal probability:
\begin{equation}
    \text{IPR}_{k,\mathcal B_A}[A]:=\sum_{x\in \mathcal B_A}p_A(x)^k\,. 
\end{equation}
This satisfies
\begin{equation}
   \frac{1}{d_A^{k-1}}\leq  \text{IPR}_{k,\mathcal B_A}[A] \leq 1\,.
\end{equation}
The marginal collision entropy is then
\begin{equation}
    S_{k,\mathcal B}:= \frac{1}{1-k}\, \log\Big( \sum_{x\in \mathcal B_A}p_A(x)^k\Big)\,. 
\end{equation}

\ 

A necessary condition for global ergodic behaviour is that every local patch of the system exhibits ergodicity. 
In other words, global thermalization requires local thermalization across all subsystems. 
This motivates the use of marginal collision entropy as a fine-grained diagnostic of ergodicity: 
for each patch $A$, we compare its collision entropy to the corresponding Haar-random value. 
If, at a given coupling $J$ and time $t$, a patch fails to approach its Haar benchmark, this signals that the full system has not reached ergodic behaviour.

Larger patches provide increasingly stringent probes of global thermalization. 
While $1$-site marginals impose only weak constraints, $2$-site and higher-order marginals successively capture more nontrivial correlations. 
Patches comprising a significant fraction of the system therefore offer the most informative diagnostics. 
Classically, however, accessing large subsystems incurs exponentially increasing computational cost (see Section~\ref{sec:Classical algorithms}).  On quantum hardware, increasing the patch size primarily increases the sampling requirements. 
The shot counts available on current QPUs are sufficient to access nontrivial subsystems — for example $3\times3$ patches within $8\times8$, $9\times9$, and $10\times10$ systems — a regime that is extremely challenging for classical simulation.

\
 
Although a physically motivated  diagnostic, the marginal collision entropy has not, to our knowledge, been systematically studied in this context. 
From a quantum information perspective, it admits a simple and transparent interpretation: it corresponds to the R\'enyi-2 entropy after complete dephasing in a fixed basis. 
Specifically, consider the dephasing quantum channel in a basis $\mathcal B$,
\begin{equation}
    \mathcal D_{\mathcal B}(\rho)
    = \sum_{\ket{i}\in \mathcal B} 
    \ket{i}\!\bra{i}\,\rho\,\ket{i}\!\bra{i}\,,
\end{equation}
which removes all off-diagonal coherences in $\mathcal B$. 
The marginal collision entropy is then precisely the R\'enyi-2 entropy of the dephased reduced state, quantifying the degree of classical randomness in that basis.
The dephasing channel removes all off-diagonal elements in the basis $\mathcal B$. Then, the marginal IPR is the dephased R\'enyi-$k$ moment of the marginal density matrix:
\begin{equation}
    \text{IPR}_{k,\mathcal B_A}[A]= \hat H_{k}[\mathcal D_{\mathcal B}(\rho_A)]\,,\qquad    S_{k,\mathcal B}[A]=  S_{k}[\mathcal D_{\mathcal B}(\rho_A)]\,.
\end{equation}
The $k$-R\'enyi moments can be obtained by taking $k$ copies of the system, $\rho^{\otimes k}$,  and measuring the expectation value of the cyclic permutation operator (for $k=2$ this is the SWAP operator between the two copies)~\cite{ekert2002direct,daley2012measuring,islam2015measuring}. It turns out that it can also be obtained by computing correlations among random measurements on a {\it single} copy \cite{vanEnk2012, Elben2018,PhysRevA.99.052323}. 

We refer to $H_{k}$ as the $k$th classical moment in the basis $\mathcal B$.  We will mostly focus on $k=2$. It is easy to see this is a unital completely positive trace-preserving (CPTP) channel.\footnote{A unital channel satisfies $\mathcal D_{\mathcal B}(\Bbb I)=\Bbb I$.} From the definitions above it is straightforward to show that~\cite{PhysRevLett.113.140401}
\begin{equation}\label{ineqZ2M2}
   \text{Tr}\,[ \rho^2] \geq  \sum_i p_i^2\,.
\end{equation}

\ 

When the initial state is an eigenstate of the measurement basis, the collision entropy starts from its minimal value and increases as the state delocalizes in that basis.

The marginal IPR offers several practical and conceptual advantages over the global IPR.
First, global thermalization necessarily implies local thermalization, so the equilibration of marginal IPRs provides a necessary condition for ergodic behavior.
Second, estimating marginal IPRs requires only $\mathcal O(d_A^{1/2})$ samples, making them accessible for moderately sized subsystems even when global diagnostics are infeasible.
As a result, marginal IPRs provide a scalable and experimentally viable probe of delocalization and ergodicity in large quantum systems.

\paragraph{A Parseval identity for the purity and marginal IPR.} Consider a system of $n_A$ qubits described by a density matrix $\rho_A$ acting on a Hilbert space of dimension $d_A=2^{n_A}$.
For qubit systems, both the purity and the marginal inverse participation ratio admit simple and useful representations in terms of expectation values of Pauli operators supported on $A$. Let $\mathcal P_A = \{I,X,Y,Z\}^{\otimes n_A}$ denote the Pauli basis on $A$, with
$|\mathcal P_A| = 4^{n_A} = d_A^2$.
Any density matrix $\rho_A$, pure or mixed, reduced or global, admits the expansion
\begin{equation}
\rho_A
=
\frac{1}{d_A}
\sum_{P \in \mathcal P_A}
\langle P_A\rangle \, P\,,
\qquad
\langle P_A \rangle := \mathrm{Tr}\,[\rho_A\, P]\,.
\end{equation}
Using the orthogonality of Pauli operators under the Hilbert-Schmidt inner product, $\text{Tr}[P\,P']=d^A \,\delta_{P,P'}$, one obtains the Parseval identity
\begin{equation}
\label{ParsevalPurity}
\mathrm{Tr}[\rho_A^2]
=
\frac{1}{d_A}
\sum_{P \in \mathcal P_A}
\langle P \rangle^2 \,.
\end{equation}
This expression holds for any density matrix $\rho_A$ and provides a Pauli-space representation of the R\'enyi-$2$ entropy (purity).

An analogous identity holds for the marginal inverse participation ratio, applying the above to the dephased density matrix $D_{\mathcal B}[\rho_A]$ in some basis $\mathcal B$. Let $\mathcal P_A^{\mathcal B}\subset \mathcal P_A$ be the subset of Pauli operators that are diagonal in that basis.
For example, in the computational $Z$-basis, $\mathcal P_A^{Z}$ consists of tensor products of $I$ and $Z$ operators only and has cardinality
$|\mathcal P_A^{Z}| = 2^{n_A} = d_A$.
The marginal IPR in basis $\mathcal B$ is then given by
\begin{equation}
\label{ParsevalIPR}
\mathrm{IPR}_{\mathcal B}[A]
=
\frac{1}{d_A}
\sum_{P \in \mathcal P_A^{\mathcal B}}
\langle P \rangle^2 \,.
\end{equation}
Thus, while the purity involves the full set of Pauli operators on $A$, the marginal IPR corresponds to a restriction to those Pauli operators that are diagonal in the chosen measurement basis.
Both quantities are therefore sums of squared expectation values of operators with full support on the subsystem, making them sensitive probes of delocalization and simultaneously challenging to compute classically as the marginal size increases. 

\subsection{Optimal collision probability estimator}
\label{sec:Collision probability estimator}

\begin{table}[]
\centering
\begin{tabular}{c c c c c c c c c c c}
\toprule
Marginal 
& $1\times1$ 
& $1\times2$ 
& $2\times2$ 
& $2\times3$ 
& $2\times4$ 
& $3\times3$ 
& $3\times4$ 
& $4\times4$ 
& $4\times5$ 
& $5\times5$ \\
\midrule
$\varepsilon$ 
& $0.01$ 
& $0.01$ 
& $0.02$ 
& $0.03$ 
& $0.04$ 
& $0.05$ 
& $0.08$ 
& $0.16$ 
& $0.32$ 
& $0.76$ \\
\bottomrule
\end{tabular}
\caption{Values of $\sqrt{2^{n_A/2}/n_S}$ for $n_S=10^4$ and various marginal sizes $n_A$.}
\label{tab:10Kshotsnoise}
\end{table}

We now describe the sampling complexity for estimating R\'enyi moments $M_k := \sum_{i=1}^D p(i)^k$. Since the quantum device provides samples from the probability distribution, our
goal is to estimate $M_k$ from these samples and determine the number of samples required to obtain a certain precision.

Let $S=\{s_1,\dots,s_{n_S}\}$ denote $n_S$ independent samples drawn
from the probability distribution $p$.
A naive approach consists of first forming the empirical frequencies
$\hat p_i = n_i / n_S$, where
$n_i$ is the number of times that bitstring $i$ was measured,
and using these to compute $\hat M_k=\sum_i \hat p_i^k$.
However, this estimator is biased and is not optimal in the number of samples required. An unbiased and statistically optimal estimator is obtained instead
by directly counting $k$-fold collisions in the sample set, which is given by \cite{SampleComplexityRenyi}
\begin{equation}
\hat{M}_k
=
\frac{1}{\binom{n_S}{k}}
\sum_i \binom{n_i}{k}\,.
\label{eq:collision_estimator_counts}
\end{equation}
This estimator is unbiased,
$\mathbb{E}[\hat{M}_k] = M_k$ and its sample complexity has been analyzed in detail in \cite{SampleComplexityRenyi,RenyiEntropyRevisited}. For integer $k>1$, estimating $M_k$ to {\it multiplicative} precision $\varepsilon$ with failure probability at most $\rho$ requires
\begin{equation}
n_S 
=
O\left(
\, \frac{1}{\varepsilon^{2}M_k^{1/k}}
\, \log\frac{1}{\rho}
\right).
\label{eq:optimal_mk_scaling}
\end{equation}
Recall that at worst, $M_k\approx D^{1-k}$ and the number of samples grows is large for an exponentially large $D$,  but can be significantly lower for lower values of $M_k$. We focus on $k=2$ and marginal distribution on a number of $n_A$ qubits. In this case, 
\begin{equation}
n_S = O\left(\frac{2^{n_A/2}}{\varepsilon^{2}} \,\log\frac{1}{\rho}\,\right)\,.
\end{equation}
Thus, while global moment estimation (for which $n_A=n$) remains exponentially hard in the total system size, estimating marginal collision probabilities scales exponentially only in the subsystem size and benefits from a square-root improvement in $D$ relative to naive frequency-based estimators. For later reference we include the explicit values for various marginals at $10^{4}$ shots in Table~\ref{tab:10Kshotsnoise}. Taking $b$ batches of such samples reduces the multiplicative error by $\varepsilon \to \varepsilon/\sqrt{b}$.

\section{Ergodicity and Random Matrix Theory}
\label{sec:Random matrix theory and ergodicity}

Random matrix theory (RMT) plays a central role in modern discussions of quantum ergodicity and thermalization~\cite{DAlessio2014}.
In many-body systems with sufficiently chaotic dynamics, long-time quantum states are expected to exhibit statistical properties that coincide with those of maximally random states, up to constraints imposed by symmetries and conservation laws. We now review some basic tools from RMT and use them to compute the  ergodic values of collision entropies and R\'enyi in the presence of a global $U(1)$ symmetry,  yielding the appropriate symmetry-resolved random-matrix ensemble for the Floquet Heisenberg systems studied in this work. 

\subsection{{Haar}-random ensemble (Page value)}

The expected value of purity in a Haar random state was first computed by Lubkin and Page \cite{lubkin1978entropy,PhysRevLett.71.1291}. Here we review this result and provide a self-contained derivation based on Weingarten calculus~\cite{weingarten1978asymptotic,collins2006integration}. For a pedagogical review of these techniques see \cite{Mele2024introductiontohaar}. 

Two fundamental identities for Haar-random states:
\begin{equation}\label{propsHaar12}
  \Bbb E_{\rm Haar} [\ket{\psi}\bra{\psi}]= \frac{\Bbb I}{d}    \,,\qquad \mathbb{E}_{\rm Haar} \left[ |\psi\rangle\langle\psi|^{\otimes 2} \right] 
    = \frac{\mathbb{I} + \mathbb{S}}{d(d+1)}\,,  
\end{equation}
where $\Bbb I$ is the identity operator, \( \mathbb{S} \) is the swap operator exchanging the two tensor factors, and $d=2^n$ is the dimension of the Hilbert space. The first equation states that the average state over the Haar measure is the maximally mixed state. The second equation follows from group-theoretical properties. An immediate consequence of the identities \cref{propsHaar12} are the well known expressions for the first and second moments of an observable $\mathcal O$ in the Haar ensemble:\footnote{To see this one uses   $\mathbb{E}_{\rm Haar} \left[ \langle \psi | O | \psi \rangle \right]
    = \text{Tr} \left[ \mathbb{E}_{\rm Haar} \left[ |\psi\rangle\langle\psi| \right] O \right] 
    = \frac{1}{d}\text{Tr} \left[  O\right]$ and  $\mathbb E_{\rm Haar}\!\left[
\langle \psi | O | \psi \rangle^2
\right]
=
\mathbb E_{\rm Haar}\!\left[
\mathrm{Tr}\!\left(
(|\psi\rangle\langle\psi|)^{\otimes 2}
\,(O \otimes O)
\right)
\right]=\frac{1}{d(d+1)}
\Big(
\mathrm{Tr}(O)^2 + \mathrm{Tr}(O^2)
\Big)$,  where we used $\mathrm{Tr}(A\otimes B)=\mathrm{Tr}(A)\mathrm{Tr}(B)$ and
$\mathrm{Tr}(\mathbb S\,A\otimes B)=\mathrm{Tr}(AB)$. } 
\begin{align}
    \mathbb{E}_{\rm Haar} \left[ \langle \psi | O | \psi \rangle \right]
    = \frac{1}{d}\text{Tr} \left[  O\right]\,, \qquad \mathbb E_{\rm Haar}\!\left[
\langle \psi | O | \psi \rangle^2
\right]=
\frac{1}{d(d+1)}
\Big(
\mathrm{Tr}(O)^2 + \mathrm{Tr}(O^2)
\Big).
\end{align}
The first equation shows the expected value of an operator in the Haar ensemble is identical to that of in the maximally mixed state.  The difference between the Haar-random ensembles is only detected at the level of fluctuations around the mean; while the maximally mixed state has none, the  Haar ensemble has fluctuations of size $\sim \sqrt{\mathrm{Tr}(O^2)}/d$ around that mean. 

The expected value of purity and marginal IPR follow directly from applying these formulas to the Parseval representations  \cref{ParsevalPurity} and \cref{ParsevalIPR}. Indeed, applying the identity above to $O = P_A \otimes \mathbb{I}_B$ and $\langle O \rangle
=
\langle \psi | (P_A \otimes \mathbb{I}_B) | \psi \rangle$ one has that for Pauli operators:
\begin{equation}
\mathbb E_{\rm Haar}[\langle P_A \rangle^2]
=
\begin{cases}
1, & P_A = \mathbb{I}_A, \\
\dfrac{1}{d_A d_B + 1}, & P_A \neq \mathbb{I}_A.
\end{cases}
\end{equation}
\paragraph{Purity.} Taking the expected value on both sides of the Parseval formula \cref{ParsevalPurity} gives
\begin{equation}
\mathbb E_{\rm Haar}[\mathrm{Tr}\,(\rho_A^2)]
=\frac{1}{d_A}
\sum_{P \in \mathcal P_A}
\mathbb E_{\rm Haar}[\langle P \rangle^2]=
\frac{1}{d_A}
\left[
1 + \frac{d_A^2 - 1}{d_A d_B + 1}
\right]
=
\frac{d_A + d_B}{d_A d_B + 1}\,,
\end{equation}
reproducing the result by Page. It is worth noting that the largest contribution comes from the identity operator  and that all other operators contribute equally. The contribution from the identity matches that of the maximally mixed state and the nontrivial Pauli operators are corrections due to the Haar random ensemble. Note the symmetry of this value under exchanging $A$ and $B$.  We denote,
\begin{equation}
    S_2^{\rm Haar}[A]:=-\log \mathbb E_{\rm Haar}[\mathrm{Tr}\,(\rho_A^2)]=-\log \left(\frac{d_A + d_B}{d_A d_B + 1}\right)\,.
\end{equation}
Note that in this definition, the ensemble average is taken inside the log. 

\paragraph{Marginal IPR.}  The analogous computation for the marginal IPR gives
\begin{equation}
    \mathbb E_{\rm Haar}[\text{IPR}_Z[A]]=\frac{1}{d_A}\sum_{P \in \mathcal P^Z_A}
\mathbb E_{\rm Haar}[\langle P \rangle^2]=\frac{1}{d_A}
\left[
1 + \frac{d_A - 1}{d_A d_B + 1}
\right]
=
\frac{1 + d_B}{d_A d_B + 1}\,.
\end{equation}
Note the difference with the purity is the number of nontrivial Pauli operators is $d_A-1$ instead of $d_A^2-1$ and that, unlike for the purity, the value is not symmetric under exchanging $A$ and $B$. 

\paragraph{Comparison to maximally mixed state.} In contrast, for a maximally mixed state, the expected value of the IPR and purity are given by: 
\begin{equation}
    \Bbb E_{T=\infty}\left[\text{IPR}_Z[A]\right]= \Bbb E_{T=\infty}\left[\mathrm{Tr}_A\,(\rho_A^2)\right]=\frac{1}{d_A}\,.
\end{equation}
Note that keeping a marginal size fixed, the difference between the value of the marginal IPR for Haar-random state and the maximally mixed state is  exponentially small in the size of the {\it full} system:
\begin{equation}
     \Bbb E_{\rm Haar}\left[\text{IPR}_Z[A]\right]-\Bbb E_{T=\infty}\left[\text{IPR}_Z[A]\right]=\frac{1}{d_A}\frac{d_A - 1}{d_A d_B + 1}\approx \frac{1}{d}\,,
\end{equation}
where $d=d_A d_B$ is the dimension of the full Hilbert space. Thus, in order to distinguish a global maximally mixed state from a Haar-random state by measuring marginals of fixed size requires an exponential precision, just like for the full IPR.

\subsection{{Haar}-random with $U(1)$ symmetry}

We consider a system of $n$ qubits with fixed total Hamming weight $k$, i.e., $\mathcal H_k = \mathrm{span}\{ |x\rangle : |x|=k \}$~\cite{vidmar2017entanglement,bianchi2019typical,PhysRevX.14.041051}. We denote the dimension by $d_k :=\dim \mathcal H_k= \binom{n}{k}$. Now, let $\ket{\psi}$ be a Haar-random state in $\mathcal H_k$.  The fundamental identity we use is now 
\begin{equation}
\mathbb E_{{\rm Haar}, U(1)}\!\left[
(\ket{\psi}\bra{\psi}|)^{\otimes 2}
\right]
=
\frac{\Pi_k\otimes \Pi_k  + \mathbb S_k}{d_k(d_k+1)},
\end{equation}
where $\Pi_k$ the projector onto $\mathcal H_k$ and
$\mathbb S_k$ the swap operator restricted to $\mathcal H_k\otimes \mathcal H_k$.

\paragraph{Purity.} Taking the Haar average of the purity and using the identities above we have
\begin{equation}
\mathbb E_{{\rm Haar}, U(1)}[\mathrm{Tr}(\rho_A^2)]
=\mathbb E_{{\rm Haar}, U(1)}[\mathrm{Tr}\!\left[
(|\psi\rangle\langle\psi|)^{\otimes 2}\,
\mathbb S_A\right]]=
\frac{
\mathrm{Tr}(\Pi_k\otimes \Pi_k\, \mathbb S_A)
+
\mathrm{Tr}(\mathbb S_k \mathbb S_A)
}{d_k(d_k+1)}\,.
\end{equation}
To evaluate this we decompose the Hilbert space  in terms of states with a fixed Hamming weight in $A$ (and hence $B$), i.e.,  $\mathcal H_k
=
\bigoplus_{h=0}^{\min(n_A,k)}
\mathcal H_{A,h} \otimes \mathcal H_{B,k-h}$. We denote the dimensions of each factor by
\begin{equation}
  d_{A,h} = \binom{n_A}{h}\,,\qquad d_{B,k-h} = \binom{n_B}{k-h}  \,.
\end{equation} 
The operators decompose then as
\begin{equation}
\Pi_k = \sum_{h=0}^{\min(n_A,k)}
\Pi_{A,h} \otimes \Pi_{B,k-h}\,,
\qquad
\mathbb S_k =\sum_{h=0}^{\min(n_A,k)} \mathbb S_{A,h} \otimes \mathbb S_{B,k-h}\,.
\end{equation}
Using this we have
\begin{align}
\mathrm{Tr}(\Pi_k\otimes \Pi_k \mathbb S_A)
=
\sum_{h=0}^{\min(n_A,k)}
d_{A,h}^2\, d_{B,k-h}\,, \qquad
\mathrm{Tr}(\mathbb S_k \mathbb S_A)
=
\sum_{h=0}^{\min(n_A,k)}
d_{A,h}\, d_{B,k-h}^2 \,.
\end{align}
Combining both contributions yields the  Haar average
\begin{equation}
\mathbb E_{{\rm Haar}, U(1)}[\mathrm{Tr}(\rho_A^2)]
=
\frac{1}{d_k(d_k+1)}
\sum_{h=0}^{\min(n_A,k)}
\Big[
d_{A,h}^2\,d_{B,k-h}
+
d_{A,h}\, d_{B,k-h}^2
\Big].
\end{equation}
This is the main result of this section, providing the asymptotic value for the purity. The R\'enyi-2 entropy of a typical state is then given by
\begin{equation}\label{S2HaarU1}
   S_{2}^{U(1)\text{-Haar}}[A]:=-\log \mathbb E_{{\rm Haar}, U(1)}[\mathrm{Tr}(\rho_A^2)]\,.
\end{equation}

\paragraph{Marginal IPR.}  The computation for the marginal IPR is similar. Let $|\psi\rangle \in \mathcal H_k$ be Haar-random within $\mathcal H_k$,
with amplitudes $\psi_{ab}$ in the computational basis.
The marginal probability distribution on subsystem $A$ is
\[
p_A(a) = \sum_b |\psi_{ab}|^2,
\]
and the marginal IPR in the $Z$ basis is
\[
\mathrm{IPR}_Z[A]
=
\sum_a p_A(a)^2
=
\sum_a \sum_{b,b'} |\psi_{ab}|^2 |\psi_{ab'}|^2 .
\]
The Haar averages of amplitude moments in $\mathcal H_k$ are
\begin{align}
\mathbb E_{{\rm Haar}, U(1)}\!\left[|\psi_x|^4\right]
=
\frac{2}{d_k(d_k+1)}\,,\qquad
\mathbb E_{{\rm Haar}, U(1)}\!\left[|\psi_x|^2 |\psi_y|^2\right]
=
\frac{1}{d_k(d_k+1)},
\qquad x\neq y .
\end{align}
Separating diagonal ($b=b'$) and off-diagonal ($b\neq b'$) contributions,
we find for a fixed $a$ with Hamming weight $h$:
\begin{align}
\mathbb E_{{\rm Haar}, U(1)}\!\left[\sum_b |\psi_{ab}|^4\right]
=
\frac{2\, d_{B,k-h}}{d_k(d_k+1)}\,,\qquad 
\mathbb E\!\left[\sum_{b\neq b'} |\psi_{ab}|^2 |\psi_{ab'}|^2\right]
=
\frac{d_{B,k-h}(d_{B,k-h}-1)}{d_k(d_k+1)} \,.
\end{align}
Summing over all $a$ with $|a|=h$ and then over all allowed $h$ yields
\begin{equation}
\mathbb E_{U(1)\text{-Haar}}\!\left[\mathrm{IPR}_Z[A]\right]
=
\frac{1}{d_k(d_k+1)}
\sum_{h=0}^{\min(n_A,k)}
d_{A,h}d_{B,k-h}
\left(
1+d_{B,k-h} 
\right)\,.
\end{equation}
This is the main result of this section, providing the asymptotic value for the marginal IPR in our system. The collision entropy of a typical state is then  given by
\begin{equation}\label{S2ZHaarU1}
S_{2,Z}^{{\rm Haar}, U(1)}[A]:=-\log \mathbb E_{{\rm Haar}, U(1)}\!\left[\mathrm{IPR}_Z[A]\right]\,.
\end{equation}

\paragraph{Comparison to maximally mixed state.} Let us contrast these to the case of a maximally mixed state in a fixed Hamming-weight-$k$ sector, $\rho^{(k)} = \frac{1}{d_k}\,\Pi_k$, where $\Pi_k$ is given above.  Tracing out subsystem $B$ yields
\begin{equation}
\rho_A^{(k)} = \mathrm{Tr}_B[\rho^{(k)}] = \frac{1}{d_k}\sum_{h=0}^{\min(n_A,k)} d_{B,k-h}\,\Pi_{A,h}\,,
\end{equation}
which is block-diagonal in the Hamming-weight sectors of $A$.
The purity follows immediately:
\begin{equation}
\mathrm{Tr}_A\big[(\rho_A^{(k)})^2\big]
=
\frac{1}{d_k^2}
\sum_{h=0}^{\min(n_A,k)}
d_{A,h}\, d_{B,k-h}^2\,.
\end{equation}
The marginal IPR in the computational  basis is identical,
\begin{equation}
\mathrm{IPR}_Z[A]
=
\sum_a p_A(a)^2
=
\frac{1}{d_k^2}
\sum_{h=0}^{\min(n_A,k)}
d_{A,h}\, d_{B,k-h}^2,
\end{equation}
reflecting the fact that for the maximally mixed state at fixed charge the reduced density matrix is
classical in the $Z$-basis. 

\ 

We note that, like in the unconstrained case, the difference between the Haar-random value and the maximally mixed state is extremely small.  Indeed, in the limit of $n\gg1$ with $n_A$ fixed (and assuming $n_A\leq k$ for simplicity) one has
\begin{equation}
    \Bbb E_{{\rm Haar}, U(1)}\left[\text{IPR}_Z[A]\right]-\Bbb E_{U(1),\, T=\infty}\left[\text{IPR}_Z[A]\right] \approx \frac{1}{d}\,,
\end{equation}
and thus detecting any difference between the two ensembles requires exponential precision on the size of the system. 

\section{The 2d Heisenberg Floquet system}
\label{sec:The 2d Heisenberg Floquet system}

In this section, we introduce the Heisenberg Floquet and describe its dynamics in different regimes of parameter space by exact simulation at small system sizes. We consider a set of diagnostics of ergodicity to argue that at any fixed system size, there is a range of the Heisenberg coupling  $J$ above which the system is ergodic and below which it is subergodic (in the sense of Hilbert space ergodicity~\cite{PhysRevLett.131.250401,PhysRevX.14.041059,PhysRevX.14.041051}). The various metric are complementary in that they access different spatial scales and time-regimes of the system and thus together provide a more complete picture of the behavior of the system.

The section is organized as follows. In \cref{sec:The Heisenberg Floquet operator} we define the model. In \cref{sec:Distribution of level spacings} we study the level spacing statistics of the Floquet operator. This reveals the behavior of the system at late times (the Heisenberg time), showing a clear crossover transition as a function of $J$. Although a powerful metric of (non)ergodicity, this requires diagonalizing an exponentially large matrix and is thus limited to small system sizes. In \cref{sec:Quantum ergodicity and the Porter-Thomas distribution} we show that deep in the ergodic regime, the system becomes ergodic much faster than the Heisenberg time. In fact, we observe that for a system of size $n=L\times L$, time $t=O(L)$ is sufficient to see convergence to RMT statistics to a large degree. This is revealing because it indicates that studying the transition to ergodicity is accessible in circuits depths that are available in current QPUs. Finally, in \cref{sec:Marginal collision entropy} we study the behavior of the  observable that is the main focus of this work, the (computational basis) marginal collision entropy $S_{2,Z}[A]$. As discussed there, these are meaningful diagnostics of global ergodicity that are sample-accessible even for large total system sizes that are outside of the reach of exact or approximate classical simulation. 

\subsection{The Heisenberg Floquet operator}
\label{sec:The Heisenberg Floquet operator}

Although we focus on Floquet evolution we motivate the choice of model by a well  studied Hamiltonian system.  The standard model for studying  ergodic vs localizing dynamics is  evolution by the Heisenberg Hamiltonian with local disorder, $H= -J\sum_{\langle i,j \rangle} \left[X_{i}X_j+Y_{i}Y_j+  Z_{i}Z_j\right] -\sum_{i} h_i Z_i$, where the sum in the first term is over the edges of some graph and the local magnetic field is chosen uniformly at random $h_i\in [-W,W]$. In a 1d chain, it has been argued that the model exhibits an ergodic/MBL transition \cite{PhysRevB.75.155111,pal2010many,PhysRevB.91.081103}. \footnote{In a 1d chain, it has been argued that the model exhibits an ergodic/MBL transition at $J/W\approx 0.067$ \cite{PhysRevB.91.081103, Luitz_2017}.To see this we set    $\Delta=1$ and $J=1/4$ in the model discussed in \cite{PhysRevB.91.081103}, where the transition is reported at $W_c\approx 3.7$. Rescaling the Hamiltonian so that the range of disorder is fixed to $h_i\in [-1,1]$ we equivalently state the transition at a value of $J/W\approx 1/(4\times 3.7)=0.067$. } Motivated by the standard Hamiltonian model, we consider a Floquet version of the  system, which we refer to as  the ``Heisenberg Floquet model.'' 

\begin{figure}[]
    \centering
    \includegraphics[width=\textwidth]{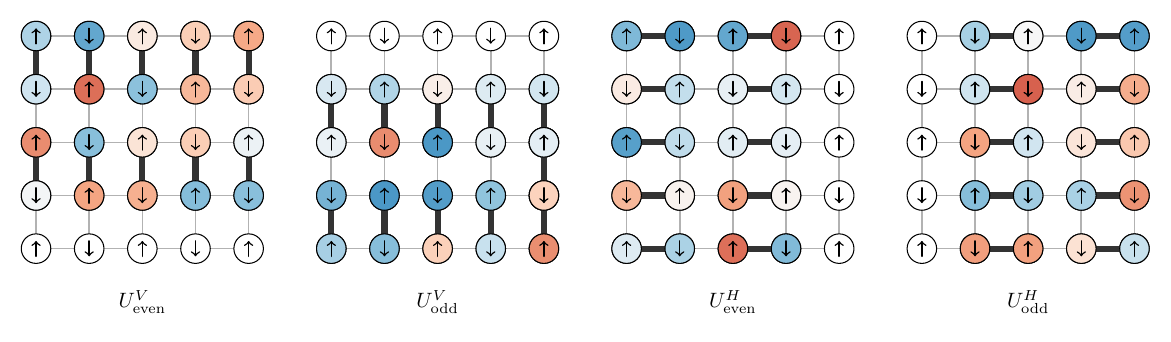}
    \caption{\textbf{Sequence of gate operations implementing one Floquet cycle.} The two-qubit gate \cref{GateuijSM} is sampled freshly in each application.  Other orderings could be chosen but we do not expect the physics to depend on this choice. }
    \label{fig:layers_choices}
\end{figure}

The basic interaction between spins $i$ and $j$ is given by
\begin{equation}\label{GateuijSM}
    U_{i,j}= e^{i J (X_i X_j+Y_iY_j +Z_iZ_j)}e^{i(h_i Z_i +h_j Z_j)}\,,
\end{equation}
where $J\in [0,\pi/4]$  is a fixed coupling constant and $h_i,h_j$ are local disorder fields acting on qubits $i$ and $j$, each drawn uniformly at random from  $[-\pi/2,\pi/2]$. Note the endpoints $J=0$ and $J=\frac{\pi}{4}$ have trivial dynamics, with the latter corresponding to the iSWAP 2-qubit gate~\cite{PhysRevLett.123.210601,PhysRevLett.126.100603,Suzuki2022computationalpower}. The gates $U
_{i,j}$ are applied in a brickwork pattern on a two-dimensional lattice to build the Floquet operator, which we take to be\footnote{It it possible to make different choices on how the horizontal and vertical terms are defined and the order in which they are applied but we do not expect this to affect the physics of the system.}
\begin{equation}\label{Floquet_layers_SM}
    U_F= U_{\rm odd}^H \, U_{\rm even}^H \, U_{\rm odd}^V \,  U_{\rm even}^V\,,
\end{equation}
where the terms are applied from right to left and each   $ U_{ \rm even, \, odd}^{H,V}$ corresponds to acting on the even/odd horizontal/vertical edges, e.g.,  
\begin{equation}
    U_{\rm even}^V= \prod_{\langle i,j \rangle \in\, \text{even vert}} U_{i,j}\,,
\end{equation}
where the product is over all vertical even bonds, and similarly for the other terms (see \cref{fig:layers_choices}). It is important that for each occurrence of $U_{i,j}$ in this expression, the gate is drawn fresh from the ensemble, i.e., the disorder is freshly resampled for every gate application, while $J$ is held fixed.  Finally, once the Floquet operator is constructed, we apply the same $U_F$ an $n_F$ number of cycles and the full time evolution operator is given by $U=(U_F)^{n_F}$.

\subsection{Level-spacing statistics}
\label{sec:Distribution of level spacings}

To obtain an initial diagnostic of whether the system exhibits ergodic or non-ergodic behavior, we examine its level-spacing statistics. 
Such spectral diagnostics are standard tools in the study of quantum chaos~\cite{wigner1951statistical,wigner1993characteristic,dyson1962statistical,mehta2004random,bohigas1984characterization,PhysRevB.75.155111,Morningstar_2022}. 
RMT predicts universal spectral correlations for complex quantum systems whose classical counterparts are chaotic. 
Different random-matrix ensembles are characterized by distinct spectral statistics~\cite{mehta2004random}. In the present Floquet setting, the relevant universality class is that of random unitary matrices, described by the Circular Unitary Ensemble (CUE), which in the quantum information community usually goes by the name of Haar-random ensemble. 

We briefly review this diagnostic, following  \cite{PhysRevB.75.155111,atas2013distribution}. 
Consider a unitary matrix $U$ with eigenvalues $\text{spec}(U)=\{e^{i\theta_n}\}$, 
where $\theta_n \in [0,2\pi)$ are ordered increasingly. 
The consecutive level spacings are defined as
\begin{equation}
  s_n = \theta_{n+1} - \theta_n \,,
\end{equation}
and the ratio of adjacent spacings, defined as \cite{PhysRevB.75.155111}:
\begin{equation}
  r_n = \frac{\min(s_n,\, s_{n-1})}{\max(s_n,\, s_{n-1})}\,.
\end{equation}
The statistics of $r_n$ provide a basis-independent diagnostic of spectral correlations and hence of quantum chaos. For an uncorrelated (Poisson) spectrum, characteristic of localized systems, the distribution of $r$ is
\begin{equation} \label{Poissonr}
  P_{\rm Poisson}(r) = \frac{2}{(1+r)^2}\,.
\end{equation}
Note the distribution is peaked at $r=0$, reflecting the absence of level repulsion. The mean gap ratio is
\begin{equation}
  \langle r \rangle_{\rm Poisson} 
  = \int_0^1 dr\, r\, P_{\rm Poisson}(r) 
  \approx 0.386\,.
\end{equation}
For chaotic spectra described by the CUE, one commonly uses the Wigner-surmise expression \cite{atas2013distribution},
\begin{equation}
  P_{\mathrm{CUE}}(r)
  = \frac{81\sqrt{3}}{2\pi}
    \frac{(r+r^2)^2}{(1+r+r^2)^4}\,.
\end{equation}
This distribution vanishes as $r\to 0$, reflecting the characteristic level repulsion of chaotic systems. The corresponding mean gap ratio is
\begin{equation}
  \langle r \rangle_{\mathrm{CUE}}
  = \int_0^1 dr\, r\, P_{\mathrm{CUE}}(r) \approx 0.603\,.
\end{equation}
More precise evaluations give $\langle r \rangle_{\mathrm{CUE}} \approx 0.5996$ (see, e.g.,~\cite{atas2013distribution,PhysRevA.107.032418,nishigaki2025distributions}), which is closer to the value obtained in our numerics.

\begin{figure}[]
    \centering

    \subpanel{0.48\linewidth}{a}{\includegraphics[width=\linewidth]{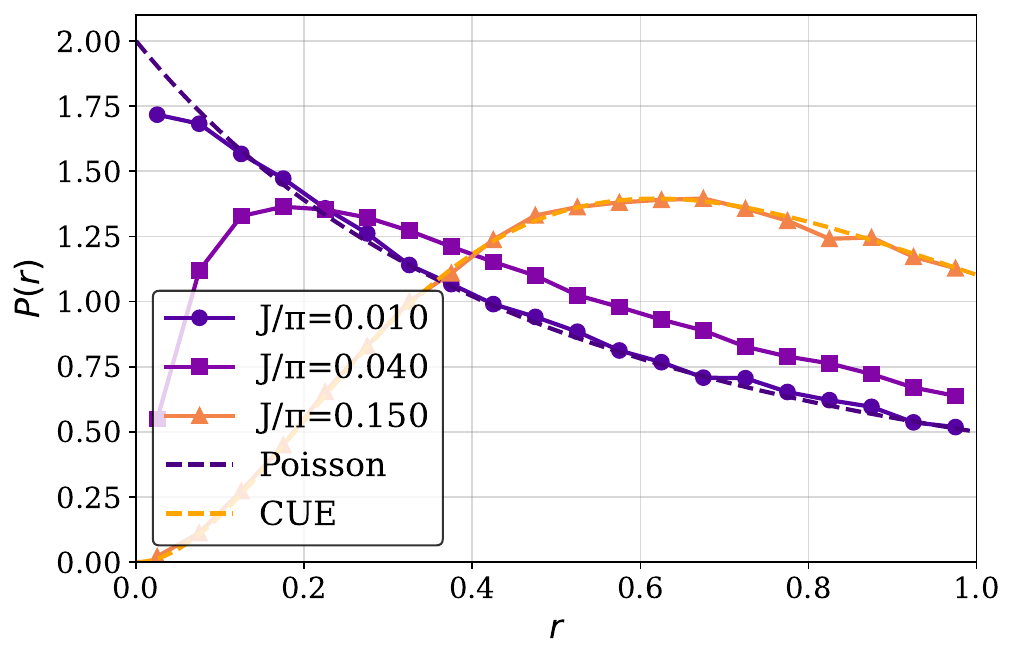}}
    \hfill
    \subpanel{0.48\linewidth}{b}{\includegraphics[width=\linewidth]{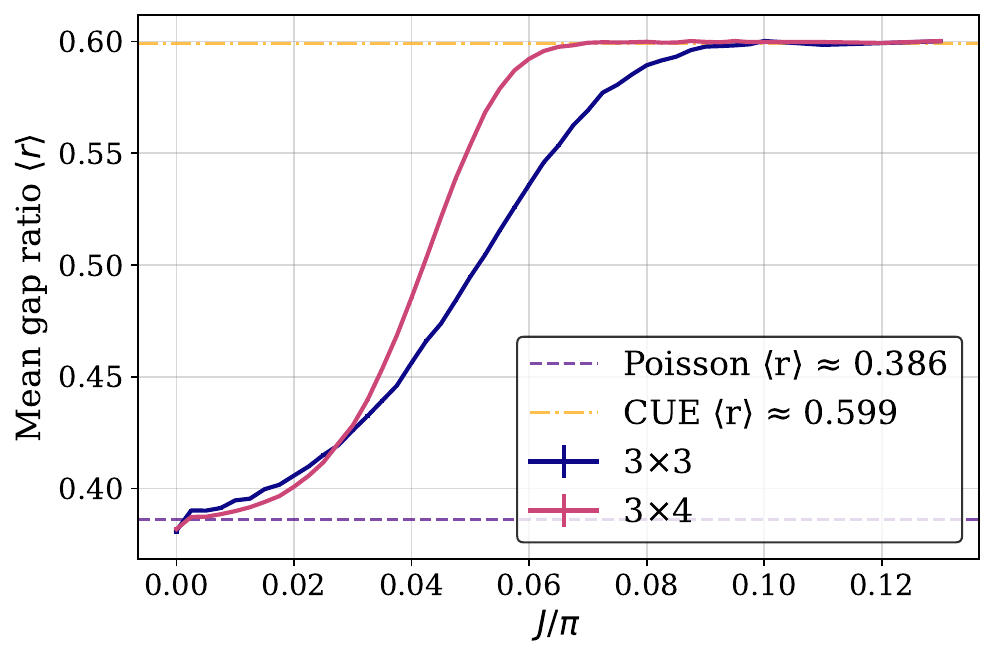}}

    \caption{
    \textbf{Gap-ratio statistics.} \textbf{a}, Probability distributions of gap ratios for the $3\times3$ system at representative values of $J$.  
    For small $J$ the distribution approaches the Poisson prediction, while for larger $J$ it converges toward the CUE form.  
    \textbf{b}, Mean gap ratio $\langle r\rangle$ as a function of the hopping parameter $J$ for several system sizes.  
    The data interpolate between the Poisson and CUE benchmarks, indicating a crossover between localized and ergodic regimes.
    }
    \label{fig:gap_statistics}
\end{figure}

To characterize the behavior of the two-dimensional Heisenberg Floquet system, we computed the exact spectrum of $U_F$ via exact diagonalization for system sizes $3\times3$ and $3\times4$, at various values of $J$, restricting to the sector of fixed maximal magnetization.\footnote{This corresponds to Hilbert space dimensions $D={9 \choose 4}=126$ for the $3\times3$ system and $D={12 \choose 6}=924$ for the $3\times4$ system.}  As shown in \cref{fig:gap_statistics}(a), for $J/\pi = 0.01$ the $3\times3$ system exhibits Poisson level statistics, indicating non-ergodic behavior. As $J$ increases, the level-spacing distribution evolves smoothly and, by $J/\pi = 0.04$, matches the CUE prediction, signaling quantum chaotic behavior in this universality class. Thus, varying $J$ drives a crossover between non-ergodic and ergodic regimes.\footnote{See \cite{regnault2016floquet} for analogous behavior in the one-dimensional model.} The $3\times4$ system displays the same qualitative crossover.

To quantify the transition more precisely, \cref{fig:gap_statistics}(b) shows the mean gap ratio $\langle r\rangle$ as a function of $J$ for both system sizes. For the $3\times3$ system, $\langle r\rangle$ approaches the CUE value for $J/\pi \gtrsim 0.10$. In the $3\times4$ system, this crossover occurs at a smaller coupling, $J/\pi \gtrsim 0.07$, indicating that the non-ergodic region contracts with increasing system size. An important question is whether this trend persists at larger $n$~\cite{suntajs2020quantum,sierant2023stability,PhysRevB.109.094207,Sierant_2025}. Addressing this via spectral diagnostics alone is computationally prohibitive, as it requires diagonalizing matrices whose dimension grows exponentially with system size. We therefore turn to alternative probes in the following sections.

We note that the level statistics discussed here diagnose the behavior of the system at the Heisenberg time scale, 
$t_H \sim 2\pi/\Delta \sim \mathcal{O}(D)$, where $\Delta$ is the mean level spacing and $D$ the Hilbert space dimension. 
Thus, for any fixed system size $n$, the non-ergodic behavior we observe is not a short-time effect: it reflects the asymptotic dynamical structure of the Floquet operator. 
In this sense, even on time scales sufficient to resolve the full spectrum, the system fails to exhibit ergodic level statistics. 
A distinct and physically important question concerns the fate of this behavior in the thermodynamic limit, where $n \to \infty$ and $t \to \infty$ are taken together. 
While this is the conventional framework for discussing thermalization, our focus here is instead on Hilbert-space ergodicity at fixed system size—namely, whether the dynamics approximate a $k$-design~\cite{brandao2016local,PhysRevLett.131.250401,PhysRevX.14.041059,PhysRevX.14.041051}. 
In this setting, it is meaningful to examine how the approach to (or deviation from) ergodicity develops as a function of time for a given finite $n$.

\subsection{Fast convergence to Porter--Thomas}
\label{sec:Quantum ergodicity and the Porter-Thomas distribution}

\begin{figure}[] 
    \centering
    \includegraphics[width=0.48\linewidth]{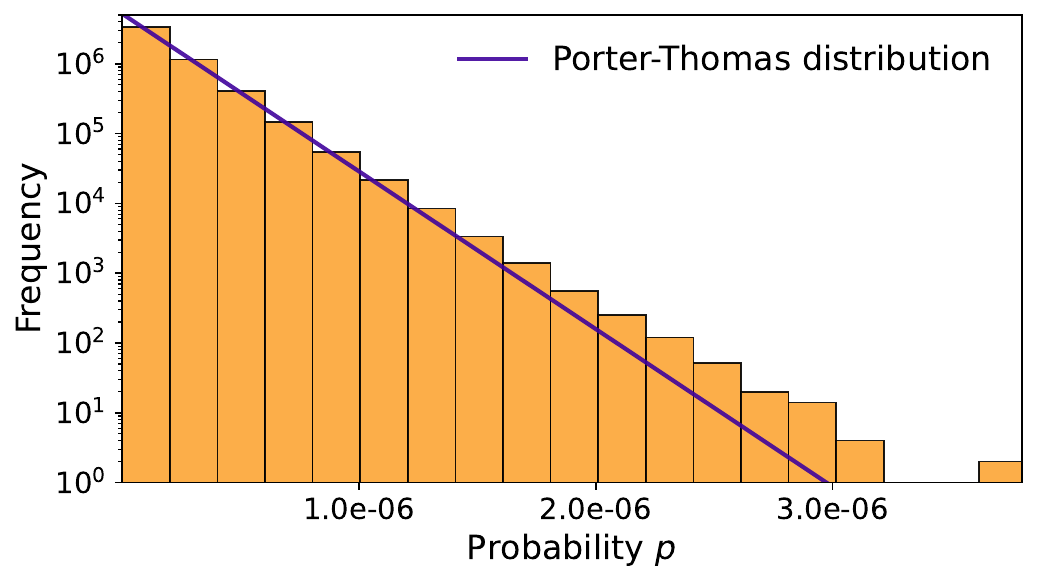}
    \hfill\includegraphics[width=0.48\linewidth]{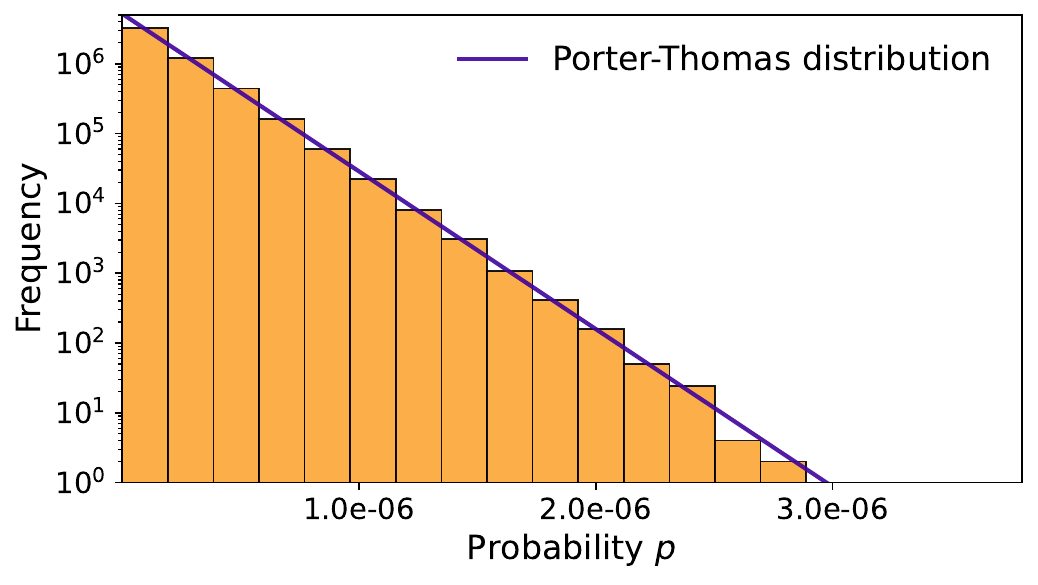}
    \caption{
    \textbf{Porter--Thomas distribution of output probabilities.} Distribution of computational-basis probabilities $P(x)=|\langle x|\psi(t)\rangle|^2$ for the $5\times5$ system at $J/\pi=0.14$ after $3$ (left) and $5$ (right) Floquet cycles, obtained by exact statevector simulation.
    The histogram is for a single disorder realization.
    After $5$ cycles the distribution is virtually indistinguishable from the Porter--Thomas form with $D=\binom{25}{12}$, indicating that in this parameter regime the Floquet dynamics produces states that are locally indistinguishable from Haar-random states within the accessible Hilbert space. Importantly, the distribution is already very close to Porter--Thomas by $t\sim \lceil L/2\rceil=3$.
    }
    \label{fig:PT}
\end{figure}

For states drawn from the Haar measure, the computational-basis probabilities 
$P(x)=|\langle x|\psi\rangle|^2$ are distributed according to the 
Porter--Thomas law~\cite{porter1956fluctuations,boixo2018characterizing,PhysRevLett.131.110601,PhysRevLett.84.3690,popescu2006entanglement}.
Precisely, if $|\psi\rangle$ is drawn uniformly from the unit sphere in a $D$-dimensional Hilbert space, then each probability $p=P(x)$ is distributed according to
\begin{equation}
    P_{\mathrm{PT}}(p)=(D-1)(1-p)^{D-2}\approx D e^{-Dp}\,,
\end{equation}
where the exponential form holds for $D\gg1$. 
This distribution provides a stringent diagnostic of quantum ergodicity. 

We studied the Floquet circuit using exact statevector simulation for system sizes up to $5\times5$ and for various values of the hopping strength $J$.  
As shown in \cref{fig:PT}, for $J=0.14$ the empirical distribution of probabilities in the $5\times5$ system after $5$ Floquet cycles is already virtually indistinguishable from the Porter--Thomas prediction with $D={25\choose 12}$.  
Remarkably, this agreement is observed for a single disorder realization, indicating that self-averaging over basis states suffices in the ergodic regime.  
This behavior provides strong evidence that, in this parameter range, the Floquet dynamics scrambles the wavefunction on time scales proportional to the linear system size and generates states that are effectively Haar-random within the symmetry sector.  

\begin{figure}[] 
    \centering
    \includegraphics[width=\linewidth]{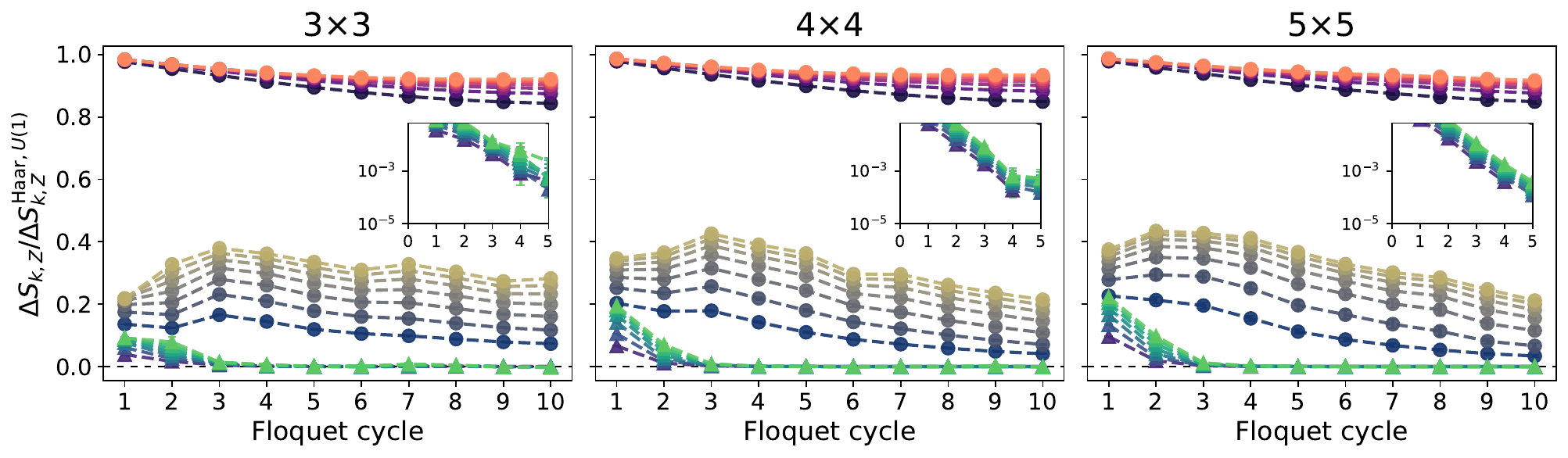}
    \caption{
    \textbf{Marginal R\'enyi-$k$ entropies and $k$-designs.} Convergence of the marginal R\'enyi-$k$ entropies toward their Haar-random values (with $U(1)$ symmetry), quantified by the normalized deviation
    $
    {\Delta S_k}/{S_k^{\mathrm{Haar},U(1)}} 
    = {(S_k^{\mathrm{Haar},U(1)} - S_k)}/{S_k^{\mathrm{Haar},U(1)}}.
    $
    This provides a probe of the convergence of the distribution moments and, equivalently, of whether the dynamics approximate a $k$-design. 
    Results are obtained via exact statevector simulation for system sizes $L \times L$ with $L = 3, 4, 5$, using $4096$, $512$, and $32$ disorder realizations, respectively. 
    The lower, middle, and upper groups of curves correspond to $J/\pi = 0.01$, $0.04$, and $0.07$. 
    Within each group, curves are shown for R\'enyi indices $k = 2, 3, \ldots, 8$. 
    For  $J$ deep in the ergodic regime, the system approaches Haar-typical behavior in  $n_F = \lceil L/2 \rceil$ number of cycles (see insets).
}\label{fig:moments_convergence}
\end{figure}

The approach to Haar-typical behavior can be quantified by tracking the R\'enyi entropies of the probability distribution as a function of the number of Floquet cycles. In~\cref{fig:moments_convergence}, we show their evolution for various system sizes, providing a direct measure of how closely the dynamics approximate a $k$-design. Deep in the ergodic regime (for sufficiently large $J$), the entropies rapidly converge to their Haar values: already by $n_F \sim \lceil L/2\rceil$, both the second moment and higher moments (shown up to the seventh) are very close to their Haar-random values. It is reasonable to expect that this fast convergence persists at larger system sizes, although direct verification becomes increasingly difficult due to the rapidly growing cost of classical simulation. Motivated by these observations, we perform experiments on the quantum device for larger systems of size $8\times8$, $9\times9$, and $10\times10$. Deep in the ergodic regime, we expect the states to be close to fully entangled already after $n_F = 4$, $ n_F=4$, and $ n_F=5$ Floquet cycles, respectively. On the theoretical front, it would be interesting to determine whether this rapid convergence scales linearly, as a sublinear power law, or logarithmically with system size, which we do not address here.

Finally, we comment briefly on the implications of these results for sampling complexity and for the time complexity of classical simulation. At small $J$, the output distribution remains structured and comparatively tractable to sample from classically. As $J$ increases, the spectrum crosses over to Haar-typical statistics, for which sampling is widely believed to be classically intractable and which underlies quantum supremacy claims in random circuit sampling experiments~\cite{GoogleRCS}. In this sense, the physical transition from non-ergodic to ergodic behavior is accompanied by a corresponding change in sampling complexity. As discussed in more detail in \cref{sec:Classical algorithms}, we also observe a sharp increase in the cost of classical tensor-network simulations across the same parameter regime. These trends reflect the same underlying evolution of the quantum state toward higher entanglement and Haar-typical structure.

\subsection{Marginal collision entropy}
\label{sec:Marginal collision entropy}

\begin{figure}[h!] 
    \centering
    \includegraphics[width=\linewidth]{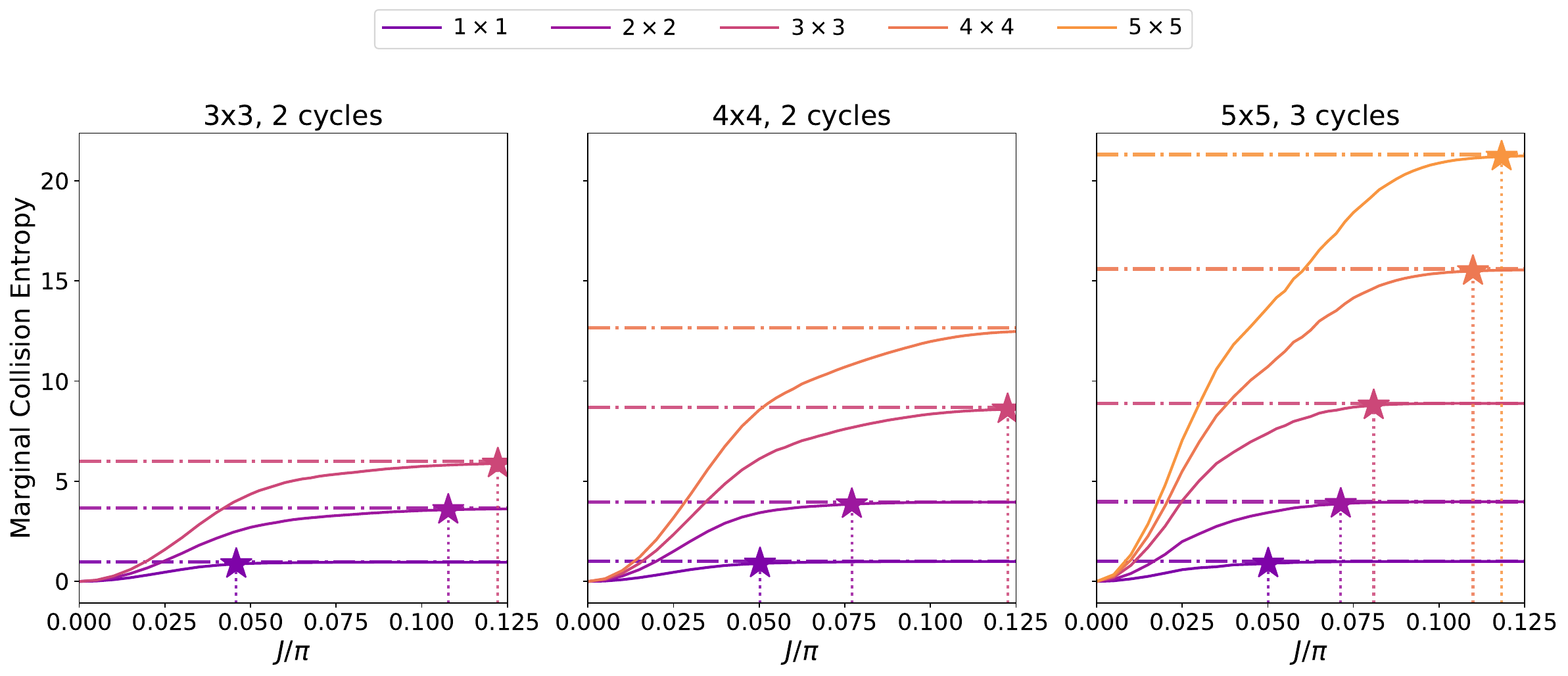}

    \caption{
     \textbf{Marginal collision entropies from exact simulations.} Marginal collision entropies $S_2$ for \emph{central} connected marginals in $3\times3$, $4\times4$, and $5\times5$ systems, computed via exact statevector simulation, averaged over 4096 disorder instances for $3\times3$ and $4\times4$, and over 256 instances for $5\times5$.
    The stars ($\bigstar$) indicate the coupling $J^\star$ at which a given marginal reaches a plateau, defined as the point where its value lies within $0.1$ of the corresponding Haar value.
    For fixed $J$, the curves are ordered by marginal size: larger marginals exhibit larger $S_{2,Z}$ values (i.e., closer to the Haar value), whereas smaller marginals lag behind.
    Consequently, marginal collision entropies provide systematic lower bounds on the global collision entropy, with the bound becoming tighter as the marginal size increases.
    }
    \label{fig:marginals_ed}
\end{figure}

\begin{figure}[h!] 
    \centering

    \includegraphics[width=\linewidth]{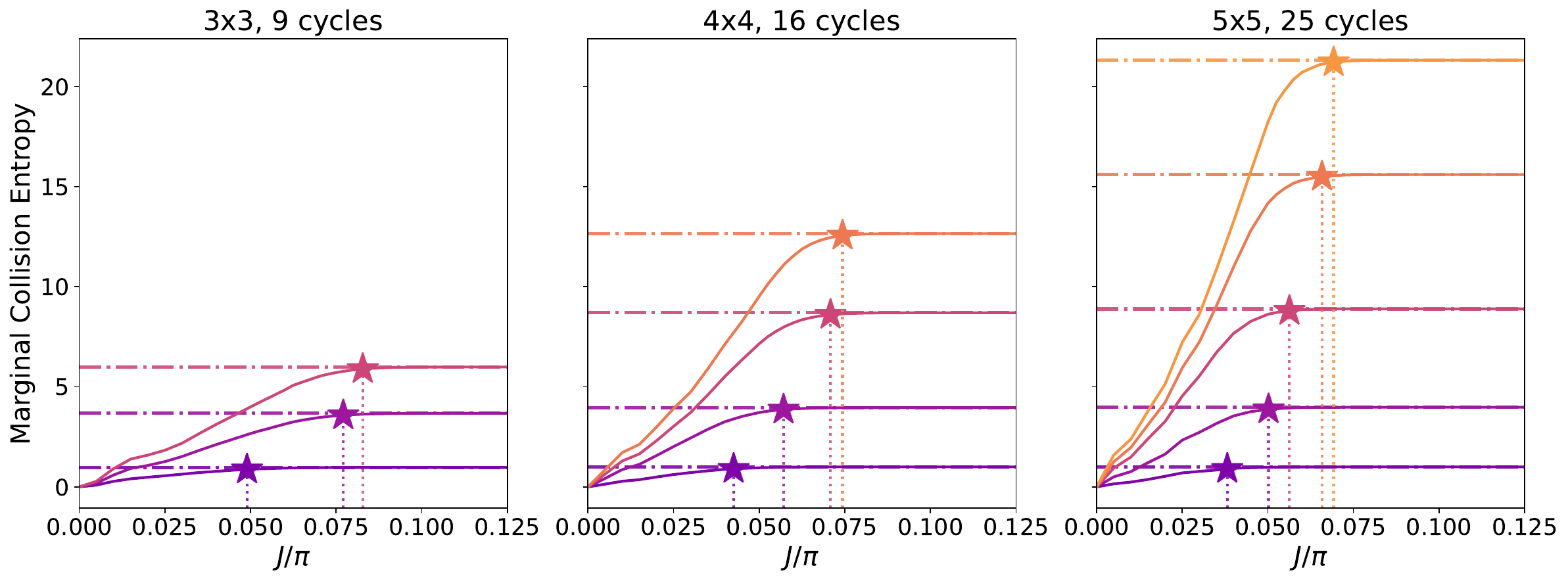}
    
    \caption{
     \textbf{Marginal collision entropies at later times.} Same quantities as in \cref{fig:marginals_ed} but shown at the later times $n_F=L^2$.
    }
    \label{fig:marginals_ed_long}
\end{figure}

\begin{figure}[h!] 
    \centering
    \includegraphics[width=0.5\linewidth]{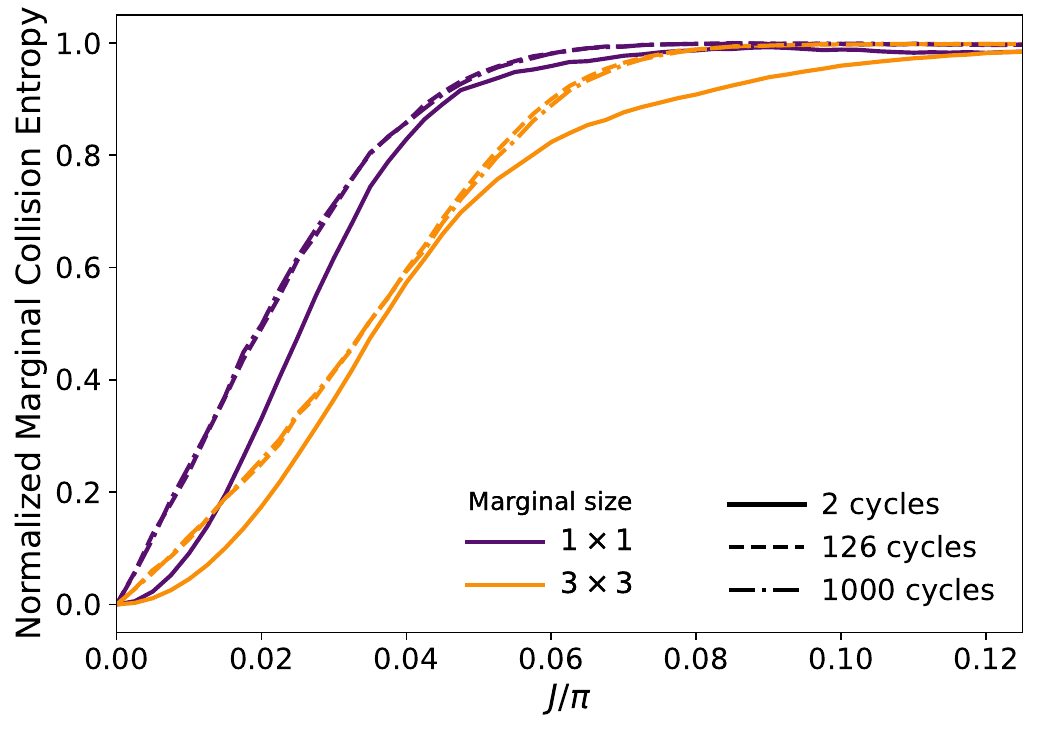}
    
    \caption{
     \textbf{Long-time limit of the collision entropy.} Long-time behavior of global and marginal collision entropy $S_{2,Z}(J)$ for the $3\times3$ system, averaged over 4096 disorder instances.
     The short-time evolution captures qualitative features of the long-time limit shape of $S_{2,Z}[A](J)$ which freezes at $\sim t_H=126$ cycles.
    }
    \label{fig:ipr_ultralong}
\end{figure}

The central object of our experimental study is the sample-accessible marginal collision entropy discussed in \cref{sec:Background}. We thus investigate its behavior in small system sizes using exact statevector simulation.  We compute the marginal collision entropies $S_{2,Z}[A]$ as functions of $J$ for systems of sizes $3\times3$, $4\times4$, and $5\times5$. We do so at two representative times, $n_F=\left\lceil \frac{L}{2}\right\rceil$ (in light of the discussion above) and $n_F=L^2$, to compare with later-time behavior and to investigate the behavior of the system at the Heisenberg time we simulate extremely long times for the $3\times 3$ system. The results are shown in \cref{fig:marginals_ed}, \cref{fig:marginals_ed_long}, and \cref{fig:ipr_ultralong}.  

The main takeaways are the following. As seen in \cref{fig:marginals_ed} the collision entropy reaches its Haar-random value for marginals of all sizes, provided $J$ is sufficiently large, already at $n_F=\left\lceil \frac{L}{2}\right\rceil$ cycles.

It is important to note that since the transition towards ergodicity seems smooth, there is no sharp value of $J$ at which the transition happens. Nonetheless, for the purposes of comparison it is useful to define a criterion at which the system is near ergodic and a simple criterion is when the collision entropy is $\epsilon$-close to its maximum. The point at which this happens are indicated with the  star markers on the plots, where we have set $\epsilon=0.1$. 

The rapid convergence of the collision entropy in the ergodic regime is consistent with the rapid convergence to Porter–Thomas statistics discussed in \cref{sec:Quantum ergodicity and the Porter-Thomas distribution}.  From the locations of the stars, we note the value of $J^\ast[A]$ increases with the size of the marginal as expected on physical grounds; larger marginals are harder to thermalize. 

An important question is how representative these short-time dynamics are of the late-time behavior. To probe this, we extend the simulations to times $n_F = L^2$. As shown in \cref{fig:marginals_ed_long}, the collision-entropy curves shift toward smaller $J$: the system reaches ergodicity at lower coupling strengths, and all extracted values of $J^\ast$ move to the left. Although the subergodic region narrows, it remains finite, and the hierarchy between marginals of different sizes persists.

\ 

One may further ask what the is the ``infinite-time'' behavior of the system. As already discussed, from the level-spacing statistics analysis in \cref{sec:Distribution of level spacings}, we expect that a finite-sized system in fact remain localized at infinite times for sufficiently small values of $J$.
To see this explicitly in terms of the collision entropy, in \cref{fig:ipr_ultralong} we show $S_{2,Z}[A](J)$ for three different evolution times: the short-time value $t=\lceil L/2\rceil=2$, the Heisenberg time $t=t_H={9 \choose 4}=126$, and an ultralong evolution of $1000$ cycles.
Consistent with the prediction from level-statistics, the subergodic region remains and the system fails to thermalize even at infinite times for small enough $J$.

\subsection{Quantum R\'enyi entropy and volume-law entanglement}
\label{sec:Quantum Renyi entropy}

We now study the behavior of the R\'enyi-2 entropy for various subsystems. This quantity probes the entanglement structure of the system and provides a basis-independent diagnostic of ergodicity. We compute the quantum R\'enyi entropy for subsystems of the $3\times 3$, $4\times 4$, and $5\times 5$ lattices. As we discuss, the quantum R\'enyi-2 entropy grows rapidly in the system, reaching maximum bipartite entanglement in as timescales accessible to the experiment. Indeed, as we show below the quantum R\'enyi entropy has a similar behavior to that of the collision entropy and thus the former is a proxy for the latter. 

\begin{figure}[h!] 
    \centering

    \includegraphics[width=\linewidth]{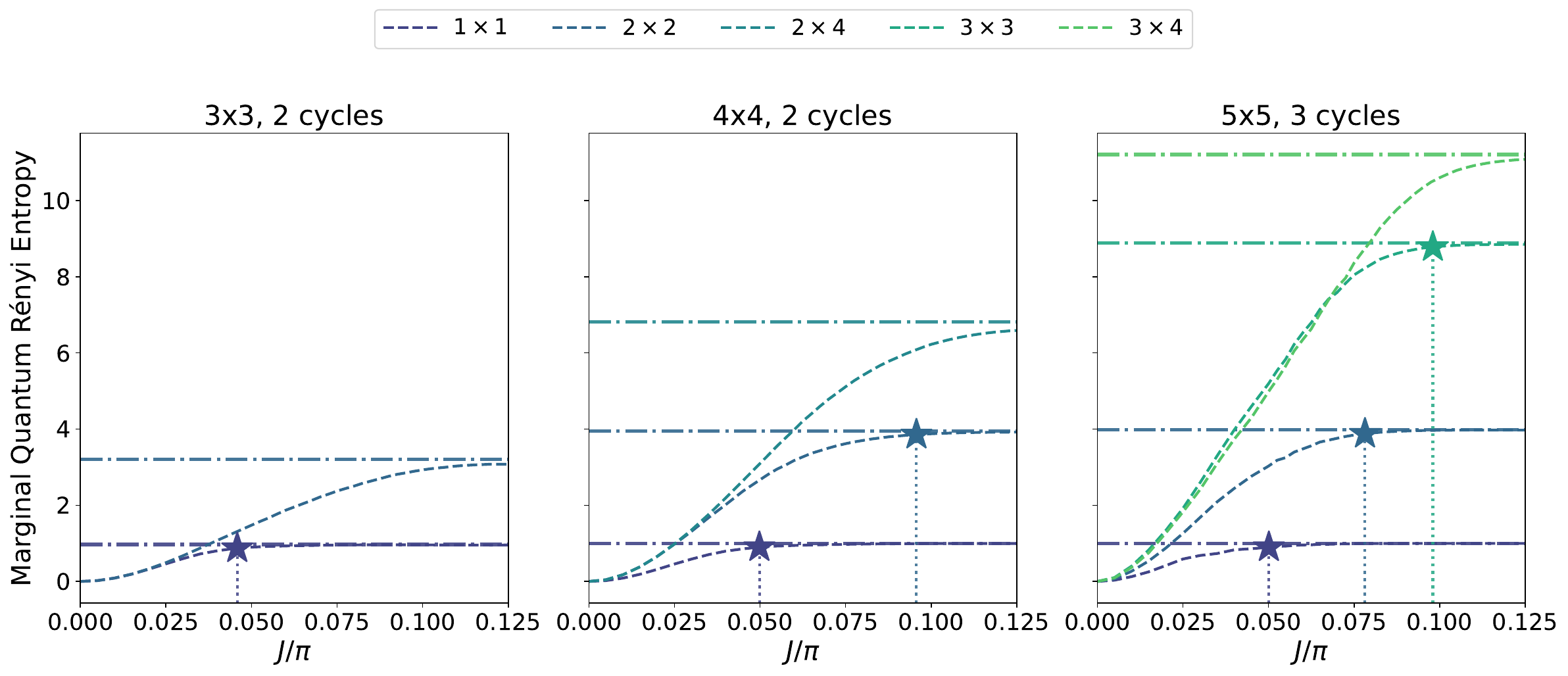}

    \includegraphics[width=\linewidth]{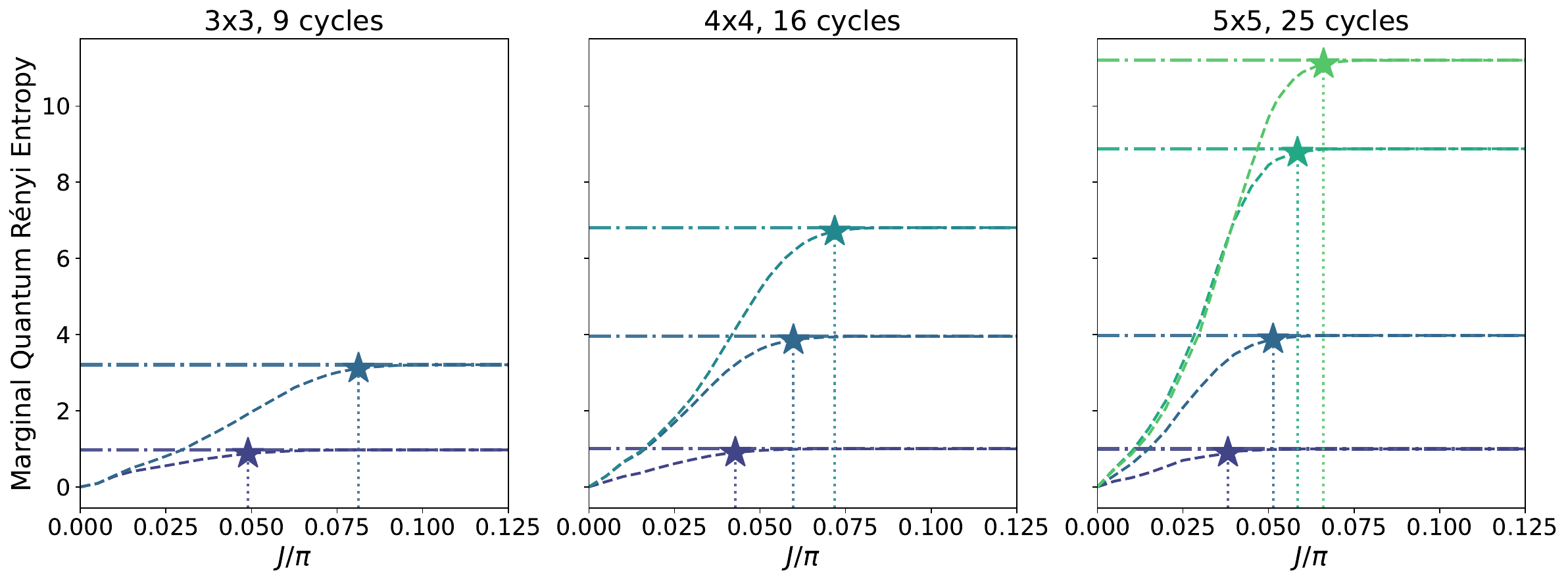}
    
    \caption{\textbf{Marginal R\'enyi-2 entropies from exact simulations.} Quantum R\'enyi-2 entropy 
for \emph{central} connected marginals in $3\times3$, $4\times4$, and $5\times5$ systems, computed via exact statevector simulation, averaged over 4096 disorder instances for $3\times3$ and $4\times4$, and over 256 instances for $5\times5$.
The stars ($\bigstar$) indicate the coupling $J^\star_P$ at which a given subsystem reaches a plateau, defined as the point where its value lies within $0.1$ of the corresponding Haar value.
For fixed $J$, the curves are ordered by marginal size: larger marginals exhibit larger $S_2$ values (i.e., closer to the Haar-random value), whereas smaller marginals lag behind.
Consequently, marginal collision entropies provide systematic lower bounds on the global collision entropy, with the bound becoming tighter as the marginal size increases.
    }
    \label{fig:marginals_ed_purity}
\end{figure}

The results are shown in \cref{fig:marginals_ed_purity}. As seen there, for sufficiently large $J$ the system attains the fully entangled Haar-random value, subject to $U(1)$ symmetry,\footnote{This value is computed in \cref{sec:Random matrix theory and ergodicity}.} on time scales as short as $n_F=\lceil L/2 \rceil$, consistent with the rapid approach to Porter–Thomas statistics discussed in \cref{sec:Quantum ergodicity and the Porter-Thomas distribution}. We denote the corresponding value by $J^\ast_P$, to distinguish it from that obtained from the collision entropy; in general $J^\ast_P > J^\ast$, as expected on physical grounds. At longer times ($n_F=L^2$), entanglement is further enhanced and full entanglement occurs at smaller values of $J$.
Recall that the collision entropy  provides an upper bound on the quantum R\'enyi entropy, $S_{2,Z}[A]\geq S_{2}[A]$. Thus, a low value of the collision entropy places an upper bound on the amount of entanglement in the system and therefore constitutes an obstruction to thermalization. We note that at values of $J$ where the collision entropy reaches its maximum, the quantum R\'enyi entropy has not yet saturated, consistent with the inequality above and the physical expectation that a stronger interactions (or longer times) are needed to scramble in all bases. 

On the other hand, we note that although the collision entropy in general only provides an upper bound on the quantum R\'enyi entropy, in the system at hand it actually provides a good proxy for it. Indeed, comparing \cref{fig:marginals_ed} to \cref{fig:marginals_ed_purity} one sees that these two quantities track each other.  It is convenient to parametrize 
\begin{equation}
    S_{2}[A]= s_A(J)\,  S_{2}^{\text{Haar}, U(1)}[A]\,,
\end{equation}
with $0\leq s_A(J)\leq 1$. Since $S_{2}^{\text{Haar}, U(1)}[A]\approx n_A$, for $n_A\leq n/2$, this represents a volume-law entanglement. This sets a fundamental obstruction to tensor networks with a low bond dimension to be a faithful representation of the quantum state of the system, even at times as short as $n_F=L/2$. See \cref{sec:Classical algorithms} for tensor network techniques used in this work and a discussion of their limitations. 

\section{Experimental details}
\label{sec:Experimental details}

In this section, we discuss specifications of IBM Quantum's Nighthawk-family QPUs and provide details on how we executed our circuits on them. 

\subsection{The quantum processor}

\begin{figure}[h]
    \centering
    \includegraphics[width=0.6\linewidth]{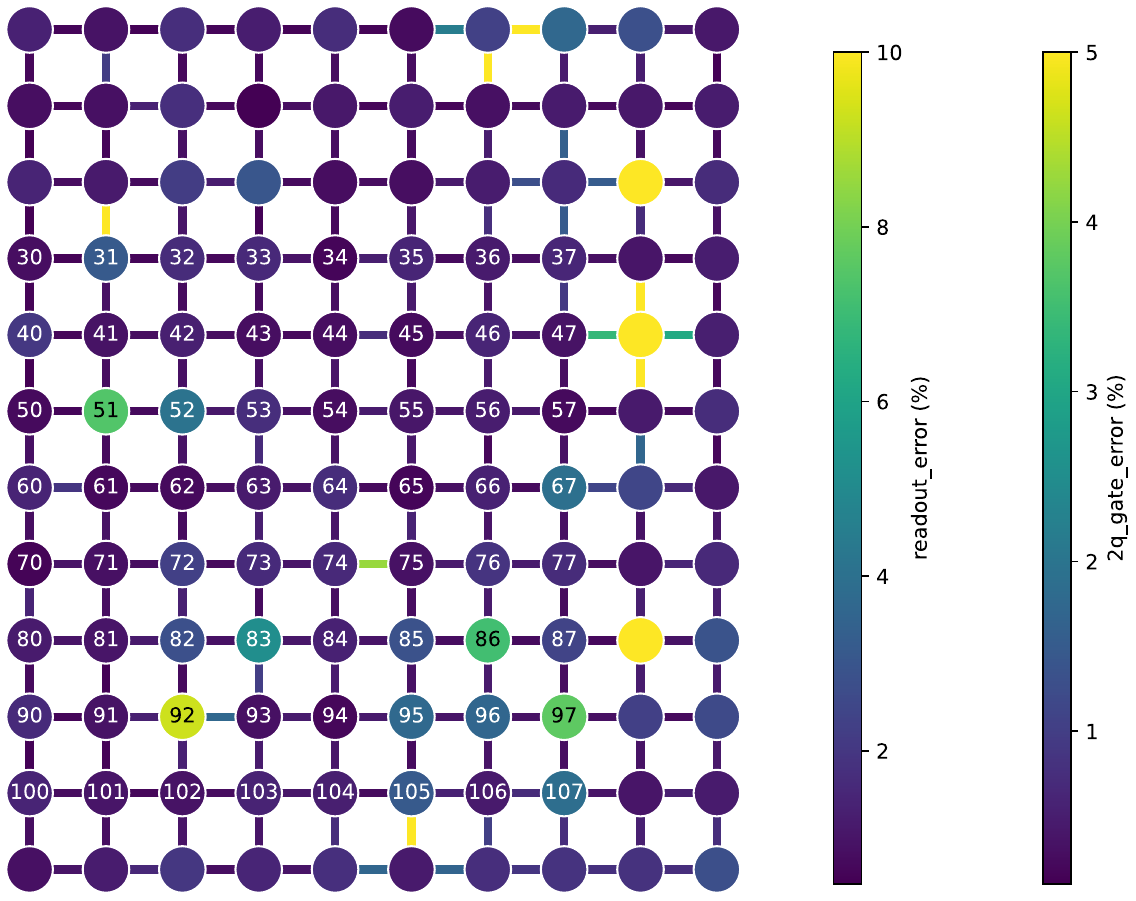}
    \caption{\textbf{Topology of IBM Quantum's \texttt{ibm\_miami} QPU.} The nodes of the graph correspond to qubits, while its edges represent two-qubit gate couplings between qubits. They are colored according to readout and $\mathrm{CZ}$ gate error rates, capped at $10\%$ and $5\%$, respectively. Labeled qubits correspond to the ones used in the experiment. As can be seen from the saturation of the colormaps, few qubits and/or couplers have large error rates. However, the qubit layout selected by Qiskit SKD's transpiler for the run does not include any of them. The figure uses calibration data taken by IBM Quantum on 6th February 2026.}
    \label{sm:fig:ibm_miami}
\end{figure}

\begin{figure}
    \centering
    \includegraphics[height = 2.7in]{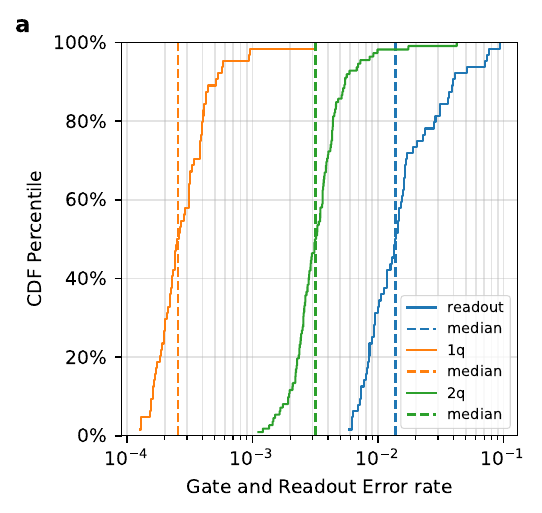}
    \hspace{0.5cm}
    \includegraphics[height = 2.7in]{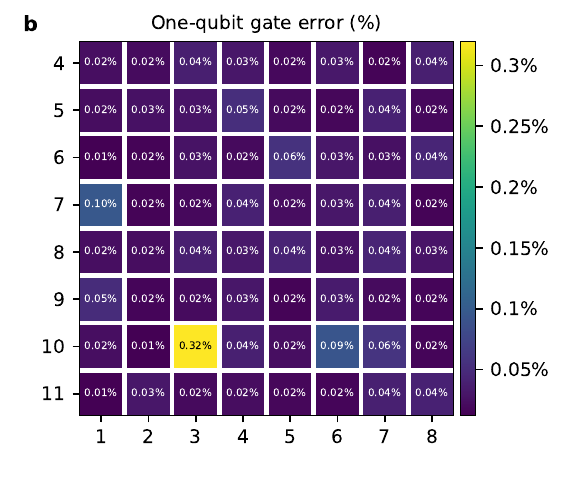} \\

    \includegraphics[height = 2.7in]{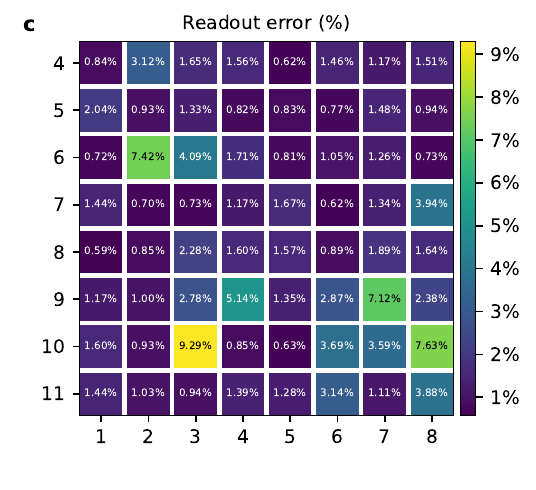} 
    \hspace{0.5cm}
    \includegraphics[height = 2.7in]{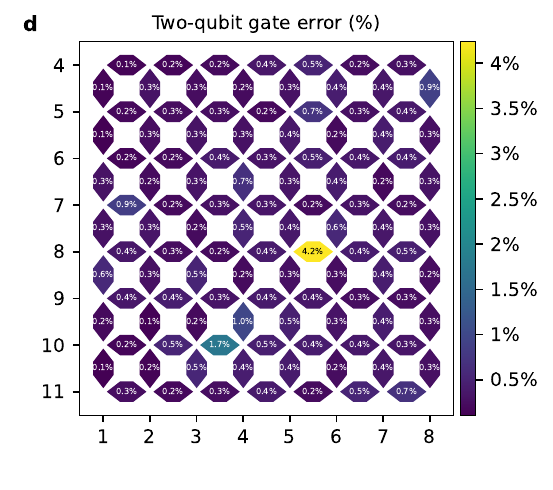} 
    
    \caption{\textbf{Device calibration data.} Calibration data for the qubits of \texttt{ibm\_miami} used in an $8\times8$ experiment on 6th February 2026. \textbf{a}, Cumulative distribution functions (solid lines) and median values (dashed lines) for readout (blue), one-qubit gate (yellow), and two-qubit gate (green) error rates. \textbf{b}, Heatmap of one-qubit ($X$ and $\sqrt{X}$) gate error rates. \textbf{c}, Heatmap of readout errors. \textbf{d}, Heatmap of two-qubit (CZ) error rates. In all panels, calibration data are shown only for qubits used in the $8\times8$ experiment. Data was obtained from backend calibration provided by IBM Quantum at the time of the experiment.}
    \label{sm:fig:app:calibration}
\end{figure}

All our experimental results have been obtained by executing circuits on \texttt{ibm\_miami} and \texttt{ibm\_berlin}, the first Nighthawk-family QPUs released by IBM Quantum, between January 2026 and July 2026. Each device has 120 superconducting qubits arranged on a $12\times 10$ square lattice (see \cref{sm:fig:ibm_miami}). 
The native single-qubit gates are $X$, $\sqrt{X}$, and (virtual) $R_Z$, while the native two-qubit gate is $
\mathrm{CZ}$. An example of \texttt{ibm\_miami}'s error rates during an $8\times 8$ experiment on 6th February 2026 is reported in \cref{sm:table:hardware_specs} while qubit-dependent calibration data are shown in \cref{sm:fig:app:calibration}. 

The results presented in \cref{sec:Quantum simulation results} were obtained from experiments that took a total QPU execution time of 22.1 hours. This excludes additional QPU usage for benchmarking device capabilities and for planning our experimental runs. We found the repetition rate of both \texttt{ibm\_miami} and \texttt{ibm\_berlin} to be approximately 4 ms. 

\begin{table}
    \centering
    \begin{tabular}{lcccc}
    \toprule
         &  Mean & Median & Min & Max\\
    \midrule
    Single-qubit gate error & 3.5E-04 & 2.6E-04 & 1.3E-04 & 3.2E-03\\
    Two-qubit gate error    & 3.9E-03 & 3.2E-03 & 1.1E-03 & 4.2E-02\\
    Readout error           & 2.0E-02 & 1.8E-02 & 5.9E-03 & 9.2E-02\\
    T1 ($\mu$s)             & 346.0 & 332.8 & 92.0 & 605.9\\
    T2 ($\mu$s)             & 249.7 & 245.8 & 105.0 & 451.5\\
    \bottomrule
    \end{tabular}
    \caption{Aggregate specifications for single-qubit, two-qubit, and readout error rates, and relaxation and dephasing times ($T_1$ and $T_2$) for \texttt{ibm\_miami} used in an $8\times8$ experiment on 6th of February, 2026.}
    \label{sm:table:hardware_specs}
\end{table}

\subsection{Compilation to native gates}
\begin{figure}
    \centering
    \includegraphics[width=\linewidth]{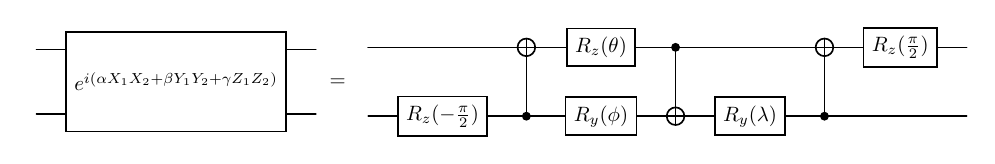}
    \caption{\textbf{Decomposition of the general Heisenberg gate $e^{i(\alpha X_1X_2+\beta Y_1Y_2+\gamma Z_1Z_2)}$ into CNOT gates.} Here, $\theta = \pi/2-2\alpha$, $\phi = 2\beta - \pi/2$, and $\lambda=\pi/2-2\gamma$. For the $XXX$ Heisenberg gate of \cref{GateuijSM}, one has $\alpha = \beta = \gamma = J $ and, therefore, $\theta = \lambda = \pi/2 - 2J$ and $\phi = 2J - \pi/2$. Each CNOT can be decomposed into CZ gates, which are the native two-qubit gates of IBM Nighthawk QPUs, via the identity $\mathrm{CNOT} = (I_1\otimes H_2)\, \mathrm{CZ}_{1,2}\,(I_1\otimes H_2)$, with $H$ the Hadamard gate.}
    \label{sm:fig:CNOT_decomp}
\end{figure}

To execute the Floquet circuit of \cref{Gateuijmain}  on the device, we must decompose the gates $U_{i,j}$ of \cref{Gateuijmain} into Nighthawk family's native gates. The evolution under local fields $e^{i h_i Z_i}$ can be implemented natively via the (virtual) $R_Z(\theta) = e^{-i\frac{\theta}{2}Z}$ gate, while the Heisenberg interaction $e^{iJ(X_1X_2+Y_1Y_2+Z_1Z_2)}$ must be decomposed in terms of CZ gates. Its standard decomposition in terms of three CNOT gates is shown in \cref{sm:fig:CNOT_decomp} (see \cite{vidal2004universal,vatan2004optimal,Smith_2019}). 
Each CNOT can be further decomposed into native CZ gates via the identity $\mathrm{CNOT}_{1,2} = (I_1\otimes H_2)\,\mathrm{CZ}_{1,2}\,(I_1\otimes H_2)$, with $H$ the Hadamard gate. Each Floquet cycle applies a number of Heisenberg gates equal to the number of edges in the lattice, given by $E=L_x(L_y-1)+L_y(L_x-1)$. Since each Heisenberg interaction is compiled into three CZ gates, the total number of CZ gates and the CZ depth of a circuit with $n_F$ Floquet cycles are given by
\begin{equation}
     \text{\# CZ}= 3 \, n_F \, E\,,\qquad\qquad  \text{CZ-depth}=12 \, n_F\,.
\end{equation}
\cref{sec:Quantum simulation results} details the gate counts and circuits depths that were necessary for our experiments. 

\subsection{Error suppression techniques and readout error calibration}
\label{sec:Error suppression techniques}
To suppress noise in our simulations, we employed an XX dynamical decoupling sequence~\cite{Viola_1998_dynamical} and Pauli twirling (PT)~\cite{Wallman_2016_noise,vandenberg2022model} on two-qubit gates and measurements as implemented in the Qiskit runtime SDK~\cite{JavadiAbhari_2024_qiskit}. 
\begin{figure}
    \centering
    \includegraphics[height=0.42\linewidth]{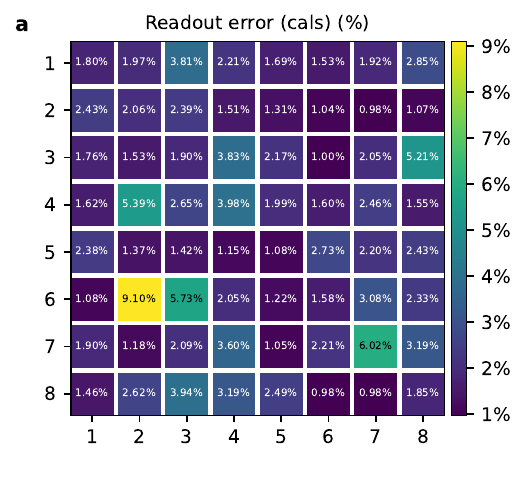}
    \hspace{0.5cm} \includegraphics[height=0.42\linewidth]{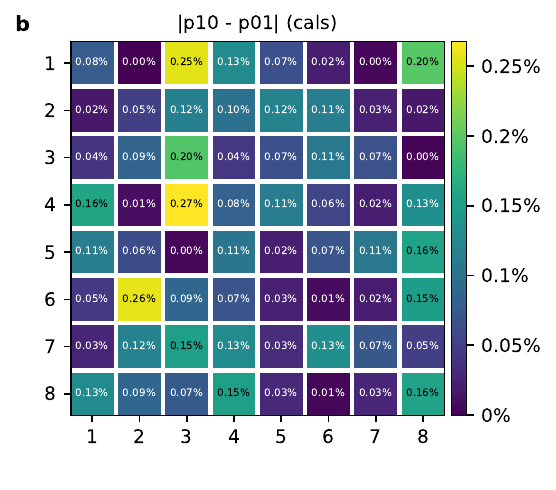}
    \caption{\textbf{Readout errors.} \textbf{a}, Average readout error $p = (p_{01} + p_{10})/2$ and \textbf{b}, difference $|p_{01} - p_{10}|$ obtained from calibration circuits for an $8\times8$ experiment on 6th February 2026.}
    \label{fig:readout_calibrations}
\end{figure}
As a consequence of the large readout error rates of the device (see 
\cref{sm:fig:app:calibration}c), we also ran readout calibration circuits during each experimental run to obtain real-time single-qubit assignment matrices $A_i$~\cite{bravyi2021mitigating},
\begin{equation}
    A_i = \begin{bmatrix}
        1-p^{(i)}_{01} & p^{(i)}_{10}\\
        p^{(i)}_{01} & 1-p^{(i)}_{10}
    \end{bmatrix}.
\end{equation}
Here, assuming uncorrelated readout errors, the readout error rates for qubit $i$,
\begin{align}
    p^{(i)}_{10} = \mathrm{Prob}(\mathrm{prepare}\  0_i\ | \ \mathrm{measure}\ 1_i)\,,\qquad 
    p^{(i)}_{01} = \mathrm{Prob}(\mathrm{prepare}\ 1_i\ | \ \mathrm{measure}\ 0_i)\,,
\end{align}
can be obtained by running two simple circuits, in which all the qubits are initialized in state $\ket{0}$ and $\ket{1}$, respectively, and then measured in the computational basis. To estimate $p^{(i)}_{01}$ and $p^{(i)}_{01}$ we used the same number of shots and readout twirling instances as in the experimental circuits. Note that, while the magnitude of readout error rates obtained from this procedure and the ones reported by IBM are comparable, there can be significant qubit-specific differences due to either readout twirling or time-dependent processes--compare \cref{sm:fig:app:calibration}c and 
\cref{fig:readout_calibrations}a. Furthermore, we have verified that readout Pauli twirling is effective in  suppressing the asymmetry between $p^{(i)}_{01}$ and $p^{(i)}_{10}$ (see Fig.~\ref{fig:readout_calibrations}b).

\section{Error mitigation}
\label{sec:Error mitigation}

\subsection{Low entanglement calibration}\label{sec:lec}

\begin{figure}[htbp]
    \centering
    \includegraphics[width=0.5\linewidth]{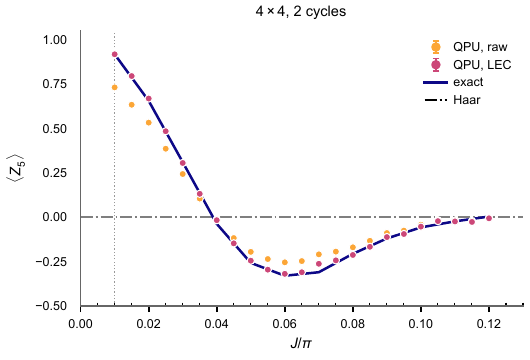}
    \caption{\textbf{Attenuation of a Z string.} $\langle Z_5\rangle$ on a central qubit of the $4\times4$ system after 2 cycles, for one disorder instance, from the QPU and from exact simulation. Noise attenuates the raw QPU values towards the Haar value $0$. LEC rescales them by a single factor, fitted at $J/\pi=0.01$ (dotted line), the smallest coupling with an exact reference.}
    \label{fig:attenuation}
\end{figure}

By comparing the experimental implementation of small system sizes against exact classical simulations, we discover that observables produced by the quantum computer are attenuated with respect to their true values. \cref{fig:attenuation} shows a $Z$ expectation value in a $4\times 4$ system. The raw measurements from the QPU capture the coherent features of the true signal, but hardware noise attenuates them: the signal above the fixed point of zero is reduced by a factor. This suggests that learning the attenuation factor induced by noise for a given experiment might allow us to recover a close approximation to the true signal. 

Note that at small $J$, the propagator of the Heisenberg interaction is close to identity. The disorder fields are single qubit operators and do not generate entanglement. As a result, the evolved state is expected to have low entanglement across every bipartition and to have an accurate and efficient representation as a classical tensor network state. For a given system, disorder instance and observable, one can learn the attenuation factor by comparing the value measured by the quantum computer at small $J$ against tensor network simulations. 

We call this rescaling of experimentally measured observables based on attenuation factors learnt from small $J$ values \textit{low entanglement calibration} or LEC. It is the primary error mitigation technique used in our work. \Cref{fig:attenuation} shows the result of applying it to a $Z$ expectation value. LEC is related to learning-based mitigation methods such as Clifford data regression~\cite{czarnik2021error} and training with fermionic linear optics~\cite{montanaro2021error}, as well as reference-circuit rescaling methods~\cite{urbanek2021mitigating,Mi2021,kim2023evidence,cai2023quantum,shtanko2025uncovering}, which learn a correction to noisy observables from circuits that can be simulated efficiently classically; in LEC these are the same circuits at small $J$.

In \cref{sec:depolarizingnoiseinversion} we motivate the multiplicative form of the calibration from a local depolarizing noise model and state the assumptions made by LEC. We then point out how these assumptions lead to a systematic bias in our results (\cref{sec:op_spread}) and describe an approach to removing the bias (\cref{sec:wplecerrormodel}). In \cref{sec:lec_variants}, we discuss a variant of LEC that can be applied directly to the IPR rather than to its constituent expectation values. 

\subsection{Inverting the action of depolarizing noise on observables}
\label{sec:depolarizingnoiseinversion}

\begin{figure}
    \centering
    \includegraphics[width=0.75\linewidth]{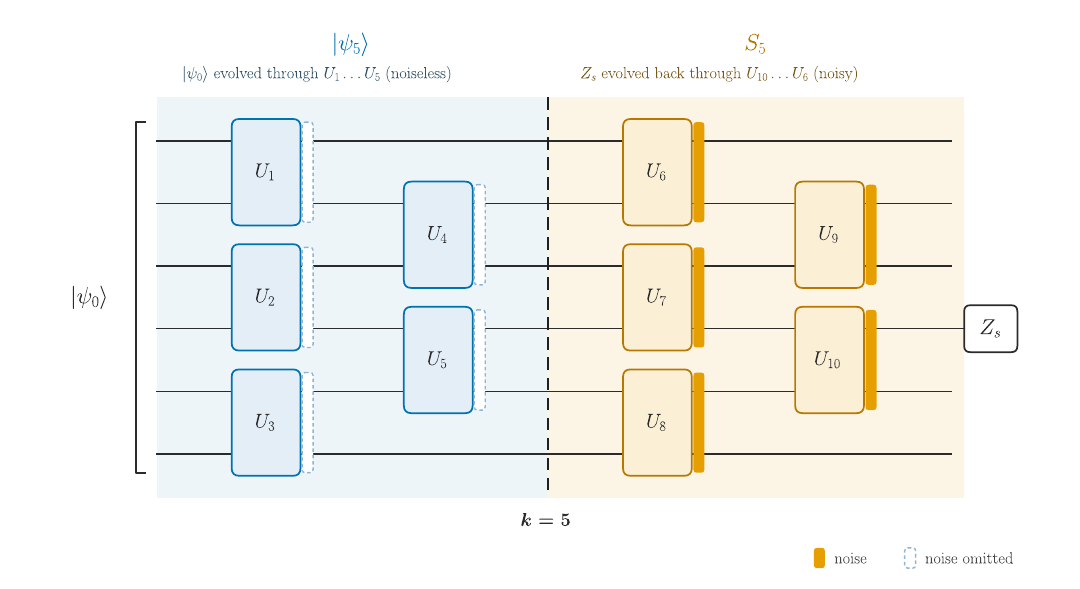}
    \caption{\textbf{Effect of noisy gates on an expectation value.} The initial state evolves
            noiselessly in the Schr\"odinger picture up to the cut (blue), giving $\ket{\psi_k}$.
            The measured $Z_s$ evolves in the Heisenberg picture through the noisy gates after it
            (orange), giving $S_k$. The expectation value is the overlap
            $\bra{\psi_k}S_k\ket{\psi_k}$, shown here for $k=5$. Moving the cut back past a gate
            switches on that gate's noise (\cref{eq:damp}): at $k=K$ the overlap is the exact
            expectation value, at $k=0$ the experimentally measured one.}
    \label{fig:circuit_cut}
\end{figure}

Consider a sequence of gates $\{U_k\}$ with $k=1,\dots,K$. To understand how noisy gates progressively dampen expectation values, we imagine a \textit{cut} in this sequence as shown in \cref{fig:circuit_cut}. The initial state evolves noiselessly in the Schrodinger picture until the cut, producing $\ket{\psi_k}$ after $k$ gates. The measured operator evolves in the Heisenberg picture through the gates to the right of the cut, each of which is accompanied by a two-qubit depolarizing channel:
\begin{align}
\mathcal{E}_g(O) = \lambda O + (1-\lambda)\frac{I_g}{4}\otimes\text{Tr}_g(O),
\end{align}
where $g$ is the subsystem on which the gate acts. The evolved operator that $U_k$ acts on is a Pauli sum: 
\begin{align}
S_k = T_k + V_k,
\end{align}
where $T_k$ is the part of the Pauli sum that has support on $g$ and $V_k$ is the part that does not. After the action of $U_k$ on $S_k$ we get: 
\begin{align}
S_{k-1} = \lambda\, U_k^{\dagger}\,T_k\,U_k + V_k.\label{eq:update}
\end{align}
Here we have used the fact that every string within $T_k$ is of the form $P_g\otimes R$, where $P_g$ is some non-identity Pauli on $g$. Therefore, $\text{Tr}_g(P_g\otimes R)=\text{Tr}_g(P_g)\otimes R=0$. Every string within $V_k$ is of the form $I_g\otimes Q$ and since $\mathcal{E}_g$ is unital, $V_k$ is left unchanged by the noisy gate. 

\begin{figure}
    \centering
    \includegraphics[width=1\linewidth]{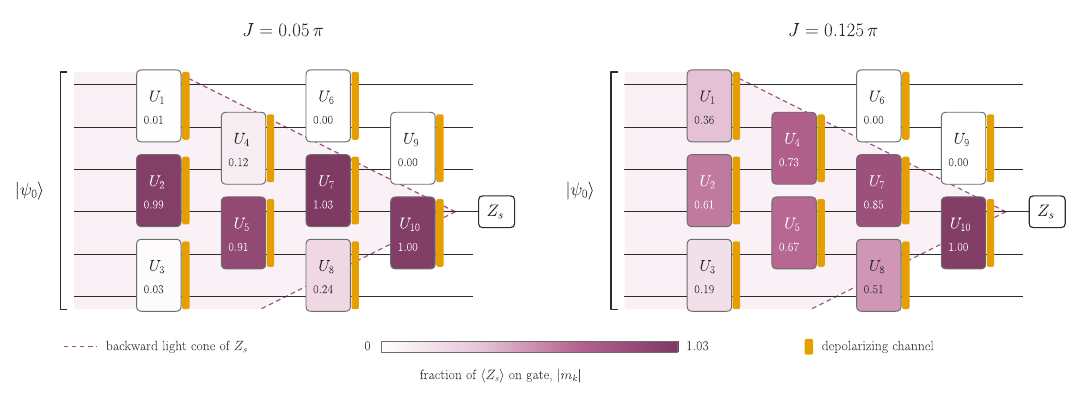}
    \caption{\textbf{The incident fractions depend on $J$.} Fraction of expectation value incident on the $k$th gate: $m_k=\bra{\psi_k}T_k\ket{\psi_k}/\bra{\psi_k}S_k\ket{\psi_k}$.  
        Computed with $J=0.05\pi$ and $J=0.125\pi$ in the absence of noise for a fixed initial product state and disorder instance. 
        The lightcones and constituent Pauli strings in $T_k$ and $S_k$ are the same at both couplings. But as $J$ grows, the fraction of expectation value carried
        to each noisy gate changes.}
    \label{fig:lec_limit}
\end{figure}

Imagine starting with the cut at $k=K$. We know that $z_{\text{exact}} \equiv \bra{\psi_K}S_K\ket{\psi_K}$ is the exact expectation value. Moving the cut back by a step, we find: 
\begin{align}
\bra{\psi_{K-1}}S_{K-1}\ket{\psi_{K-1}}
&= \bra{\psi_{K-1}}U_K^\dagger\big[\lambda\,T_K+V_K\big]U_K\ket{\psi_{K-1}} \nonumber\\
&= \lambda\,\bra{\psi_K}T_K\ket{\psi_K}+\bra{\psi_K}V_K\ket{\psi_K} \nonumber\\
&= \bra{\psi_K}S_K\ket{\psi_K}-(1-\lambda)\,\bra{\psi_K}T_K\ket{\psi_K} \nonumber\\
&= \big[1-(1-\lambda)\,m_K\big]\,\bra{\psi_K}S_K\ket{\psi_K},\label{eq:damp}
\end{align}

where $m_k$ is defined for every $k$ as: 
\begin{align}
m_k=\frac{\bra{\psi_k}T_k\ket{\psi_k}}{\bra{\psi_k}S_k\ket{\psi_k}}.
\end{align}
\cref{eq:damp} tells us that the effect of a noisy $U_k$ is to dampen the expectation value by a factor that depends on the noise parameter as well as $m_k$, which quantifies the fraction of the expectation value that is incident on $U_k$. 

Note that $z_{\text{noisy}}\equiv \bra{\psi_0}S_0\ket{\psi_0}$ represents the expectation value measured in an experiment where every gate is subject to noise. \cref{eq:damp} defines a recurrence relation that we can use to uncover a relationship between $z_\text{noisy}$ and $z_\text{exact}$: 
\begin{align}
    z_\text{noisy} = z_\text{exact}\prod_{k=1}^K\left[1-(1-\lambda)m_k\right].\label{eq:scale}
\end{align}
In practice, we cannot compute the scaling factor in \cref{eq:scale} by finding $\lambda$ and $\{m_k\}$ explicitly. The first would require an accurate characterization of gate-level noise. The second would require classical simulation of the full circuit, which is not possible for the large scale circuits in our experiments at all values of the Heisenberg coupling $J$. However, if we assume that $\lambda$ and $\{m_k\}$ are independent of $J$, we can \textit{empirically} extract the scaling factor at small values of $J$, which remain classically simulable, and apply it to all values of $J$. This is the central idea of LEC.

We do not expect $\lambda$ to depend on $J$. The coupling enters our circuits only through single-qubit rotation angles, so the two-qubit gates, and where they sit in the circuit, are identical at every $J$. The noise properties of a fixed two-qubit gate is set primarily by its duration relative to qubit coherence times. Although the Pauli strings that form $S_k$ are the same at every $J>0$, their coefficients vary with $J$. Therefore, the fraction of expectation value incident on each noisy gate, $\{m_k\}$, also depends on $J$. This is the fundamental limitation of LEC, illustrated in \cref{fig:lec_limit}. LEC's failure to capture this $J$-dependence leads to a systematic bias that we attempt to understand next. 

\subsection{Operator spreading and LEC}
\label{sec:op_spread}

\Cref{eq:scale} is exact, but $m_k$ is a fraction of a signed expectation value in a
particular state: it can lie outside $[0,1]$, and on its own it does not tell us in
which direction LEC errs. A state-independent counterpart follows from tracking the
norm of the evolved operator rather than its expectation value. Writing
$S_k=\sum_i a_i P_i$ and $\|S_k\|^2=\sum_i a_i^2$ for the normalized Hilbert--Schmidt
norm, we define the \textit{operator weight} on the noisy gate $U_k$ as
\begin{align}
w_k=\frac{\|T_k\|^2}{\|S_k\|^2},
\end{align}
the fraction of the evolved operator that is incident on $U_k$. Since $T_k$ and $V_k$ share no Pauli strings and conjugation by
$U_k$ preserves the norm, \cref{eq:update} gives
\begin{align}
\|S_{k-1}\|^2=\lambda^2\|T_k\|^2+\|V_k\|^2=\big[1-(1-\lambda^2)\,w_k\big]\,\|S_k\|^2,
\end{align}
and, with $\|S_K\|=1$,
\begin{align}
\|S_0\|^2=\prod_{k=1}^K\big[1-(1-\lambda^2)\,w_k\big].\label{eq:normdecay}
\end{align}
This shows how the effect of noise appears in the Heisenberg picture. Unitary evolution preserves operator norm. Noise shrinks it at each gate in proportion to
the operator weight incident on that gate. Note that the norm of an operator sets the typical size of its expectation values. For a traceless operator $O$ on $n$ qubits, averaging over random
pure states gives $\mathbb{E}_\psi\bra{\psi}O\ket{\psi}^2=\|O\|^2/(2^n+1)$.
\Cref{eq:normdecay} is therefore the state-averaged analogue of \cref{eq:scale} with $w_k$ in place of $m_k$.

The operator weight connects LEC to the literature on operator
spreading~\cite{nahum2018operator,Khemani_2018}, where the same quantity is defined on
a single site. It also makes the direction of the bias clear. At the calibration coupling
$J_0$ the evolved string has barely spread, so $w_k\approx1$ on the gates that touch the
support of $Z_s$ and $w_k\approx0$ elsewhere. As $J$ grows, operator weight spreads onto
gates in the backward light cone that the string did not originally touch. 

The calibration point is thus the least damped, and on
average the scaling factor learnt at $J_0$ underestimates the one needed at larger $J$.
The mitigated $Z_s^2$ are then too small, so the IPR is underestimated and the entropy
overestimated, increasingly so at greater depth, where the light cone holds more gates
for the operator to spread onto. We model and mitigate this bias in \cref{sec:wplecerrormodel}.

\subsection{Correcting the residual bias of LEC}
\label{sec:wplecerrormodel}

LEC finds its calibration factors from the smallest coupling $J_0$ and applies them
unchanged at every other $J$. Compared against classical references, the LEC entropies
agree at $J_0$ but on average overestimate the entropy at larger $J$, by an amount that
grows with circuit depth (\cref{tab:bias_model_holdout}). \Cref{sec:op_spread}
attributes this to operator spreading: the damping of each string is set by the operator
weight incident on each noisy gate, which grows with $J$, while LEC fixes the damping at
its value at $J_0$. In this section we turn this into a correction, fit it to the runs
that have a trusted classical reference, and attach to the corrected entropies an error
bar that is itself fitted and validated on held-out runs.

By \cref{eq:parseval}, the IPR of a marginal $A$ of $n_A$ qubits is
\begin{equation}
\text{IPR}_{2,Z}[A] = 2^{-n_A}+\Delta Q\,,\qquad
\Delta Q=2^{-n_A}\sum_{s\neq I} z_s^2\,,
\end{equation}
where $\Delta Q$ is the contribution of the nontrivial $Z$ strings on $A$, and
$S_{2,Z}[A]=-\log_2\text{IPR}_{2,Z}[A]$. Suppose that at coupling $J$ each string is
damped by an additional factor $e^{-\delta}$ relative to the calibration point $J_0$.
Every string boosted by LEC is then too small by $e^{-\delta}$, and $\Delta Q$, a sum of
squared strings, by $e^{-2\delta}$. The correction undoes this,
\begin{equation}
\Delta Q_\textrm{corr}=e^{2\delta}\,\Delta Q_\textrm{LEC}\,,
\label{eq:bias_correction_signal}
\end{equation}
boosting the part of the IPR above its maximally mixed value. The corresponding change in entropy,
\begin{equation}
\Delta S = S_\textrm{LEC}-S_\textrm{corr}
=\log_2\frac{2^{-n_A}+e^{2\delta}\Delta Q_\textrm{LEC}}{2^{-n_A}+\Delta Q_\textrm{LEC}}\,,
\label{eq:bias_correction_size}
\end{equation}
is at most $2\delta/\ln 2$ bits when $\Delta Q_\textrm{LEC}\gg2^{-n_A}$ and vanishes for
a maximally mixed marginal ($\Delta Q=0$). A marginal
whose measured $\Delta Q_\textrm{LEC}$ is slightly negative because of shot noise is
corrected in the same way rather than clipped, so that spatial averages remain unbiased.

\subsubsection{Form of $\delta$.}

In \cref{eq:bias_correction_signal}, $e^{-2\delta}$ is the factor by which $\Delta Q$ remains damped after LEC, i.e. the additional damping at $J$ relative to $J_0$.
\Cref{eq:normdecay} tells us what the damping factors look like as a function $J$ on average. Motivated by this and with $\lambda$ independent of $J$,
we can write to the first order in $1-\lambda^2$,
\begin{align}
e^{-2\delta}
&=\frac{\|O_0(J)\|^2}{\|O_0(J_0)\|^2}
 =\prod_{k=1}^{K}\frac{1-(1-\lambda^2)\,w_k(J)}{1-(1-\lambda^2)\,w_k(J_0)}\,, \nonumber\\
\delta
&=-\tfrac12\sum_{k=1}^{K}\Big\{\ln\!\big[1-(1-\lambda^2)\,w_k(J)\big]
 -\ln\!\big[1-(1-\lambda^2)\,w_k(J_0)\big]\Big\} \nonumber\\
&\simeq\tfrac12\,(1-\lambda^2)\sum_{k=1}^{K}\big[w_k(J)-w_k(J_0)\big].
\label{eq:delta_from_weights}
\end{align}
The additional damping is thus set by the operator weight that the noisy gates gain between
$J_0$ and $J$. This holds string by string. The single $\delta$ of a marginal is an
average over its strings, weighted by their signal. 

\Cref{eq:delta_from_weights} fixes the qualitative
behaviour of $\delta$. It vanishes at $J_0$ by construction and is non-negative on
average, since the calibration point is the least damped (\cref{sec:op_spread}). It
starts at second order in $J$: the part of the operator that a gate spreads onto a new
qubit has an amplitude of order $J$, and its operator weight is the square of that
amplitude. It grows with depth, as each cycle spreads more of the operator onto new
gates. Finally, it saturates once the operator has spread over its backward light cone,
since no operator weight can exceed one. If each of the $G$ two-qubit gates in the light
cone gains the same weight $\Delta w$ at saturation, then $\delta\to\alpha G$ with
$\alpha=\tfrac12(1-\lambda^2)\,\Delta w\le\tfrac12(1-\lambda^2)$. The simplest smooth
form with these properties is
\begin{equation}
\delta=\alpha\,G\left[e^{-(n_FJ_0/x_0)^2}-e^{-(n_FJ/x_0)^2}\right],
\label{eq:bias_model_delta}
\end{equation}
where $x_0$ sets the accumulated coupling at which the operator fills its light cone.
The onset, growth and saturation follow from \cref{eq:delta_from_weights}; the use of
$n_FJ$ as the single variable controlling the spreading, and the Gaussian form, are
heuristic choices. The disorder fields rotate the parts of the operator spread in
successive cycles relative to each other, so these do not add up fully coherently. We
test the form against the data in \cref{fig:bias_model_delta}. We compute $G$ exactly
for each marginal by tracing the backward light cone of its qubits through the logical
Floquet circuit, so that it depends on the marginal's shape and position, on the lattice
size and on $n_F$.

\subsubsection{Fitting the error model}

We fit $\alpha$ and $x_0$ by least squares of $\text{IPR}_\textrm{corr}$ against the reference IPR, over every marginal (of shapes $1\times1$ to $3\times3$) and every coupling of the 20 runs with a trusted classical reference, listed in \cref{tab:bias_model_holdout}. The references are MPS simulations for $L\le6$ and BP-PEPS for $L\ge8$, each restricted to the range of $J$ in which it is converged (\cref{sec:reliability_map}); for most of the large lattices this is $J\le0.02\pi$. Fitting in IPR space rather than in entropy space means that the correction is linear in the measured signal, and near-thermal marginals, whose IPR is close to $2^{-n_A}$ for any $\delta$, carry little weight without any selection or flooring of points. The fit gives $\alpha=1.58\times10^{-4}$ per gate and $x_0=0.148\pi$.

\begin{figure}[htbp]
    \centering
    \includegraphics[width=\linewidth]{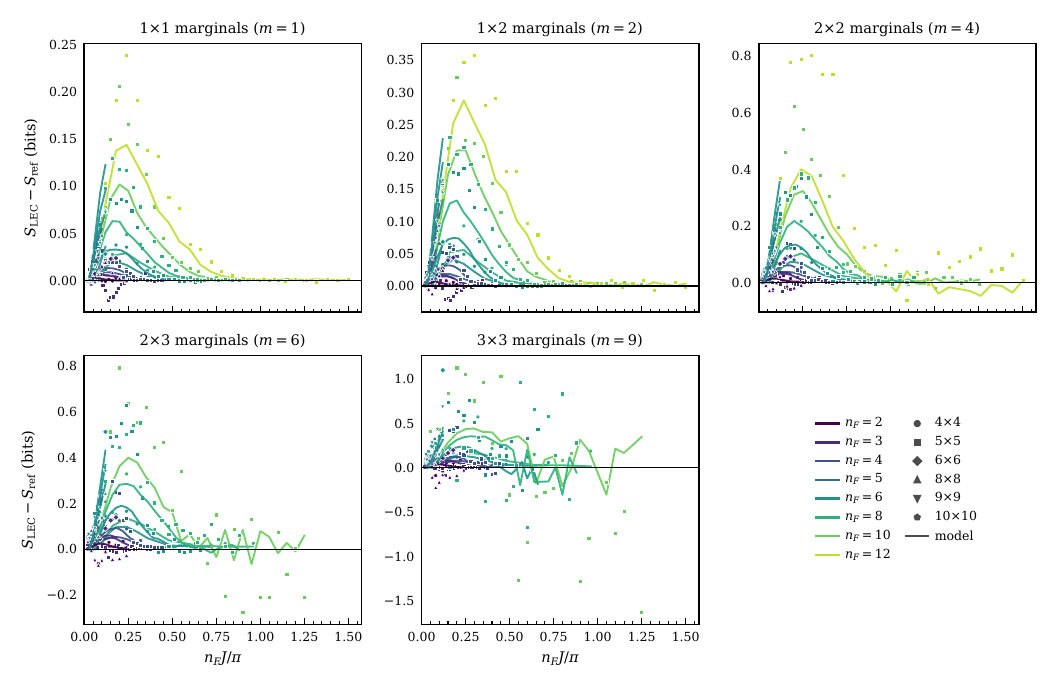}
    \caption{\textbf{Test of the form of $\delta$.} Error of weight-resolved LEC against the classical reference, $S_\textrm{LEC}-S_\textrm{ref}$ (points), and the error predicted by the bias model, $\Delta S$ of \cref{eq:bias_correction_size} (lines), against the accumulated coupling $n_FJ$, for marginals of each shape from $1\times1$ ($m=1$) to $3\times3$ ($m=9$). Each point is the mean over all marginals of that shape in one run at one coupling. Colours give the number of cycles $n_F$ and markers the lattice size The model is fitted to all 20 runs of \cref{tab:bias_model_holdout}, each compared with its classical reference only at couplings where that reference is converged. If \cref{eq:bias_model_delta} held exactly, the points of each run would lie on its line. For every shape the model captures the onset of the error and its decay at large $n_FJ$, where the marginals thermalize and leave no signal for the bias to act on. For the larger marginals at large $n_FJ$, the points are dominated by shot noise magnified by the mitigation, and the predicted error fluctuates with them, since $\Delta S$ depends on the measured signal.}
    \label{fig:bias_model_delta}
\end{figure}

\Cref{fig:bias_model_delta} tests the form of \cref{eq:bias_model_delta} directly. For every marginal that still carries signal we compute the additional damping required to match its reference, $\delta_\textrm{obs}=\tfrac12\ln(\Delta Q_\textrm{ref}/\Delta Q_\textrm{LEC})$. If \cref{eq:bias_model_delta} held exactly, all runs rescaled by $\alpha G$ would collapse onto the single curve $1-e^{-(n_FJ/x_0)^2}$. The quadratic onset at small $n_FJ$ is captured. At intermediate $n_FJ$ most runs lie above the curve, the measured damping continuing to grow after the ansatz has saturated, before returning towards it at larger $n_FJ$, where few marginals still carry enough signal for $\delta_\textrm{obs}$ to be determined. The amplitude scales roughly as $A\alpha G$: fitted run by run it varies by about half its value, with no systematic trend in $G$ or depth. Both deviations are absorbed into the error bars described next, and part of them remains as residual bias.

The corrected entropy of marginal $i$ has a systematic error, which we model as a zero-mean Gaussian of width
\begin{equation}
\sigma_i=\sqrt{(\sigma\,n_{A,i})^2+(\eta\,\Delta S_i^{\kappa})^2}\,,
\label{eq:bias_model_width}
\end{equation}
added in quadrature to its shot noise. The two terms have different origins. The first is the error of LEC itself. All strings of a given weight on a marginal share one factor $R_k$, although they act on different qubits that may be damped differently, so LEC fits the strings of a marginal only on average. This error is present even at $J_0$, where the correction vanishes, and grows with the number of strings sharing each factor; we take it proportional to $n_A$. The second is the uncertainty of the correction. \Cref{eq:bias_model_delta} is an approximate form whose amplitude varies between runs and marginals, so the shift $\Delta S_i$ of \cref{eq:bias_correction_size} is itself uncertain, by an amount that should vanish when nothing is corrected and grow with the size of the correction. We do not assume the rate of this growth but fit it through the exponent $\kappa$. In evaluating \cref{eq:bias_model_width} we use $|\Delta Q_\textrm{LEC}|$ in \cref{eq:bias_correction_size}, so that marginals measured just above and just below the maximally mixed value receive the same error bar.

With $\alpha$ and $x_0$ fixed, the residual $r_i=S_{\textrm{corr},i}-S_{\textrm{ref},i}$ of each training point is the sum of two independent zero-mean Gaussian errors: the shot noise of known width $s_i$ (from the experiment and the reference, in quadrature) and the systematic error of width $\sigma_i$. We therefore fit $\sigma$, $\eta$ and $\kappa$ by maximum likelihood,
\begin{equation}
\min_{\sigma,\eta,\kappa}\ \sum_i\left[\ln\!\big(\sigma_i^2+s_i^2\big)+\frac{r_i^2}{\sigma_i^2+s_i^2}\right],
\label{eq:bias_model_likelihood}
\end{equation}
which removes the known shot noise from the observed scatter, so that points dominated by shot noise constrain the systematic width only weakly. The few points whose corrected IPR is negative have no defined entropy and are excluded. The fit gives $\sigma=0.007$ bits per qubit, $\eta=0.69$ and $\kappa=0.58$: the uncertainty of the correction grows roughly as the square root of its size, so small corrections carry large relative errors and large corrections smaller ones. The floor $\sigma\,n_A$ is necessary for the fit to be well defined, since the second term vanishes at $J_0$ while the error of LEC does not.

On the training runs, the resulting intervals widen with the light-cone gate count and close once the marginals thermalize. They are too narrow at intermediate $J$, where part of the bias remains on the deep $5\times5$ runs, and for the largest marginals.

Spatial averages are formed in IPR space, as for the uncorrected data: the entropy of an average is $-\log_2$ of the mean corrected IPR over the marginals of a given shape. The error of each marginal is converted to an IPR error, $\ln2\,\text{IPR}_{\textrm{corr},i}\,(\sigma_i^2+s_i^2)^{1/2}$, these are combined as independent errors, the standard error of the mean over marginals is added in quadrature, and the result is propagated back through the logarithm.

We validate the model by leaving each of the 20 runs out in turn, refitting all five parameters on the remaining 19, and predicting the left-out run (\cref{tab:bias_model_holdout}). Over the held-out runs, the correction reduces the median bias of the spatial averages from $0.077$ to $0.026$ bits, removing about two thirds of it. The 68\% intervals cover a median of 82\% of the held-out spatial averages and 79\% of the held-out single marginals (99\% and 96\% for the 95\% intervals), with a mean absolute error of $0.062$ and $0.071$ bits respectively. The error bars are therefore close to their nominal coverage overall, somewhat conservative on the small lattices and too narrow on the large ones, most of all on $10\times10$ at 6 cycles.

\begin{table}[htbp]
\centering
\begin{tabular}{lcccc}
\toprule
Held-out run & $J/\pi$ & LEC bias (bits) & Corrected bias (bits) & 68\% coverage \\
\midrule
$4\times4$, $n_F=2$ & 0.010--0.120 (12) & $+0.005$ & $-0.001$ & 98\% \\
$4\times4$, $n_F=4$ & 0.010--0.120 (9) & $+0.042$ & $+0.021$ & 100\% \\
$4\times4$, $n_F=6$ & 0.010--0.120 (9) & $+0.062$ & $+0.028$ & 98\% \\
$4\times4$, $n_F=8$ & 0.010--0.120 (9) & $+0.100$ & $+0.058$ & 91\% \\
$5\times5$, $n_F=3$ & 0.005--0.125 (25) & $+0.006$ & $-0.011$ & 90\% \\
$5\times5$, $n_F=4$ & 0.005--0.110 (22) & $+0.053$ & $+0.022$ & 96\% \\
$5\times5$, $n_F=6$ & 0.005--0.110 (22) & $+0.130$ & $+0.075$ & 68\% \\
$5\times5$, $n_F=8$ & 0.005--0.110 (22) & $+0.110$ & $+0.036$ & 94\% \\
$5\times5$, $n_F=10$ & 0.005--0.125 (25) & $+0.050$ & $-0.054$ & 84\% \\
$5\times5$, $n_F=12$ & 0.005--0.125 (25) & $+0.126$ & $+0.051$ & 87\% \\
$6\times6$, $n_F=3$ & 0.010--0.060 (6) & $+0.058$ & $+0.028$ & 83\% \\
$6\times6$, $n_F=6$ & 0.010--0.020 (2) & $+0.230$ & $+0.180$ & 60\% \\
$8\times8$, $n_F=2$ & 0.010--0.120 (9) & $-0.014$ & $-0.022$ & 58\% \\
$8\times8$, $n_F=4$ & 0.005--0.020 (4) & $+0.052$ & $+0.019$ & 70\% \\
$8\times8$, $n_F=6$ & 0.005--0.020 (4) & $+0.130$ & $+0.026$ & 80\% \\
$9\times9$, $n_F=4$ & 0.005--0.020 (4) & $+0.065$ & $+0.026$ & 80\% \\
$9\times9$, $n_F=5$ & 0.005--0.020 (4) & $+0.089$ & $+0.009$ & 80\% \\
$9\times9$, $n_F=6$ & 0.005--0.020 (4) & $+0.193$ & $+0.056$ & 70\% \\
$10\times10$, $n_F=5$ & 0.005--0.020 (4) & $+0.158$ & $+0.071$ & 80\% \\
$10\times10$, $n_F=6$ & 0.005--0.020 (4) & $+0.169$ & $-0.009$ & 50\% \\
\midrule
Median & & $+0.077$ & $+0.026$ & 82\% \\
\bottomrule
\end{tabular}
\caption{\textbf{Leave-one-run-out validation of the bias correction.} For each run with a trusted classical reference, the model is refitted on the other 19 runs and applied to the held-out one. The $J/\pi$ column gives the range of couplings compared, with the number of values in brackets. LEC bias and corrected bias are the mean of $S_\textrm{LEC}-S_\textrm{ref}$ and $S_\textrm{corr}-S_\textrm{ref}$ over the run's spatial averages; coverage is the fraction of corrected spatial averages whose 68\% interval contains the reference. References are MPS for $L\le6$ and BP-PEPS for $L\ge8$, each restricted to the range of $J$ in which it is converged.}
\label{tab:bias_model_holdout}
\end{table}

The model assumes a single $\delta$ for all strings of a marginal, whereas higher-weight strings may be damped more strongly, and it has no lattice- or device-specific terms: the plateau depends only on $G$ and the onset only on $n_FJ$. The run-to-run variation of the amplitude and the deviation from \cref{eq:bias_model_delta} at intermediate $n_FJ$ (\cref{fig:bias_model_delta}) enter the error bars only through the fitted width, and part of them remains as bias. The results of \cref{fig:TDL_main} at $J/\pi=0.03$ and $0.04$ on the largest lattices are an extrapolation of the model. These circuits have the largest light cones and therefore the largest corrections, while their references are trusted only up to $J/\pi=0.02$, and the held-out coverage is lowest for the largest lattices. The error bars of spatial averages are also likely optimistic, since an error in $\delta$ is shared by all marginals of a run while the averaging treats the marginal errors as independent. 

For most of the results in this work the bias correction is small: it shifts weight-resolved LEC only slightly (\cref{fig:mitigation_comparison}). It matters most for the approach to the thermodynamic limit (\cref{sec:tdl_sm}), where six cycles on lattices up to $10\times10$ give the largest light cones in this work, and where the differences between lattice sizes are comparable to the bias; without the correction the entropy at fixed $J$ would appear to increase with system size, opposite to the classical simulations. The interplay between operator spreading and error mitigation is an active direction of recent work which will be interesting to continue exploring~\cite{suchsland2026quantum}.

\subsection{Applying LEC to the IPR}
\label{sec:lec_variants}

By \cref{ParsevalIPR}, the IPR of a marginal $A$ of $n_A$ qubits in the computational
basis is a sum over $Z$ strings,
\begin{equation}
\text{IPR}_{2,Z}[A]=2^{-n_A}\sum_{s\subseteq A}z_s^2\,,\qquad z_s=\langle Z_s\rangle\,,
\label{eq:parseval_zstrings}
\end{equation}
where $Z_s$ is the product of $Z$ operators on the qubits in $s$, $z_\emptyset=1$, and
the weight $|s|$ is the number of qubits in $s$. All $2^{n_A}$ expectation values are
estimated from the same computational-basis shots that give the IPR, so LEC can act at
three levels of granularity without additional measurements. In each case the factors
are fitted for each marginal at $J_0$ against the classical reference and applied
unchanged at every other $J$.

\paragraph{IPR-level LEC.} A single factor per marginal rescales the part of the IPR
above its maximally mixed value,
\begin{equation}
\Delta Q^\textrm{LEC}(J)=R\,\Delta Q^\textrm{exp}(J)\,,\qquad
R=\frac{\Delta Q^\textrm{exact}}{\Delta Q^\textrm{exp}}\bigg|_{J=J_0}\,,
\label{eq:lec_ipr}
\end{equation}
with $\Delta Q=\text{IPR}_{2,Z}[A]-2^{-n_A}$. This form exactly inverts a depolarizing
channel acting on the marginal as a whole,
\begin{equation}
\mathcal{E}_{p}(\rho_{A}) = (1 - p)\, \rho_{A} + p\,\frac{\mathbb{I}}{2^{n_{A}}}\,.
\end{equation}
Since $\text{IPR}_{2,Z}[A]=\sum_x\bra{x}\rho_A\ket{x}^2$,
\begin{equation}
\text{IPR}_{2,Z}[A]\big(\mathcal E_p(\rho_A)\big)
=(1-p)^2\,\text{IPR}_{2,Z}[A](\rho_A)+\frac{p(2-p)}{2^{n_A}}\,,
\qquad\text{so}\qquad
\Delta Q^\textrm{noisy}=(1-p)^2\,\Delta Q^\textrm{exact}\,,
\end{equation}
and $R=(1-p)^{-2}$. Fitting $R$ independently for each marginal, we find that the
effective depolarizing rate increases with marginal size. We use the same single-factor
form to mitigate the purity, applied to the estimated purity of each marginal as a
whole, with the fixed point $2^{-n_A}$ replaced by the purity of the maximally mixed
state within the magnetization sector of the initial state, reduced to the marginal.

\paragraph{String-resolved LEC.} Each nontrivial string receives its own factor,
\begin{equation}
z_s^\textrm{LEC}(J)=R_s\,z_s^\textrm{exp}(J)\,,\qquad
R_s=\frac{z_s^\textrm{exact}}{z_s^\textrm{exp}}\bigg|_{J=J_0}\,,
\label{eq:lec_string}
\end{equation}
and the IPR is reconstructed from \cref{eq:parseval_zstrings} with the rescaled
expectation values, each $z_s^2$ replaced by its unbiased estimator from the measured
shots.

\paragraph{Weight-resolved LEC.} The strings are grouped by weight, and all strings of
weight $\ell$ share one factor,
\begin{equation}
z_s^\textrm{LEC}(J)=R_{|s|}\,z_s^\textrm{exp}(J)\,,\qquad
R_\ell=\frac{\sum_{|s|=\ell}z_s^\textrm{exact}\,z_s^\textrm{exp}}
{\sum_{|s|=\ell}\big(z_s^\textrm{exp}\big)^2}\bigg|_{J=J_0}\,,
\label{eq:lec_weight}
\end{equation}
which is the least-squares factor matching the $\binom{n_A}{\ell}$ strings of weight
$\ell$ to the reference at $J_0$. The IPR is reconstructed as for string-resolved LEC.

\paragraph{Comparison.} String-resolved LEC is the natural choice for gate-level noise
(\cref{sec:depolarizingnoiseinversion}). At small $J$ a measured string $Z_s$ remains on
its own support, since the $ZZ$ part of each coupling and the $Z$ rotations from the
disorder fields commute with it. Each string is therefore damped once for every noisy
gate that touches its support (\cref{eq:scale} with $m_k$ equal to $1$ on those gates
and $0$ elsewhere). The number of such gates is set mainly by the weight of the string,
but also by its position, through boundary effects and through error rates that vary
across the device, so each string has its own damping factor. This is also why a single
depolarizing channel on the whole marginal is too crude, and why its effective rate
grows with marginal size.

IPR-level LEC cannot resolve this. Because the IPR is quadratic in the expectation
values, its single factor $R$ is a weighted mean of the squared string factors $R_s^2$,
weighted by the signal $(z_s^\textrm{exact})^2$ each string carries at $J_0$. Near $J_0$
the state is close to a computational-basis product state, every string carries
comparable signal, and the mean is dominated by the most numerous weights. As $J$
increases and correlations build up, the distribution of the signal over weights changes,
and a factor calibrated with the weights at $J_0$ no longer matches the damping of the
signal that remains. IPR-level LEC therefore acquires a $J$-dependent bias of its own,
on top of the one corrected in \cref{sec:wplecerrormodel}.

String-resolved LEC removes this bias but is expensive in shots. Each $R_s$ is the ratio
of a measured and a reference value at a single coupling, so the shot noise of
$z_s^\textrm{exp}(J_0)$ enters $R_s$ directly and is then applied, as a fixed systematic
error, at every $J$. This error is largest for the high-weight strings, which are the
most strongly damped and whose measured expectation values are therefore smallest
relative to the shot noise. A $3\times3$ marginal has 511 such factors, and determining
each of them precisely would require many more shots at $J_0$ than the measurement of the
IPR itself.

Weight-resolved LEC is a middle ground between the two. It keeps the dominant dependence
of the damping, on the weight of the string, while each of the $n_A$ factors is fitted to
all strings of that weight. In \cref{eq:lec_weight} the strings with a clear signal
dominate the fit, and the shot noise of individual strings averages out. The cost is that
strings of the same weight in different positions share one factor, so LEC matches the
reference at $J_0$ only on average over each weight. This error is present at every $J$,
and it is the origin of the floor $\sigma n_A$ in the error bars of
\cref{sec:wplecerrormodel}. We use weight-resolved LEC for all collision entropies in
this work, and in the rest of this section LEC refers to the weight-resolved form unless
stated otherwise.

\subsection{Inferring and inverting errors from the Hamming spread}\label{sec:mitigationbrief}

A natural first approach to assessing noise in a number-conserving circuit like ours, where the Hamming weight in the computational basis should be the same at the beginning and end of the ideal evolution is to compute the \emph{retention rate}: the fraction of output samples that fall in the expected Hamming-weight sector. A closely related strategy is to post-select on those samples~\cite{bonet2018low,mcardle2019error,rosenberg2024dynamics}. A first drawback of this approach is that post-selection discards a large fraction of the data, thereby increasing statistical uncertainty. More importantly, a second drawback is that at the system sizes and error rates relevant to our QPU experiments, many samples that land in the correct Hamming-weight sector still contain an even number of bit-flip errors whose net effect cancels in the total Hamming weight, so the post-selected ensemble can remain substantially corrupted.

In this work we go beyond the retention rate and post-selection toolbox and make use of the \emph{full} distribution of output Hamming weights. We map the observed spread into a simple error model with a single non-tunable parameter and then invert the corresponding attenuation in a principled way. This yields mitigated estimates for any $r$-qubit observable that is diagonal in the computational basis, including the collision probability and collision entropy of $r$-qubit subsystems.

\begin{figure}[hbtp]
    \centering
    \subpanel{0.49\linewidth}{a}{\includegraphics[width=\linewidth]{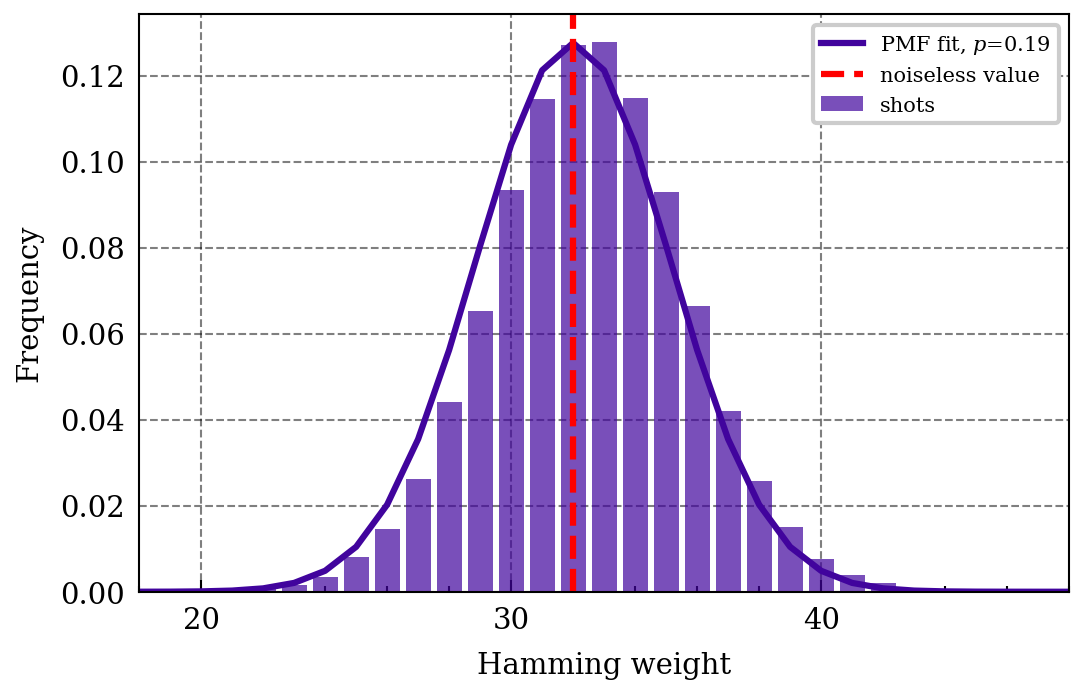}}
    \hfill
    \subpanel{0.49\linewidth}{b}{\includegraphics[width=\linewidth]{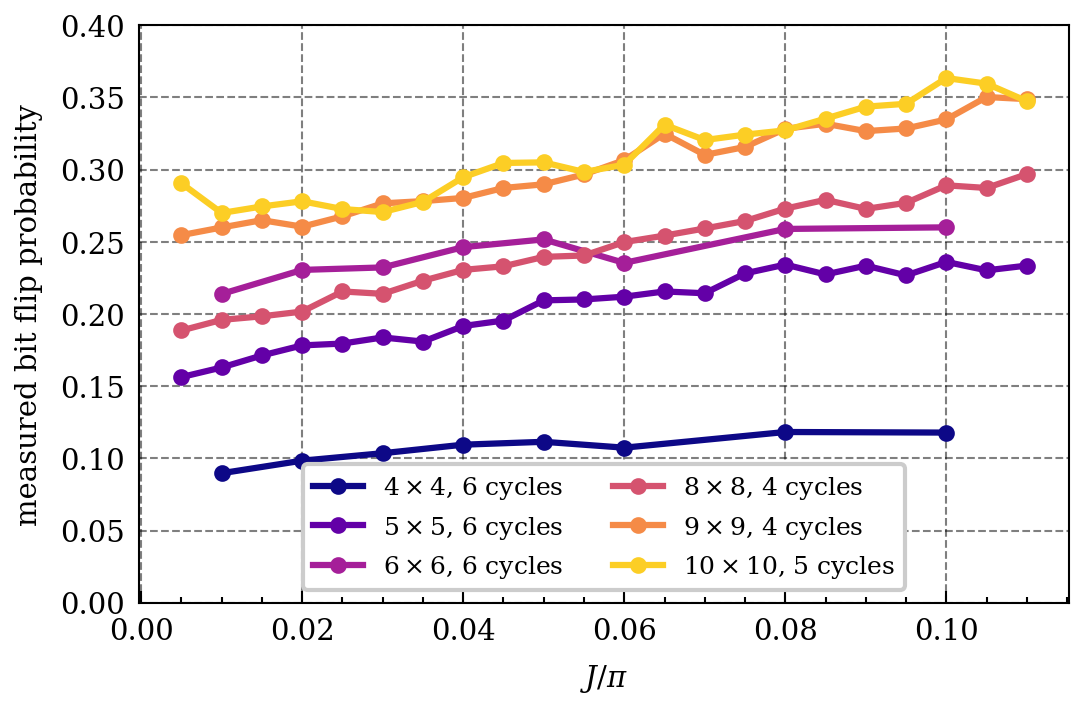}}

    \caption{
    \textbf{Bit-flip noise in the raw data.} \textbf{a}, Distribution of Hamming weights in raw experimental data. We fit a bit flip noise model to the observed data to extract the bit-flip probability. The data used is from an experiment with system size $8\times 8$, 4 Floquet cycles, $J/\pi=0.06$ and 40k shots.
    \textbf{b}, Probability of bit flips across system sizes and values of $J$. We observe that the probability grows with both system size and $J$, but that smaller system sizes at later $J$ are often as noisy as larger system sizes at earlier $J$.
    }
    \label{fig:pflip_noise_diagnostics}
\end{figure}

Motivated by the empirical distribution of Hamming weight frequencies, as seen in \cref{fig:pflip_noise_diagnostics}, we assume a simple error model: each qubit experiences a bit flip error with total probability $p$ during the course of the circuit.  The probability mass function for the output Hamming weight random variable $h$ derived from this model is not exactly a binomial distribution, however it can be derived by applying the convolution theorem to a sum of binomial random variables. The result is
\begin{equation}\textrm{Pr}(h) = \sum_{d = \max\{0,m-h\})}^{\min\{m,n-h\}} {m \choose d} {n- m \choose h - m + d} p^{h - m +2 d} ( 1- p)^{n + m - h - 2 d}\,, \label{eq:hspmf}\end{equation}
where $n$ is the number of qubits and $m = \lfloor n/2\rfloor$ is the Hamming weight of the initial state--see \cref{sec:hsadditional} for the calculation.  The exact variance from \cref{eq:hspmf} is $\sigma^2 = n p(1-p)$.  Therefore a quick approach is to fit the value of $p$ from the sample variance of the output Hamming weights; instead, we fit $p$ according to a least squares minimization between the empirical output Hamming weight frequencies and the probability mass function (which is equivalent to a fit over all moments of the distribution), and we apply that choice throughout all applications of the mitigation technique in this work.  

Since $p$ is the total bit-flip probability \emph{per qubit} over the full circuit, it is an intensive quantity and is naturally normalized for comparisons across circuit sizes and depths. Nevertheless, we examine the dependence of $p$ on system size in \cref{fig:pflip_noise_diagnostics} and find that $p$ generally increases for larger systems. One reason is that the QPU compiler typically selects qubit layouts to avoid the noisiest qubits and couplers identified during calibration; as the target system size grows, the embedding is increasingly forced to include less well-performing regions of the device. A second reason is that the inferred $p$ reflects not only \emph{local} flip events but also their propagation through entangling gates. For example, a bit flip on the control qubit prior to a $\mathrm{CZ}$ can spread to correlated flips on both qubits after the gate. Thus, even at fixed circuit depth, increasing the system size can increase the effective spread of a single fault if the corresponding lightcone would otherwise be truncated by the smaller system.

In \cref{fig:pflip_noise_diagnostics}, we also observe a clear dependence of the modeled error rate $p$ on the coupling strength $J$. This illustrates how the method provides circuit-level benchmarking information in real time: along each curve, the circuits have the same gate counts (the same numbers of one- and two-qubit gates), and only the strength of the two-qubit coupling is varied, yet the inferred flip probability changes measurably. Finally, \cref{fig:pflip_noise_diagnostics} shows that a smaller system run at larger depth can exhibit a higher per-qubit error rate than a larger system at shorter depth.\footnote{When comparing $p$ across data sets, it is also relevant that device performance can fluctuate from day to day, consistent with other time-dependent calibration metrics.} This is useful for interpreting the impact of errors on the output: it allows comparisons between circuits inside and outside the classically simulable regime that nevertheless share similar inferred values of $p$.  

Once we fit the parameter $p$ according to the empirical Hamming weight frequencies, we assume that the output samples are as if they passed through an ideal version of the circuit followed by a tensor product of binary symmetric channels all with the same error parameter $p$.   In other words we consider the single bit flip channel $\mathcal{E}_p$
\begin{equation}
\mathcal{E}_p(\rho) = (1-p)\rho + p X \rho X
\end{equation}
and then mitigate the experimental data in a principled way according to the assumption that the measured samples were obtained by passing the output of the ideal circuit through this channel acting on each qubit,
\begin{equation}
\textrm{Assume: } p_\textrm{measured}(x) = \langle x |\rho_\textrm{device} | x\rangle = \langle x |  \mathcal{E}_p^{\otimes n} \left( U |x_\textrm{in}\rangle \langle x_{\textrm{in}} U^\dagger\rangle \right)|x\rangle \quad , \quad x \in \{0,1\}^n\,. \label{eq:hsassumption}
\end{equation}
In particular, we can compute the action of this assumed error model on expectations of any observable in the $Z$ basis.  In the single-qubit case,
\begin{equation}
\mathcal{E}_p(Z) = (1-p)Z + p X Z X = (1-2p)Z
\end{equation}
and in the general case of a Pauli $Z$ string on a subset of qubits $R$, $Z_R = \prod_{i\in R} Z_i$, the attenuation caused by the assumed action of the error model is
\begin{equation}
\langle Z_R\rangle_\textrm{measured} = (1-2p)^{|R|} \langle Z_R\rangle_\textrm{ideal}\,,
\end{equation}
where $\langle Z_R\rangle_\textrm{ideal}$ indicates the ideal result under the assumption of \cref{eq:hsassumption}.  Therefore we solve this equation and regard the inflated measured result as a mitigated prediction,
\begin{equation}
\langle Z_R\rangle_\textrm{mitigated} = (1-2p)^{-|R|} \langle Z_R\rangle_\textrm{measured}\,.
\label{eq:hsZmitigation}
\end{equation}
Of course, this assumption is radical in its simplicity as mid-circuit bit flip errors do not commute with the remainder of the circuit.  Yet $p$ still takes some of this spreading into account by aggregating the probability of bit flip errors over the full circuit.  We apply randomized Pauli twirling so that the noise in the circuit can be analyzed in terms of an approximate Pauli channel, but the method does not attempt to treat phase errors.  Ultimately it is a heuristic that is best justified by its empirical performance even in the \emph{highly scrambling} circuits considered in this work.

When we estimate the collision probability for small marginals of $1-4$ qubits, statistical errors are low enough that we can mitigate each of the expectation values of $Z$ strings using \cref{eq:hsZmitigation} in the Parseval form \cref{eq:parseval}. As the size of the marginal increases, statistical errors begin to dominate such an approach, and we instead mitigate the collision probability of marginals of size above 4 qubits with the form
\begin{equation}
\alpha = (1-p)^2 + p^2 \quad , \quad \textrm{IPR}_{2,R, \textrm{mitigated}} = 2^{-|R|} + \alpha^{-|R|} (\textrm{IPR}_{2,R,\textrm{measured}} - 2^{-|R|})\,.
\label{eq:hsrank1}
\end{equation}
Some intuition for this form can be gained by thinking of the collision estimator as formed from two samples at a time; $\alpha$ is the probability that a bit flip error strikes a single qubit in either both samples or neither (both events leaving the sample collision unchanged). See \cref{sec:hsadditional} for a more complete derivation.
\begin{figure}[!t]
    \centering
    \includegraphics[width=1\linewidth]{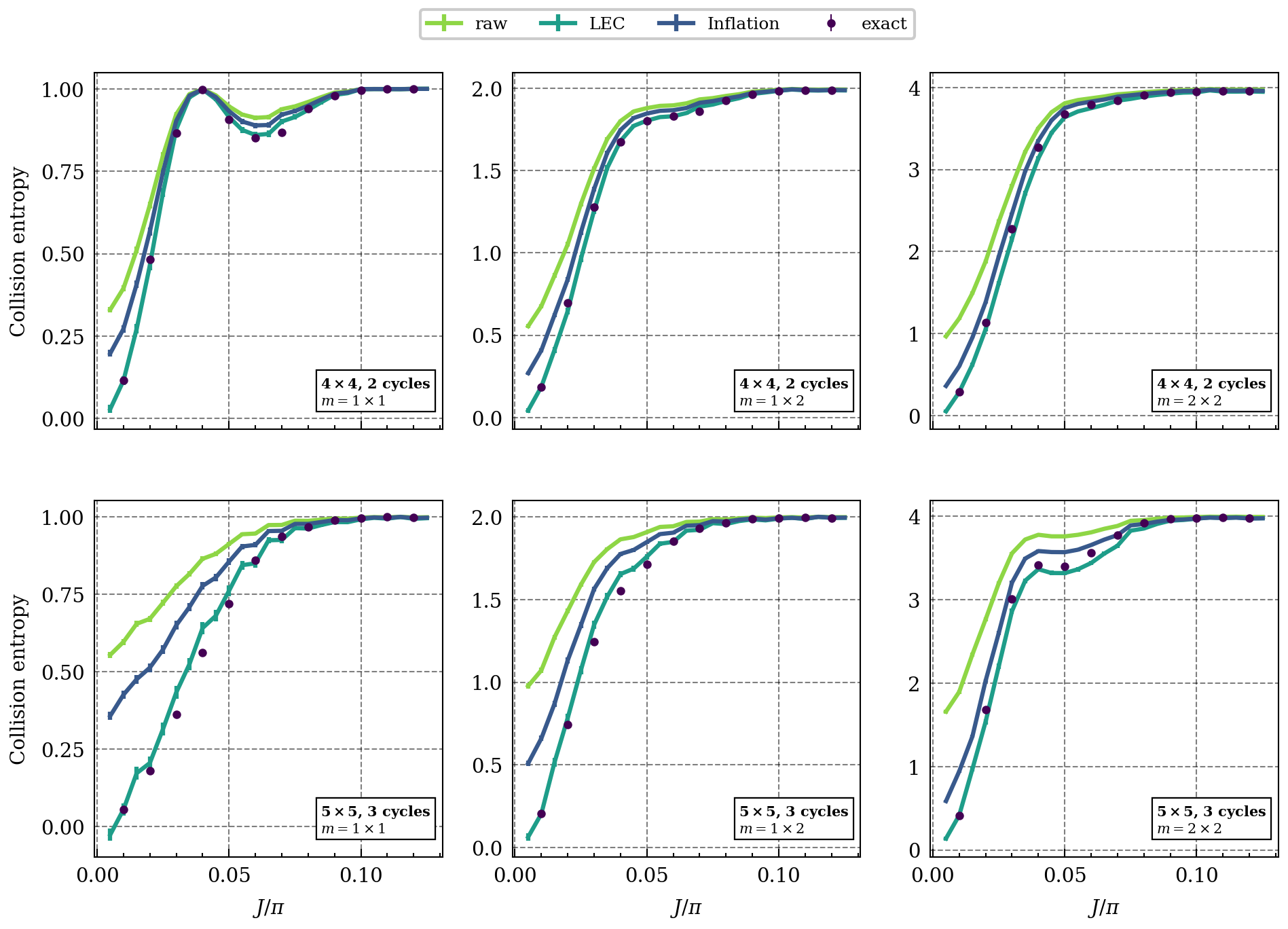}
    \caption{\textbf{Comparison of LEC and bit-flip inflation.} Using central marginals of different shapes, we applied the two mitigation strategies discussed in this section: Low Entanglement Calibration as defined in \cref{eq:LECdefinitionmain} and inflation  via \cref{eq:hsZmitigation} and \cref{eq:parseval}.  
    }
    \label{fig:bitflip_mitigation_comparison}
\end{figure}
From \cref{eq:hsZmitigation} we see the prediction that the inflation factor $(1-2p)^{-|R|}$ grows exponentially with the size of the subsystem $R$. In our experiments, we consider marginals up to size $3\times 3 = 9$ qubits, and since the collision probability depends on the square of the $Z$ expectation values, we apply inflation factors as large as $10^7$! Whether using \eqref{eq:hsrank1} or Low Entanglement Calibration (LEC) method in \eqref{eq:LECdefinitionmain}, these inflation factors amplify the statistical uncertainty of the raw collision probability estimated by samples, and this amplified statistical uncertainty is reflected in error bars throughout this work.  

To summarize this discussion, in \cref{fig:bitflip_mitigation_comparison} we compare the raw data with the mitigated results and the comparison to exact classical simulations at system size $4\times 4$ and $5\times 5$.   

\subsection{Additional Details for Hamming Spread Error Mitigation}\label{sec:additional_hs}
This section contains supporting calculations for the Hamming spread error mitigation method described in \cref{sec:mitigationbrief}: deriving the probability mass function under the assumed error model, and our method for approximating the error inversion directly on the optimal collision estimator.   
\label{sec:hsadditional}
\paragraph{Distribution of Hamming Weights in the 1-parameter error model} %
 Let $n$ be the number of qubits and $m$ be the Hamming weight of the initial state.  The core assumption of our model is that each qubit experiences a bit flip error with independent total probability $p$ during the course of the circuit.  For initial filling $m$ and bit flip probability for qubit  $p$ model, the Hamming weight $h = h(m,p)$ of output bit strings is a random variable, and we have
\begin{equation}
h(m,p) = m - d(m,p) + u(m,p)\,,
\end{equation}
where $d(m,p)$ is a random variable that describes the number of flips $1 \rightarrow 0$, at a given filling $m$ and probability $p$, and similarly $u(m,p)$ is a random variable for the number of flips $0 \rightarrow 1$.  The variable $d(m,p)$ is binomially distributed  $d(m,p) \sim \textrm{binom}(m,p)$--there are $m$ ones which each have a probability $p$ of flipping--, and similarly there are $n-m$ zeroes, so $u(m,p) \sim \textrm{binom}(n-m,p)$.  

Therefore we have expressed our random variable of interest, the Hamming weight of output strings, as a sum of two random variables, each of which has a known probability distribution.   For a sum of two independent integer valued random variables $A,B$, the convolution theorem gives the distribution for the sum $C = A + B$ as
\begin{equation}
    \textrm{Pr}(C = c) = \sum_{\textrm{all } a} \textrm{Pr}(A = a) \textrm{Pr}(B = c -a)\,,
\end{equation}
which sums over all ways that $A + B = C$ can occur the product of the probabilities for each event since they are independent. In our setting, the probability of $d = d(m,p)$ or $u = u(m,p)$ is given by the binomial distribution,
\begin{equation}
\textrm{Pr}(d) = {m \choose d} p^d (1-p)^{m - d}  \quad , \quad \textrm{Pr}(u) = {n- m \choose u} p^u ( 1- p)^{n-m - u}\,. 
\end{equation}
Now using the convolution formula and $u = h - m + d$,
\begin{equation}
\begin{aligned}
\mathrm{Pr}(h)
&= \sum_d 
\binom{m}{d} p^d (1-p)^{m-d}
\binom{n-m}{h-m+d} p^{h-m+d} (1-p)^{n-m-(h-m+d)} \\
&= \sum_d 
\binom{m}{d} \binom{n-m}{h-m+d}
p^{h-m+2d} (1-p)^{n+m-h-2d}.
\end{aligned}
\end{equation}
Lastly, we think about the range of the sum: $d$ cannot be below $0$ or below $m - h$, and it cannot be above $m$ or $n - h$, so the probability distribution over Hamming weights is
\begin{equation}\textrm{Pr}(h) = \sum_{d = \max\{0,m-h\})}^{\min\{m,n-h\}} {m \choose d} {n- m \choose h - m + d} p^{h - m +2 d} ( 1- p)^{n + m - h - 2 d}\,.\end{equation}

\paragraph{Mitigation of the Collision Estimator} %
The collision probability for a subsystem with density matrix $\rho$ can be expressed as an observable on two copies of the system,
$$C_R= \sum_{s =0,1} \sum_{i\in R}|s,s\rangle \langle s,s|_i  \,, \qquad C_{2,R}(\rho) = \textrm{tr}[C_R \;  \rho \otimes \rho]$$
where $|s,s\rangle \langle s,s|_i$ projects on states where the $i$-th bit of both copies is equal to $s$.  Therefore, we seek to analyze  two copies of the error channel acting in tensor product on two copies of the system.  Our error model assumes that some bit masks $e,e'$ are applied to the system, a bit mask being the list of bits that flip a string $x$ into $x\oplus e$.  The notation $X^e$ refers to a Pauli $X$ string that acts with $X$ on bit $i$ with $e_i = 1$, and 0 otherwise, so $(X^e)(X^{e'}) = X^{e\oplus e'}$.   With this we can express the trace as
\begin{equation}
    C_{\textrm{measured}} := C_{2,R}(\mathcal{E}(\rho))= \textrm{Tr}[C_R \mathcal{E}(\rho \otimes \rho)] = \mathbb{E}_{e,e'} \left[ \textrm{Tr}[C_R (X^e \rho X^e)(X^{e'}\rho \,,X^{e'})]\right]
\end{equation}
where the subscript ``measured'' refers to the value the collision estimator would converge to, assuming this error model. Expanding the trace in the computational basis,
\begin{equation}
     \textrm{Tr}[C_R (X^e \rho X^e)(X^{e'}\rho X^{e'})] = \sum_{x}  \langle x|(X^e \rho X^e) |x \rangle  \langle x |(X^{e'}\rho X^{e'}) |x \rangle = \sum_x p(x \oplus e) p(x \oplus e')\,.
\end{equation}
Therefore, by defining $u := e\oplus e'$ and relabeling the sum we have
\begin{equation}
    \textrm{C}_\textrm{measured}= \mathbb{E}_{u}   \sum_x p(x) p(x\oplus u) \,.
\end{equation}
So far, everything we have done is exact, and we could also have derived this final expression by considering the collision estimator.  Define $\alpha = (1-p)^2 + p^2$ so that $\alpha^r$ is the probability of the event $u = 0$, then
 \begin{equation}
     C_\textrm{measured}= \alpha^r \left(\sum_x p(x)^2 \right) + \sum_{u \neq 0} P(u) \sum_{x} p(x) p(x\oplus u)\,,
 \end{equation}
where $P(u)$ is the probability that the error mask was $e\oplus e' = u$.   This is still exact, and the first term in parentheses is the collision probability of the subsystem without the error channel applied.  Now we make an approximation,
\begin{equation}
    \textrm{for all } u\neq 0 \textrm{ we assume} \sum_{x} p(x) p(x\oplus u)  = \frac{1}{D}\,,
\end{equation}
where $D = 2^r$ is the dimension of the marginal's Hilbert space.   Note that for any distribution $p$, the average of this sum over all bit masks $u$ is always $1/D$, 
\begin{equation}
    \frac{1}{D} \sum_u  \sum_x p(x) p(x\oplus u) = \frac{1}{D} \sum_x p(x) \sum_u p(x\oplus u) = \frac{1}{D}\,.
\end{equation}
Therefore, the replacement of this sum by the fixed term $1/D$ assumption of non-correlation between amplitudes on distinct bit strings together with typicality of the  distribution.  Another perspective is that this can be regarded as a rank-1 approximation that treats the error channel with a two-dimensional subspace: ideal or not.   The subspaces are ``no error'' or ``any non-zero error bit mask $u$ acts like an average error bit mask and reduces the bit-shift autocorrelation to its average value."   

Making this replacement, 
\begin{equation}
    C_\textrm{measured}= \alpha^r \left(\sum_x p(x)^2 \right) + \frac{1}{D}\sum_{u \neq 0} P(u)  = \alpha^r C_{2,\textrm{mitigated}} + \frac{(1-\alpha^r)}{D}\,.
\end{equation}
Therefore, we solve for the mitigated value predicted under the model in terms of the measured value,
\begin{equation}
    C_{2,\textrm{mitigated}} = 2^{-r} + \alpha^{-r}(C_{2,\textrm{measured}} - 2^{-r})\,,
\end{equation}
  and we apply this correction in the mitigation of all marginal collision probabilities above $r > 4$ qubits.  A final note is that the sum  $A(u) = \sum_{x} p(x) p(x\oplus u) $ is an autocorrelation function in a general sense (how correlated is $p$ with itself after a shift by $u$?). In principle, this information can also be estimated from noisy samples and the method can be extended.  Such corrections are most relevant at low values of $J$ when the $Z$ expectation values are highly polarized and the bit-shift autocorrelation function is less than $1/D$ (e.g., $p(1-p) \ll 1/2$ when $p$ is near 0 or 1).

\section{Quantum simulation on IBM Nighthawk QPUs}
\label{sec:Quantum simulation results}
\begin{figure}[hbtp]
    \centering
    \includegraphics[width=1\linewidth]{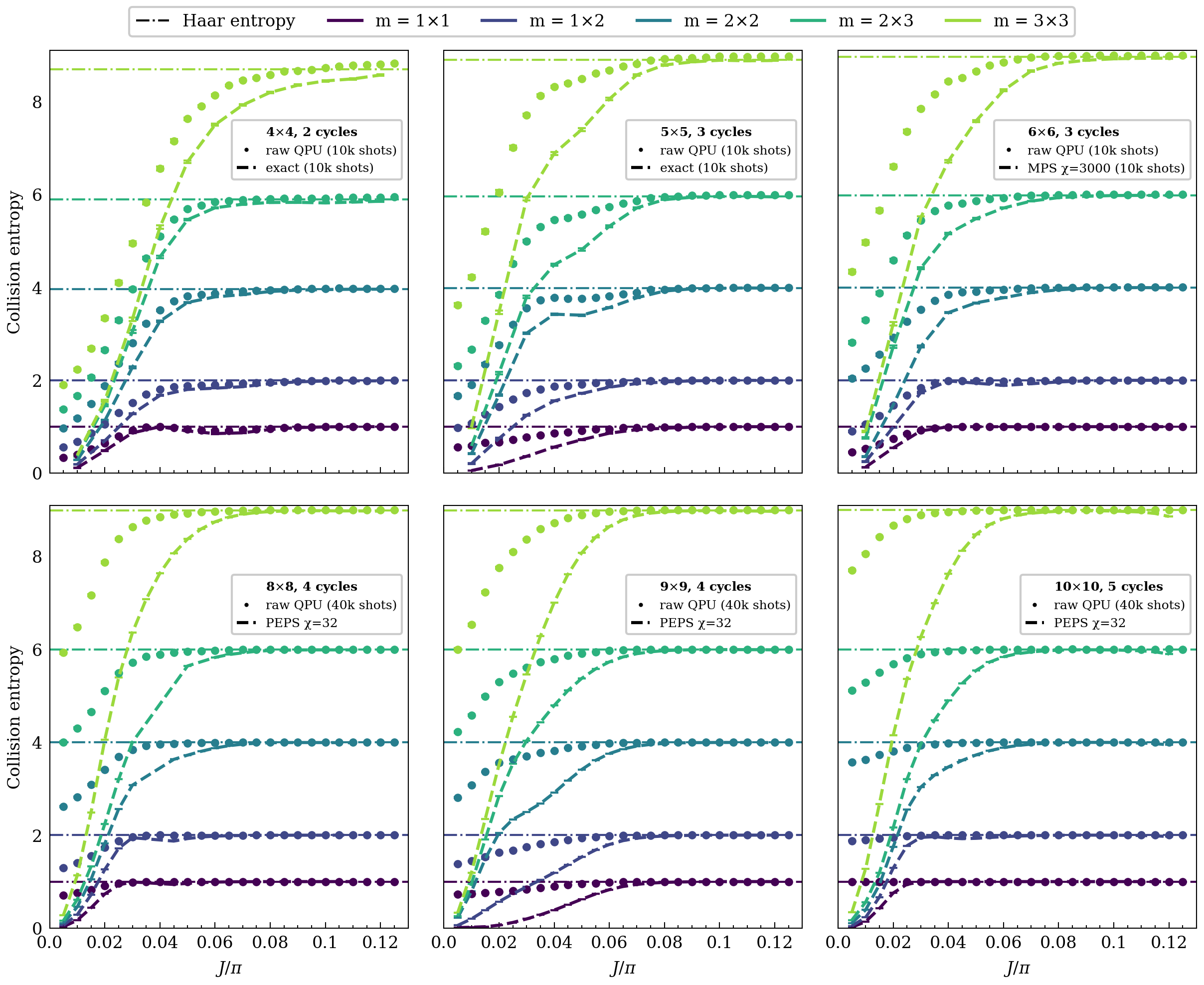}
    \caption{\textbf{Entropy from raw data.} Collision entropy of central marginals estimated from raw experimental samples. As the system and marginal sizes increase, the raw data retains the qualitative trends of the classical simulations, but mitigation is required for quantitative agreement. }
    \label{fig:raw_small_system}
\end{figure}
In this section we include additional quantum simulation results obtained from IBM Quantum's \verb|ibm_miami| and \verb|ibm_berlin| QPUs for the Floquet circuit ensemble described  in~\cref{Gateuijmain}. In \cref{sec:rawdata} we present the raw experimental data, and see that it exhibits the correct qualitative trends but has limited accuracy at system size $4\times 4$, that rapidly decreases further with system size.  In \cref{sec:additionaldata} we compare the error mitigation methods of \cref{sec:Error mitigation} on the central marginal of every system, and show the bias-corrected results for individual marginals and for spatial averages.

\subsection{Comparison of classical simulations to raw experimental data}
\label{sec:rawdata}

For the smallest system sizes ($4\times4$ and $5\times5$), exact classical simulations allow us to directly quantify error in the experimental results. For $6\times6$ and larger systems, we compare to approximate tensor-network simulations, which provide both a benchmark for experimental signal quality and a consistency check at small values of $J$, where the dynamics remain weakly entangled and are accurately captured by low--bond-dimension tensor network simulations. Results for the collision entropy of several marginal sizes, estimated from unmitigated (raw) experimental samples, are shown in \cref{fig:raw_small_system}. As expected, device noise generically suppresses the collision probability, thereby increasing the measured collision entropy across all $J$. We also observe a rapid degradation of the raw results with increasing system size: at sizes of $6\times6$ and above, the unmitigated QPU data systematically shifts the apparent crossover and underestimates the value of $J^\star$ that we aim to extract.

The unmitigated QPU results can be viewed as probing the same crossover in an effective open-system (noisy) analogue of the Heisenberg Floquet circuit. This perspective explains why the raw data is qualitatively similar to the ideal model while remaining quantitatively biased. In this work, rather than analyze the effective noisy analogue directly, we apply mitigation procedures to infer the trends expected for an idealized digital quantum simulation.
\begin{figure}[hbtp]
    \centering
    \includegraphics[width=1\linewidth]{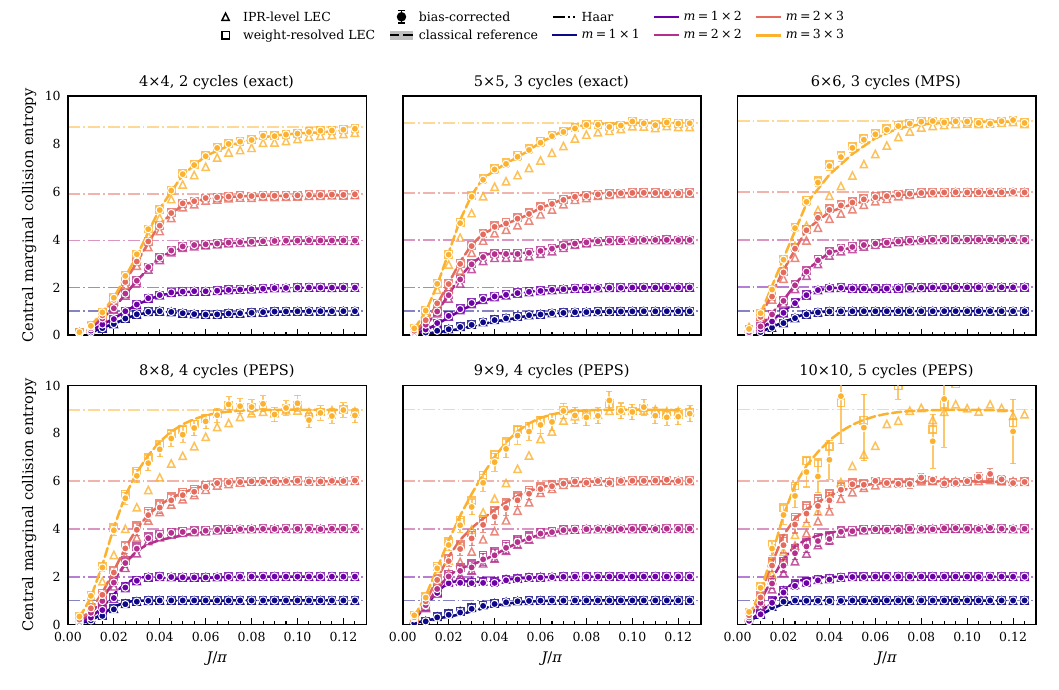}
    \caption{\textbf{Comparison of mitigation methods.} Collision entropy of the central marginal of each shape, for one disorder instance per system size, mitigated with IPR-level LEC (open triangles), weight-resolved LEC (open squares) and weight-resolved LEC with the bias correction of \cref{sec:wplecerrormodel} (filled points, with 68\% intervals of the total error). The two uncorrected methods are shown without error bars. Dashed lines and bands: classical reference, named in each panel title; dash-dot lines: Haar values.}
    \label{fig:mitigation_comparison}
\end{figure}

\subsection{Mitigated experimental data across a range of system sizes}\label{sec:additionaldata}

\Cref{fig:mitigation_comparison} compares the mitigation methods of \cref{sec:lec_variants,sec:wplecerrormodel} on the central marginal of each system. For single-qubit marginals the IPR contains a single string, and the IPR-level and weight-resolved forms of LEC coincide. For larger marginals IPR-level LEC underestimates the entropy at intermediate $J$, increasingly so with marginal size and most visibly for the $3\times3$ marginals of $6\times6$ to $9\times9$, as expected from the $J$-dependent bias of a single calibration factor discussed in \cref{sec:lec_variants}. Weight-resolved LEC follows the classical reference closely for these circuits, and the bias correction, which grows with depth and with the size of the light cone, shifts it only slightly. The error of the mitigated results increases with the size of the marginal, since the mitigation magnifies statistical fluctuations along with the signal; the $3\times3$ marginal of $10\times10$ at $J/\pi\gtrsim0.04$ is the clearest example.

\begin{figure}[hbtp]
    \centering
    \includegraphics[width=0.75\linewidth]{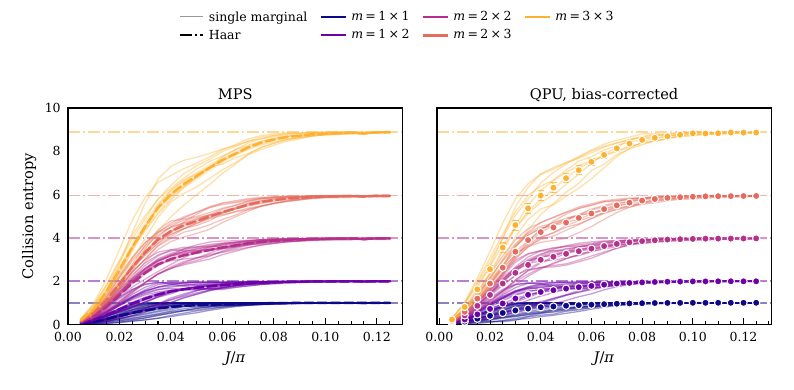}
    \caption{\textbf{Variation between marginals of the same shape.} Entropy for different choices of marginals of a fixed shape on $5\times5$ (3 cycles). Thin lines: every placement of each marginal shape, excluding marginals that contain a broken qubit, from MPS (left) and from bias-corrected weight-resolved LEC (right). Dashed lines and points: their spatial averages, taken in IPR space, with 68\% intervals of the total error on the QPU averages; dash-dot lines: Haar values. We see from both classical and (mitigated) quantum simulations that there is a variance in the thermalization of each marginal, justifying our choice to use spatial averaging to study disorder averaged physics.}
    \label{fig:variance_over_marginals}
\end{figure}

Turning to the measurement of disorder averaged properties of the ensemble of Floquet circuits, these provide a relaxed target since agreement between the QPU and exact results for fixed instances is sufficient but not necessary for accuracy in averages taken with respect to the disordered ensemble. While there is in principle no obstacle to averaging over disorder instances on a QPU, due to resource limitations we present marginal collision probabilities averaged over spatial translations of each marginal of a given size and shape within a single fixed disorder instances in \cref{fig:variance_over_marginals}.  While the entropy of neighboring marginal patches within a fixed instance is not independent (they have some non-zero covariance), this does not bias the ensemble average (it only reduces our convergence to it with the number of samples, since they are not independent samples).  

The spatially averaged results for the larger systems are shown in \cref{fig:entropy_large_systems} of the main text. Spatial averaging is often considered alongside disorder averaging in studies of thermalization, and the spatial average and disorder average of a local quantity are equivalent in the thermodynamic limit. Naturally we are far from this limit, so these results serve primarily to assess the readiness of the QPU to resolve questions about the disorder averaged ensemble.

In light of residual errors in the experimental results and resource limitation that prevent sufficient averaging over the disordered ensemble, we refrain from extracting quantitative finite-sized trends about the fate of the early-time crossover with increasing system size.   However, we find that applying a fixed constant relative threshold between the marginal collision entropy and the asymptotic value predicted for an ergodic state produces qualitative agreement with the expected trend of the crossover point $J^*(L)$ for a fixed marginal sized decreasing as the overall $L\times L$ system size increases.

\subsection{Approaching the thermodynamic limit}
\label{sec:tdl_sm}

In \cref{fig:TDL_main} we fixed the number of Floquet cycles to $n_F=6$ and compared the bulk-averaged collision entropy of small marginals across lattice sizes. Here we show how this comparison depends on the bias correction, and compare it with a second, independent classical method.

\begin{figure*}[!htbp]
    \centering
    \includegraphics[width=\textwidth]{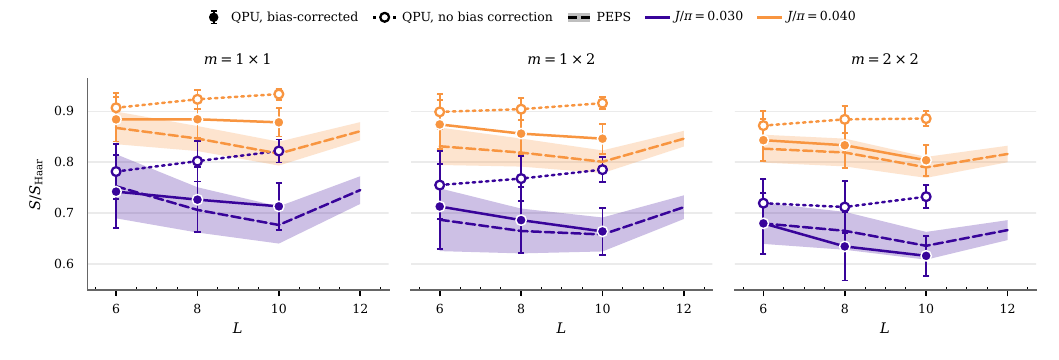}
    \caption{\textbf{Effect of the bias correction on the approach to the thermodynamic limit.} Subsystem-averaged collision entropy after 6 cycles, normalized by its Haar value, against the linear size $L$ of the lattice at $J/\pi=0.030$ and $0.040$, as in \cref{fig:TDL_main}. Open points (dotted): QPU data mitigated with weight-resolved LEC, with statistical error bars. Filled points (solid): the same data with the bias correction of \cref{sec:wplecerrormodel}, with 68\% intervals of the total error. Dashed lines: BP-PEPS at $\chi=32$, with the uncertainty of the subsystem average shaded. Averages run over the subsystems of each shape that do not touch the lattice boundary, and over those that also avoid broken couplers in the quantum device, replicated in the classical simulations.}
    \label{fig:tdl_debias}
\end{figure*}

\Cref{fig:tdl_debias} shows the QPU data with and without the bias correction. Without it, weight-resolved LEC overestimates the entropy, and the overestimate grows with $L$: at $J/\pi=0.030$ the $1\times1$ value rises from $0.78$ at $L=6$ to $0.82$ at $L=10$, whereas BP-PEPS falls from $0.75$ to $0.68$. The uncorrected data would therefore suggest that the entropy at fixed $J$ increases with system size, opposite to the classical trend. The bias correction removes this discrepancy. It is largest on the largest lattice, lowering $S/S_{\rm Haar}$ by $0.05$--$0.12$ at $L=10$, because the backward light cone of a bulk marginal after six cycles contains more gates on a larger lattice, and the residual bias of LEC grows with the number of gates in the light cone (\cref{sec:wplecerrormodel}). After the correction the QPU entropies are in broad agreement with BP-PEPS.

\begin{figure*}[!htbp]
    \centering
    \includegraphics[width=\textwidth]{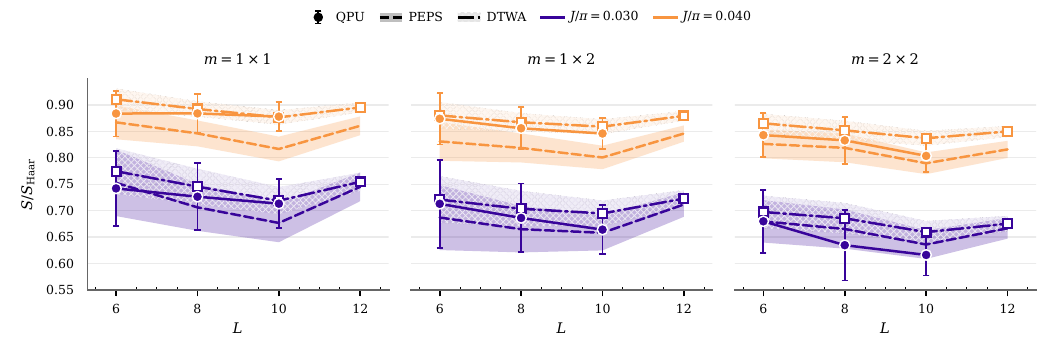}
    \caption{\textbf{Approach to the thermodynamic limit with cluster DTWA.} As \cref{fig:tdl_debias}, showing the bias-corrected QPU data (filled points, solid lines, 68\% intervals of the total error) together with BP-PEPS at $\chi=32$ (dashed lines, shaded bands) and cluster DTWA with $4\times4$ clusters (open squares, dash-dotted lines, hatched bands), both averaged over all bulk subsystems.}
    \label{fig:tdl_dtwa}
\end{figure*}

\Cref{fig:tdl_dtwa} adds cluster DTWA (\cref{app:dtwa:cluster}) for the same circuits and disorder realizations. DTWA lies slightly above BP-PEPS, consistent with the faster thermalization it shows in the benchmarks of \cref{app:dtwa}, and the difference is larger at $J/\pi=0.040$ than at $0.030$. The two classical methods nevertheless show the same dependence on $L$: both decrease from $L=6$ to $L=10$ and rise again at $L=12$. Since they approximate the dynamics in different ways, this common dependence supports attributing it to the disorder instances probed rather than to the approximation of either method. The 68\% interval of every bias-corrected QPU point overlaps the band of at least one of the two classical methods, showing all methods are in broad agreement.

\section{Tensor Network simulations}
\label{sec:Classical algorithms}

In this section, we outline the classical tensor-network simulation algorithms used in our study of the 2D Heisenberg Floquet system. Although these techniques can be powerful for small coupling $J$ and small marginal sizes $m$ and system size $n$, they become increasingly challenging as $J$ approaches the crossover transition.
The goal of these classical simulations is thus twofold: to validate the (mitigated) experimental results in regions of the parameter space beyond the capabilities of full-state vector simulation where tensor-network methods can be trusted, and to characterise the intrinsic limits of classical simulation itself.

To study the Heisenberg Floquet system, we use both Matrix Product States (MPS) and Belief-Propagation Tensor Network States (in this work dubbed BP-PEPS for projected entangled pair states on the square lattice)~\cite{tindall2023gauging}. We find MPS effective for small and medium system sizes, up to $8\times8$ at the largest bond dimensions, while BP-PEPS simulations enable exploration of larger lattices but only for moderate values of $J$, where entanglement remains sufficiently weak for controlled bond dimensions. In both approaches, as $J$ approaches the crossover region and marginal collision entropies $S_{2,Z}[A]$ of increasing size $n_A=\abs{A}$ are targeted, we find that the required bond dimension for their accurate computation at fixed precision increases rapidly and controlled convergence is lost due to increasing computational costs.

We  begin by describing the matrix product state (MPS) techniques used in this work. MPS wavefunctions at a given bond dimension inherently encode area-law entangled states in one-dimensional systems~\cite{eisert2010colloquium,schollwock2011}, thus these are known to be in general computationally inefficient for time-dynamics simulations of two-dimensional quantum systems~\cite{stoudenmire2012studying}. Nonetheless, up to $8\times8$ they provide the most accurate reference available to us, and we use them both to benchmark the BP-PEPS simulations and for some of the results of the main text.

\subsection{Methods: MPS and Belief-Propagation PEPS}
\label{sec:tn_methods}

\paragraph{MPS techniques.} We tested time-evolving block decimation (TEBD)~\cite{vidal2004efficient}, the time-dependent variational principle (TDVP)~\cite{haegeman2011time,haegeman2016unifying} and the density matrix renormalisation group for quantum circuits~\cite{ayral2023density}, and found the best compromise between performance and accuracy in a hybrid TEBD-TDVP scheme, with TEBD for the first Floquet cycle and two-site TDVP for the subsequent cycles. The lattice is mapped onto the chain in row-major order, so that horizontal bond gates act on chain neighbours while vertical ones connect sites $N_y$ apart. TEBD applies each vertical gate through a ladder of $N_y-1$ SWAP gates and back, each truncated to the same bond dimension as a physical gate, so that on an $N\times N$ lattice SWAPs make up a fraction $(N-1)/N$ of the two-site operations of a cycle, $87.5\%$ on $8\times8$. In the TDVP stage we instead apply each bond term as an MPO with the local-TDVP scheme of Ref.~\cite{sander2025quantum}, which restricts the sweep to the $N_y+1$ chain sites between the two qubits and needs no SWAP network; this is the main reason why the combined scheme outperforms TEBD alone. The maximum bond dimensions we could simulate are listed in Table~\ref{tab:tn_sims}. At bond dimension $\chi$ the memory of an MPS of $N$ qubits grows as $N\chi^2$ and the time of each two-site update as $\chi^3$.
All MPS values of the marginal IPR$_m$ reported below are obtained by exact contraction of the MPS on the sites of the marginal, so that they carry no statistical error; the $10^4$ samples of the state that we also draw after each Floquet cycle are not used for these values. Some of the $6\times6$ results at $\chi=4096$ shown in the main text were instead estimated from $10^4$ samples of the state, which carries a relative statistical error on the $3\times3$ IPR of typically $1$--$2\%$ and of up to $5\%$ for individual patches, as we have checked against the exact contraction of the same states.

\paragraph{PEPS techniques.}

In tensor network states beyond one dimension, a quantum state is represented as a tensor network whose geometry mirrors that of the underlying lattice, which in our case corresponds to a square lattice~\cite{verstraete2004renormalization,verstraete2008matrix}. As for MPS, the entanglement supported by the PEPS ansatz is controlled by the dimension of the virtual bonds connecting neighbouring tensors.
Contractions of PEPS to \emph{exactly} compute norms and observable expectation values are however computationally infeasible~\cite{schuch2007computational}, and therefore several heuristic contraction schemes have been proposed to circumvent this problem. In this work, we used the belief-propagation (BP) approximation~\cite{alkabetz2021tensor,tindall2023gauging}, which has been recently successfully used to simulate several
quantum experiments~\cite{tindall2024efficient,tindall2025dynamics}.

Time evolution is implemented through application of the Floquet-layer operator (see Fig.~\ref{fig:layers_choices}) via the simple update BP algorithm~\cite{jiang2008accurate,alkabetz2021tensor,tindall2023gauging}. In this approach, following each two-site gate application, a truncated singular value decomposition is performed with truncation precision set to $10^{-12}$ and maximum dimension $\chi$, conditioned on message-environment tensors providing a rank-1 approximation of the environment network calculated with the BP algorithm. On the square lattice each tensor has four virtual bonds, so that the memory of the PEPS grows as $N\chi^4$ and the time of each gate update as $\chi^5$. In our simulations, BP is performed without damping factors,
using a maximum number of iterations equal to 25 and a convergence tolerance of $10^{-8}$.

While the BP approximation is exact on tree tensor networks, it often remains accurate for lattices containing short loops when the correlation length is limited or the quantum state exhibits moderate amount of entanglement. In our approach, after the evolution of each Floquet layer, we computed length$-m$ Pauli string expectation values using the BP-based contraction method with same precision parameters used above. From these, the marginal IPR$_m$ is constructed using the Parseval formula, Eq.~\eqref{ParsevalIPR} (or Eq.~\eqref{eq:parseval} of the main text).
We find BP-contraction computationally more efficient than the boundary-MPS contraction method of Ref.~\cite{tindall2025dynamics}. Although the square lattice contains short loops, we found the computation of expectation values to build IPR$_m$
without loop corrections already computationally very expensive, and therefore for marginals with large support we employed a cumulant expansion approximation to evaluate the expectation values required for the IPR (see Sec.~\ref{sec:cumulant_expansion} below).

The BP-PEPS implementation used in this work is based on the open-source Julia package \verb|TensorNetworkQuantumSimulator.jl|\cite{tindall_tnqs_2025}, which enables tensor-network simulations on lattices of arbitrary topology and is built upon the \verb|ITensorNetworks.jl|\cite{fishman_itensornetworks_2026} library.

\subsection{Results}
\label{sec:tn_results}

MPS and BP-PEPS tensor-network simulations enable us to probe different system sizes and entanglement regimes within controlled approximations. We first define the criteria used to assess convergence, and then discuss the MPS results, their comparison with BP-PEPS, and the BP-PEPS results on large lattices.

\begin{figure}[tbp]
    \centering
    \includegraphics[width=\linewidth]{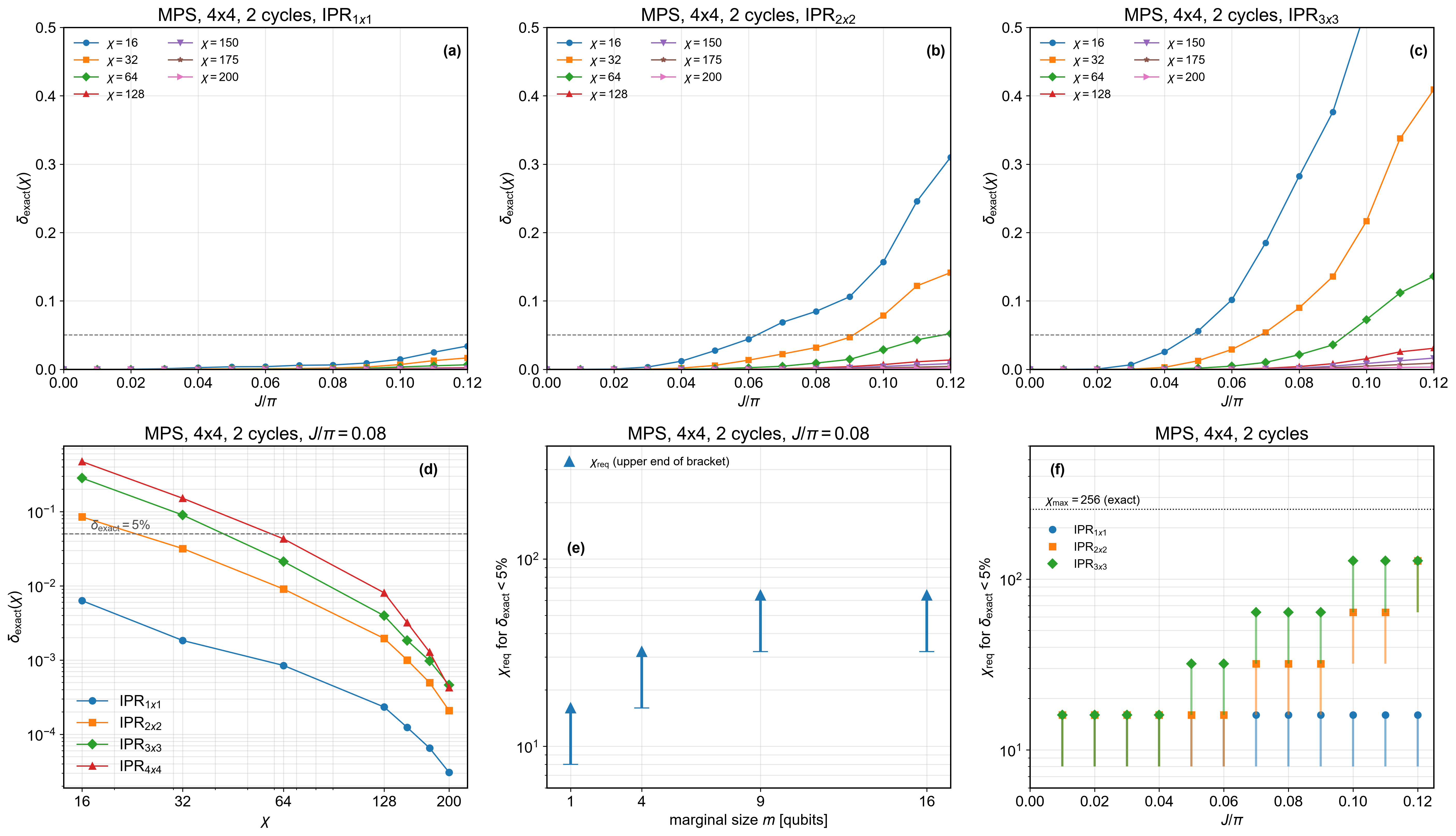}
    \caption{\textbf{MPS accuracy and required bond dimension on $4\times 4$ (2 cycles), where $\chi_{\textrm{max}}=256$ is exact.}
    \textbf{a}--\textbf{c}, relative error $\delta_{\rm exact}$ (Eq.~\eqref{eq:delta_exact}) on marginal IPR$_m$ with respect to the exact state vector, which coincides with MPS at $\chi_{\textrm{max}}=256$, as a function of $J/\pi$ for different MPS bond dimensions. All IPRs are obtained by exact contraction of the MPS and averaged over the translated patches. We show marginal IPR$_{1\times1}$ ($1\times 1$ block) (panel (a)), IPR$_{2\times2}$ (panel (b)), and IPR$_{3\times3}$ (panel (c)) for 2 Floquet cycles. The dashed line marks $\delta_{\rm exact}=0.05$, which corresponds to $\delta S_{2,Z}\simeq0.07$ bits. \textbf{d}, $\delta_{\rm exact}$ as a function of $\chi$ for the $1\times1$ to $4\times4$ marginals at $J/\pi=0.08$. \textbf{e}, bond dimension $\chi_{\rm req}$ required for $\delta_{\rm exact}<0.05$ at $J/\pi=0.08$ as a function of the marginal size $m$. \textbf{f}, $\chi_{\rm req}$ as a function of $J/\pi$ for the $1\times1$, $2\times2$ and $3\times3$ marginals. Here $\chi_{\rm req}$ is the smallest bond dimension of our grid for which $\delta_{\rm exact}<5\%$ holds at that and at every larger bond dimension. Since the grid is discrete, the true threshold lies between $\chi_{\rm req}$ (symbols) and the next smaller bond dimension of the grid, and the vertical bars span this interval.}
    \label{fig:MPS_4x4}
\end{figure}

\paragraph{Criterion for convergence.} Except on $4\times4$ and $5\times5$, convergence is defined operationally, since an exact reference is not available at the sizes studied here. Extrapolations in $1/\chi\rightarrow0$, often used for this purpose, proved unreliable in the crossover regime at the bond dimensions we could reach. We therefore increase the bond dimension along a sequence, in practice often doubling it at each step, and quote the successive refinement
\begin{equation}\label{eq:delta_chi}
\Delta_\chi=|\mathrm{IPR}_m(\chi)-\mathrm{IPR}_m(\chi')|/\mathrm{IPR}_m(\chi),
\end{equation}
where $\chi'$ is the bond dimension preceding $\chi$ in the sequence, and, where an exact reference is available, the true error
\begin{equation}\label{eq:delta_exact}
\delta_{\rm exact}=|\mathrm{IPR}_m-\mathrm{IPR}_m^{\rm exact}|/\mathrm{IPR}_m^{\rm exact}.
\end{equation}
On $4\times4$, where MPS at $\chi_{\textrm{max}}=2^{n/2}=256$ is exact, Fig.~\ref{fig:MPS_4x4} shows that $\delta_{\rm exact}$ grows with $J/\pi$ and with the size of the marginal, so that at fixed precision increasingly large bond dimensions are required. We classify each calculation and coupling as
\begin{equation}
  \Delta_\chi<0.5\%\ \text{(reliable)},\qquad
  0.5\%\le\Delta_\chi<5\%\ \text{(borderline)},\qquad
  \Delta_\chi\ge5\%\ \text{(uncontrolled)} ,
  \label{eq:classification}
\end{equation}
shaded green, orange and red in the figures. Beyond the coupling $J^*$ at which the refinement of a calculation is largest, the refinement decreases again because the marginals struggle to approach their random-state values, not because the calculation converges, and couplings beyond $J^*$ with a refinement below $0.5\%$ are marked grey (stable but unconverged) in the reliability map (Fig.~\ref{fig:reliability}). When the largest refinement available is uncontrolled we do not extrapolate. Since $S_{2,Z}=-\log_2\mathrm{IPR}$, a relative error $\varepsilon$ on the IPR corresponds to an error $\varepsilon/\ln2$ on $S_{2,Z}$, so that $5\%$ amounts to about $0.07$ bits of collision entropy.

We remark, however, that the stability of a result under bond-dimension refinement does not guarantee its accuracy. The BP-PEPS results are affected by three approximations,
\begin{equation}
  \delta\,\mathrm{IPR}_m \;\sim\; \delta_{\rm state} + \delta_{\rm env} + \delta_{\rm cum},
  \label{eq:taxonomy}
\end{equation}
namely the truncation of the state at finite $\chi$, the belief-propagation approximation of the environment, and the truncation of the cumulant expansion of Sec.~\ref{sec:cumulant_expansion}. We use Eq.~\eqref{eq:taxonomy} as a classification of error sources rather than as a rigorous additive bound. Of these, only $\delta_{\rm state}$ is probed by $\Delta_\chi$. The convergence of the BP messages ensures that a fixed point has been found, but not that this fixed point is accurate, and a refinement in $\chi$ at fixed cumulant order does not provide any information on the cumulant truncation. For the MPS results only the first term is present, since the MPS values of the IPR are obtained by exact contraction.

\begin{figure}[tbp]
    \centering
    \includegraphics[width=\linewidth]{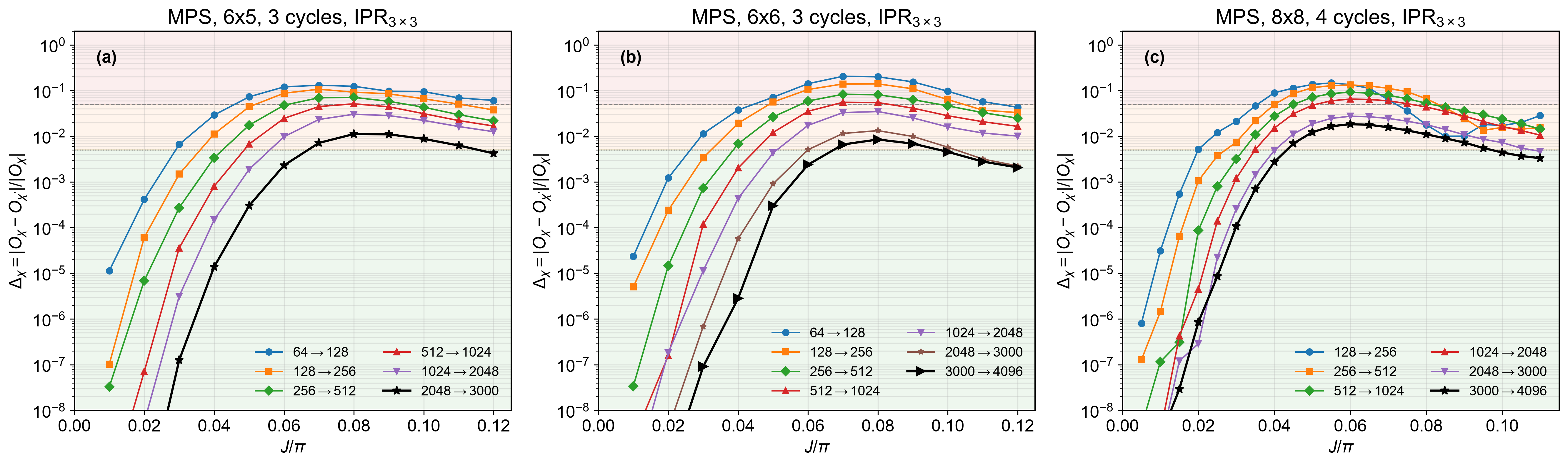}
    \caption{\textbf{MPS computational complexity at a fixed marginal size ($3\times 3$ at $[L/2]$ cycles, for linear system size $L$ and total number of qubits $n=L\times L$), varying the total system size $n$.}
    \textbf{a}--\textbf{c}, successive refinements $\Delta_\chi$ (Eq.~\eqref{eq:delta_chi}) of the marginal IPR$_9$ ($3\times 3$ block) as a function of $J/\pi$ for (a) 6$\times$5, (b) 6$\times$6, and (c) 8$\times$8 systems. Each curve compares two successive bond dimensions of the sweep, up to $\chi_{\textrm{max}}=3000$ for 6$\times$5 and $\chi_{\textrm{max}}=4096$ for 6$\times$6 and 8$\times$8; the last refinement is drawn in black. All IPRs are obtained by exact contraction of the MPS. For 6$\times$5 and 6$\times$6 the IPR is averaged over all translated $3\times3$ patches, and for 8$\times$8 it is the patch with lower-left site $(4,4)$. The fields of 6$\times$5 are those of a $5\times6$ block of the $8\times8$ field realisation. Shading marks the bands of Eq.~\eqref{eq:classification}.}
    \label{fig:MPS_system_size}
\end{figure}

With this in mind, we studied the computational scaling of MPS results as a function of system size in Fig.~\ref{fig:MPS_system_size}. Panels (a--c) show the successive refinements $\Delta_\chi$ of the $3\times 3$ marginal as a function of $J/\pi$ for different system sizes and bond dimensions.
In particular, we simulated $6\times 5$ up to $\chi_{\text{max}}=3000$ and $6\times 6$ and $8\times8$ up to $\chi_{\text{max}}=4096$.
At weak coupling the refinements decrease rapidly at each step of the sweep, and for $J/\pi\le0.03$ the refinement $1024\to2048$ is at most $0.12\%$ on all three lattices. Near the crossover the refinements decrease more slowly. On $8\times8$ the refinement $1024\to2048$ reaches $6.5\%$ at $J/\pi=0.06$, the next one, $2048\to3000$, $2.8\%$, and the last one, $3000\to4096$, $1.8\%$; the last refinement is at least $1\%$ for $0.05\le J/\pi\le0.08$ and below $1\%$ elsewhere. For $J/\pi\ge0.025$ the MPS values decrease monotonically as $\chi$ grows. We remark that the last two steps increase the bond dimension by factors of $1.46$ and $1.37$ rather than $2$, so that they understate the change that a further doubling would produce. On $6\times5$ the refinement $2048\to3000$ remains below $1.2\%$, and on $6\times6$ the last refinement $3000\to4096$ remains below $0.9\%$ at every coupling $J/\pi\le0.12$. On both lattices the refinements are largest at $J/\pi\simeq0.08$ and decrease again at stronger coupling, as on $8\times8$. On these two lattices the $3\times3$ IPR is therefore reliable or borderline in the sense of Eq.~\eqref{eq:classification} over the whole range of couplings examined, and on $8\times8$ it is reliable up to $J/\pi=0.04$ and borderline from $J/\pi=0.045$ to $0.09$.
As expected, MPS simulations become increasingly difficult with system size, as the bond dimension required for a given accuracy grows exponentially with the linear size of the system for generic two-dimensional states.

\begin{table}[t]
\centering
\renewcommand{\arraystretch}{1.15}
\begin{tabular}{@{}l l l l l@{}}
\toprule
\textbf{Method} & \textbf{Lattice} & \textbf{Cycles} & \textbf{Bond dimensions} $\chi$ & $J/\pi$ \textbf{(step)}\\
\midrule
MPS & $4\times4$ & $2$ & $16$--$256$ & $0$--$0.25$ ($0.01$)\\
MPS & $5\times5$ & $2$, $4$, $6$ & $64$--$4096$ & $0$--$0.25$ ($0.01$)\\
MPS & $6\times5$ & $3$ & $64$--$3000$ & $0$--$0.12$ ($0.01$)\\
MPS & $6\times6$ & $3$, $6$ & $64$--$4096$ & $0$--$0.12$ ($0.01$)\\
MPS & $8\times8$ & $4$, $6$ & $128$--$4096$ & $0$--$0.11$ ($0.005$)\\
BP-PEPS & $4\times4$ & $2$, $4$ & $4$--$64$ & $0$--$0.25$ ($0.01$)\\
BP-PEPS & $8\times8$ & $2$, $4$--$6$ & $4$--$32$, $48$, $64$ & $0.005$--$0.12$ ($0.005$)\\
BP-PEPS & $9\times9$ & $2$, $4$--$6$ & $4$--$32$ & $0.005$--$0.12$ ($0.005$)\\
BP-PEPS & $10\times10$ & $2$, $4$--$6$ & $16$, $32$, $48$ & $0.005$--$0.12$ ($0.005$)\\
BP-PEPS & $12\times12$ & $2$--$6$ & $32$ & $0.005$--$0.125$ ($0.005$)\\
\bottomrule
\end{tabular}
\caption{Tensor-network simulations of this section. On $4\times4$ and $5\times5$ the largest MPS bond dimension is exact. The $8\times8$ MPS calculation at six cycles covers $J/\pi=0.01$--$0.04$ (step $0.01$) with $\chi=128$--$3000$. On $8\times8$, $9\times9$ and $10\times10$ the BP-PEPS calculations at $\chi=32$ cover two, four, five and six cycles; on $10\times10$, $\chi=48$ was run for $J/\pi=0.025$ and $0.05$ after $2$ to $6$ cycles, for marginals of up to $2\times2$ sites. On $8\times8$ BP-PEPS reaches $\chi=48$ and $64$ for marginals of up to $2\times3$ sites, and $\chi=48$ for the $3\times3$ marginal for $0.04\le J/\pi\le0.08$; marginals of up to $2\times2$ sites were also computed at $\chi=4$--$48$ after $2$ to $6$ cycles on $6\times6$ and $2$ to $8$ cycles on $8\times8$. On $12\times12$ two disorder realisations were simulated, the second for $J/\pi\le0.1$; the $6\times6$ MPS calculation at six cycles enters the approach to the thermodynamic limit of the main text. The stability of each calculation under bond-dimension refinement is summarised in Fig.~\ref{fig:reliability}.}
\label{tab:tn_sims}
\end{table}

Since the bond dimension fixes the computational resources rather than the accuracy of the simulation, we have also recorded the discarded weight $\epsilon_i$ of every truncation, and the product $F_{\rm proxy}=\prod_i(1-\epsilon_i)$ commonly used as an estimate of the fidelity~\cite{zhou2020limits,ayral2023density}. On $4\times4$, where the true fidelity is accessible, the marginal IPRs remain accurate long after the global fidelity has decayed: at $J/\pi=0.12$ and $\chi=64$ the fidelity falls to $0.07$ after four cycles, while the collision entropies of the $1\times1$, $2\times2$ and $3\times3$ marginals remain within $0.003$, $0.04$ and $0.1$ bits of the exact values, partly because these marginals approach their maximal entropy within one or two cycles at this coupling. $F_{\rm proxy}$, on the other hand, underestimates the true infidelity by a median factor of five, so that it is only an upper bound on the fidelity. On $8\times8$ the TEBD layer dominates the loss of fidelity, $76\%$ to $87\%$ of its discarded weight coming from the SWAP networks.

\begin{figure}[tbp]
    \centering
    \includegraphics[width=\linewidth]{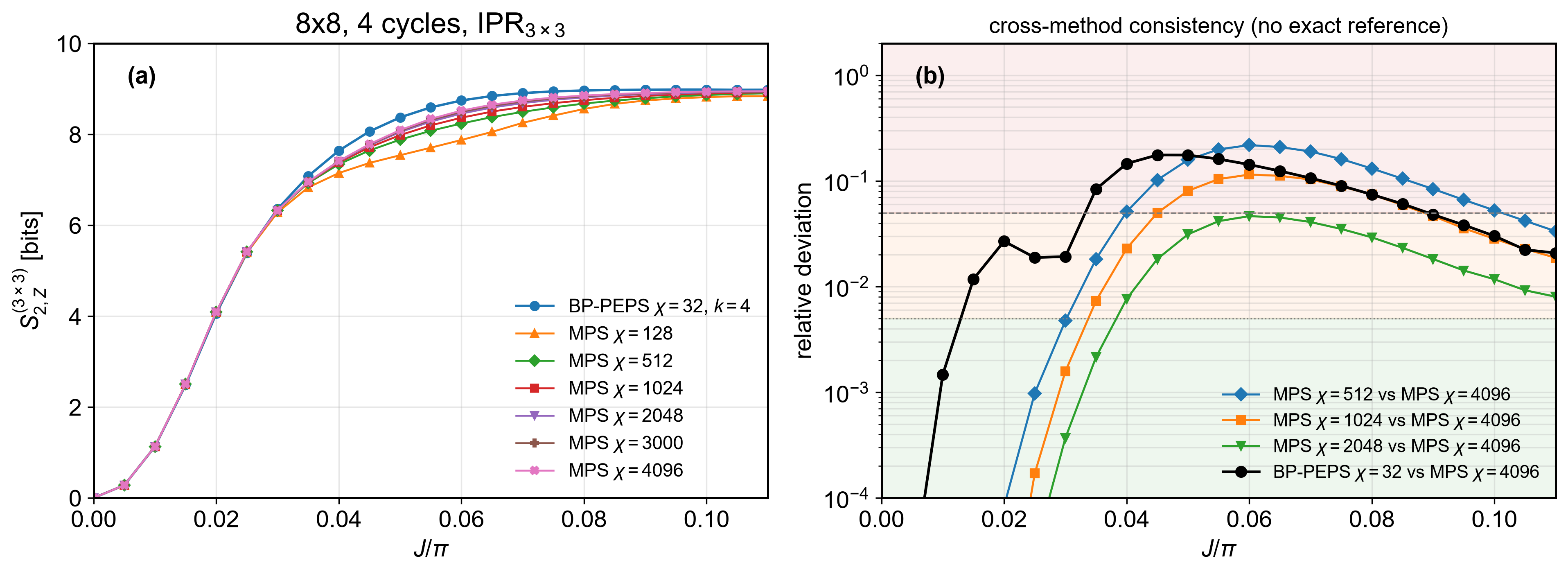}
    \caption{\textbf{BP-PEPS compared with MPS on $8\times8$.} \textbf{a}, collision entropy of the $3\times3$ marginal with lower-left site $(4,4)$ on $8\times8$ after four Floquet cycles, computed with BP-PEPS at $\chi=32$ with a fourth-order cumulant expansion and with MPS at $\chi=128$, $512$, $1024$, $2048$, $3000$ and $4096$. \textbf{b}, relative deviations underlying panel (a), all measured with respect to MPS at $\chi=4096$. The 8$\times$8 MPS values are obtained by exact contraction of the MPS. Since neither method is exact at $8\times8$, panel (b) is a consistency check between two approximate calculations; the comparison of BP-PEPS with the exact result on $4\times4$ is shown in \cref{sm:fig:MPS_vs_PEPS}. Shading marks the bands of Eq.~\eqref{eq:classification}.}
    \label{fig:MPS_vs_PEPS}
\end{figure}

\begin{figure}[tbp]
    \centering
    \includegraphics[width=0.9\linewidth]{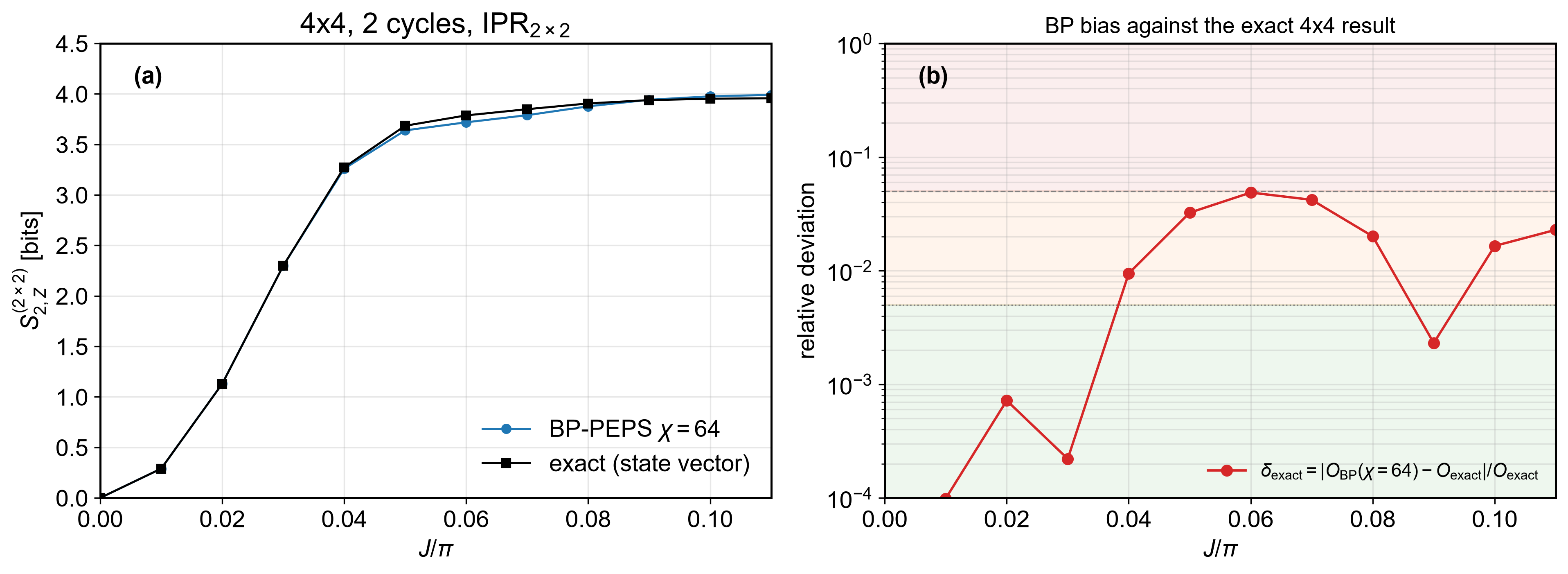}
    \caption{\textbf{BP-PEPS against the exact result on $4\times4$.} \textbf{a}, collision entropy of the $2\times 2$ marginal as a function of $J$ after two Floquet cycles, computed using BP-PEPS at $\chi=64$ and a state-vector simulation of the same circuit, which coincides with MPS at $\chi=256$. \textbf{b}, relative error $\delta_{\rm exact}$ (Eq.~\eqref{eq:delta_exact}) of the BP-PEPS result. The corresponding comparison on $8\times8$, where neither method is exact, is shown in Fig.~\ref{fig:MPS_vs_PEPS}. Shading marks the bands of Eq.~\eqref{eq:classification}.}
    \label{sm:fig:MPS_vs_PEPS}
\end{figure}

Figs.~\ref{sm:fig:MPS_vs_PEPS} and~\ref{fig:MPS_vs_PEPS} compare MPS and BP-PEPS on $4\times4$ and $8\times8$, for the $2\times2$ and $3\times3$ marginal collision entropies, respectively, as a function of $J$. On $4\times4$, where the reference is exact, BP-PEPS can be validated over a wide range of $J/\pi$. On $8\times8$ neither method is exact, and the comparison is a cross-check between two independent approximations: on the patch of Fig.~\ref{fig:MPS_vs_PEPS}(a) the two agree within $3\%$ for $J/\pi\le0.03$, but on a second patch they already differ by $16\%$ at $J/\pi=0.03$ (see below), and the MPS results at increasing bond dimensions ($\chi=128$ to $4096$) drift towards BP-PEPS (Fig.~\ref{fig:MPS_vs_PEPS}(b)). For $0.035\le J/\pi\le0.08$, however, the MPS IPR at $\chi=4096$ exceeds the BP-PEPS one by $8\%$ to $21\%$, with the largest gap at $J/\pi\simeq0.045$--$0.05$, whereas the MPS value changes by at most $1.8\%$ between $\chi=3000$ and $4096$. The gap closes again at stronger coupling, to $2.1\%$ at $J/\pi=0.11$, as both IPRs approach their ergodic value.
Figure~\ref{sm:fig:MPS_vs_PEPS}(a) also shows that, even at the smallest system size where MPS provides the ground truth, the $2\times2$ marginal IPR computed with BP-PEPS at $\chi=64$ displays a noticeable discrepancy with respect to the exact result (MPS at $\chi=256$), which reaches $4.9\%$ in the IPR near $J/\pi=0.06$ (Fig.~\ref{sm:fig:MPS_vs_PEPS}(b)).

We have repeated the $8\times8$ comparison on a second patch, with lower-left site $(3,3)$. At $J/\pi=0.03$ the MPS results change by at most $0.2\%$ between $\chi=1024$ and $3000$ on both patches; they agree with BP-PEPS to $2\%$ on the patch $(4,4)$ but lie $16\%$ above it on the patch $(3,3)$. For $J/\pi\ge0.04$ the MPS results are still decreasing as $\chi$ grows, and at $\chi=4096$ they lie $15\%$ to $17\%$ above BP-PEPS at $J/\pi=0.04$ and $0.06$ on both patches, and $8\%$ to $9\%$ above at $J/\pi=0.08$, while the BP-PEPS values are stable under $\chi=32\to48$. These gaps bound the combined error of the two methods on $8\times8$, without assigning it to either of them, and they are considerably larger than the contraction error found on $4\times4$. On the same runs TEBD alone is three to ten times less accurate than the hybrid TEBD-TDVP scheme at equal bond dimension, and the hybrid scheme reaches a given accuracy in about a third of the computing time.

\begin{figure}[tbp]
    \centering
    \includegraphics[width=\linewidth]{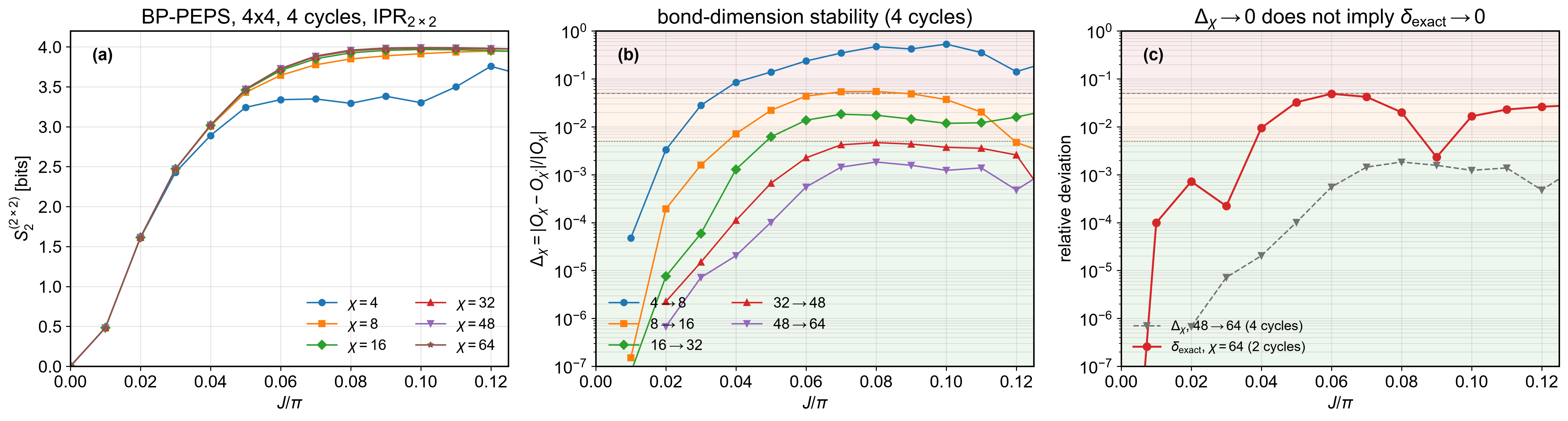}
    \caption{\textbf{BP-PEPS convergence in bond dimension.} \textbf{a}, collision entropy of the $2\times 2$ marginal as a function of $J$ on $4\times4$ after four Floquet cycles, computed using BP-PEPS at $\chi=4,\dots,64$. \textbf{b}, successive refinements $\Delta_\chi$ (Eq.~\eqref{eq:delta_chi}) of the same data; for the last step $48\to64$ we find $\Delta_\chi\le1.8\times10^{-3}$ over the whole range $J/\pi\le0.125$. \textbf{c}, the same last refinement (grey, dashed) compared with the relative error $\delta_{\rm exact}$ of BP-PEPS at $\chi=64$ with respect to the exact result, which is available for the two-cycle data set of \cref{sm:fig:MPS_vs_PEPS}(a). The residual BP bias reaches $4.9\%$ near $J/\pi=0.06$, more than an order of magnitude above $\Delta_\chi$.}
    \label{fig:conv_bp_4x4}
\end{figure}

Figure~\ref{fig:conv_bp_4x4} shows that on $4\times4$ after four Floquet cycles BP-PEPS requires a relatively large bond dimension to become stable under refinement at the level of $10^{-3}$.
The comparison with the exact $4\times4$ result in Fig.~\ref{fig:conv_bp_4x4}(c) shows that stability and accuracy have to be assessed separately: while the BP-PEPS result becomes essentially independent of the bond dimension, it retains a systematic bias of a few percent with respect to the exact result, which is largest where the dynamics approach the crossover. In this range, after two cycles, BP-PEPS overestimates the IPR, i.e.\ underestimates the collision entropy. On the square lattice a bias of this kind is expected, since the short four-site loops are precisely the correlations that belief propagation neglects.

\begin{figure}[tbp]
  \centering
  \includegraphics[width=\linewidth]{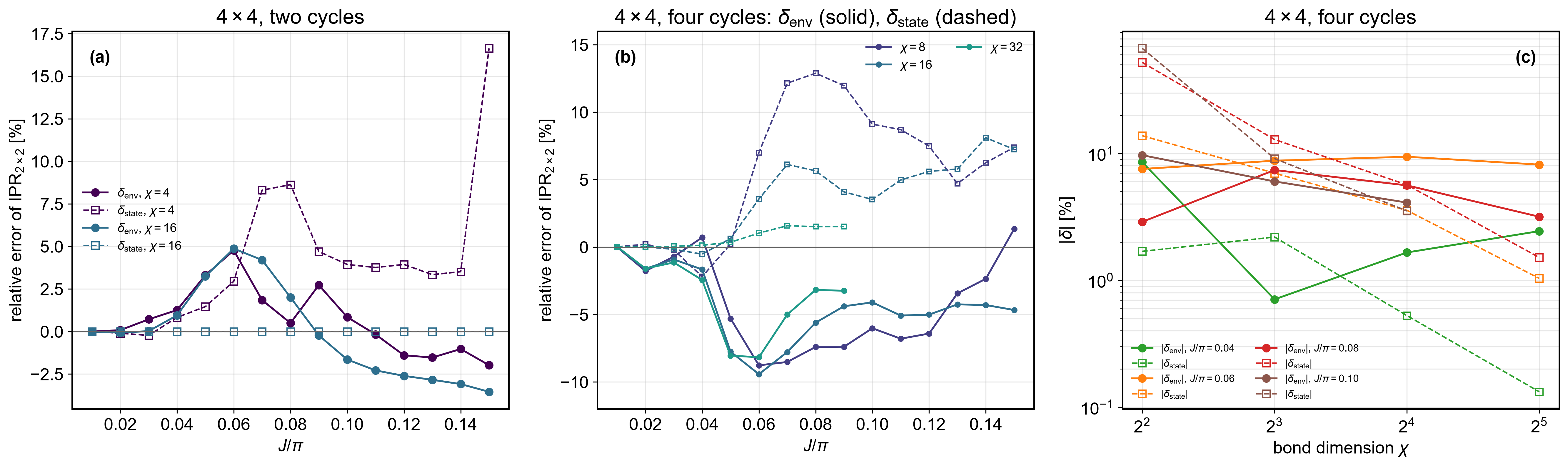}
  \caption{\textbf{Contraction and truncation errors of BP-PEPS on $4\times4$.}
  Relative error of the central $2\times2$ IPR, split into the error of the BP contraction, $\delta_{\rm env}$, and of the state truncation, $\delta_{\rm state}$, by contracting each BP-PEPS state exactly. \textbf{a}, two cycles; for $\chi\ge16$ the state is not truncated and $\delta_{\rm state}$ vanishes. \textbf{b}, four cycles, $\chi=8$--$32$ ($\chi=32$ for $J/\pi\le0.09$), with $\delta_{\rm env}$ solid and $\delta_{\rm state}$ dashed. \textbf{c}, $|\delta_{\rm env}|$ and $|\delta_{\rm state}|$ after four cycles as a function of $\chi$ at fixed $J/\pi$.}
  \label{fig:bp_env_state}
\end{figure}

This bias does not originate from the truncation of the state. Indeed, after two cycles on $4\times4$ the BP-PEPS state is not truncated for $\chi\ge16$: the bond dimension saturates at $16$ and the largest discarded weight of any gate is of order $10^{-31}$. Contracting this state exactly into a state vector reproduces the exact IPR to machine precision, so that the whole deviation in Fig.~\ref{sm:fig:MPS_vs_PEPS}(b) comes from the BP contraction used to evaluate the Pauli strings. We have also checked that the exact reference, a state-vector simulation of the circuit, agrees with the $\chi=256$ MPS result at every coupling within the statistical error of its $10^4$ samples.
After four cycles the state is truncated, and the two contributions of Eq.~\eqref{eq:taxonomy} can be separated by contracting the same BP-PEPS state exactly: with $\mathrm{IPR}_{\rm state}$ evaluated on the exactly contracted state, $\delta_{\rm env}=(\mathrm{IPR}_{\rm BP}-\mathrm{IPR}_{\rm state})/\mathrm{IPR}_{\rm exact}$ and $\delta_{\rm state}=(\mathrm{IPR}_{\rm state}-\mathrm{IPR}_{\rm exact})/\mathrm{IPR}_{\rm exact}$ (Fig.~\ref{fig:bp_env_state}). The state truncation decreases steadily with $\chi$, and at $J/\pi=0.06$ it is $7.0\%$, $3.6\%$ and $1.0\%$ for $\chi=8$, $16$ and $32$. The contraction error instead remains approximately constant: at the same bond dimensions it is $-8.8\%$, $-9.4\%$ and $-8.2\%$, of opposite sign to the two-cycle value, and it dominates the total error for every $\chi\ge16$ in the crossover region. The two contributions can also cancel each other. At $J/\pi=0.08$ and $\chi=16$ they are $-5.6\%$ and $+5.7\%$ and the total error is only $0.05\%$, so that the agreement with a reference at a single bond dimension cannot be taken as evidence of accuracy. Finally, the fidelity of the truncated state is not a good indicator of either contribution: at $J/\pi=0.08$ and $\chi=32$ it is $0.82$, while $\delta_{\rm state}=1.5\%$.

\begin{figure}[tbp]
    \centering
    \includegraphics[width=\linewidth]{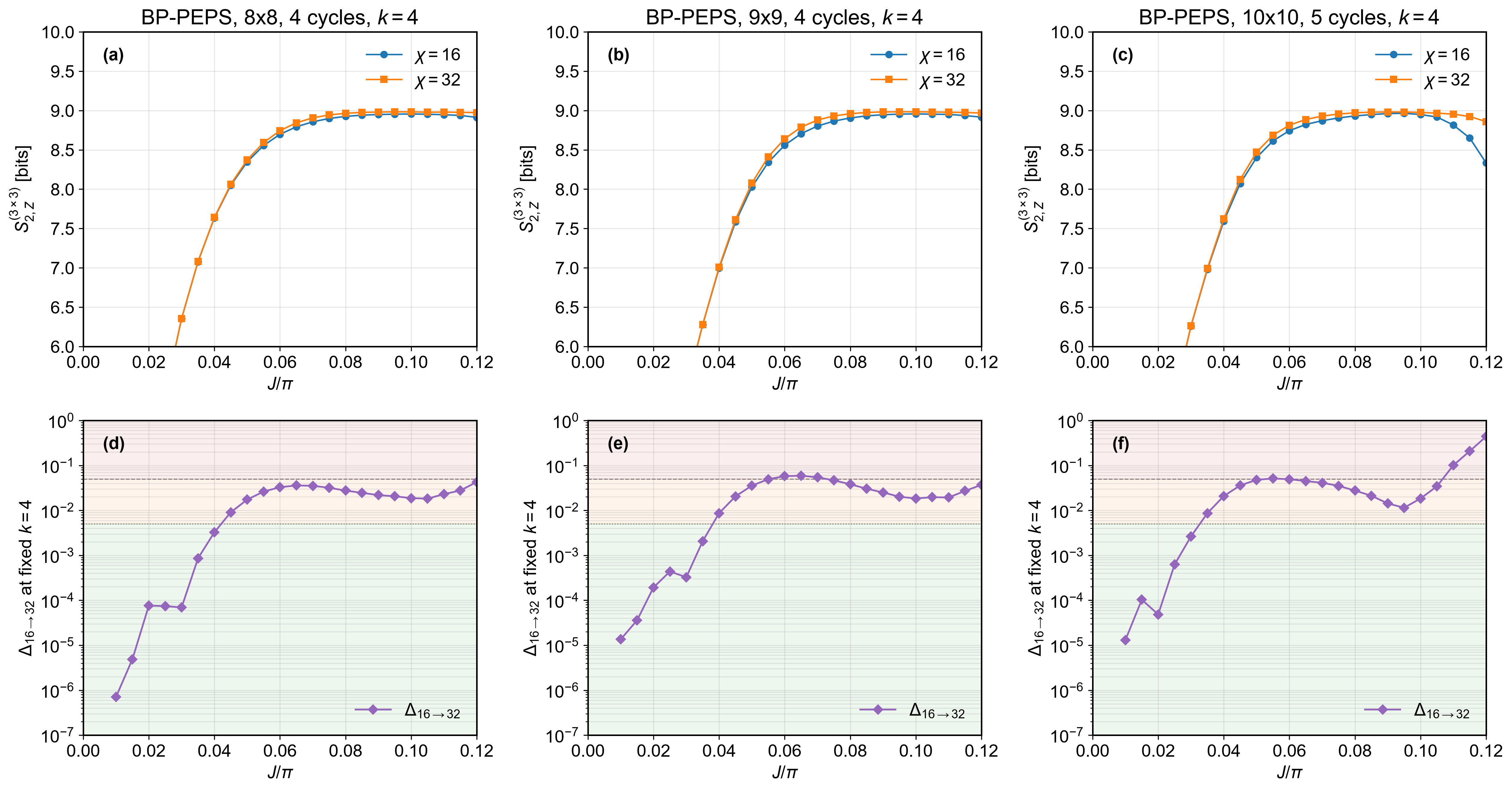}
    \caption{\textbf{BP-PEPS convergence in bond dimension at large lattice sizes.} \textbf{a}--\textbf{c}, collision entropy of the $3\times 3$ marginal as a function of $J$ computed using BP-PEPS with a fourth-order cumulant expansion at $\chi=16$ and $\chi=32$ (on $8\times8$ for the patch with lower-left site $(4,4)$), for $8\times 8$ and $9\times 9$ system sizes after $4$ Floquet cycles, and $10\times 10$ after $5$ Floquet cycles. \textbf{d}--\textbf{f}, the corresponding refinement $\Delta_{16\to32}$. The refinement stays below $1\%$ for $J/\pi\le0.045$ on $8\times8$, $J/\pi\le0.04$ on $9\times9$ and $J/\pi\le0.035$ on $10\times10$.}
    \label{fig:PEPS_bond_dim2}
\end{figure}

We then investigate the convergence of a $3\times 3$ marginal IPR collision entropy at large lattice sizes ($8\times 8$ and $9\times 9$ for 4 Floquet cycles, $10\times 10$ for 5 Floquet cycles) in Fig.~\ref{fig:PEPS_bond_dim2}, comparing $\chi=16$ and $\chi=32$. The refinement $\Delta_{16\to32}$ is below $0.01\%$ for $J/\pi\le0.03$ on $8\times8$ and for $J/\pi\lesssim0.015$ on $9\times9$ and $10\times10$, and below $1\%$ up to $J/\pi=0.045$, $0.04$ and $0.035$, respectively. At larger couplings it reaches $4.3\%$ on $8\times8$ (at $J/\pi=0.12$), $5.9\%$ on $9\times9$ and, for the five-cycle $10\times10$ data, $44\%$. The refinement grows with the size of the marginal and, more weakly, with the size of the system, and so does the bond dimension required for a given refinement (Fig.~\ref{fig:methods_complexity}).

\begin{figure}[tbp]
  \centering
  \includegraphics[width=\linewidth]{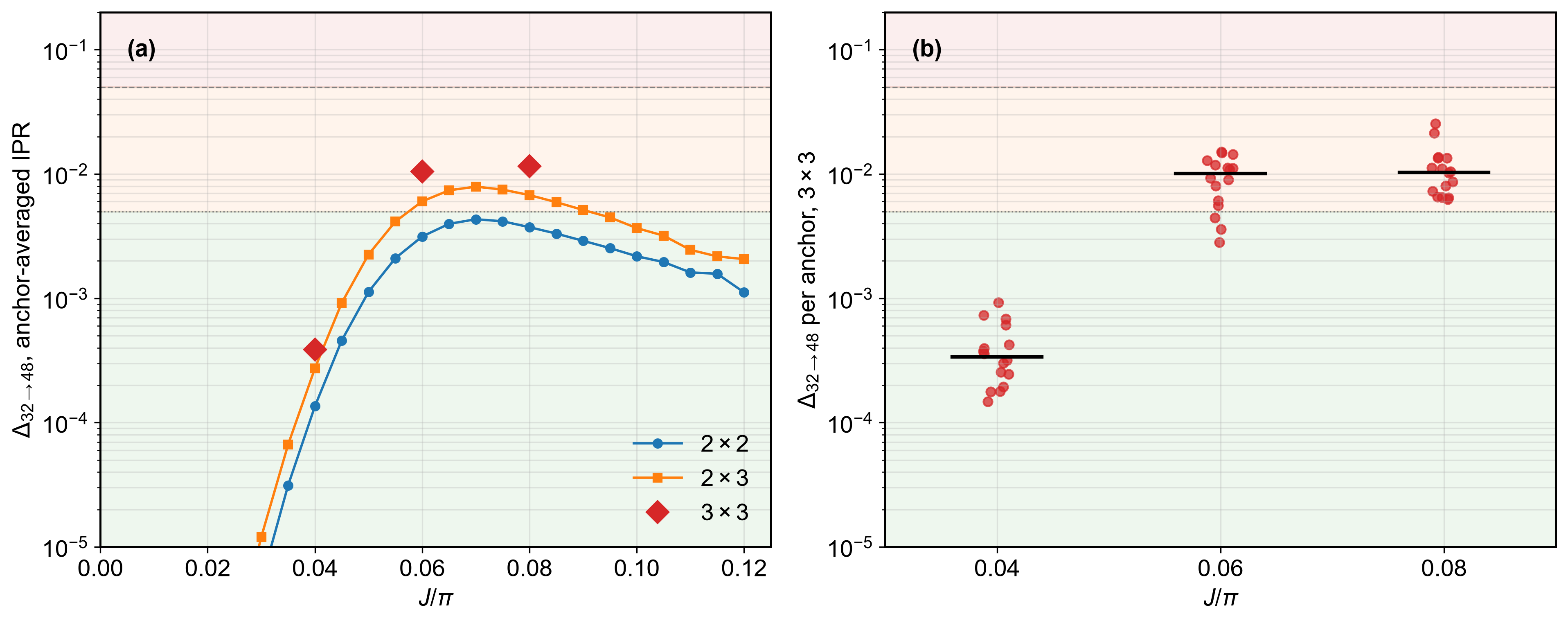}
  \caption{\textbf{Refinement $\chi=32\to48$ of the $3\times3$ marginal on $8\times8$.}
  Four Floquet cycles, cumulant order $k=4$, at the $16$ positions of the production data. \textbf{a}, spatially averaged $\Delta_{32\to48}$ for the $2\times2$ and $2\times3$ marginals over the full coupling range, and for the $3\times3$ marginal (diamonds) at $J/\pi=0.04$, $0.06$ and $0.08$. \textbf{b}, $\Delta_{32\to48}$ of each of the $16$ patches for the $3\times3$ marginal; bars mark the median. Shading marks the bands of Eq.~\eqref{eq:classification}.}
  \label{fig:bp_chi48_3x3}
\end{figure}

Since the $3\times3$ production data contain a single refinement, we have tested it with larger bond dimensions on $8\times8$, where runs at $\chi=48$ and $64$ completed four Floquet cycles, produced the $2\times2$ and $2\times3$ marginals over the whole coupling range and stored the evolved states. For these marginals the refinements $\chi=32\to48\to64$ remain below $1\%$ throughout, peaking near $J/\pi\simeq0.07$ at $0.43\%$ and $0.17\%$ for $2\times2$ and at $0.79\%$ and $0.31\%$ for $2\times3$, and they decrease by a factor between $2.3$ and $3.5$ at each step, as expected when the truncation error decreases systematically with $\chi$. From the stored $\chi=48$ states we have also evaluated the $3\times3$ marginal at $J/\pi=0.04$, $0.06$ and $0.08$, at the $16$ positions of the production data (Fig.~\ref{fig:bp_chi48_3x3}). The spatially averaged refinement $\Delta_{32\to48}$ is $0.04\%$, $1.05\%$ and $1.16\%$ at the three couplings, no individual patch moves by more than $2.6\%$, and at $J/\pi=0.06$ the refinement roughly doubles with each increase of the marginal, from $0.31\%$ for $2\times2$ to $0.60\%$ for $2\times3$ and $1.05\%$ for $3\times3$. The $3\times3$ collision entropies of the main text are therefore stable to about one percent under the refinement $\chi=32\to48$, also in the crossover region. This test probes only $\delta_{\rm state}$: it is insensitive both to the BP bias of Fig.~\ref{fig:conv_bp_4x4}(c), which does not decrease with $\chi$, and to the cumulant truncation, which is held fixed at $k=4$.

The largest discarded weight of the simple-update truncations, recorded after each Floquet cycle, measures the state truncation directly. For $\chi\ge16$ no truncation occurs during the first two cycles at any system size, after which the discarded weight grows rapidly with $J$. At $J/\pi=0.06$ the step $\chi=32\to48$ reduces it by a factor $2.8$, in line with the refinements above, whereas for $J/\pi\gtrsim0.09$ the reduction is only a factor $1.1$--$1.5$, the singular-value spectra being nearly flat on the scale of the accessible bond dimensions.

Evaluating expectation values of \emph{local} observables with BP-PEPS without loop corrections costs at best $O(n\chi^{z+1})$ time~\cite{tindall2025dynamics}, where $z$ is the coordination number of the lattice and $n$ the system size. $Z$ strings of large support, as required for large marginals, are more expensive, since their region is contracted exactly (Sec.~\ref{sec:tns_complexity}). For marginals of more than four sites we therefore resort to the cumulant expansion described in the next section.

\subsection{Cumulant expansion}
\label{sec:cumulant_expansion}

The connected correlators, or cumulants, $\langle O_{i_1}\cdots O_{i_k}\rangle_c$ of a set of operators $\{O_1,\dots,O_n\}$, with $\langle A\rangle=\mathrm{Tr}(\rho A)$, are defined by the moment--cumulant relation~\cite{KuboCumulant}
\begin{equation}
\langle O_1\cdots O_n\rangle
=
\sum_{\pi \in \mathcal{P}([n])}
\;\prod_{B\in \pi}
\Big\langle \prod_{i\in B} O_i \Big\rangle_c ,
\label{eq:moment_cumulant}
\end{equation}
where $\mathcal{P}([n])$ is the set of partitions of $\{1,\dots,n\}$ and each block $B$ keeps the operator order of the original expression. In systems with decaying correlations the higher-order cumulants are suppressed, and a truncation at order $k$ retains only the cumulants of at most $k$ operators: $k=1$ gives the mean-field product $\prod_i\langle O_i\rangle$, $k=2$ adds all pairings, and so on. We apply this truncation, with $k=4$, to each $Z$ string of the Parseval expression, Eq.~\eqref{ParsevalIPR}, so that only strings of weight up to four have to be contracted, and the individual contributions can be computed in parallel. The approximation deteriorates as the ratio $k/n_A$ decreases, and since the Parseval expansion of a marginal of $n_A$ sites contains $2^{n_A}-2$ nontrivial strings, each with its own truncation error, a fixed precision would require $k$ to grow with $n_A$. For the six-site $2\times3$ marginal of $8\times8$ BP-PEPS at $J/\pi=0.065$ and $\chi=32$, the truncation at $k=4$ reproduces the directly contracted IPR to $3\times10^{-6}$.

\begin{figure}[tbp]
  \centering
  \includegraphics[width=\linewidth]{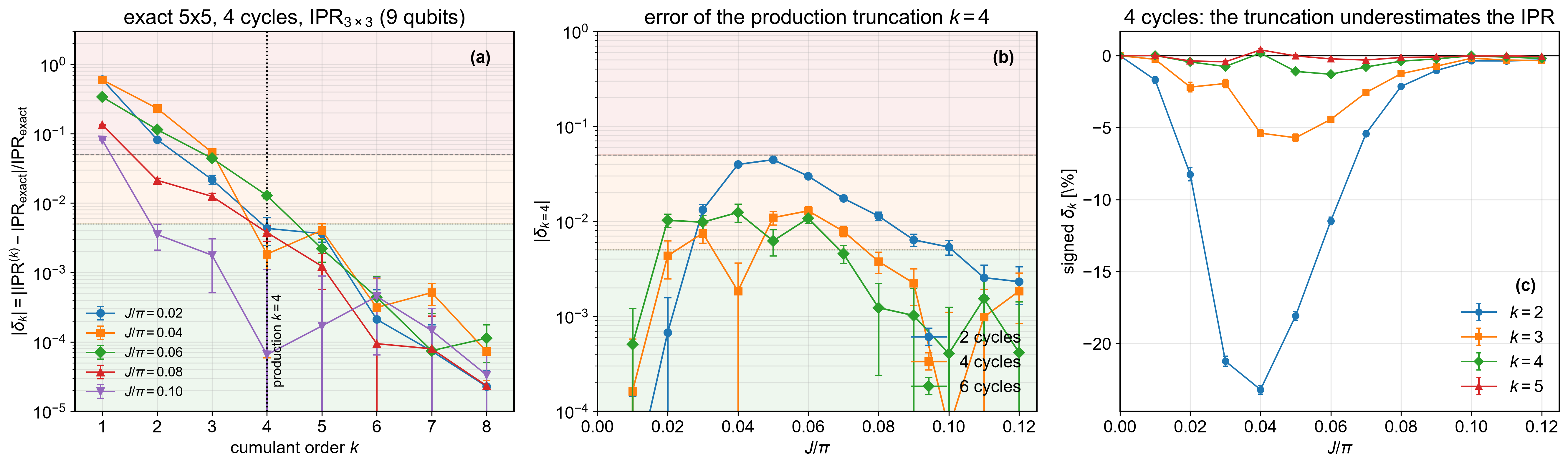}
  \includegraphics[width=\linewidth]{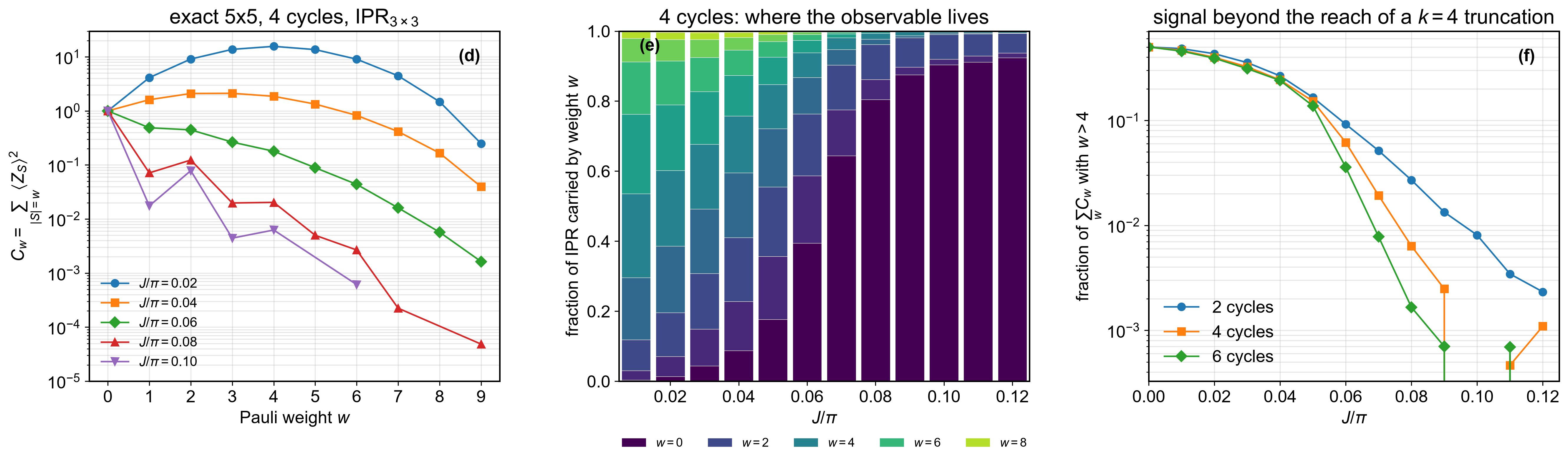}
  \caption{\textbf{Cumulant truncation at the $3\times3$ marginal size.}
  Exact $5\times5$ MPS data ($\chi=4096$), averaged over the nine translated $3\times3$ patches.
  \textbf{a}, $|\delta_k|$ (Eq.~\eqref{eq:delta_k}) as a function of the cumulant order $k$ after four cycles; the dotted line marks $k=4$. \textbf{b}, $|\delta_{k=4}|$ as a function of $J/\pi$ after two, four and six cycles. \textbf{c}, signed $\delta_k$.
  Error bars are obtained with a delete-one-block jackknife over the $10^4$ samples, with $\delta_k$ evaluated as a correlated difference so that the sampling fluctuations common to $\mathrm{IPR}^{(k)}$ and $\mathrm{IPR}^{(9)}$ cancel. This is necessary because the plug-in Parseval estimator carries a sampling bias of order $1/N_s$, about $3\%$ at $N_s=10^4$ samples, comparable to or larger than $\delta_k$ itself.
  \textbf{d}, contributions $C_w=\sum_{|S|=w}\langle Z_S\rangle^2$ to $\mathrm{IPR}_9=2^{-9}\sum_w C_w$ as a function of the Pauli weight $w$ after four cycles. \textbf{e}, fraction of the IPR carried by each weight as a function of $J/\pi$. \textbf{f}, fraction carried by weights $w>4$.}
  \label{fig:cumulant_exact}
\end{figure}

For the nine-site marginal used in the main text, whose exact Parseval expansion contains connected correlators of order five to nine, a truncation at $k=4$ discards much more, and we tested it directly at this size. For $5\times5$, with the fields of a $5\times5$ block of the $8\times8$ field realisation, the largest MPS bond dimension, $\chi=4096$, saturates the maximal Schmidt rank of a $25$-site chain, so that the MPS samples are drawn from the exact Floquet state. Applying to the $3\times3$ marginals of this state the same cumulant scheme used for BP-PEPS gives the truncation error
\begin{equation}
  \delta_k=\bigl(\mathrm{IPR}^{(k)}-\mathrm{IPR}^{(9)}\bigr)/\mathrm{IPR}^{(9)}
  \label{eq:delta_k}
\end{equation}
for every order $k=1,\dots,9$, where $\mathrm{IPR}^{(9)}$ is the untruncated Parseval expansion. Figure~\ref{fig:cumulant_exact} shows $\delta_k$ averaged over the nine translated $3\times3$ patches of the lattice. We find that $|\delta_k|$ decreases roughly geometrically with $k$ at every coupling, as expected of a systematically improvable scheme, but that at $k=4$ it is not negligible near the crossover. After two Floquet cycles $|\delta_{k=4}|$ reaches $4.4\pm0.1\%$ at $J/\pi=0.05$ ($4.0\%$ at $J/\pi=0.04$ and $3.0\%$ at $J/\pi=0.06$) and drops below $0.6\%$ for $J/\pi\ge0.09$; after four and six cycles it peaks at $1.3\%$ and $1.1\%$, respectively, at $J/\pi=0.06$. Where it is appreciable the truncation underestimates the IPR, i.e.\ it overestimates the collision entropy. Near the crossover the cumulant truncation is therefore comparable to the error of the BP contraction after two cycles, and smaller than it, but not negligible, after four. Its sign is opposite to that of the contraction error on $4\times4$ after two cycles and the same after four cycles (Fig.~\ref{fig:bp_env_state}), so that at the cycle count of the production runs the two errors add.

Figure~\ref{fig:cumulant_exact}(d)--(f) shows which correlators contribute most to the IPR. Writing $\mathrm{IPR}_m=2^{-m}\sum_w C_w$, with $C_w=\sum_{|S|=w}\langle Z_S\rangle^2$ the contribution of the Pauli strings of weight $w$, we find that at weak and intermediate coupling the IPR is dominated by high-weight correlators: at $J/\pi=0.02$ after four cycles, weights $w=3$ to $6$ account for about $72\%$ of $\sum_w C_w$, and the identity term for only $1.4\%$. As the coupling increases the marginal approaches the maximally mixed state and the identity term takes over ($80\%$ at $J/\pi=0.08$, $92\%$ at $J/\pi=0.12$). The fraction of the IPR carried by weights above four decreases by three orders of magnitude across the range and follows $|\delta_{k=4}|$ closely. The cumulant truncation is thus least accurate at intermediate coupling, where the IPR is carried by correlators of weight comparable to the marginal size. This is also where the bond-dimension refinement and the BP bias are largest.

\subsection{Computational cost}
\label{sec:tns_complexity}

We ran the tensor-network calculations on a server with an AMD Ryzen Threadripper PRO 7965WX processor ($24$ cores, $503$\,GB of memory) and on E2 virtual machines of Google Cloud Platform with $4$ to $16$ vCPUs and up to $128$\,GB of memory, and the cluster DTWA calculations on a laptop with an AMD Ryzen AI 9 HX 370 processor ($12$ cores). The timings quoted below refer to one Floquet cycle on four cores for MPS, to a single core for BP-PEPS and to eight cores for cluster DTWA.

For MPS we first extract the bond dimension required for the MPS simulations to evaluate a marginal IPR$_m$ within a given relative deviation from the result at the largest bond dimension $\chi_{\textrm{max}}$ of the sweep. On $4\times4$, where $\chi_{\textrm{max}}=256$ is exact, this is the true error $\delta_{\rm exact}$ of Eq.~\eqref{eq:delta_exact}.
Figure~\ref{fig:MPS_4x4}(d) shows the relative error as a function of bond dimension at a fixed system size ($4\times 4$), for several marginal sizes (from 1 to 16 qubit marginals), from which we extract the bond dimension $\chi_{\rm req}$ required for $\delta_{\rm exact}=0.05$ (Fig.~\ref{fig:MPS_4x4}(e)). Since our bond dimensions form a discrete grid, $\chi_{\rm req}$ is only resolved up to one grid step: at $J/\pi=0.08$ we find $\chi_{\rm req}\le16$ for the $1\times1$ marginal, $\chi_{\rm req}\in(16,32]$ for $2\times2$ and $\chi_{\rm req}\in(32,64]$ for both $3\times3$ and $4\times4$. Figure~\ref{fig:MPS_4x4}(f) finally shows that $\chi_{\rm req}$ grows steeply as $J/\pi$ increases, reaching $\chi=128$ for the $3\times3$ marginal for $J/\pi\ge0.10$, half of the exact $\chi_{\textrm{max}}=256$.

\begin{figure*}[t]
  \centering
  \includegraphics[width=\textwidth]{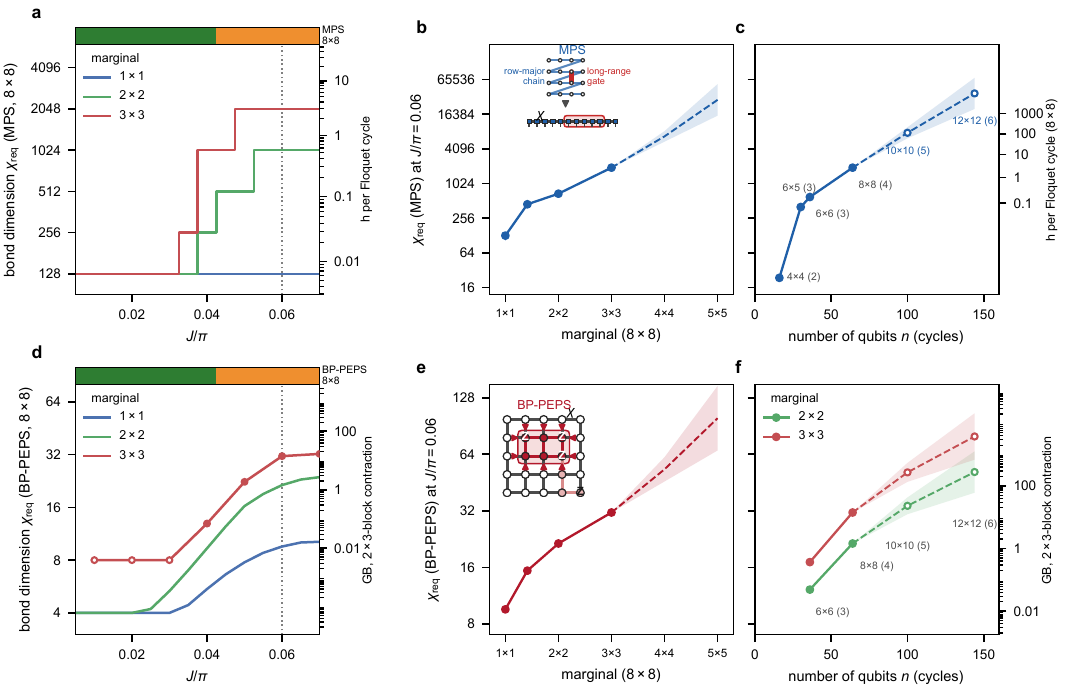}
  \caption{\textbf{Classical simulation methods and their cost.}
  \textbf{a}, MPS on $8\times8$ after four cycles: bond dimension $\chi_{\rm req}$ required for a refinement below $5\%$ of the $1\times1$, $2\times2$ and $3\times3$ marginal IPRs; the right axis gives the measured wall-clock time of one TDVP Floquet cycle at that bond dimension on a $4$-vCPU GCP E2 machine.
  \textbf{b}, $\chi_{\rm req}$ at $J/\pi=0.06$ (dotted line in \textbf{a}) as a function of the size of the marginal on $8\times8$ ($1\times1$, $1\times2$, $2\times2$, $3\times3$), extrapolated to $4\times4$ and $5\times5$ marginals; the shaded band spans an exponential growth with the number of sites and with the linear size of the marginal.
  \textbf{c}, $\chi_{\rm req}$ of the $3\times3$ marginal at $J/\pi=0.06$ as a function of the number of qubits (number of cycles in brackets), extrapolated to $10\times10$ and $12\times12$ (band as in \textbf{b}, exponential in the number of qubits and in the linear size). Right axis: wall-clock time of one Floquet cycle on $8\times8$, extrapolated beyond $\chi=4096$ with the measured power law. In \textbf{a}--\textbf{c}, $\chi_{\rm req}$ is measured against the largest bond dimension of each sweep, and in \textbf{b}, \textbf{c} it is interpolated between the bond dimensions of the sweep.
  \textbf{d}--\textbf{f}, the same for BP-PEPS, with a refinement threshold of $1\%$ measured against $\chi=48$ (cumulant-truncated IPR, averaged over all positions of the marginal): \textbf{d}, $8\times8$ after four cycles, from bond dimensions $\chi=4$, $8$, $16$, $32$ and $48$ ($1\times1$, $2\times2$) and $\chi=8$, $16$, $32$ and $48$ at the $16$ positions used in production ($3\times3$, with $\chi=32$ as reference for $J/\pi\le0.03$; open symbols: refinement below $1\%$ already at $\chi=8$); \textbf{e}, at $J/\pi=0.06$ as a function of the size of the marginal, extrapolated as in \textbf{b}; \textbf{f}, $2\times2$ marginal at $J/\pi=0.06$ on $6\times6$ and $8\times8$ after $L/2$ cycles, extrapolated to $10\times10$ and $12\times12$ (band between a linear and an exponential growth with the number of cycles; after the same number of cycles $6\times6$, $8\times8$ and $12\times12$ require nearly the same bond dimension), and the same for the $3\times3$ marginal; open symbols mark the extrapolated $10\times10$ and $12\times12$ values. Right axes of \textbf{d} and \textbf{f}: largest intermediate tensor (complex double precision) of the exact contraction of a $2\times3$ block at that bond dimension (\cref{fig:bp_contraction_cost}).
  The strips above \textbf{a} and \textbf{d} give the stability class of the corresponding $8\times8$ tensor-network calculation (green: reliable, orange: borderline, red: uncontrolled, grey: stable but unconverged, refinement below $0.5\%$ beyond its maximum; \cref{fig:reliability}).
  Insets of \textbf{b} and \textbf{e}: sketches of the methods. MPS: the lattice is mapped onto a chain in row-major order, vertical bonds become long-range gates, applied by TEBD in the first cycle and by local TDVP on a window of $N_y+1$ sites afterwards. BP-PEPS: the expectation value of a $Z$ string (sites marked Z) is obtained from the smallest lattice region that contains it; its loop core (dark red) is contracted exactly with the belief-propagation messages entering it (arrows), its loop-free branches (light red) as on a tree.}
  \label{fig:methods_complexity}
\end{figure*}

On the larger lattices the reference $\chi_{\textrm{max}}$ is not exact, and a bond dimension required to come within $5\%$ of the result at $\chi_{\textrm{max}}$ cannot exceed $\chi_{\textrm{max}}$ itself, so that it is only a lower bound. With this caveat, for the $3\times3$ marginal at $J/\pi=0.06$ it grows from $\chi\approx24$ on $4\times4$ to $\chi\approx400$, $600$ and $1900$ on $6\times5$, $6\times6$ and $8\times8$ (Fig.~\ref{fig:methods_complexity}(c)); on $8\times8$ it is largest for $0.05\lesssim J/\pi\lesssim0.08$ and decreases again at stronger coupling, as the marginal approaches its ergodic value.

Fig.~\ref{fig:methods_complexity} extrapolates these trends at $J/\pi=0.06$. For MPS on $8\times8$, with a refinement threshold of $5\%$, a $4\times4$ marginal would require $\chi\approx5{,}000$--$8{,}000$, i.e.\ $2$--$6$ days per Floquet cycle (the $2\%$ threshold of Fig.~\ref{fig:classical_cost_main}b gives the slightly lower $\chi\approx5{,}000$--$7{,}000$, because at $2\%$ the $3\times3$ value, $\chi\approx2{,}900$, lies close to the reference $\chi=4096$, which flattens the extrapolated slope), and a $5\times5$ marginal $\chi\approx15{,}000$--$55{,}000$; for the $3\times3$ marginal the required bond dimension extrapolates to $\chi\approx6{,}000$--$10{,}000$ on $10\times10$ and $\chi\approx20{,}000$--$70{,}000$ on $12\times12$. The bond dimension that BP-PEPS requires for a refinement below $1\%$ is much smaller but follows the same trends. On $8\times8$ it grows from $\chi\approx10$ for the $1\times1$ marginal to $\chi\approx21$ and $31$ for the $2\times2$ and $3\times3$ ones; extrapolated, a $4\times4$ marginal would require $\chi\approx45$--$60$ and a $5\times5$ one $\chi\approx65$--$150$, for which the contraction of a single $2\times3$ block would involve intermediate tensors of $0.1$--$0.8$\,TB and $1$--$160$\,TB, respectively. With the size of the lattice it grows from $\chi\approx12$ on $6\times6$ to $\chi\approx21$ on $8\times8$ for the $2\times2$ marginal, and from $\chi\approx17$ to $\chi\approx31$ for the $3\times3$ one, which extrapolates to $\chi\approx60$--$105$ on $12\times12$. Since $6\times6$, $8\times8$ and $12\times12$ require nearly the same bond dimension after the same number of cycles, this growth reflects the number of cycles, $L/2$, rather than the size of the lattice. These bond dimensions are measured against the reference $\chi=48$ and are therefore lower bounds when they approach it.

\begin{figure}[tbp]
  \centering
  \includegraphics[width=\linewidth]{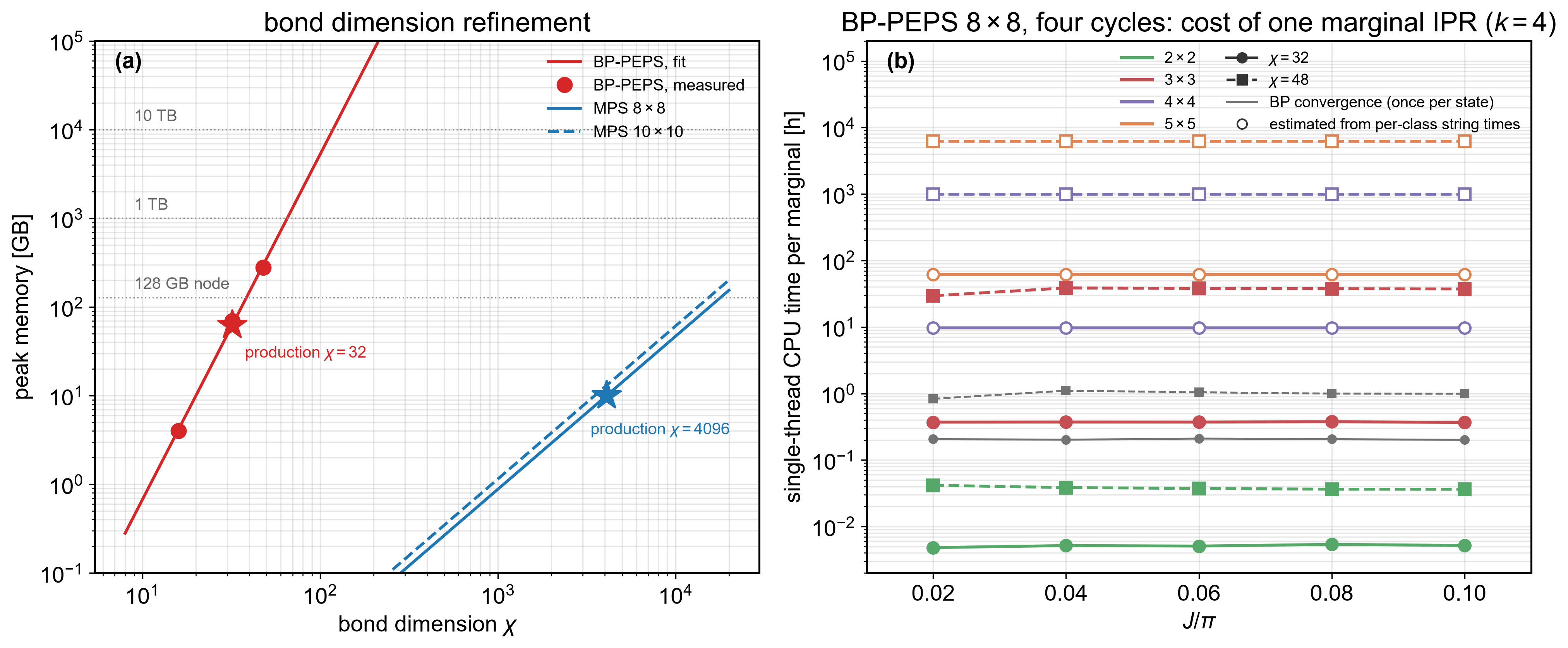}
  \caption{\textbf{Memory and computing time of the tensor-network simulations.}
  \textbf{a}, peak memory as a function of bond dimension. BP-PEPS points are measured; the MPS curves are the fit $M\simeq4.9\times10^{-7}\,n^{0.60}\chi^{1.73}$\,GB of the peak memory of the hybrid TEBD-TDVP runs on $6\times5$, $6\times6$ and $8\times8$, measured for $\chi\le2048$ and $n\le64$, so that the $10\times10$ curve and the range $\chi>2048$ are extrapolations. Stars mark the production settings.
  \textbf{b}, single-thread CPU time, measured on a Google Cloud e2-highmem-8 machine, of the fourth-order cumulant IPR of the central $m\times m$ marginal of the $8\times8$ BP-PEPS state after four cycles, as a function of $J/\pi$, for $m=2$ to $5$ at $\chi=32$ (solid lines) and $\chi=48$ (dashed lines). Filled symbols are measured directly; open symbols are obtained from the measured cost of each class of contraction region (loop-free regions, plaquettes and larger loops) multiplied by the number of regions of that class in the marginal; at $\chi=48$, where the $2\times3$ loops were not timed individually, their cost is the measured $3\times3$ marginal minus its other regions. Grey: convergence of the belief-propagation messages, needed once per state (measured).}
  \label{fig:resource_horizon}
\end{figure}

With the hybrid scheme of Sec.~\ref{sec:tn_methods}, the four Floquet cycles of an $8\times8$ calculation at $\chi=3000$ take between $23$ and $47$ hours on a cloud virtual machine with two physical cores (four virtual CPUs), of which $60\%$ to $80\%$ are spent in the time evolution and the remainder in the exact contraction of the patch IPRs and in sampling, while three cycles on $6\times6$ at $\chi=4096$ take between $13$ and $20$ hours. On $8\times8$ a single TDVP cycle takes about $0.5$, $3$, $8$ and $19$ hours at $\chi=1024$, $2048$, $3000$ and $4096$, so that its cost grows approximately as $\chi^{2.5}$ in this range, somewhat more slowly than the asymptotic $O(n\chi^3)$ scaling of the two-site update.
If we were to double the bond dimension to converge larger marginals in a wider interval of values of $J/\pi$ \emph{at the same level of precision}, each doubling would multiply the computing time by about $6$, so that already at $\chi=8192$ a single Floquet cycle on $8\times8$ would take almost five days on the same hardware.
Memory, on the other hand, is not the limiting factor of these MPS simulations. The peak memory of our implementation, which includes the cached bond MPOs of the TDVP stage, the exact contraction of the patch IPRs and the sampling, grows as $\chi^{1.7}$ at fixed system size and only weakly with $n$ (a joint fit on $6\times5$, $6\times6$ and $8\times8$ gives $M\simeq4.9\times10^{-7}\,n^{0.60}\chi^{1.73}$\,GB), and it is $3.3$\,GB on $8\times8$ at $\chi=2048$. The fit gives about $10$\,GB at $\chi=4096$ and $33$\,GB at $\chi=8192$ on $8\times8$ (Fig.~\ref{fig:resource_horizon}(a)), which would fit on a single GPU, so that faster architectures such as GPUs or TPUs could reduce the computing time without running into the memory limit, although the scaling of the cost with $\chi$ would remain.

For BP-PEPS the time evolution is cheap, and the cost lies in the contraction of the $Z$ strings entering the marginal IPRs. In our production code the peak memory of this stage grows as $\chi^{3.9}$ ($M\simeq8.5\times10^{-5}\,\chi^{3.90}$\,GB for marginals of up to nine sites on $8\times8$), so that the $3\times3$ marginal on $8\times8$ requires $62$\,GB at $\chi=32$ and would require about $0.3$\,TB at $\chi=48$. This growth reflects the way each Steiner region is contracted rather than the BP approximation. For the $\chi=48$ data of Fig.~\ref{fig:bp_chi48_3x3} we contract the same networks in the same order, but whenever the largest intermediate tensor would exceed a fixed memory budget we sum explicitly over the internal bond that dominates it, and we reuse the denominator, which depends only on the region, for all Pauli strings that share it. This reproduces the production $\chi=32$ values to a relative accuracy of $10^{-13}$, and at $\chi=48$ the peak memory is $30$\,GB, with about $1.3$ hours per patch on $16$ cores. These costs depend on our implementation rather than on belief propagation itself, and our bond dimensions are comparable to those of other recent studies: a study of the relaxation dynamics of a two-dimensional quantum magnet on a $256$-qubit processor~\cite{mendessantos2026exploring} used BP at $\chi=40$ and snake-ordered MPS at $\chi=1000$ for $L=8$, and likewise found that the tensor-network methods lose control in the dynamical regime of interest.

\begin{figure*}[tbp]
  \centering
  \includegraphics[width=\linewidth]{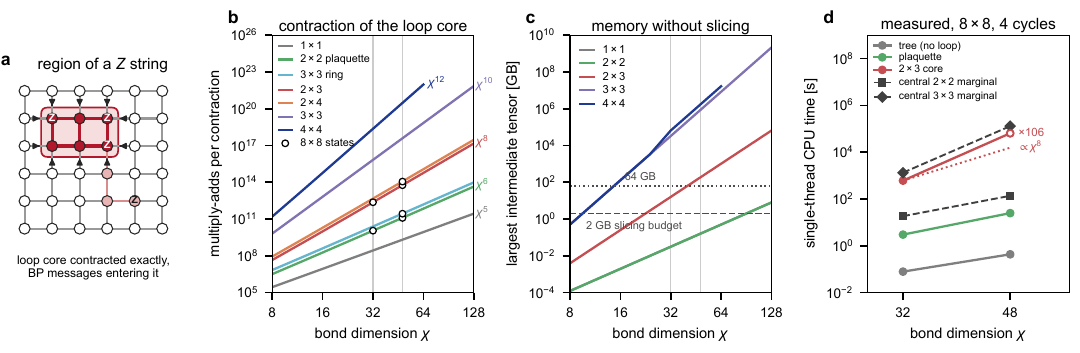}
  \caption{\textbf{Cost of the exact contractions in the BP-PEPS expectation values.}
  \textbf{a}, the expectation value of a $Z$ string (sites marked Z) is obtained from the smallest lattice region that contains it. Repeatedly removing the sites attached by a single bond leaves its loop core (dark red); the removed sites (light red) are contracted site by site into the core, which is then contracted exactly with the belief-propagation messages entering it (arrows).
  \textbf{b}, complex multiply-adds of the exact contraction of a loop core in the bulk of the lattice (ket, operator and bra layers, with a $\chi\times\chi$ environment on every bond leaving the core) as a function of the bond dimension $\chi$, with the contraction order used in production (exhaustive search; for the $3\times3$ core the order found at $\chi=4$, for $4\times4$ a simulated-annealing order); labels give the exponent of $\chi$ fitted for $16\le\chi\le32$. A $3\times3$ ring is a loop of eight sites around a site that is not part of the region. Circles: multiply-adds of the contraction regions of the $8\times8$ calculations at $\chi=32$ and $48$.
  \textbf{c}, largest intermediate tensor of the same contractions (complex double precision). Intermediates larger than a budget of $2$\,GB are computed in slices, by summing explicitly over an internal bond.
  \textbf{d}, measured single-thread CPU time (GCP e2-highmem-8) of one $Z$ string whose region is loop-free, contains a plaquette or contains a $2\times3$ core, and of the central $2\times2$ and $3\times3$ marginal IPRs of the $8\times8$ BP-PEPS state after four cycles, at $\chi=32$ and $48$, averaged over $J/\pi=0.02$--$0.1$. The $2\times3$-core time at $\chi=48$ (open circle) is the measured $3\times3$ marginal minus its other strings; dotted, the growth $\propto\chi^8$ of its multiply-adds, the remaining factor of about four coming from the slicing.}
  \label{fig:bp_contraction_cost}
\end{figure*}

In practice, the region of a $Z$ string is the smallest connected set of sites that contains it, together with all the lattice bonds between these sites, so that it contains a loop whenever it contains a $2\times2$ block of sites. We contract it in two steps (Fig.~\ref{fig:bp_contraction_cost}(a)). First, we repeatedly remove the sites attached to the rest of the region by a single bond: what remains is the loop core, and the removed sites form loop-free branches attached to it. Each branch is contracted from its end towards the core, one site at a time as in a belief-propagation sweep, at a cost of about $\chi^5$ per site, and enters the core as an additional message. The core is then contracted exactly as a general tensor network, together with the belief-propagation messages on the bonds leaving the region. This last step dominates the cost: for a core of $w\times\ell$ sites with $w\le\ell$ the number of operations grows as $\chi^{2w+4}$, i.e.\ as $\chi^8$ for the $2\times3$ and $2\times4$ cores that occur in marginals of up to $5\times5$ sites. The split changes only the order of the contraction, not its result, which agrees with the direct contraction of the whole region to round-off.

Fig.~\ref{fig:resource_horizon} summarises the cost of extending these calculations to larger couplings, which for the system sizes of interest are stable under bond-dimension refinement only up to $J/\pi\simeq0.035$--$0.06$ (Sec.~\ref{sec:reliability_map}). Panel (a) gives the memory cost of moving this limit. For BP-PEPS, $\chi=128$ would require about $14$\,TB. For MPS, the $8\times8$ run at $\chi=4096$ occupies about $10$\,GB, $\chi=8192$ would require about $33$\,GB and $10\times10$ at $\chi=4096$ about $13$\,GB, so that for MPS the cost of refinement is set by the computing time discussed above rather than by memory. Panel (b) gives the single-thread computing time of one marginal for BP-PEPS. At fixed bond dimension it is nearly independent of the coupling, since all bonds of the $\chi=32$ states are saturated already at $J/\pi=0.02$, but it grows steeply with the size of the marginal, from $19$\,s for the central $2\times2$ marginal to $22$ minutes for the $3\times3$ one, and to about $10$ hours and $2.6$ days for the $4\times4$ and $5\times5$ ones. We obtain the last two from the measured cost of each class of contraction region multiplied by the number of regions of that class, an estimate that reproduces the directly measured $3\times3$ time to within $10\%$. The time is dominated by the few Pauli strings whose region contains a block of at least $2\times3$ sites, i.e. two or more adjacent plaquettes (the elementary $2\times2$ squares of the lattice), whose exact contraction grows approximately as $\chi^8$: the single such string of the $3\times3$ marginal accounts for about $1100$ of its $1300$\,s. At $\chi=48$ the convergence of the BP messages takes about one hour instead of $12$ minutes, and the marginals of $3\times3$ and more sites become about $100$ times more expensive, the $3\times3$ one taking about $36$ hours: the multiply-adds of the $2\times3$ and $2\times4$ loops grow by $(48/32)^8\simeq26$, and the slicing needed to keep their intermediate tensors ($17$\,GB at $\chi=32$, $200$\,GB at $\chi=48$) within a budget of $2$\,GB adds a further factor of about four (Fig.~\ref{fig:bp_contraction_cost}).

For BP-PEPS a larger bond dimension alone would, however, not improve the accuracy near the crossover, since of the three terms in Eq.~\eqref{eq:taxonomy} only $\delta_{\rm state}$ decreases with $\chi$. At $J/\pi\simeq0.06$ the error of the BP contraction measured on $4\times4$ after four cycles is $8.2\%$ (Fig.~\ref{fig:bp_env_state}), the cumulant truncation at four cycles is $1.3\%$, and neither depends on $\chi$. Extrapolating the decay of the refinements $\chi=32\to48\to64$ of Sec.~\ref{sec:tn_results}, $\delta_{\rm state}$ drops below the contraction error already at $\chi\simeq23$, and going from $\chi=32$ to $\chi=128$ would multiply the memory by about $220$ while reducing the extrapolated $\delta_{\rm state}$ of the $3\times3$ marginal on $8\times8$ from $3.3\%$ to below $0.1\%$, leaving the other two contributions unchanged. The same extrapolation gives $\delta_{\rm state}\simeq1.1\%$ at $\chi=48$, close to the spatially averaged $\Delta_{32\to48}=1.05\%$ measured directly for the $3\times3$ marginal at this coupling (Fig.~\ref{fig:bp_chi48_3x3}). For BP-PEPS at these system sizes, the accuracy near the crossover is thus limited by the environment approximation and the cumulant truncation rather than by $\chi$.

For MPS the bond dimension remains the limiting factor: on $8\times8$ at $J/\pi=0.06$ the refinement decreases from $6.5\%$ for $1024\to2048$ to $2.8\%$ for $2048\to3000$ and $1.8\%$ for $3000\to4096$, while the computing time per cycle grows by a factor of about $35$ between $\chi=1024$ and $4096$.

\subsection{Entanglement of the simulated states}
\label{sec:tn_entanglement}

\begin{figure*}[tbp]
  \centering
  \includegraphics[width=\linewidth]{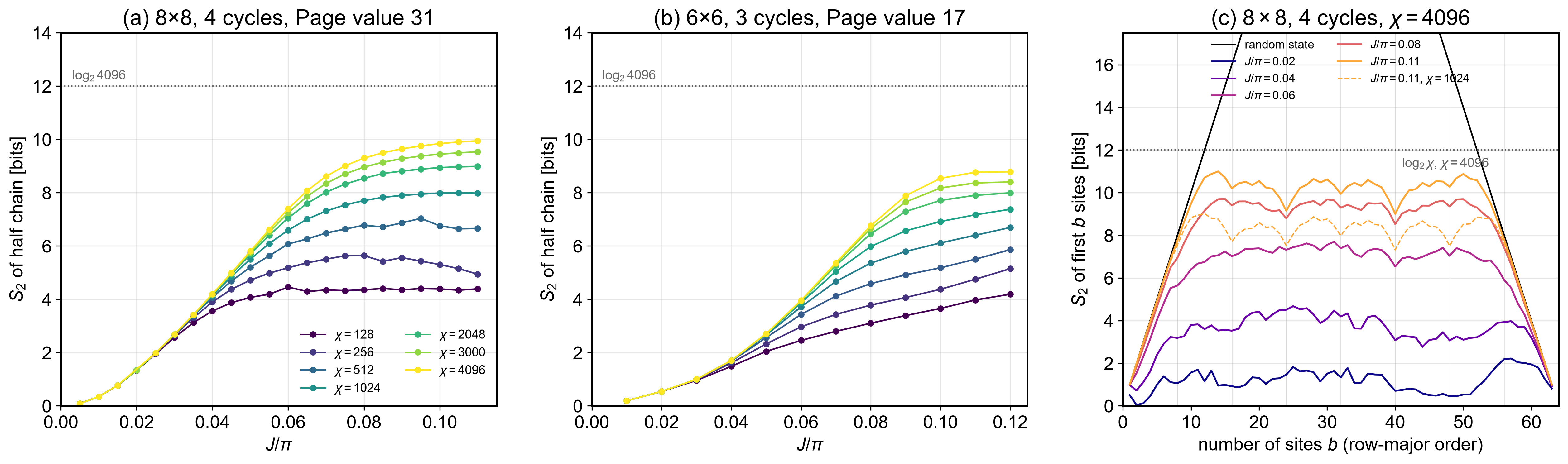}
  \caption{\textbf{Bipartite entanglement of the MPS states.}
  R\'enyi-2 entropy $S_2=-\log_2P_2$, with $P_2=\textrm{Tr}\,\rho_b^2$ the purity, of the first $b$ sites of the MPS chain (row-major order), from the Schmidt values of the hybrid TEBD-TDVP production runs (one disorder realisation).
  \textbf{a}, half chain ($b=32$) of $8\times8$ after four cycles as a function of $J/\pi$, for $\chi=128$ to $4096$.
  \textbf{b}, half chain ($b=18$) of $6\times6$ after three cycles, for $\chi=64$ to $4096$.
  \textbf{c}, $S_2$ as a function of $b$ on $8\times8$ after four cycles at $\chi=4096$ ($\chi=3000$ for $J/\pi=0.02$), and at $\chi=1024$ for $J/\pi=0.11$ (dashed); vertical lines mark complete rows. Black: average R\'enyi-2 entropy of $b$ qubits of a random state of $64$ qubits. Dotted: capacity $\log_2\chi$ of an MPS at $\chi=4096$; the Page values $31$ and $17$ in the titles of (a) and (b) are the half-chain entropies of a random state.}
  \label{fig:mps_entanglement}
\end{figure*}

\paragraph{Bipartite entanglement of the MPS states.} Besides the marginal IPRs, after every Floquet cycle our MPS simulations record the purity $\textrm{Tr}\,\rho_b^2$ of the first $b$ sites of the chain for every bond $b$, obtained from the Schmidt values of the MPS in mixed-canonical form. Since the chain follows the lattice in row-major order, $\rho_b$ is the reduced density matrix of a block of complete rows plus part of the next one, starting at a corner of the lattice. Figure~\ref{fig:mps_entanglement} shows the corresponding R\'enyi-2 entropy $S_2=-\log_2P_2$, with $P_2=\textrm{Tr}\,\rho_b^2$, for the single disorder realisation of the MPS production runs. On $8\times8$ after four cycles the entropy of the half chain is independent of $\chi$ up to $J/\pi\simeq0.04$, where it reaches $4.2$ bits, and grows from there with each increase of the bond dimension: at $\chi=4096$ it is $7.4$ bits at $J/\pi=0.06$ and $9.9$ bits at $J/\pi=0.11$, still $0.4$ bits above the value at $\chi=3000$ (Fig.~\ref{fig:mps_entanglement}(a)). Beyond $J/\pi\simeq0.05$ these values are therefore lower bounds, which lie well below both the capacity $\log_2\chi=12$ of the MPS and the value $31$ of a random state. The same holds on $6\times6$ after three cycles, where the half chain reaches $8.8$ bits at $J/\pi=0.12$ against $17$ for a random state (Fig.~\ref{fig:mps_entanglement}(b)).

Figure~\ref{fig:mps_entanglement}(c) shows how the entropy depends on the size of the block. For blocks of up to one row the entropy is close to its maximal value $b$ for $J/\pi\gtrsim0.08$: the first row of eight sites carries $7.7$ bits at $J/\pi=0.11$, against $3.2$ bits at $J/\pi=0.04$. On the scale of a few qubits the MPS states are thus nearly maximally entangled, consistently with the exact results of Sec.~\ref{sec:Quantum Renyi entropy}. For $b\gtrsim12$, on the other hand, the entropy at strong coupling hardly depends on $b$ and stays $1$--$3$ bits below $\log_2\chi$, and this plateau moves up with the bond dimension (from about $9$ bits at $\chi=1024$ to about $11$ bits at $\chi=4096$ for $J/\pi=0.11$). The plateau is thus set by $\chi$, and it cannot distinguish an entropy growing with the number of sites of the block (volume law) from one growing with the length of its boundary (area law), which is nearly the same for all the cuts between complete rows. The steep growth of $\chi_{\rm req}$ discussed above does not discriminate between the two either, since in two dimensions an MPS requires a bond dimension exponential in the linear size of the lattice already for area-law states. Our MPS data therefore establish that the entanglement grows rapidly with $J$ and, for $J/\pi\gtrsim0.05$, exceeds what can be represented at $\chi=4096$, and that subsystems of up to about eight qubits approach maximal entanglement, while volume-law entanglement of extensive subsystems at these system sizes remains beyond their reach.

\paragraph{Entanglement of small patches of the BP-PEPS states.} The entropies of the MPS chain are limited by the bond dimension already at moderate coupling, and they only probe blocks that start at a corner of the lattice. We have therefore also evaluated the R\'enyi-2 entropy $S_2[A]=-\log_2P_2[A]$, with $P_2[A]=\textrm{Tr}\,\rho_A^2$ the purity, of small patches $A$ in the bulk of the $8\times8$ BP-PEPS states after four cycles. The reduced density matrix $\rho_A$ is obtained by contracting exactly the PEPS tensors of the patch and their conjugates, with the physical indices of the patch left open, together with the converged BP messages that enter the patch from its boundary. This is the same environment that enters our BP evaluation of the $Z$-parity strings supported on $A$, and $\rho_A$ is positive by construction. The cost of this contraction grows quickly with the size of the patch: at $\chi=32$ we have computed strips of up to eight sites along the fourth row of the lattice and $2\times2$ and $2\times3$ patches, while the $3\times3$ patch is only accessible at $\chi=8$ (at $\chi=32$ it would require about $10^{18}$ multiply-add operations). From the converged messages we have also obtained the entanglement of every bond of the PEPS, as the R\'enyi-2 entropy of the eigenvalues of the product of the two messages exchanged across the bond, which are the Schmidt weights of the bond in the BP (Vidal) gauge~\cite{alkabetz2021tensor,tindall2023gauging}.

\begin{figure*}[tbp]
  \centering
  \includegraphics[width=\linewidth]{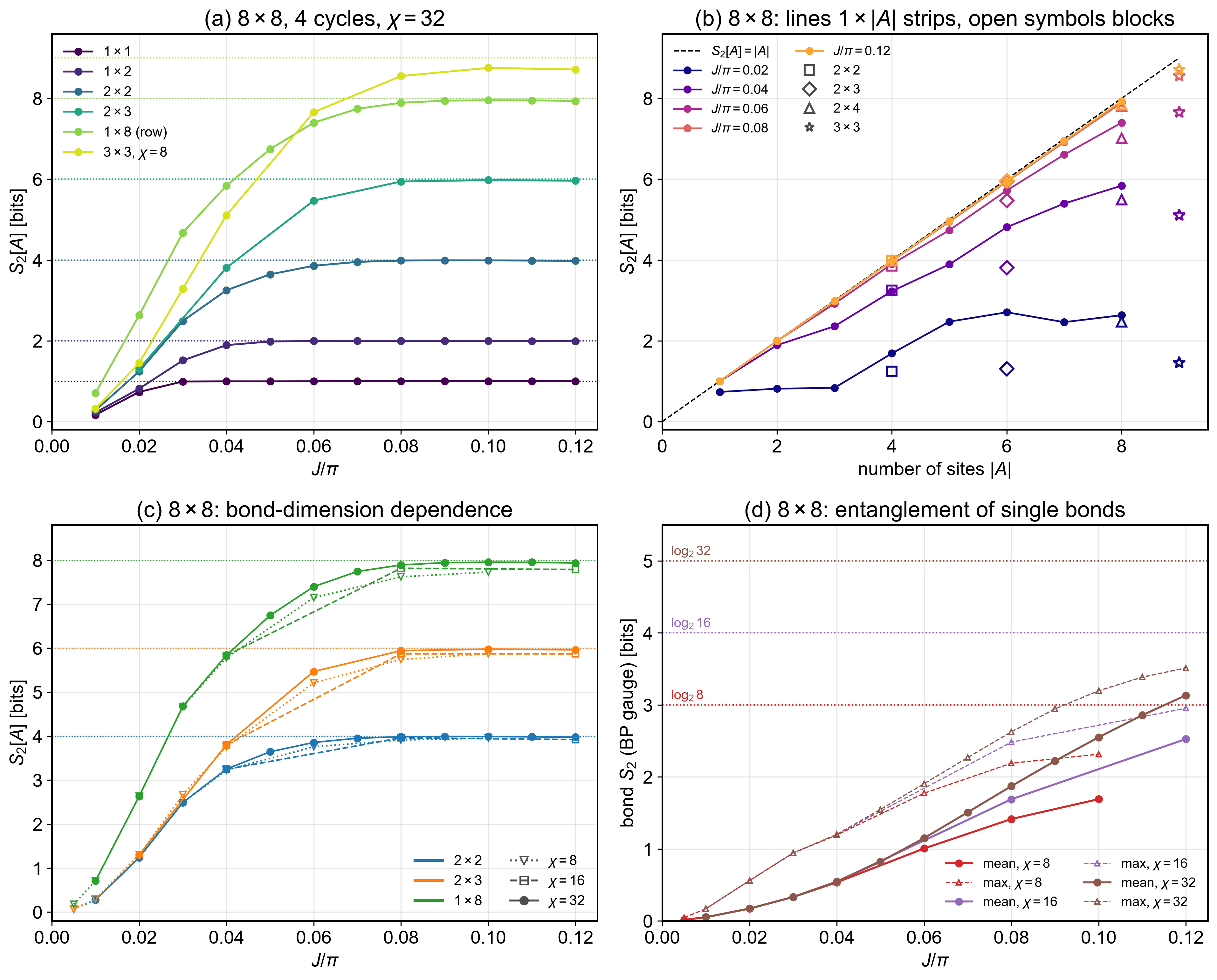}
  \caption{\textbf{Entanglement of small patches of the BP-PEPS states.}
  R\'enyi-2 entropy $S_2[A]=-\log_2P_2[A]$, with $P_2[A]=\textrm{Tr}\,\rho_A^2$ the purity and $\rho_A$ the reduced density matrix of patch $A$ contracted exactly with the converged BP messages entering it, after four cycles.
  \textbf{a}, $8\times8$ at $\chi=32$ ($\chi=8$ for the $3\times3$ patch) as a function of $J/\pi$; dotted lines mark the maximal value $|A|$.
  \textbf{b}, the same as a function of the number of sites $|A|$, for strips of $1\times|A|$ sites along the fourth row (lines) and for $2\times2$, $2\times3$, $2\times4$ and $3\times3$ patches (open symbols, at the largest bond dimension available); dashed: $S_2[A]=|A|$.
  \textbf{c}, $2\times2$, $2\times3$ and $1\times8$ patches of $8\times8$ at $\chi=8$, $16$ and $32$.
  \textbf{d}, mean and maximum over the $112$ bonds of $8\times8$ of the R\'enyi-2 entropy of the Schmidt weights in the BP gauge; dotted: $\log_2\chi$.
  The check of this estimate against the exact state on $4\times4$ is shown in \cref{sm:fig:bp_entanglement}.}
  \label{fig:bp_entanglement}
\end{figure*}

Fig.~\ref{fig:bp_entanglement}(a) shows that the entropy of every patch grows with $J$ towards its maximal value $|A|$, the number of qubits in $A$. At $J/\pi=0.02$ the $2\times2$ patch carries $1.2$ bits and the full row of eight sites $2.6$ bits, at $J/\pi=0.06$ these become $3.9$ and $7.4$ bits, and for $J/\pi\gtrsim0.08$ all the patches we computed at $\chi=32$ are within $0.1$ bit of $|A|$ (the $2\times3$ patch carries between $5.94$ and $5.98$ bits and the row between $7.89$ and $7.95$ bits). The $3\times3$ patch reaches $8.71$ bits at $J/\pi=0.12$ already at $\chi=8$. Fig.~\ref{fig:bp_entanglement}(b) shows the same data as a function of $|A|$: at strong coupling the entropy grows linearly with the number of sites with unit slope, independently of the shape of the patch, whereas at $J/\pi=0.02$ it stays well below $|A|$ and hardly grows beyond five sites. Up to $J/\pi=0.04$ the entropies obtained at $\chi=8$, $16$ and $32$ agree within $0.06$ bits, while at stronger coupling they increase with $\chi$ towards $|A|$ (Fig.~\ref{fig:bp_entanglement}(c)): for the row at $J/\pi=0.08$ we find $7.62$, $7.82$ and $7.89$ bits at $\chi=8$, $16$ and $32$. The truncation of the bond dimension thus lowers the entanglement of the patches, as we also find on $4\times4$ for the exact contraction of the same PEPS, and the entropies of the $3\times3$ patch at $\chi=8$ ($8.55$ to $8.76$ bits for $J/\pi\geq0.08$) are likely to underestimate those of the converged state. The bond entropies of Fig.~\ref{fig:bp_entanglement}(d), on the other hand, grow steadily with $J$ and with $\chi$ (from a mean of $2.5$ bits at $\chi=16$ to $3.1$ bits at $\chi=32$ for $J/\pi=0.12$), but remain below the capacity $\log_2\chi$ of the bonds.

\begin{figure*}[tbp]
  \centering
  \includegraphics[width=0.85\linewidth]{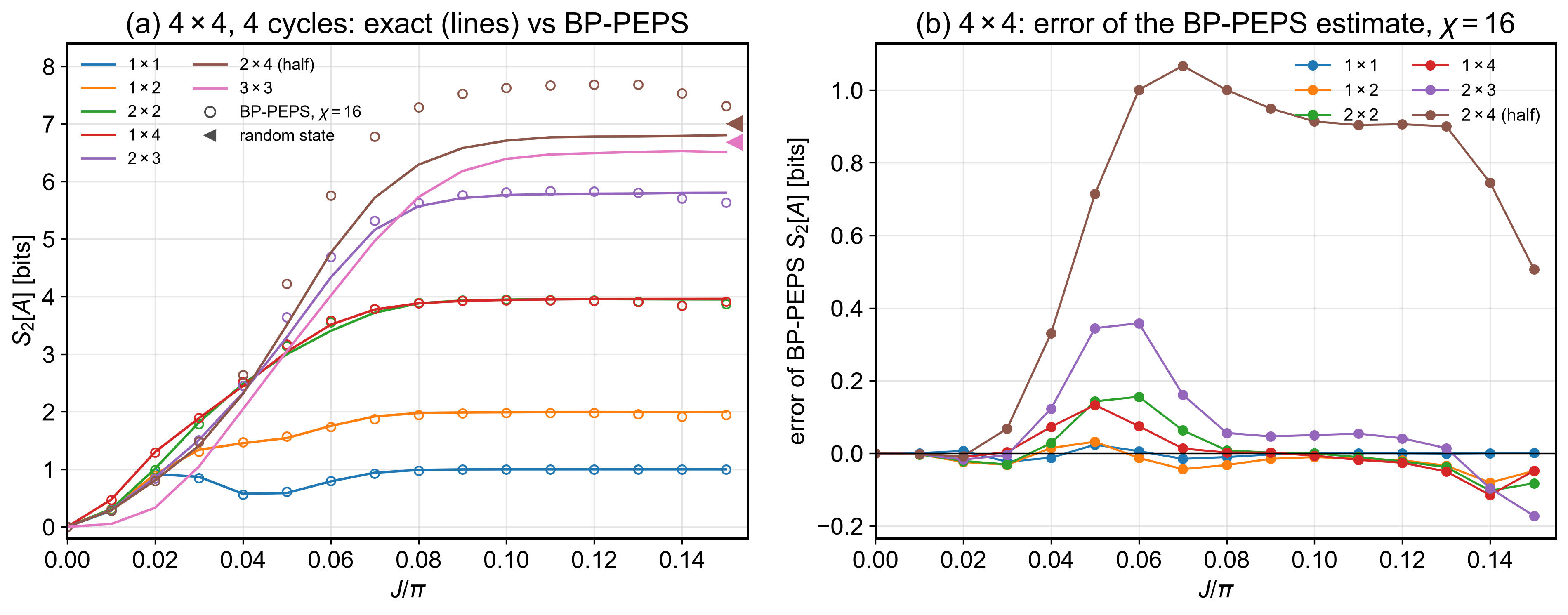}
  \caption{\textbf{Patch entanglement of BP-PEPS against the exact state on $4\times4$.}
  R\'enyi-2 entropy $S_2[A]=-\log_2P_2[A]$ after four cycles, with $\rho_A$ contracted exactly with the converged BP messages entering patch $A$ as in Fig.~\ref{fig:bp_entanglement}.
  \textbf{a}, exact $S_2[A]$ (lines) and BP-PEPS estimate at $\chi=16$ (circles); triangles: average R\'enyi-2 entropy of $8$ and $9$ qubits of a random state of $16$ qubits.
  \textbf{b}, difference between the BP-PEPS estimate and the exact value.}
  \label{sm:fig:bp_entanglement}
\end{figure*}

We have checked this estimate of $S_2[A]$ on $4\times4$, where the exact state is available (Fig.~\ref{sm:fig:bp_entanglement}). For patches of up to four sites the BP-PEPS estimate at $\chi=16$ agrees with the exact value within $0.16$ bits for all couplings up to $J/\pi=0.15$, and for the $2\times3$ patch within $0.36$ bits, the largest deviations occurring around $J/\pi\simeq0.06$. For the $2\times4$ patch, i.e. half of the lattice, the BP-PEPS estimate instead exceeds the exact value by up to $1.1$ bits, an overestimate that we attribute to the correlations, not captured by the BP environment, between the two disconnected rows that make up the complement of the patch. On $8\times8$ the patches contain at most nine of the $64$ sites, and we expect them to be in the regime where the estimate is accurate. At strong coupling the exact entropies on $4\times4$ approach those of a random state also for the largest subsystems: at $J/\pi=0.11$ the $2\times4$ patch carries $6.77$ bits and the $3\times3$ patch $6.47$ bits, against $7.00$ and $6.68$ for a random state of $16$ qubits.

These results thus show that, on the scale of the patches that we can access, the states at strong coupling are nearly maximally entangled, as expected from volume-law entanglement, consistently with the MPS results above and with the exact results of Sec.~\ref{sec:Quantum Renyi entropy}. Establishing the scaling of the entropy with the size of the subsystem for extensive subsystems of the $8\times8$ lattice is, however, beyond the reach of both our MPS and BP-PEPS simulations.

\subsection{Summary of numerical reliability}
\label{sec:reliability_map}

\begin{figure}[tbp]
  \centering
  \includegraphics[width=\linewidth]{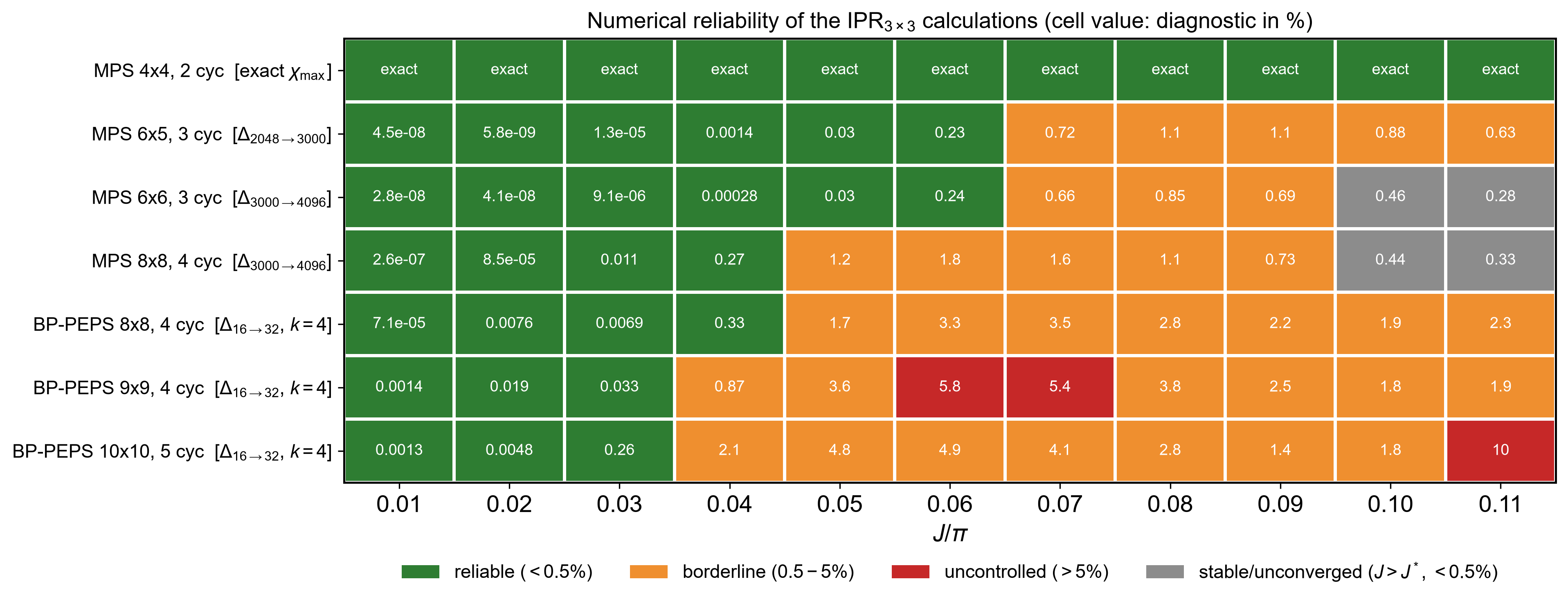}
  \caption{\textbf{Numerical reliability of the $3\times3$ marginal IPR.}
  Each row corresponds to one calculation and each column to one value of $J/\pi$. Cells are classified according to Eq.~\eqref{eq:classification}, using $\delta_{\rm exact}$ for $4\times4$ MPS, the largest refinement performed for the other MPS calculations, and $\Delta_{16\to32}$ at fixed $k=4$ for BP-PEPS. Grey cells, stable but unconverged, lie beyond the first maximum $J^*$ of the refinement of that row, where the refinement decreases because the marginal approaches its random-state value rather than because the calculation converges: there, a refinement below $0.5\%$ no longer indicates a reliable result, while borderline and uncontrolled cells keep their class. Cell values give the diagnostic in percent. The map stops at $J/\pi=0.11$, the largest coupling of the $8\times8$ MPS calculation. For $8\times8$ MPS the diagnostic is $\Delta_{3000\to4096}$.}
  \label{fig:reliability}
\end{figure}

Fig.~\ref{fig:reliability} summarises the diagnostics of this section for the $3\times3$ marginal, classifying each calculation and coupling with the best diagnostic available: $\delta_{\rm exact}$ for $4\times4$, the largest refinement performed for the other MPS calculations, and $\Delta_{16\to32}$ at $k=4$ for BP-PEPS. On a finer grid of couplings, the largest available refinement first exceeds $5\%$ at $J/\pi=0.06$ on $9\times9$ and $0.055$ on $10\times10$ for BP-PEPS, and stays below $5\%$ for BP-PEPS on $8\times8$ up to $J/\pi=0.12$. For MPS it stays below $5\%$ on all three lattices over the whole range examined, $J/\pi\le0.12$ on $6\times5$ and $6\times6$ and $J/\pi\le0.11$ on $8\times8$, and it exceeds $1\%$ only on $8\times8$ for $0.05\le J/\pi\le0.08$ and on $6\times5$ at $J/\pi=0.08$ and $0.09$. On $8\times8$, where both methods are available, the MPS result for the same marginal exceeds the BP-PEPS one by $8\%$ to $21\%$ for $0.035\le J/\pi\le0.08$ (Fig.~\ref{fig:MPS_vs_PEPS}), so that at least one of the two calculations carries an error well beyond its last refinement. For the system sizes of interest the classical calculations are therefore stable under bond-dimension refinement up to $J/\pi\simeq0.035$--$0.06$, depending on method and lattice, below the ergodic crossover.

As the dynamics approach the ergodic crossover, the bond-dimension refinement, the BP bias and the cumulant truncation deteriorate together, in the same window $J/\pi\simeq0.02$--$0.06$ in which the patch-to-patch variation is largest. In this regime the available tensor-network results are approximations of unknown accuracy. A similar conclusion was reached in Ref.~\cite{mendessantos2026exploring} for a different model and platform, where belief-propagation, matrix-product-state and tree-tensor-network simulations all lose control in the regime of slow relaxation, and the more general question of how to validate quantum estimates in the absence of a controlled classical reference is addressed in Ref.~\cite{barron2026observable}.

\begin{figure}[tbp]
  \centering
  \includegraphics[width=\linewidth]{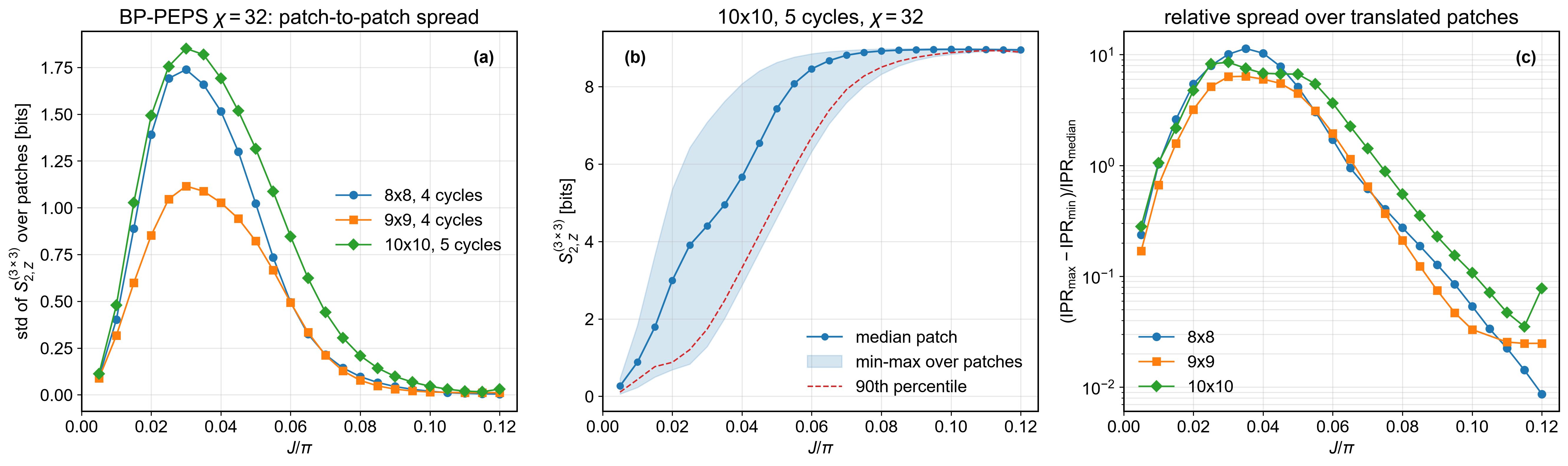}
  \caption{\textbf{Patch-to-patch variation of the $3\times3$ marginal.}
  BP-PEPS at $\chi=32$ and $k=4$, over the $16$ translated positions of the marginal.
  \textbf{a}, standard deviation of the collision entropy $S_{2,Z}$ over patches, for $8\times8$ and $9\times9$ after four cycles and $10\times10$ after five.
  \textbf{b}, median (solid), $90$th percentile (dashed) and min--max envelope over patches, for $10\times10$.
  \textbf{c}, relative spread $(\mathrm{IPR}_{\max}-\mathrm{IPR}_{\min})/\mathrm{IPR}_{\rm median}$ of the same data.}
  \label{fig:patch_stats}
\end{figure}

The collision entropies reported in the main text are averages over translated patches, whereas the convergence analysis of Sec.~\ref{sec:tn_results} was carried out on a single $3\times3$ marginal. Figure~\ref{fig:patch_stats} shows the spread over the $16$ translated positions computed with BP-PEPS at $\chi=32$ and $k=4$. The patch-to-patch standard deviation of the $3\times3$ collision entropy $S_{2,Z}$ peaks at $J/\pi\simeq0.03$, where it reaches $1.74$ bits on $8\times8$, $1.12$ bits on $9\times9$ and $1.85$ bits on $10\times10$ after five cycles, with the most extreme patch lying $3.4$, $2.3$ and $3.3$ bits from the mean, respectively. At this coupling the IPRs of the largest and smallest patch differ by up to a factor of $90$ on $8\times8$. The spread then decreases rapidly, falling below $0.1$ bits at $J/\pi=0.09$ and below $0.05$ bits at $J/\pi=0.10$, as all patches approach the maximally mixed value.

This spread is physical rather than numerical: the field realisation is fixed and strongly site dependent, so that different patches experience different local environments. It is relevant, however, for the interpretation of Fig.~\ref{fig:PEPS_bond_dim2}. The marginal used there, the patch with lower-left site $(4,4)$, is not one of the $16$ positions; it lies between the $12$th and $31$st percentile of the distribution of IPRs over the $16$ positions on $8\times8$ at every coupling, and at $J/\pi=0.03$ its IPR is about a third of the median. It is therefore among the more strongly entangled patches, and among the harder ones to converge, which makes the refinement test conservative; its entropy, however, is not the spatially averaged quantity shown in the main text. The refinement of the individual patches is shown in Fig.~\ref{fig:bp_chi48_3x3}(b): at $J/\pi=0.06$ and $0.08$ the $16$ patches move by between $0.3\%$ and $2.6\%$ under $\chi=32\to48$, and none leaves the borderline band.

The separation of the BP-PEPS error into its environment and state contributions, which we have carried out exactly on $4\times4$, cannot be repeated by exact contraction at the larger sizes. It could be obtained there by evaluating the same states with one of the loop and cluster corrections to belief propagation developed recently~\cite{evenbly2026loop,midha2025beyond,gray2025tensor,Park2025,evenbly2025partitioned,rudolph2025simulating,guo2023block}, which would test whether the environment error found on $4\times4$ persists on the lattices of the main text.

\clearpage

\section{Discrete truncated Wigner approximation}
\label{app:dtwa}

In this appendix, we describe the discrete truncated Wigner approximation
(DTWA)~\cite{polkovnikov2010phase,schachenmayer2015many} and the cluster DTWA~\cite{wurtz2018cluster} approaches for simulating the two-dimensional
Floquet Heisenberg dynamics, and we benchmark them against both MPS and BP-PEPS simulations.

The DTWA replaces the quantum evolution of the many-body wavefunction by a stochastic
average over classical trajectories sampled from the Wigner distribution of the
initial state. In this approach, each spin is represented by a classical vector
$\mathbf{s}=(s^x,s^y,s^z)$ whose components take values $\pm1$, and the initial
ensemble is chosen so that the discrete sampling reproduces the exact
single-site expectation values of the magnetisation and magnetisation square.
For a spin polarised along $+z$ (as for all the qubits in one sublattice in the
N\'eel state) the longitudinal component is fixed, $s^z=+1$, while the two
transverse components are sampled independently and uniformly,
$s^x,s^y\in\{+1,-1\}$ each with probability $1/2$; this gives
$\langle s^a\rangle=\delta_{a,z}$, matching the polarised state, while endowing
the transverse directions with ``quantum'' fluctuations
$\overline{(s^{x,y})^2}=1$.

Each trajectory is then propagated by the classical equations of
motion generated by the Floquet-Heisenberg Hamiltonian.
For example, the single-site disorder rotation
$\exp(-\mathrm{i}\,h\,\sigma^z)$ acts as a rotation of the transverse plane by
angle $2h$,
\begin{equation}
  \begin{pmatrix} s^x\\ s^y\end{pmatrix}
  \longmapsto
  \begin{pmatrix} \cos 2h & \sin 2h\\ -\sin 2h & \cos 2h\end{pmatrix}
  \begin{pmatrix} s^x\\ s^y\end{pmatrix},
  \label{eq:zrot}
\end{equation}
and the two-spin Heisenberg gate
$\exp(-\mathrm{i}\,J\,\boldsymbol{\sigma}_i\!\cdot\!\boldsymbol{\sigma}_j)$
maps to the \emph{exact} classical two-spin precession that conserves the total
vector $\mathbf{m}=\mathbf{s}_i+\mathbf{s}_j$: the relative vector
$\mathbf{d}=\mathbf{s}_i-\mathbf{s}_j$ rotates about $\hat{\mathbf{m}}$ by an
angle $-2J\,|\mathbf{m}|$, after which
$\mathbf{s}_{i,j}=\tfrac12(\mathbf{m}\pm\mathbf{d})$.
Because of this mapping from gate unitary dynamics to classical rotations,
the approximation in DTWA resides in (i) the
factorised sampling of the initial Wigner function and (ii) the mean-field
factorisation implicit in evolving one classical configuration per trajectory.

Observables are finally reconstructed from the trajectory ensemble. The $z$-basis
marginal $p_S(\mathbf{x})$ of a patch $S$, i.e.\ the probability of the bit-string
$\mathbf{x}=(x_i)_{i\in S}$, $x_i\in\{+1,-1\}$, is obtained from the sampled $s^z$ values.
Its inverse participation ratio $\mathrm{IPR}_S=\sum_{\mathbf{x}} p_S(\mathbf{x})^2$, which gives
the \emph{collision entropy} $S_{2,Z}=-\log_2\mathrm{IPR}_S$, is
estimated with a split-batch estimator. Each trajectory yields an estimate
of the marginal from its sampled longitudinal components,
\begin{equation}
  \hat p_S(\mathbf{x})=\prod_{i\in S}\tfrac12\big(1+x_i\,s^z_i\big),
  \label{eq:patch-marginal}
\end{equation}
so that the batch average $\overline{\hat p_S(\mathbf{x})}$ converges to
$p_S(\mathbf{x})$. The estimator is evaluated for all $2^{|S|}$ bit-strings,
and $s^z_i\in[-1,1]$ is the continuous longitudinal component of the
trajectory (equal to $\pm1$ only in the initial N\'eel state); the factor
$\tfrac12(1+x_i s^z_i)$ is the probability that a $z$ measurement of that
classical spin returns $x_i$. Because $\mathrm{IPR}_S$ is \emph{quadratic} in the
distribution, squaring a single batch average would bias it upwards by the
sampling variance, $\overline{\hat p_S(\mathbf{x})}^{\,2}=p_S(\mathbf{x})^2+
\mathcal{O}(1/N_{\mathrm{traj}})$. We therefore split the trajectories into two
independent batches, form the two marginal estimates $\hat p_S^{(1)}$ and
$\hat p_S^{(2)}$, and take
\begin{equation}
  \mathrm{IPR}_S \;=\; \sum_{\mathbf{x}}
     \hat p_S^{(1)}(\mathbf{x})\,\hat p_S^{(2)}(\mathbf{x}),
  \label{eq:split-batch}
\end{equation}
whose expectation is exactly $\sum_{\mathbf{x}} p_S(\mathbf{x})^2$ because the
two batches are independent and each unbiased. In practice, expanding Eq.~\eqref{eq:patch-marginal} turns Eq.~\eqref{eq:split-batch} into a sum
over the $z$ correlators of all subsets of $S$, the analogue of the Parseval expression, Eq.~\eqref{ParsevalIPR},
with each squared expectation value replaced by the product of its two batch averages.
For single-site DTWA we use $8192$ trajectories, split into two batches of $4096$, and on each bond we apply
the exchange gate before the two single-site disorder rotations, as in the MPS and BP-PEPS
circuits.

\begin{figure}[t]
  \centering
  \includegraphics[width=\linewidth]{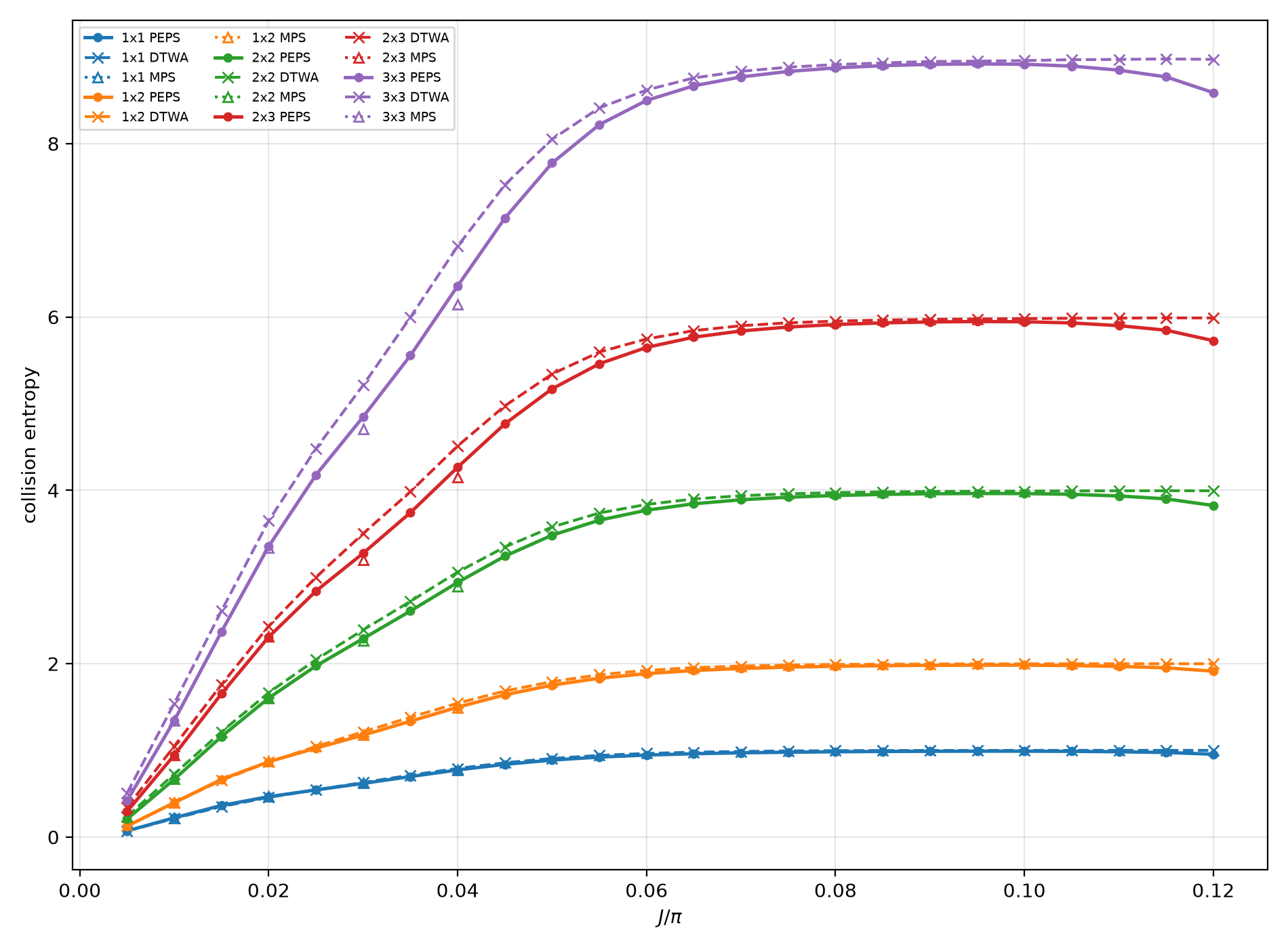}
  \caption{%
    \textbf{DTWA against tensor networks on $8\times8$.} Collision entropy as a function of $J/\pi$, at Floquet cycle $6$ for the
    $8\times8$ lattice and a fixed disorder realisation. Colours label the
    patch size and shape ($1\times1$, $1\times2$, $2\times2$, $2\times3$, $3\times3$);
    for each shape we show DTWA (crosses, dashed), belief-propagation PEPS at
    bond dimension $\chi=32$ (circles, solid) and MPS
	at $\chi=3000$ (open triangles, dotted; $J/\pi\le0.04$). }
  \label{fig:dtwa-benchmark}
\end{figure}

Figure~\ref{fig:dtwa-benchmark} shows that DTWA tracks both BP-PEPS at $\chi=32$ and MPS at
$\chi=3000$ closely for all patch shapes and sizes up to $3\times 3$.
The discrepancy grows with the size of the patch and with $J$ up to $J/\pi\simeq0.05$, and is largest for the $2\times3$ and $3\times3$ patches,
signalling that DTWA overscrambles the Floquet dynamics, i.e.\ thermalises faster as $J$
increases.

\subsection{An elementary introduction to cluster DTWA}
\label{app:dtwa:idea}

\begin{figure}[t]
  \centering
  \includegraphics[width=\linewidth]{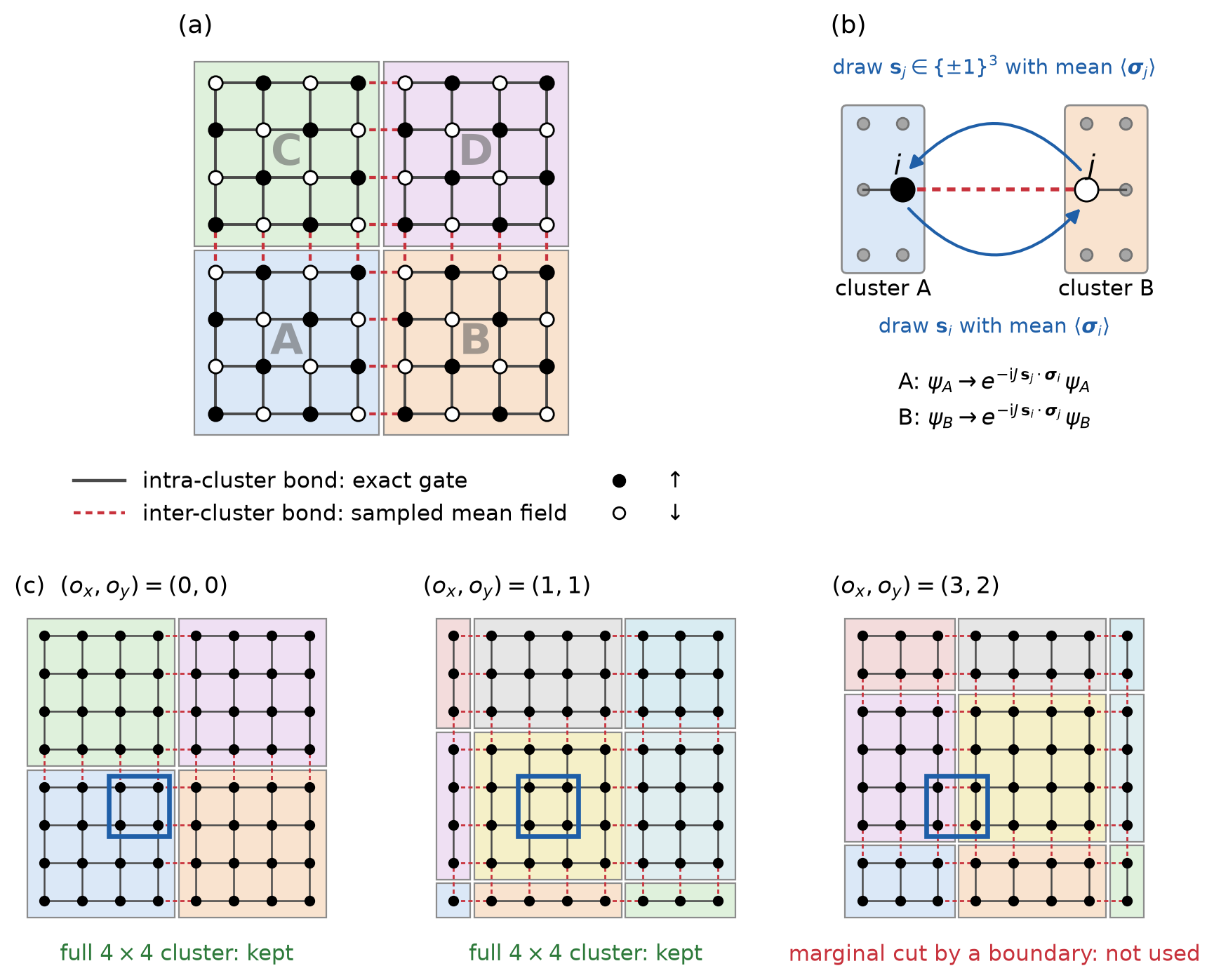}
  \caption{%
    \textbf{Cluster DTWA with $4\times4$ clusters.}
    \textbf{a}, The $8\times8$ lattice in the N\'eel state (filled: up, open: down), cut into the four
    clusters A--D. Bonds within a cluster (solid) are applied as exact gates, bond gates between clusters
    (dashed) are treated with sampled classical spins.
    \textbf{b}, A bond $(i,j)$ between clusters A and B: a spin $\mathbf{s}_j$ with components $\pm1$ is
    drawn with mean equal to the Bloch vector $\langle\boldsymbol{\sigma}_j\rangle$ and rotates
    qubit $i$, and vice versa.
    \textbf{c}, A $2\times2$ patch (blue square) for three shifts $(o_x,o_y)$ of the cluster boundaries.
    The patch is used only for the shifts in which it lies inside a complete $4\times4$ cluster.}
  \label{fig:cdtwa-schematic}
\end{figure}

We explain here the idea behind cluster DTWA and our implementation, using the $8\times8$ lattice as an
example. An exact simulation of the circuit stores $2^N$ complex amplitudes for $N=L^2$ qubits, which is
easily done for $L=4$ but would require about $3\times10^{20}$ bytes for $L=8$ ($16$ bytes per complex amplitude). Single-site DTWA goes to
the opposite extreme: every qubit is replaced by a classical spin, so that the cost grows only
linearly with $N$, but no quantum correlation between any two qubits is retained, and we have seen
in Fig.~\ref{fig:dtwa-benchmark} that this leads to an overestimate of the collision entropy.
Cluster DTWA interpolates between these two limits. The lattice is cut into small blocks, the
clusters, and each cluster is treated exactly, while only the bond gates connecting different clusters
are treated with classical spins. With $4\times4$ clusters the $8\times8$ lattice consists of the
four clusters A, B, C and D of Fig.~\ref{fig:cdtwa-schematic}(a), each stored as a statevector of
$2^{16}$ amplitudes, so that the whole simulation requires $4\times2^{16}$ amplitudes instead of
$2^{64}$.

In every Floquet cycle the circuit applies the two-qubit gate
$\exp(-\mathrm{i}J\,\boldsymbol{\sigma}_i\!\cdot\!\boldsymbol{\sigma}_j)$ to each bond $(i,j)$ of the
lattice, followed by the disorder rotations $\exp(-\mathrm{i}h\,\sigma^z)$ of its two sites. In
cluster DTWA the bond gates fall into two classes, shown in Fig.~\ref{fig:cdtwa-schematic}(a). When both
sites belong to the same cluster (solid lines), the gate acts only on the statevector of that cluster
and is applied exactly, as in a full simulation. On $8\times8$ with $4\times4$ clusters this is the
case for $96$ of the $112$ bond gates. When the two sites belong to different clusters (dashed lines),
the gate cannot be applied in the same way without merging the two clusters into a single statevector
of $2^{32}$ amplitudes, which is precisely what we want to avoid. For these $16$ bond gates the partner
qubit is therefore replaced by a classical spin.

The simplest way to do so would be to replace the operator $\boldsymbol{\sigma}_j$ by its expectation
value, the Bloch vector $\mathbf{b}_j=\langle\boldsymbol{\sigma}_j\rangle$, computed from the state of
the cluster that contains $j$. The two-qubit gate then reduces to the single-qubit rotation
$\exp(-\mathrm{i}J\,\mathbf{b}_j\!\cdot\!\boldsymbol{\sigma}_i)$ of qubit $i$ about the direction of
$\mathbf{b}_j$, and symmetrically for $j$. This is the usual mean-field approximation, and for our
initial state it fails completely. In the N\'eel state every Bloch vector points along $\pm z$, and
since the gates inside a cluster conserve the magnetisation, the Bloch vectors remain along $z$ at all
later times. The mean-field rotation is then a rotation about the $z$ axis, which does not change the
populations of qubit $i$, and the clusters never exchange magnetisation. What is missing are the
fluctuations of the transverse components: in the N\'eel state
$\langle\sigma^x_j\rangle=\langle\sigma^y_j\rangle=0$, but
$\langle(\sigma^x_j)^2\rangle=\langle(\sigma^y_j)^2\rangle=1$. As in single-site DTWA, these
fluctuations are restored by sampling. Instead of $\mathbf{b}_j$ we use a random spin $\mathbf{s}_j$
whose three components take the values $\pm1$, with probabilities
$\Pr(s^a_j=\pm1)=\tfrac12(1\pm b^a_j)$, so that the average of $\mathbf{s}_j$ over many draws is
$\mathbf{b}_j$ (Fig.~\ref{fig:cdtwa-schematic}(b)). In the N\'eel state, for instance, $s^z_j$ is
fixed by the orientation of qubit $j$, while $s^x_j$ and $s^y_j$ are $+1$ or $-1$ with equal
probability. Qubit $i$ is then rotated by $\exp(-\mathrm{i}J\,\mathbf{s}_j\!\cdot\!\boldsymbol{\sigma}_i)$,
which has transverse components and therefore transfers magnetisation between the clusters, and qubit
$j$ by $\exp(-\mathrm{i}J\,\mathbf{s}_i\!\cdot\!\boldsymbol{\sigma}_j)$, with $\mathbf{s}_i$ drawn in the
same way from the Bloch vector of $i$. Both spins are drawn from the states before the update, and new
spins are drawn every time a bond between two clusters is applied.

A single run of this procedure is called a trajectory. Each cluster is prepared in its part of the N\'eel
state, the bond gates of every Floquet cycle are applied in the order of the circuit, exactly inside a
cluster and with the sampled spins between clusters, each followed by the disorder rotations of its two
sites, and after the required number of cycles, six in the results below, the probabilities
$|\psi(\mathbf{x})|^2$ of the statevector $\psi$ of each cluster give the marginal $V_t(\mathbf{x})$ of
each patch $S$ inside it.
Different trajectories differ only in the random spins drawn at the cluster boundaries, and we repeat
the procedure for $T$ trajectories. Within one trajectory the state of the lattice is a product of the
four cluster states, but since the trajectories differ from each other, their average describes a
statistical mixture of such product states. Cluster DTWA therefore retains all the entanglement within
a cluster, while between different clusters it only retains classical correlations, which arise
because the random spins applied to a cluster depend on the state of its neighbours.

The marginal distribution of a patch is the trajectory average
$p_S(\mathbf{x})=\frac1T\sum_t V_t(\mathbf{x})$, and its IPR is $\sum_{\mathbf{x}}p_S(\mathbf{x})^2$.
Squaring this average, however, includes the product of each trajectory with itself, which adds the
statistical fluctuations of the individual trajectories to the IPR, and therefore underestimates the
collision entropy for a finite number of trajectories. We therefore only multiply marginals of distinct
trajectories, $t\neq t'$,
\begin{equation}
  \mathrm{IPR}_S \;=\; \frac{1}{T(T-1)}\sum_{t\neq t'}\sum_{\mathbf{x}}
     V_t(\mathbf{x})\,V_{t'}(\mathbf{x}),
  \label{eq:allpairs}
\end{equation}
which is the average of the split-batch estimator of Eq.~\eqref{eq:split-batch} over all pairings of
the trajectories, and is unbiased for the same reason.

Finally, the result for a given patch depends on its position with respect to the cluster boundaries.
A patch in the middle of a cluster is surrounded by bond gates that are treated exactly, whereas a patch next
to a boundary is coupled to some of its neighbours only through the random spins. Since the position of
the boundaries is arbitrary, we shift them by $(o_x,o_y)$ with $o_x,o_y\in\{0,\dots,3\}$, as shown in
Fig.~\ref{fig:cdtwa-schematic}(c), and for each patch we average the IPR over the shifts in which the
patch lies inside a complete $4\times4$ cluster. Shifts in which the patch is cut by a boundary, such as
$(o_x,o_y)=(3,2)$ in Fig.~\ref{fig:cdtwa-schematic}(c), are not used.

The size of the clusters controls both the accuracy and the cost of the method. If a single cluster
covers the whole lattice, there are no bond gates between clusters and cluster DTWA reduces to the exact
simulation of the circuit, which is the case for $L=4$ with $4\times4$ clusters. If every cluster
contains a single site, every bond is treated with random spins and no quantum correlation between
different sites is retained. Larger clusters therefore give more accurate results, but the cost of a
$c_x\times c_y$ cluster grows as $2^{c_xc_y}$: a $4\times4$ cluster requires $2^{16}$ amplitudes, about $1$\,MB,
whereas a $5\times5$ cluster would require $2^{25}$ amplitudes, about $0.5$\,GB, in each trajectory. Since the random spins do not conserve the magnetisation of a cluster, we evolve the full cluster
Hilbert space rather than a sector of fixed magnetisation. We have also implemented a version in which
each cluster is evolved as an MPS, and verified that it gives the same results within its statistical
error.
At fixed cluster size, however, the cost grows only linearly with the number of clusters, i.e.\ with the
size of the lattice, which allows us to reach $12\times12$ in Sec.~\ref{app:dtwa:size}.

\subsection{Benchmark of cluster DTWA}
\label{app:dtwa:cluster}

\begin{figure}[tbp]
  \centering
  \includegraphics[width=\linewidth]{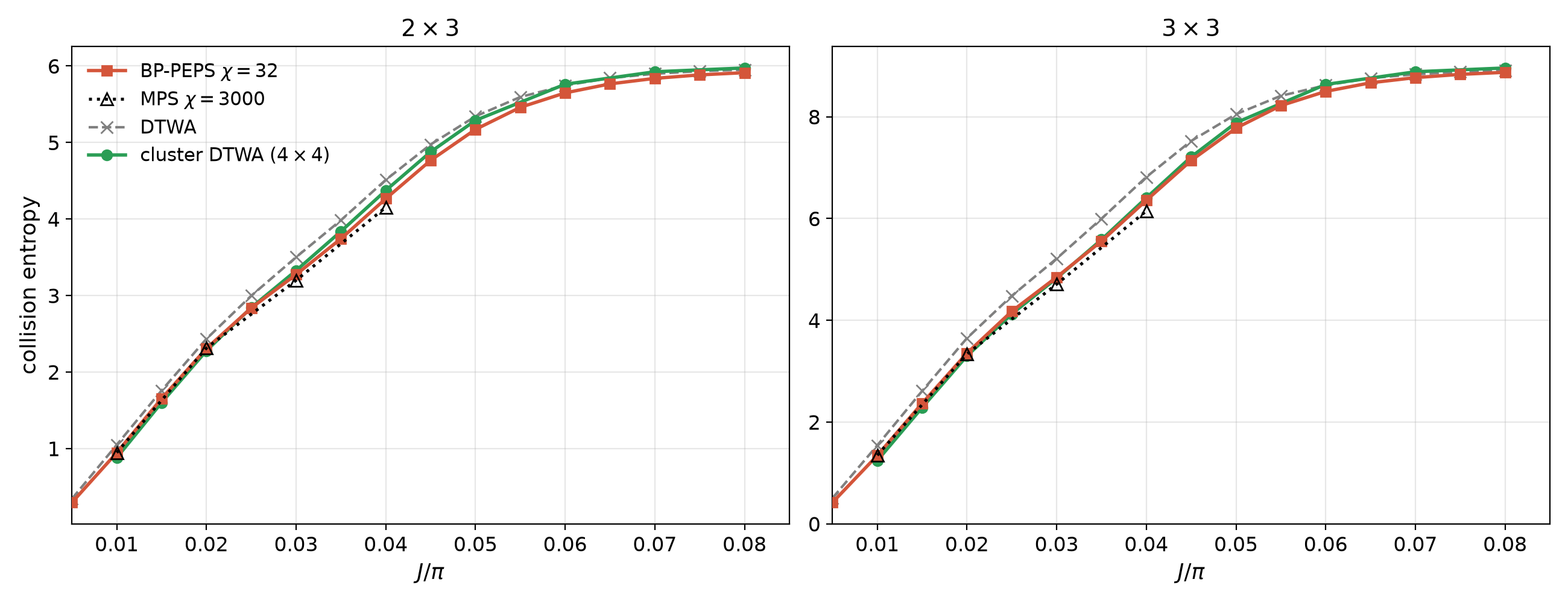}
  \caption{%
    \textbf{Cluster DTWA against single-site DTWA and tensor networks.} Collision entropy at Floquet cycle $6$ for the $8\times 8$ system as a function of $J/\pi$ for
    the two largest patch shapes, $2\times3$ (left) and $3\times3$ (right), averaged over all the
    patches of the lattice, for the disorder realisation of Fig.~\ref{fig:dtwa-benchmark}.
    Cluster DTWA with $4\times4$ clusters (green circles; $400$ trajectories per tiling, averaged
    over tilings) is compared with single-site DTWA (grey crosses), BP-PEPS at $\chi=32$
    (squares) and MPS at $\chi=3000$ (open triangles; $J/\pi\le0.04$).}
  \label{fig:cluster-benchmark}
\end{figure}

Figure~\ref{fig:cluster-benchmark} shows the $4\times4$ cluster DTWA results ($400$ trajectories per tiling, averaged over tilings and patches)
for the two most demanding patch shapes, $2\times3$ and $3\times3$, at cycle $6$ for the
$8\times 8$ system. While single-site DTWA overestimates the collision entropy, by $5$--$11\%$
for $2\times3$ and $7$--$15\%$ for $3\times3$ in the range $0.01\le J/\pi\le0.04$, the $4\times4$ cluster DTWA is close to the BP-PEPS
results over most of the range of $J/\pi$: for $0.02\le J/\pi\le0.08$ the two collision entropies differ by at most $2.6\%$
for $2\times3$ and $1.6\%$ for $3\times3$. Only at $J/\pi=0.01$, where BP-PEPS and MPS agree,
cluster DTWA underestimates the collision entropy, by $7$--$8\%$. The agreement with MPS is
limited to $J/\pi\lesssim0.02$. At $J/\pi=0.03$ and $0.04$ the MPS results lie below both BP-PEPS and
cluster DTWA, by up to $3.4\%$ and $5.3\%$ respectively, but at these couplings the $8\times8$ MPS at
cycle $6$ is itself only borderline in bond dimension, with a last refinement $2048\to3000$ of
$2\%$ for the $3\times3$ marginal at $J/\pi=0.04$.
These results thus establish
cluster DTWA as a competitive heuristic algorithm to classically
compute collision entropies in Floquet dynamics for larger systems: it reproduces BP-PEPS within $2\%$
for $0.02\le J/\pi\le0.08$, while the comparison with MPS at $J/\pi=0.03$ and $0.04$ is limited by the
convergence of the latter.

\subsection{System-size dependence with cluster DTWA}
\label{app:dtwa:size}

Since the cost of the statevector implementation grows only linearly with the number of
clusters, we can apply cluster DTWA to lattices up to $12\times12$. We use $100$ trajectories
per tiling. For $L=4$ the only cluster is the whole lattice, and cluster DTWA reduces to the
exact simulation of the circuit.
We consider only bulk marginals, i.e.\ those not touching the edge of the lattice, and report the
entropy of the spatially averaged IPR, $S_{2,Z}=-\log_2\overline{\mathrm{IPR}_S}$, normalised by its
value for a Haar-random state in the $U(1)$ sector of the N\'eel state,
\begin{equation}
  S_{2,Z}^{\rm Haar} \;=\; -\log_2\!\left[\frac{1}{D(D+1)}\sum_{r=0}^{n_S}
     \binom{n_S}{r}\, d_r\,(d_r+1)\right],
  \qquad d_r=\binom{N-n_S}{N_\uparrow-r},
  \label{eq:haar_u1}
\end{equation}
with $N=L^2$, $N_\uparrow=N/2$, $D=\binom{N}{N_\uparrow}$ and $n_S=\abs{S}$.

\begin{figure}[t]
  \centering
  \includegraphics[width=\linewidth]{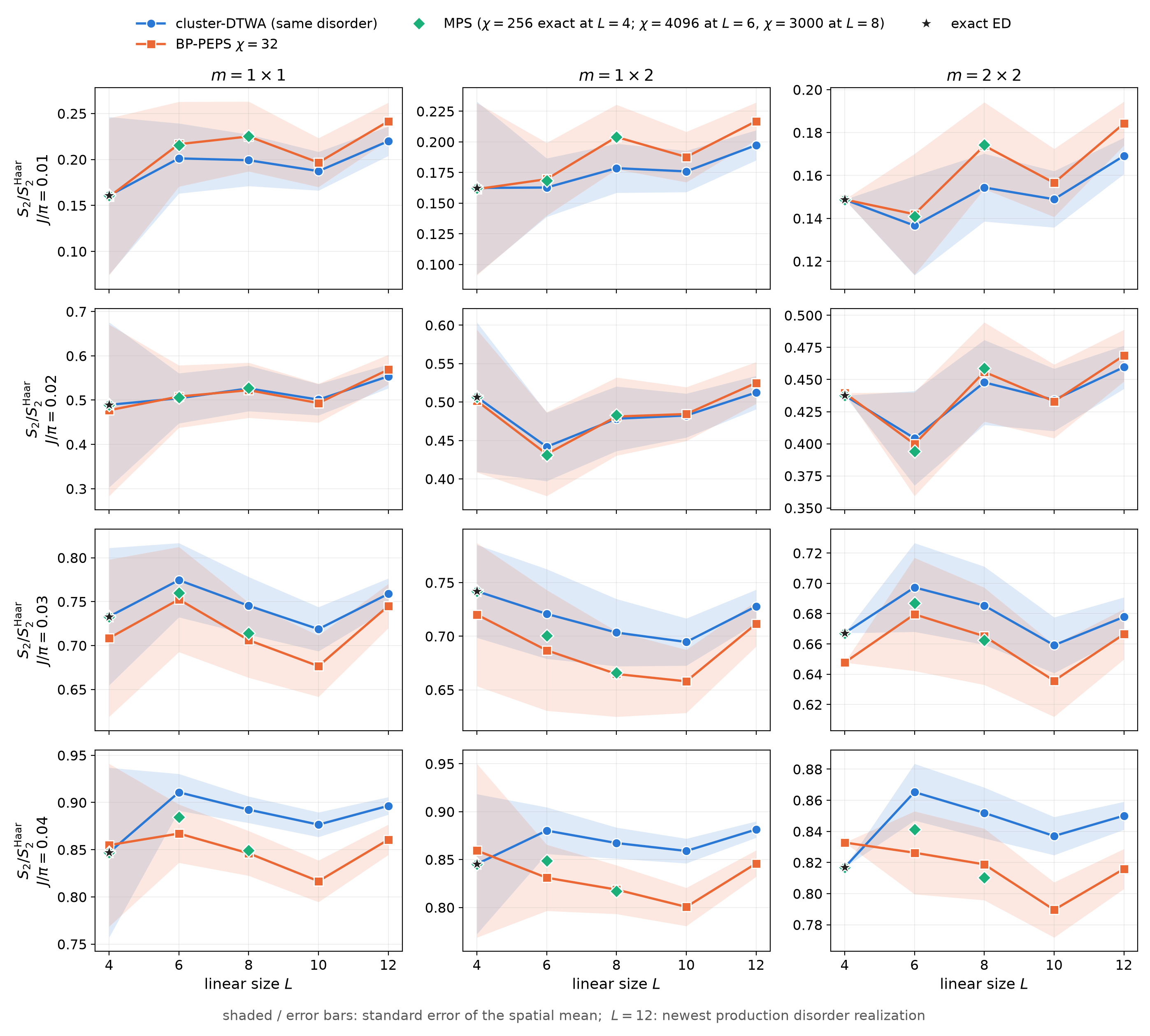}
  \caption{%
    \textbf{Cluster DTWA against tensor networks as a function of system size.}
    Normalised collision entropy $S_{2,Z}/S_{2,Z}^{\rm Haar}$ of the bulk $1\times1$, $1\times2$ (vertical
    pairs) and $2\times2$ marginals (columns) at Floquet cycle $6$ for $L\times L$ lattices,
    $L=4,\dots,12$, and $J/\pi=0.01$--$0.04$ (rows). At each $L$ all methods use the same
    production disorder realisation (at $L=12$ the second one of Fig.~\ref{fig:ced-12x12}).
    Cluster DTWA with $4\times4$ clusters (circles), BP-PEPS at $\chi=32$ (squares), MPS
    (diamonds; $\chi=256$, i.e.\ exact, at $L=4$, $\chi=4096$ at $L=6$ and $\chi=3000$ at $L=8$) and exact
    state-vector results at $L=4$ (stars). Shaded regions show the standard error of the
    spatial mean.}
  \label{fig:ced-benchmark-L}
\end{figure}

Figure~\ref{fig:ced-benchmark-L} compares cluster DTWA with BP-PEPS at $\chi=32$, with MPS and,
at $L=4$, with the exact state-vector results, for $J/\pi\le0.04$ and the same disorder
realisation at each $L$. At $L=4$ cluster DTWA coincides with the exact result by construction,
and so does MPS, whereas BP-PEPS deviates from it by at most $0.025$. For $L\ge6$ the three methods agree within
the spatial standard error at $J/\pi=0.02$, while at $J/\pi=0.01$ cluster DTWA lies below both
tensor-network results by up to $0.026$. At $J/\pi=0.03$ and $0.04$ cluster DTWA instead
overestimates the collision entropy. On $8\times8$ at $J/\pi=0.04$, it exceeds the MPS values
by $0.043$, $0.050$ and $0.042$ for the $1\times1$, $1\times2$ and $2\times2$ marginals, i.e.\
by $5$--$6\%$, whereas BP-PEPS and MPS differ by at most $0.009$; on $6\times6$ the excess over MPS
is $0.024$--$0.031$. Since the $L=4$ point is exact, this deviation must be due to the mean-field
treatment of the inter-cluster bond gates, and it has the same sign as the overscrambling of
single-site DTWA in Fig.~\ref{fig:dtwa-benchmark}. The non-monotonic dependence on
$L$, e.g.\ the dip at $L=10$, is shared by cluster DTWA and BP-PEPS: each $L$ has its own disorder
realisation, so that at fixed realisation the curves do not represent a finite-size scaling.

\begin{figure}[t]
  \centering
  \includegraphics[width=\linewidth]{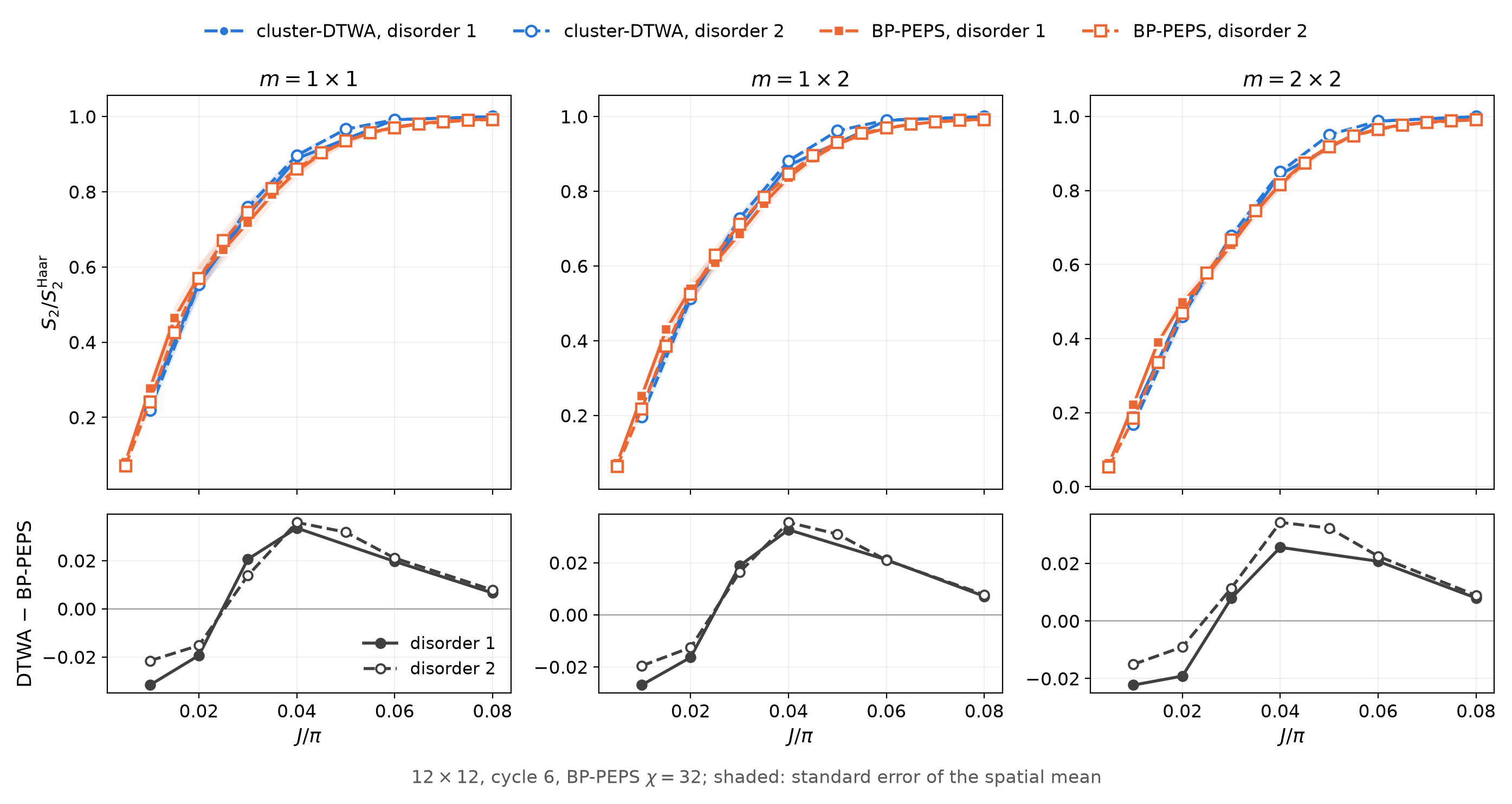}
  \caption{%
    \textbf{Cluster DTWA against BP-PEPS on $12\times12$.}
    Top row: $S_{2,Z}/S_{2,Z}^{\rm Haar}$ of the bulk $1\times1$, $1\times2$ and $2\times2$ marginals at
    Floquet cycle $6$ as a function of $J/\pi$, for cluster DTWA with $4\times4$ clusters (circles)
    and BP-PEPS at $\chi=32$ (squares), for the two production disorder realisations (filled and
    open symbols). Bottom row: difference between cluster DTWA and BP-PEPS for each realisation.
    Shaded regions show the standard error of the spatial mean.}
  \label{fig:ced-12x12}
\end{figure}

On $12\times12$ only BP-PEPS is available as a reference. Figure~\ref{fig:ced-12x12} shows both
methods for the two production disorder realisations up to $J/\pi=0.08$. The difference between
cluster DTWA and BP-PEPS has the same $J$ dependence for both realisations: it is negative for
$J/\pi\le0.02$, between $-0.032$ and $-0.009$, changes sign near $J/\pi\simeq0.025$, reaches
$0.026$--$0.035$ at $J/\pi=0.04$, and then decreases as both methods approach the Haar value,
falling below $0.01$ at $J/\pi=0.08$. Since the two realisations give the same difference to
within about $0.01$, the discrepancy is systematic between the two methods rather than a property of a
particular realisation, and it follows the same pattern as on the smaller lattices of
Fig.~\ref{fig:ced-benchmark-L}. Cluster DTWA saturates the Haar value already at
$J/\pi\simeq0.06$, with $S_{2,Z}/S_{2,Z}^{\rm Haar}=0.987$--$0.992$, where BP-PEPS gives
$0.965$--$0.972$.

\begin{figure}[t]
  \centering
  \includegraphics[width=\linewidth]{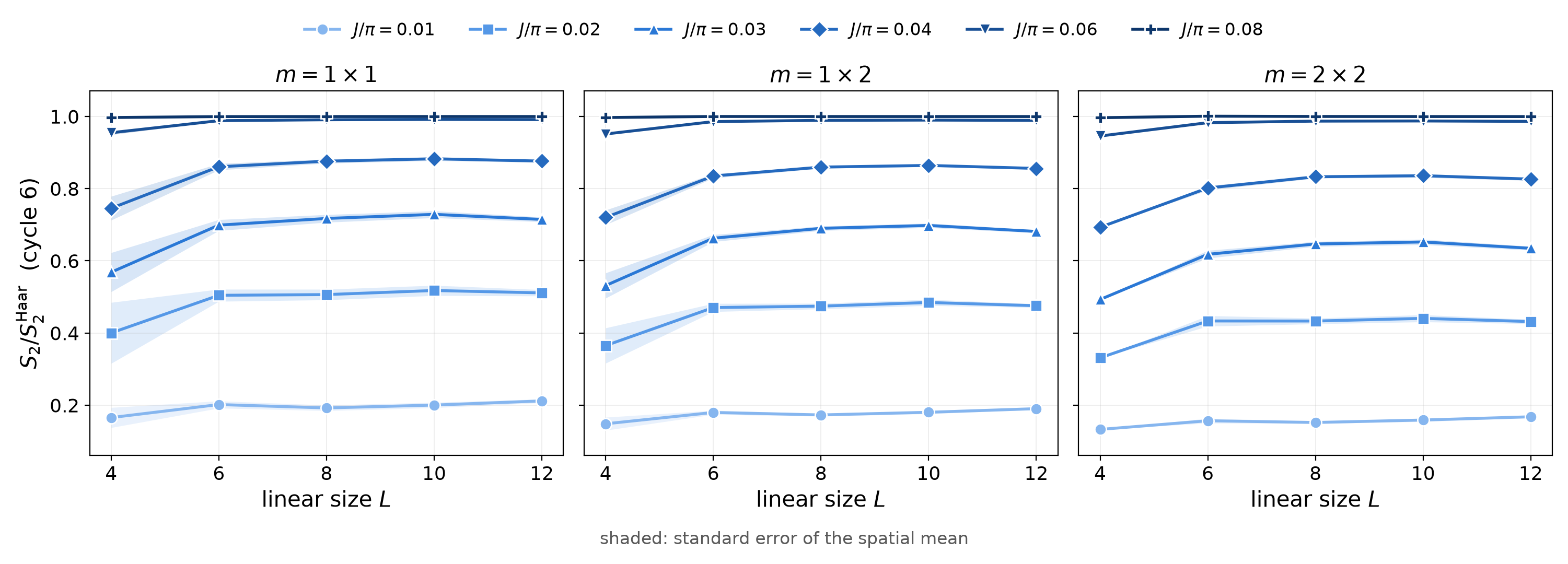}
  \includegraphics[width=\linewidth]{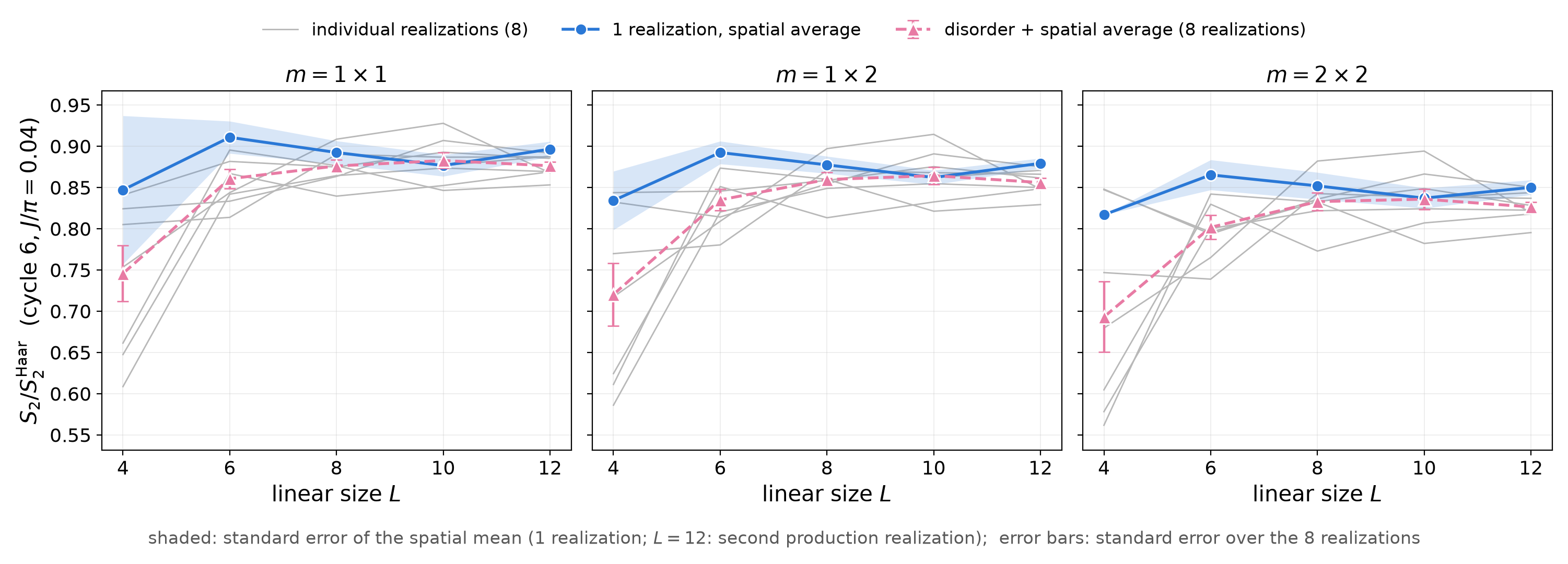}
  \caption{%
    \textbf{Disorder-averaged cluster DTWA as a function of system size.}
    Top: $S_{2,Z}/S_{2,Z}^{\rm Haar}$ of the bulk $1\times1$, $1\times2$ (both orientations) and $2\times2$
    marginals at Floquet cycle $6$ as a function of the linear size $L$, for
    $J/\pi=0.01$, $0.02$, $0.03$, $0.04$, $0.06$ and $0.08$, from cluster DTWA with $4\times4$ clusters averaged over all the
    tilings that contain each marginal and over $8$ disorder realisations. Shaded regions show the
    standard error of the spatial mean. Bottom: at $J/\pi=0.04$, the production disorder realisation
    (circles, at $L=12$ the second one of Fig.~\ref{fig:ced-12x12}; shaded: standard error of the spatial
    mean) against the average over $8$ realisations (triangles; error bars: standard error over the
    realisations); grey lines show the individual realisations.}
  \label{fig:ced-disavg-L}
\end{figure}

Keeping these deviations in mind, we use cluster DTWA to study the system-size dependence of the
bulk marginal entropies over the whole coupling range, averaging over disorder. At each $L$ we take the
production realisation together with seven further realisations of the fields drawn from the same
distribution. The top panels of Fig.~\ref{fig:ced-disavg-L} show the resulting $S_{2,Z}/S_{2,Z}^{\rm Haar}$ as a function
of $L$. For $L\ge6$ the entropies depend only weakly on the system size: between $L=8$ and
$L=12$ they change by at most $0.02$ at every coupling and marginal size, and for
$J/\pi\ge0.06$ they lie within $2\%$ of the Haar value. The $4\times4$ values are considerably
lower, e.g.\ $0.746$ against $0.861$ on $6\times6$ for the $1\times1$ marginal at $J/\pi=0.04$.
Part of this step is due to the approximation itself, since the $L=4$ point is exact whereas for
$L\ge6$ cluster DTWA overestimates the entropy near $J/\pi=0.04$. However, at $J/\pi=0.02$, where
cluster DTWA agrees with MPS and BP-PEPS, the step between $L=4$ and $L=6$ is still about $0.10$,
so that it is mostly a genuine finite-size effect of the $4\times4$ lattice.

The bottom panels of Fig.~\ref{fig:ced-disavg-L} show, at $J/\pi=0.04$, how representative a single
realisation is of the disorder average. On $4\times4$ the realisations differ strongly, with a
standard deviation of $0.10$--$0.12$ over the eight of them, and the production realisation lies
at the upper end of the distribution. The standard deviation decreases with $L$, to
$0.014$--$0.017$ at $L=12$, and for $L\ge8$ the production realisation and the disorder average
agree to within $0.025$. A single field realisation per system size, as used in the
tensor-network simulations of Sec.~\ref{sec:Classical algorithms}, is therefore representative of
the disorder average at the level of a few percent for $L\ge8$.

\bibliography{References}

\end{document}